\documentclass[LaM, oneside, english, noexaminfo]{sapthesis}
\usepackage{microtype}
\usepackage[english]{babel}
\usepackage[utf8]{inputenc}
\usepackage{imakeidx}
\usepackage{subfigure}
\usepackage{graphicx}
\usepackage{array}
\usepackage{ragged2e}
\usepackage[table]{xcolor}
\usepackage{hyperref}
\usepackage{blindtext}
\usepackage{setspace}
\usepackage{scrextend}
\usepackage{amsfonts}
\usepackage{mathtools}
\usepackage{dsfont}
\usepackage{amssymb}
\usepackage{slashed}
\usepackage{tikz}
\usepackage{float}
\usetikzlibrary{arrows.meta,decorations.markings,decorations.pathmorphing}

\newcommand{\ra}[1]{| #1 \rangle}
\newcommand{\la}[1]{\langle #1 |}
\newcommand{\lar}[1]{\langle #1 \rangle}
\newcommand{\m}{^{\mu}}
\newcommand{\p}{_{\perp}}
\newcommand{\widebar}[1]{\overline{#1}}
\renewcommand{\refeq}[1]{eq.~\eqref{#1}}

\usepackage[linguistics]{forest} 

\newcommand*{\belowrulesepcolor}[1]{%
  \noalign{%
  \kern-\belowrulesep \begingroup \color{#1}%
  \hrule height\belowrulesep \endgroup }%
}
\newcommand*{\aboverulesepcolor}[1]{%
  \noalign{%
  \begingroup \color{#1}%
  \hrule height\aboverulesep \endgroup \kern-\aboverulesep }%
}

\usepackage[sort]{natbib}
\setcitestyle{numbers,open={},close={}}
\makeatletter
\newcommand{\citeall}[1]{%
  [%
  \def\citeall@sep{}%
  \@for\citeall@key:=#1\do{%
    \citeall@sep\cite{\citeall@key}%
    \def\citeall@sep{, }%
  }%
  ]%
}
\makeatother

\makeatletter
\newcommand{\customadvisorlabel}[1]{%
  \AtBeginDocument{%
    \renewcommand{\SAP@ThesisAdvisorLabel}{#1}%
    \renewcommand{\SAP@ThesisAdvisorsLabel}{#1}%
  }%
}
\newcommand{\customcoadvisorlabel}[1]{%
  \AtBeginDocument{%
    \renewcommand{\SAP@ThesisCoAdvisorLabel}{#1}%
    \renewcommand{\SAP@ThesisCoAdvisorsLabel}{#1}%
  }%
}
\newcommand{\customdirectorlabel}[1]{%
  \AtBeginDocument{%
    \renewcommand{\SAP@ThesisDirectorLabel}{#1}%
    \renewcommand{\SAP@ThesisDirectorsLabel}{#1}%
  }%
}
\newcommand{\thesistype}[1]{%
  \AtBeginDocument{\renewcommand{\SAP@LaMthesistype}{#1}}%
}
\makeatother

\makeatletter
\def\@makechapterhead#1{%
  \vspace*{30\p@}%
  {\parindent \z@ \raggedright \normalfont
    \ifnum \c@secnumdepth >\m@ne
      \huge\bfseries \@chapapp\space \thechapter
      \par\nobreak
      \vskip 12\p@
    \fi
    \interlinepenalty\@M
    \LARGE \bfseries #1\par\nobreak
    \vskip 28\p@
  }}
\def\@makeschapterhead#1{%
  \vspace*{30\p@}%
  {\parindent \z@ \raggedright
    \normalfont
    \interlinepenalty\@M
    \LARGE \bfseries #1\par\nobreak
    \vskip 28\p@
  }}
\makeatother

\hypersetup{
  pdftitle={Gravitational scattering amplitudes in collinear and high-energy limits},
  pdfauthor={Lorenzo Pietrini},
  citebordercolor={0 1 0}
}
\title{Gravitational scattering amplitudes in collinear and high-energy limits}
\author{Lorenzo Pietrini}
\IDnumber{2024656}
\course[override]{Master's Degree in Physics}
\courseorganizer{Faculty of Mathematics, Physics and Natural Sciences}
\AcademicYear{2025/2026}
\advisor{Prof. Paolo Pani}
\advisor{Dr. Vittorio Del Duca}
\authoremail{pietrini.2024656@studenti.uniroma1.it}
\copyyear{2026}
\thesistype{Master's Thesis in Gravitational Waves, Compact Stars and Black Holes}

\makeindex
\begin{document}
\frontmatter
\maketitle
\dedication{A mia madre, \\
che porto con me in ogni momento
}

\begin{abstract}
In this thesis we study the high-energy and collinear behaviour of
gravitational scattering amplitudes and their relation to classical scattering
and radiation observables. Multi-Regge kinematics, characterised by a strong
rapidity ordering, provides the factorised building blocks governing the
leading logarithmic structure. Relaxing this ordering to next-to-multi-Regge
kinematics allows two neighbouring particles to have comparable rapidities
and provides the real emission building blocks entering at
next-to-leading logarithmic accuracy. We
extract the next-to-multi-Regge impact factor from tree-level MHV
amplitudes for pure gravitons and derive its gravitational splitting functions. We then use BCFW
recursion relations to reconstruct the five-point tree amplitude for two massive scalars
scattering with the emission of one graviton. Its high-energy limit yields the
massive scalar--graviton impact factor and, through the KMOC formalism, a
representation of the leading high-energy collinear waveform. Finally, by
sewing the relevant vertices across unitarity cuts, we construct the elastic
\(H\) kernel, its radiative extension and a classical next-to-multi-Regge
kernel. The latter reproduces the standard \(H\) kernel in the multi-Regge
limit and the universal Weinberg current in the soft limit.
\end{abstract}

\tableofcontents
\listoffigures

\mainmatter
\chapter{Introduction}
\label{cap:Introduction}

The direct observation of gravitational waves has opened a new way of
studying the Universe. The first evidence that gravity carries energy through
radiation came indirectly from the orbital decay of the Hulse--Taylor binary
pulsar \citeall{Hulse:1975ju}. Four decades later, the observation of GW150914
by the two Advanced LIGO interferometers provided the first direct detection
of a gravitational wave and the first observation of the coalescence of two
black holes \citeall{LIGOScientific:2016aoc}. Since then, the LIGO--Virgo--KAGRA
network has turned isolated detections into a steadily growing catalogue of
compact binary mergers. Gravitational waves' astronomy now probes populations
of black holes and neutron stars, the behaviour of matter at supranuclear
density and general relativity in its dynamical, strong-field, regime.

\begin{figure}[!b]
    \centering
    \includegraphics[width=0.6\textwidth]
    {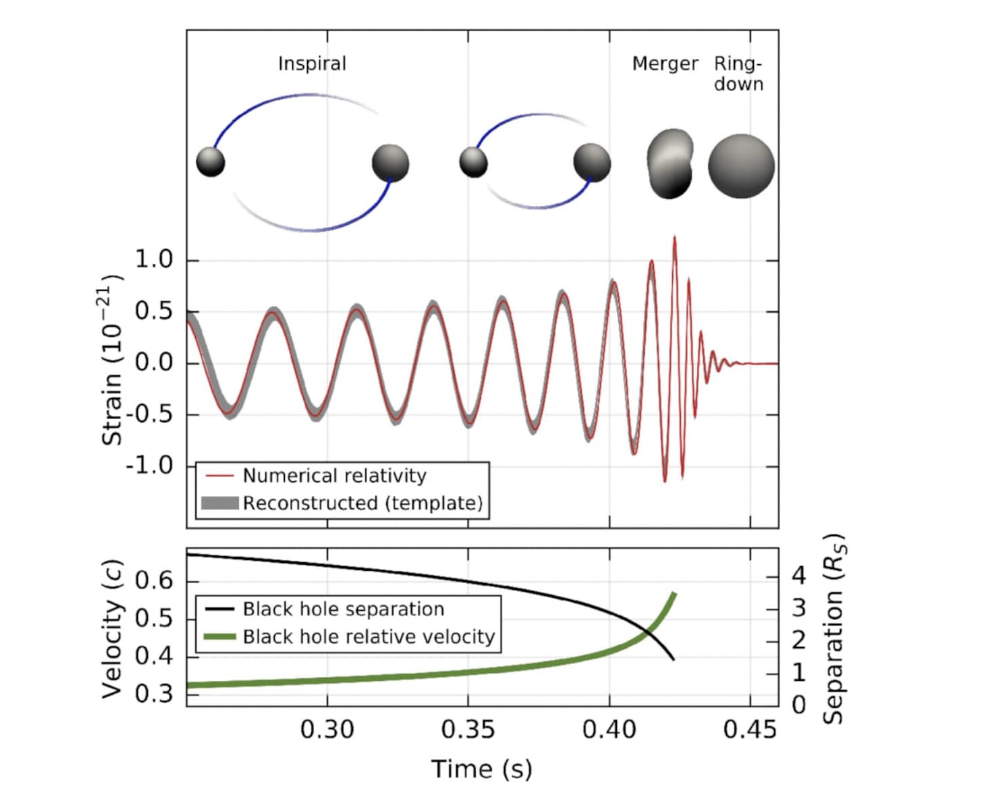}
    \caption{The GW150914 signal observed at the Hanford detector. The
    inspiral, merger and ringdown stages of the coalescence are indicated,
    together with the evolution of the separation and relative velocity of
    the two black holes. Figure adapted from
    Ref.~\protect\citeall{Abbott_2016}.}
    \label{fig:Cap1:GW150914}
\end{figure}

A gravitational waves' detector measures
the relative deformation, or strain, produced as the wave passes through
the interferometer. The physical parameters of the binary must therefore be
inferred by comparing the measured strain with theoretical templates. Their
phase and amplitude encode the individual masses and spins of the bodies, the orbital
geometry, the energy carried by radiation and possible departures from general
relativity. This makes waveform modelling an integral part of the observation.

The characteristic signal of a binary black hole coalescence is shown in
fig.~\ref{fig:Cap1:GW150914}. During the inspiral regime, the orbital time scale is
long compared with the radiation reaction time while the frequency increases
gradually. The subsequent merger is intrinsically nonlinear, while the final
ringdown is governed by the relaxation of the final black hole. A faithful theoretical
waveform model must connect these qualitatively different regimes without
losing information and accuracy during the matching.

This demand for theoretical precision will become even more stringent as the
sensitivity and frequency coverage of gravitational waves' observatories
increase. The advanced LIGO and Virgo interferometers and KAGRA established
the present ground-based network \citeall{Akutsu:2020his,Aasi:2015iha,Acernese:2014hva}.
The next generation includes the LISA mission and proposed
ground-based facilities such as Cosmic Explorer and the Einstein Telescope
\citeall{LISA:2017pwj,Punturo:2010zz,Reitze:2019iox}. These instruments will
observe a wider range of masses and length scales, and will require waveform
templates whose theoretical errors remain below the increasingly small
experimental uncertainties. The central theoretical problem is therefore not
only to solve Einstein's equations, but to organise their solutions in a form
that is accurate, systematically improvable and sufficiently efficient for
data analysis \citeall{buonanno2022snowmasswhitepapergravitational}.

The relativistic two-body problem is difficult precisely because it combines
several physical scales with the nonlinearity of general relativity. No single
expansion covers the complete evolution of a generic compact binary. For instance, the
post-Newtonian (PN) approximation expands in the orbital velocity and is naturally
suited to the early inspiral. Gravitational self-force (GSF) methods organise the
dynamics in the mass ratio, whereas numerical relativity (NR) treats the
strong-field merger non perturbatively. The effective one body (EOB) framework
combines and resums the informations from these descriptions into a model of the
entire coalescence. The post-Minkowskian (PM) expansion provides a complementary
organisation in powers of Newton's constant,
\begin{equation}\label{eq:Cap1:PM_expansion}
    \mathcal O
    =\sum_{n\geq1}G^n\,\mathcal O_{n\mathrm{PM}} \, ,
\end{equation}
without imposing a small velocity expansion. It is therefore particularly
well adapted to relativistic scattering and supplies information at all
orders in the velocity at each fixed PM order.

\begin{figure}[t]
    \centering
    \includegraphics[width=0.6\textwidth]
    {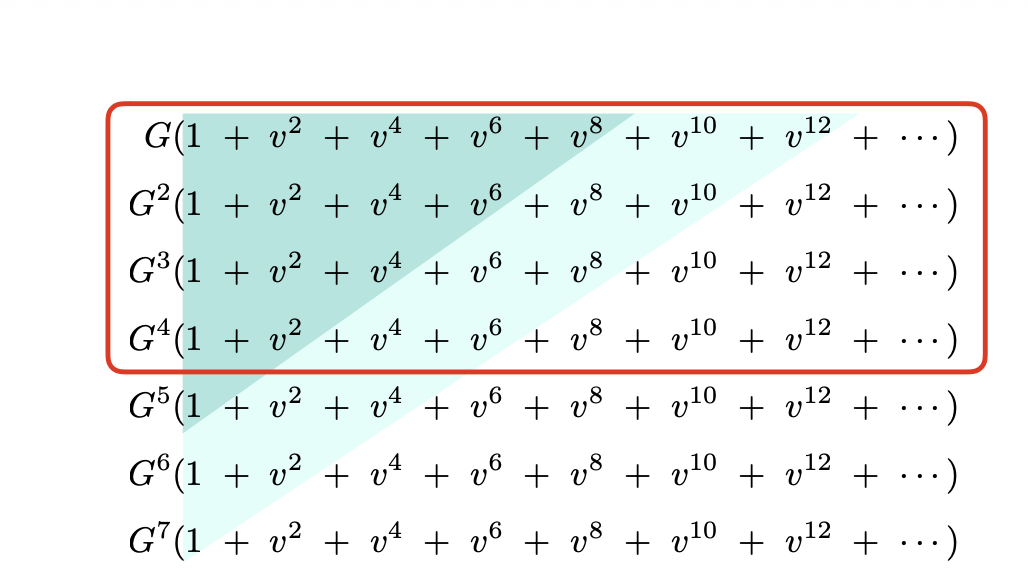}
    \caption{Schematic map of the perturbative expansion in Newton's
    constant and in the relative velocity; a fixed PM order contains an
    infinite tower of PN terms. Figure adapted from
    Ref.~\protect\citeall{buonanno2022snowmasswhitepapergravitational}.}
    \label{fig:Cap1:PM_expansion_and_strong_field}
\end{figure}

PN theory proceeds
horizontally in powers of the relative velocity, at a fixed range of powers of
\(G\), whereas an \(n\)PM result fixes a power of \(G\) while retaining the
complete velocity dependence. Their overlap provides sanity checks and
allows the informations obtained from amplitudes to be incorporated into the
traditional two body description. This complementary organisation is
illustrated in  fig.~\ref{fig:Cap1:PM_expansion_and_strong_field}.

\begin{figure}[t]
    \centering
    \includegraphics[width=1.\textwidth]
    {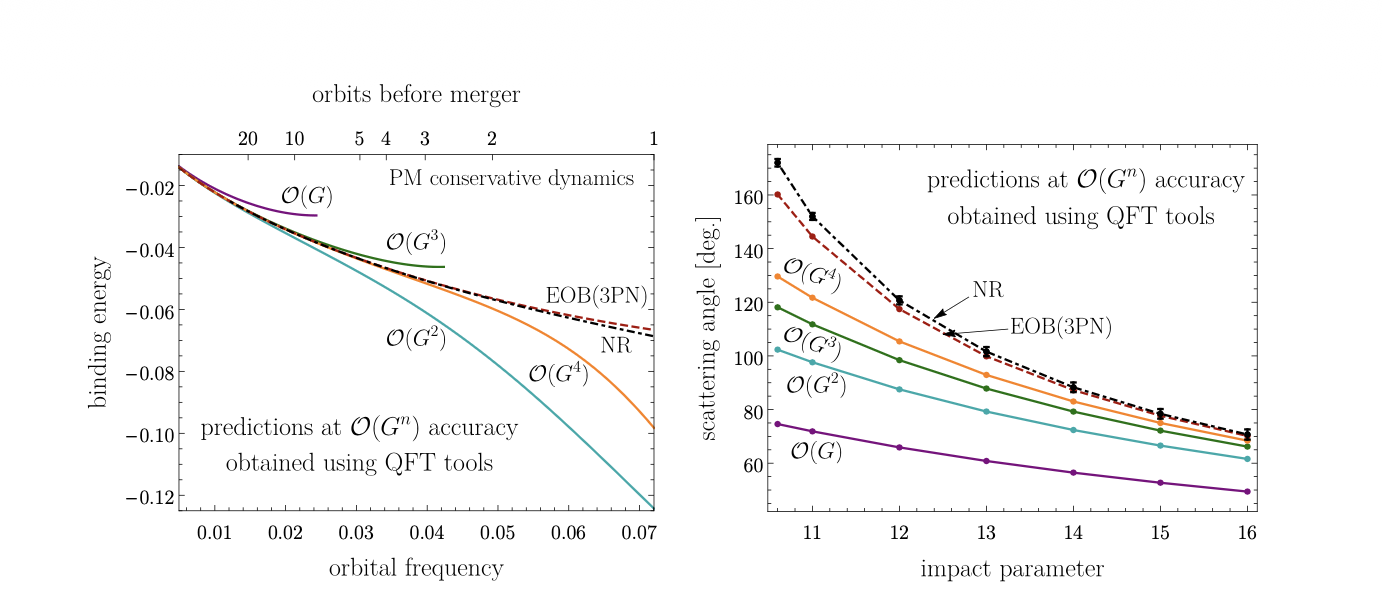}
    \caption{Binding energy for a quasi-circular binary (left) and scattering
    angle for a hyperbolic encounter (right). Increasing orders in the PM
    expansion, obtained using quantum-field-theory tools, are compared with
    effective-one-body and numerical-relativity results. Figure adapted from
    Ref.~\protect\citeall{buonanno2022snowmasswhitepapergravitational}.}
    \label{fig:Cap1:PM_NR_comparison}
\end{figure}

Scattering observables provide a useful meeting point for these approaches.
For example, the deflection angle of two compact objects 
can be extracted from numerical relativity simulations and can also be
computed perturbatively from the elastic amplitude. Through appropriate relations between scattering observables and bound-state dynamics, the same information can be used to constrain bound orbits and effective one body Hamiltonians.
 Figure~\ref{fig:Cap1:PM_NR_comparison}
illustrates how successive PM results obtained with quantum field theory
methods approach the numerical relativity and effective one body benchmarks.
The comparison also displays an important limitation: fixed order PM
predictions improve the weak field regime, but their convergence deteriorates
close to the strong field transition, where the perturbative regime starts to break down.

The modern scattering amplitudes programme offers a particularly schematic
route to the PM expansion. Individual Feynman diagrams contain gauge dependent
and redundant information, while an on-shell amplitude packages the physical
content of many diagrams into a compact, gauge-invariant, object. For instance, the
Parke--Taylor formula for maximally helicity-violating gluon amplitudes was an
early demonstration of this simplification \citeall{PT_formula}. Spinor-helicity
variables, on-shell recursion relations and unitarity methods subsequently made it
possible to construct tree and loop amplitudes directly from their poles,
factorisation channels and other building blocks rather than via the standard Lagrangian approach
\citeall{Britto_2005,BCFW_Tree_Level_Recursion,elvang2014scatteringamplitudes}.
These methods are especially effective in gravity, where conventional
perturbation theory rapidly produces very large intermediate expressions.

\begin{figure}[t]
    \centering
    \includegraphics[width=1.\textwidth]
    {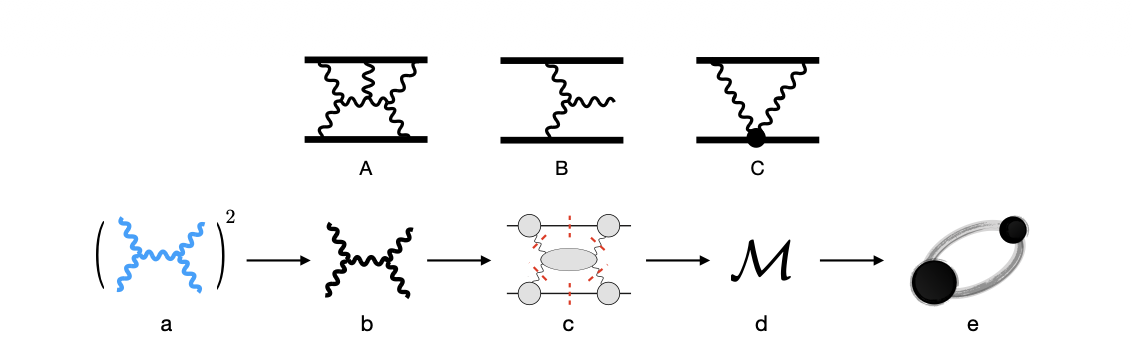}
    \caption{Schematic route from perturbative gravitational amplitudes to
    classical two-body observables. On-shell factorisation replaces a large
    set of gauge-dependent diagrams by compact amplitudes, whose appropriate
    classical limit determines the dynamics and radiation of the binary.
    Figure adapted from
    Ref.~\protect\citeall{buonanno2022snowmasswhitepapergravitational}.}
    \label{fig:Cap1:amplitudes_pipeline}
\end{figure}

An additional simplification follows from colour--kinematics duality. Once a
gauge theory amplitude is arranged so that its kinematic numerators obey the
same algebraic relations as its colour factors, replacing colour by a second
kinematic numerator yields a gravitational amplitude. Schematically,
\begin{equation}\label{eq:Cap1:double_copy_schematic}
    \text{gauge theory}\times\text{gauge theory}
    \sim \text{gravity} \, .
\end{equation}
This double-copy structure is not just a device for shortening
calculations, but instead it reveals that gauge theory and gravity share universal
on-shell building blocks and factorisation properties
\citeall{bern2019dualitycolorkinematicsapplications,Bern:2010ue}. The typical pattern from Yang--Mills amplitudes to classical gravitational observables is illustrated in fig.~\ref{fig:Cap1:amplitudes_pipeline}. One of the
goals of this thesis is to make this relation explicit in a special kinematic regime
where amplitudes acquire a defined ladder structure.

To turn an amplitude into a classical prediction, one must nevertheless take
the classical limit with care. The loop expansion is not identical to an
expansion in \(\hbar\), indeed loop amplitudes may contain classical terms. In the scattering problem the relevant limit is characterised by
large masses and angular momentum compared with the corresponding quantum
scales, together with a momentum transfer much smaller than the hard incoming
momenta. The KMOC formalism implements this limit at the level of 
expectation values of wave packets and provides a direct expression for the impulse, radiated
momentum and radiation field in terms of on-shell amplitudes
\citeall{Kosower:2018adc,kosower2022sagexreviewscatteringamplitudes}. In
particular, the elastic four-point amplitude determines the conservative
scattering informations, whereas the five-point amplitude with an emitted
on-shell graviton supplies the leading radiation kernel and hence the  waveform \citeall{Cristofoli:2022waveforms}.

In this thesis we will focuse on the high-energy, or Regge, regime of that programme.
For an elastic process it is defined by the condition
\begin{equation}\label{eq:Cap1:Regge_limit}
    s\gg |t|\, ,
    \qquad t\ \text{fixed} \, ,
\end{equation}
 The hierarchy in \refeq{eq:Cap1:Regge_limit} produces large rapidity
intervals and logarithms of \(s/|t|\). Moreover, it is also closely related to the
classical forward scattering limit: the hard particles carry large momenta,
while the exchanged momentum is predominantly transverse and much smaller.
The dominant exchange is then a Glauber graviton, and amplitudes with
additional emissions factorise into impact factors, exchanged propagators and
central emission vertices (CEVs).

In gauge theories, this strong rapidity ordering leads to the multi-Regge
factorisation of amplitudes, and sewing real emissions with virtual Regge
trajectories gives the BFKL equation, where its kernel generates the leading
high-energy evolution. Gravity retains the same kinematic factorisation and
the graviton CEV is a double copy of its gluonic counterpart. Its physical
power counting is, however, different. The dimensionful coupling enhances
eikonal exchange by the energy of the centre of mass, whereas the graviton Regge
trajectory is suppressed by \(|t|/s\) and is quantum in the \(\hbar\)
expansion. The leading classical high-energy logarithms are instead selected
from connected multi-\(H\) configurations, obtained by sewing tree amplitudes
through successive three-particle unitarity cuts
\citeall{alessio2026gravitationalamplitudesregge,barcaro2025centralemissionvertexgravitons,rothstein2026systematiclagrangianformulationquantum}.

The relation with the classical two-body problem can be understood directly
from the scattering geometry, indeed two compact objects prepared far apart approach
one another with an impact parameter \(b\), exchange a comparatively small
momentum \(q\) and emerge with deflected trajectories. The elastic four-point
amplitude encodes this momentum transfer. After a Fourier transform from
\(q\) to \(b\), its classical part determines the eikonal phase and hence the
impulse and scattering angle. If an additional on-shell graviton of momentum
\(k\) is present, the corresponding five-point amplitude contains the leading
radiative information. Its expectation value in a suitably localised incoming
state determines the radiation field measured at large distance. 

The Regge limit sharpens this picture because it separates the longitudinal and
transverse dynamics. The large energy in the centre of mass fixes the energetic
motion along the beam directions, while the deflection is controlled by the
smaller transverse momenta. Rapidity ordering then isolates universal
emission vertices between successive Glauber exchanges, while the  dependence on the
external particles is confined to the impact factors.  This separation makes it possible to compare pure graviton
amplitudes with amplitudes involving massive sources and to identify which
parts survive the classical limit. It also explains why the same CEV enters
both the high-energy kernel and the construction of scattering waveforms.

The central goal of the thesis is to develop these building blocks at
tree level and assemble the kernels needed for classical high-energy
scatterings and radiations. Particular attention is paid to next-to-multi-Regge
kinematics, where two particles have comparable rapidities. This region is
needed both to go beyond leading logarithmic accuracy and to connect
high-energy factorisation with collinear limits. The pure graviton analysis is
complemented by amplitudes involving massive scalars, which provide a simple
on-shell description of compact objects without spin. The resulting
five-point amplitude simultaneously contains the massive impact factor, its
collinear behaviour and the radiation kernel entering the leading waveform.

The organisation of the thesis is as follows. Chapter~\ref{Cap2:scatteringamp}
introduces the representation of the Poincaré group and fixes the
spinor-helicity and polarisation conventions used throughout the work. These tools are
then applied in
Chap.~\ref{Cap3:Tree_level_scattering_amplitudes}, where the tree-level gluon
and graviton MHV amplitudes, colour--kinematics duality and BCFW recursion are
reviewed, together with their universal soft and collinear factorisation properties.

Chapter~\ref{Cap4:Regge_limit_and_BFKL} develops high-energy factorisation
first in QCD and then in gravity. It introduces impact factors, the one- and
two-particle CEVs, the BFKL kernel and the distinction between the quantum
Reggeised ladder and the classical multi-\(H\) sequence. Those supply the tree-level building blocks
required if we are interested to study the leading and the next-to-leading logarithmic accuracy.

In Chap.~\ref{cap:Gravitational_amp_in_the_coll_limit} we will study the overlap of
the high-energy and collinear regimes. The five-graviton MHV amplitude is used
to obtain the NMRK impact factor and the gravitational splitting amplitudes.
The same analysis is then repeated for massive external particles: BCFW
recursion relations reconstruct the tree-level amplitude for two massive scalars
scattering with one graviton emission, from which the corrected impact factor, its
classical limit and the high-energy collinear waveform are extracted.

Finally, in Chap.~\ref{cap:Gravity_kernels} we sew the CEVs across unitarity cuts.
We construct the elastic \(H\) kernel, its radiative extension and the
classical NMRK kernel associated with the massive impact factor. The
transverse integration of the elastic kernel recovers the logarithmic
two-loop \(H\)-diagram contribution originally identified in the
high-energy gravitational scattering programme, including the infrared pole
fixed by the Weinberg soft current.

The appendices collect the Poincaré
conventions, the restoration of \(\hbar\), the KMOC extraction of classical
observables and waveforms, the five-point kinematic regions and the NMRK
phase-space reduction used in the kernel calculation.

Taken together, these results connect the on-shell structure of gauge and
gravity amplitudes, their factorisation in rapidity and the classical
observables of gravitational scattering and radiation, while separating classical dynamics from quantum Reggeisation.

\chapter{Scattering amplitudes and representation of the Poincaré group}\label{Cap2:scatteringamp}
\begin{center}
\begin{minipage}{0.8\textwidth}
\centering\itshape
``Our job in physics is to see things simply, to understand a great many
complicated phenomena in a unified way, in terms of a few simple principles.''\\[0.7em]
--- Steven Weinberg, \textit{Nobel Lecture} (December 8, 1979)
\end{minipage}
\end{center}

\hspace{0.2cm}

According to Weinberg, the task of theoretical physics is to explain complex phenomena in a unified way, in order to make predictions for observables that experimentalists can reasonably hope to verify.

Amplitudes are the mathematical objects that allow us to make a bonding between the ingoing states, prepared by the experimentalists in the laboratory much time before the interaction, and the outgoing states, those appearing in the detectors at the end of the scattering process. Our \textit{``few simple principles''} are nothing but the postulates of special relativity and the principles of quantum mechanics. 

In this chapter a basic introduction to modern helicity amplitudes is provided. Starting from the characterisation of the quantum one-particle states, compatible with causal structure of the Minkowskian spacetime, I introduce the spinor-helicity formalism, which proves particularly useful when dealing with processes involving only specific combinations of quantum numbers, like the helicity for massless particles. This new formalism allows one to reduce the length of the computations of diagrams that must be evaluated, by using a more compact form which makes the Lorentz structure more transparent and allows different amplitudes to be related through discrete symmetries such as parity and charge conjugation.

\section{Little group and one-particle states}
In this section I will show how to briefly characterize the quantum one-particle states. The historical and pedagogical work related on
the one-particle irreducible states is presented on Ref.~[\cite{Weinberg}], while for the introductory notions of group and representation theory  I will follow Ref.~\citeall{Bargmann,BargmannWigner,Tung}.

According to Bargmann and Wigner, the state $\psi$ of a physical system is described by a \textbf{ray} in the Hilbert space $\mathcal{H}$. If we define a vector $\ra{\psi} \in \mathcal{H}$, a ray [$\psi$] is an element in the quotient space, also called projective Hilbert space $\mathbb{P}\mathcal{H}\cong \frac{\mathcal{H}}{\sim}$, identified by the following equivalence relation
\begin{equation}
    \psi \sim \widetilde{\psi} : \exists \lambda \in \mathbb{C} \setminus \{0\} : \psi = \lambda \widetilde{\psi} \ .
\end{equation}

The transformations are represented, in this mathematical framework, as transformations $\hat T$ acting on $\mathbb{P}\mathcal{H}$, i.e. $[\psi] \mapsto [\psi ']=\hat{T}[\psi] $, and the transition amplitudes $\mathcal{P}([\psi] \to [\psi '])$,
which our experimental colleagues can actually measure, are described by $\mathcal{P}([\psi] \to [\psi ']) = \frac{|\langle \psi | \psi ' \rangle |^2}{||\psi||^2 \, ||\psi '||^2}$. A symmetry transformation is nothing but a projective automorphism, a linear map acting on $\mathbb{P}\mathcal{H}$ 
which is bijective and preserves the transition probability.

If we are interested to look at transformations on the projective Hilbert space induced by a group $G$, such as the Lorentz or the Poincaré group, we must deal with \textbf{projective representations}, i.e. group homomorphisms that associate, for each element of $g \in G$, a projective automorphism $\hat{T}_g \in \bf{AUT}(\mathbb{P}\mathcal{H})$, which forms 
a group under the composition.
As Wigner has shown, every projective automorphism acting on $\mathbb{P}\mathcal{H}$ is either induced by a unitary and linear operator $\mathcal{U}$ or by an anti-unitary and anti-linear operator.
In particular, in dealing with continuous transformations connected to $\mathds{1}$, we can show that we can pick unitary operators $\mathcal{U}$ acting on the rays, for instance, as 
\begin{equation}\label{eq:cap2_projectiveautom}
   \hat{T}_\mathcal{U}[\psi] = [\mathcal{U}\psi] \in \mathbb{P}\mathcal{H} \, .
\end{equation}

Notice that eq. (\ref{eq:cap2_projectiveautom}) is defined up to a phase, indeed
\begin{equation}\label{eq:cap2_projectiveautom_2}
 \hat{T}_\mathcal{U}\cong \hat{T}_\mathcal{\eta U} \, , \qquad \eta \in U(1) \, .   
\end{equation}

For generic projective unitary representations the homomorphism's properties, namely $\hat{T}_{g \, h}=\hat{T}_g \circ \hat{T}_h$, imply that, under the operation of the group, there exists  $\omega(g, h) \in U(1)$ such that $\mathcal{U}(g*h) = \omega(g, h) \, \mathcal{U}(g)\cdot \mathcal{U}(h)$. If this phase factor $\omega(g, h)$, via a $U(1)$ redefinition, \refeq{eq:cap2_projectiveautom_2}, can be 
recalibrated to the identity, then the projective unitary representation is actually a group homomorphism, i.e. the group's product is preserved, and it is called a (true) unitary representation. 

As said in the introduction, my \textit{``few simple principles''} are nothing but special relativity and quantum mechanics, so that if we would like to characterize the one-particle states
in a relativistic framework,  my symmetry transformations will be nothing but the Poincaré transformations.

Under a generic Poincaré transformation a four vector $x^{\mu}$ transforms as $x'^{ \mu}= \Lambda^{\mu}_{\ \nu}x^{\nu} + a^{\mu}$. From a QFT side, we are interested in transformations continuously connected to the identity, which preserve the time's orientation, so, rather than the full $O(1,3)$ Lorentz group, which is the set of $\Lambda \in \mathrm{GL}(4, \mathbb{R})$ such that $\Lambda^{t}\eta\,\Lambda=\eta $, we can focus on his subgroup, characterized by the condition $det\Lambda = 1$ and, in matrix form, $\Lambda^{0}_{\ 0} \geq 1$, so the group at which we are interested is just $
\mathcal{P}^{\uparrow}_{+}
= \mathbb{R}^{1,3} \rtimes SO^{+}(1,3)
$, the proper-orthochrouns Poincaré group. Keeping in mind the Poincaré algebra (see Appendix~\ref{app:poincare_algebra}), we can use the momentum components as my continuous degrees of freedom, while, for the discrete degrees of freedom we can introduce $\sigma$ that can, for instance, represent the helicity or other discrete quantum numbers.

We can therefore label the one-particle states as $\ra{p, \sigma}$, with $P\m\ra{p, \sigma} = p\m\ra{p, \sigma}$.

Notice that $P\m$ is an operator acting on the Hilbert space $\mathcal{H}$ which, by construction, belongs to a finite four-vector representation of the Lorentz group, then the transformation's rule for $P\m$  is just
\begin{equation}\label{eq:cap2_P_transf_rule}
    P^\mu \xrightarrow{ \Lambda \in SO^{+}(1,3) } 
    P'^\mu \equiv 
    \mathcal{U}(\Lambda)P^\mu \mathcal{U}^{-1}(\Lambda) 
    = (\Lambda^{-1})^{\mu}_{\ \nu}P^{\nu} 
    = \Lambda_{\nu}^{\ \mu}P^{\nu} \, ,
\end{equation}
where the unitary and linear operator $\mathcal{U}(\Lambda)$ acts on the one-particle state $\ra{p, \sigma}$ changing its momentum eigenvalue from $p\m$ to $(\Lambda p)\m$. This can be formalized by applying $P\m$ on $\mathcal{U}(\Lambda)\ra{p, \sigma}$ using \refeq{eq:cap2_P_transf_rule},
\begin{equation}
    P\m\mathcal{U}(\Lambda)\ra{p, \sigma}=\mathcal{U}(\Lambda)\left(\mathcal{U}^{-1}(\Lambda)P\m\mathcal{U}(\Lambda)\right)\ra{p, \sigma}=\mathcal{U}(\Lambda)
(\Lambda^{-1})_{\nu}^{\  \mu}P^{\nu}\ra{p, \sigma} = \Lambda^{\mu}_{\ \nu}p^{\nu} \mathcal{U}(\Lambda)\ra{p, \sigma} \, .  
\end{equation}

So $\mathcal{U}(\Lambda)\ra{p, \sigma}$ must be a linear combination of Lorentz-transformed one-particle states,
\begin{equation}\label{eq:cap2_Lorentz_1}
    \mathcal{U}(\Lambda)\ra{p, \sigma} = C_{\sigma' \sigma}(\Lambda, p)\ra{\Lambda p, \sigma'} \, .
\end{equation}
The issue now is to express the coefficients $C_{\sigma' \sigma}(\Lambda, p)$ in terms of irreducible representations of the Poincaré group. We could follow the so called \textbf{method of induced representation}. Starting by the Casimirs of my theory, $P^2$ and, in absence of supersymmetry generators, $W_{\mu}W\m$, the square of the Pauli-Lubański vector, we have chosen a complete set of commuting observables which led us to the one-particle states defined by the four vector eigenstates and helicity (or spin) eigenstates. The next step is the identification of the \textbf{Wigner Little Group} (L.G.). This Poincaré subgroup is identified by the following condition
\begin{equation}\label{eq:cap_2_LG}
    \text{L.G.}(p) = \left\{\Lambda \in SO^+(1,3) \, : \, \Lambda^{\mu}_{\ \nu}p^{\nu}=p\m\right\} \, .
\end{equation}

Now,  for each $p^2$, we can choose a reference momentum $k\m$, such that $p\m = L^{\mu}_{\ \nu}(p;k)k^{\nu}$ and we can define the one-particle states as
\begin{equation}
  \ra{p, \sigma}= \mathcal{U}\left(L(p;k)\right)\ra{k, \sigma}  .
\end{equation}

We can see how $\ra{p, \sigma}$ transforms under general Lorentz transformations,
\begin{equation}\label{eq:Cap_2_state_transf}
    \mathcal{U}(\Lambda)\ra{p, \sigma} = \mathcal{U}(\Lambda)\mathcal{U}\left(L(p;k)\right)\ra{k, \sigma} = \mathcal{U}\left(L(\Lambda p;k)\right)\mathcal{U}\left(L^{-1}(\Lambda p;k)\Lambda L(p;k)\right)\ra{k, \sigma} \, .
\end{equation}

Now, $W = L^{-1}(\Lambda p;k)\Lambda L(p;k)$ first maps $k$ to $p=L(p;k) \, k$, then to $\Lambda p$, then back to $k$, leaving at the end $k$ invariant,
\begin{equation}
    W^{\mu}_{\ \nu}k^{\nu}=k\m \, .
\end{equation}

Very similarly to \refeq{eq:cap2_Lorentz_1}, the action of $W$  on the reference state $\ra{k, \sigma}$ yields a linear combination of reference states,
\begin{equation}\label{eq:cap2_Lorentz_2}
   \mathcal{U}\left(W(\Lambda;p;k)\right)\ra{k, \sigma} = \mathcal{D}_{\sigma' \sigma}\left(W(\Lambda;p;k)\right)\ra{k, \sigma'} \, ,
\end{equation}
where the coefficients $\mathcal{D}_{\sigma' \sigma}$ provide a representation of the Wigner Little Group.

We see that we have shifted our focus from finding the coefficients $C_{\sigma' \sigma}(\Lambda, p)$ to evaluating the coefficients $\mathcal{D}_{\sigma' \sigma}\left(W(\Lambda;p;k)\right)$ which determine a representation of the little group. In order to clarify this point, we can use \refeq{eq:cap2_Lorentz_2} inside \refeq{eq:Cap_2_state_transf}, reaching 
\begin{equation}\label{eq:Cap2_LGtransf}
  \mathcal{U}(\Lambda)\ra{p, \sigma}=  \mathcal{D}_{\sigma' \sigma}\left(W(\Lambda;p;k)\right)\mathcal{U}\left(L(\Lambda p;k)\right)\ra{k, \sigma '} = \mathcal{D}_{\sigma' \sigma}\left(W(\Lambda;p;k)\right)\ra{\Lambda p, \sigma '} \, ,
\end{equation}
which, therefore, states that a one-particle state is labelled by its momentum and transforms under a representation of the little group.

Thus, an $n$-point scattering amplitude $\mathcal{M}_n$ is then labelled by the one-particle states $\ra{p_i, \sigma_i}$, with $i=1,...,n$, and Poincaré invariance, in $D$ spacetime,  implies that 
\begin{align}
    \mathcal{M}_n(p_i; \sigma_i) &= \delta^{D} \left(\sum_{i=1}^{n}p_i^{\mu}\right)M_n(p_i;\sigma_i) \, , \\
    M^{\Lambda}_{ n}(p_i;\sigma_i) &= \prod_{i=1}^n\mathcal{D}_{\sigma'_i \sigma_i}\left(W\right)M_n\left((\Lambda p)_i;\sigma_i\right) \, ,
\end{align}
so the amplitude transforms covariantly by products of little group coefficients.

For massive particles, a convenient choice of reference momentum is
\begin{equation}
    k^\mu=(m,0,0,0).
\end{equation}
The Lorentz transformations leaving this momentum invariant are precisely the
spatial rotations. Therefore, in four spacetime dimensions, the little group of a
massive particle is \(SO(3)\). At the quantum level one usually works with its
double cover \(SU(2)\), since
\begin{equation}
SO(3)\cong SU(2)/\mathbb{Z}_2,
\end{equation}
and \(SU(2)\) admits both integer and half-integer spin representations.

For massless particles, instead, one may choose a reference momentum along a
fixed spatial direction, for instance
\begin{equation}\label{eq:Cap2_ref_mom_massless}
k^\mu=(E,0,0,E).
\end{equation}
The subgroup of the Lorentz group leaving this null momentum invariant is the
two-dimensional Euclidean group
\begin{equation}
ISO(2)=SO(2)\ltimes \mathbb{R}^2.
\end{equation}
Here \(SO(2)\cong U(1)\) corresponds to rotations around the direction of
motion, while the two additional generators form the euclidean translational part of
\(ISO(2)\).

More generally, in \(D\) spacetime dimensions, the little group is \(SO(D-1)\)
for massive particles, while for massless particles it is
\begin{equation}
ISO(D-2)=SO(D-2)\ltimes \mathbb{R}^{D-2}.
\end{equation}

In light-cone coordinates, a generic four momentum in $D=4$ can be expanded as $p\m = \left(p^+,p^-;p^1,p^2\right)$, with $p^{\pm}=p^0\pm p^3$ so, for massless particles, \refeq{eq:Cap2_ref_mom_massless} becomes $k^\mu=(2E,0;0,0)$. Through  this change of basis, the generators of the Lorentz group $M^{\mu \nu}$ can be expanded in terms of $M^{j \pm}$, for $j=1,2$,
\begin{align}
    M^{j +} &= -i\left(x^j\partial^+-x^+\partial^j\right) = M^{j0} + M^{j 3} \, , \\
    M^{j -} &= -i\left(x^j\partial^--x^-\partial^j\right) = M^{j0} - M^{j 3} \, ,
\end{align}
with $\partial^{\pm} = \frac{\partial}{\partial x_0}\pm \frac{\partial}{\partial x_3}$. More details on the Lorentz and Poincaré algebra are reported in Appendix~\ref{app:poincare_algebra}. 

Moreover, the infinitesimal version of a Lorentz transformation takes the form $\Lambda^{\mu}_{\ \nu}= \delta^{\mu}_{\ \nu} + \omega^{\mu}_{\ \nu}$, then the little group condition takes the infinitesimal form $\omega^{\mu}_{\ \nu} k^{\nu} = 0$, which reduces the independent components of the little group to three real parameters.

Therefore, we can choose as my independent generators $M^{12}$, $M^{1-}$, $M^{2-}$, which fulfill the following commutation relations
\begin{equation}\label{eq:Cap2_Comm_rels_M}
    \left[M^{12}, M^{1-}\right] = iM^{2-}, \qquad \left[M^{12}, M^{2-}\right] = -iM^{1-}, \qquad \left[M^{1-}, M^{2-}\right]=0 \, .
\end{equation}
Since $M^{1-}$ and $M^{2-}$ commute among themselves, we can simultaneously diagonalize them by states $\ra{k, a, b}$, with $a$ and $b$ eigenvalues of $M^{1-}$ and $M^{2-}$ respectively. However a Lorentz rotation, looking at the commutation relations (\ref{eq:Cap2_Comm_rels_M}), would mix the eigenstates, hence leading to a tower of continuous states, also called continuous spin representations, which we think to be unphysical. Then we can assume in generality that $a=b=0$, and in this subspace all the three generators actually commute among themselves (obviously, because the translations are vanishing), thus we can introduce another quantum number $h$ related to the spectrum of $M^{12}$, so a general L.G. transformation on a massless one-particle state yields
\begin{equation}
    \mathcal{U\left(W(\theta;\alpha;\beta)\right)}\ra{k, h,\Vec{0}}=e^{i\theta M^{12}}e^{i\alpha M^{1-}+i\beta M^{2-}}\ra{k, h,\Vec{0}} = e^{i\theta h}\ra{k, h,\Vec{0}} \, .
\end{equation}

Finally, $M^{12}$ generates a rotation around the beam axis, so that we can identify $M^{12}=J_z$, and taking the angular component in the direction of the beam we find
\begin{equation}\label{eq:Cap2_helicity}
    \frac{\Vec{k}\cdot \Vec{J}}{|\Vec{k}|}\ra{k, h} = h\ra{k, h} \, ,
\end{equation}
hence its eigenvalue is exactly the helicity. So, if we compare the previous expression with \refeq{eq:Cap2_LGtransf}, we find that the little group coefficients for massless one-particle states are nothing but
\begin{equation}
 \mathcal{D}_{h' h}\left(W(\theta)\right) = e^{i\theta h}\delta_{h' h} \, ,
\end{equation}
thus helicity is Lorentz invariant, and we can label massless particles using momentum and helicity, keeping in mind that, for parity invariant theories, opposite helicity states must be related by a parity transformation.

Now, for massive particles, as we said above, we usually work with the double cover of $SO(3)$, namely $SU(2)$, which admits integer and half-integer spin representations. For massless particles we have a similar feature. In order to show that helicity is an integer or half-integer quantum number we have to study the topological nature of the Lorentz group. This can be done as follows. Firs of all let us introduce the Pauli matrices,
\begin{equation}
\sigma^1 =
\begin{pmatrix}
0 & 1 \\
1 & 0
\end{pmatrix} \, ,
\quad
\sigma^2 =
\begin{pmatrix}
0 & -i \\
i & 0
\end{pmatrix}\, ,
\quad
\sigma^3 =
\begin{pmatrix}
1 & 0 \\
0 & -1
\end{pmatrix}\, ,
\end{equation}
such that we can define two four-vectors
\begin{equation}
    \sigma\m=(\mathds{1},\Vec{\sigma}) \, , \qquad  \widebar{\sigma}\m=(\mathds{1},-\Vec{\sigma}) \, .
\end{equation}

Now we can build a map $\rho: M_4 \! \to \!H$, defined from the four Minkowskian momenta space to the space of hermitian complex matrices $H$, 
\begin{equation}
    \rho : p\m \to p\m\sigma_{\mu}=p_0\mathds{1} - \Vec{p}\cdot \Vec{\sigma} \, .
\end{equation}
It's easy to check that the matrix $p\m\sigma_{\mu}$ is Hermitian and the Hermiticity condition is also preserved under conjugation via the $GL(2, \mathbb{C})$ group, $p\m\sigma_{\mu} \to  \tau p\m\sigma_{\mu} \tau^{\dagger}$, with $\tau \in GL(2, \mathbb{C})$. Furthermore, $det(p\m\sigma_{\mu})= p^2$, so the determinant is a Lorentz invariant quantity, which is also preserved by considering  $\tau \in SL(2, \mathbb{C}) \le GL(2, \mathbb{C})$. Let's call $\Gamma$ this conjugation by $SL(2, \mathbb{C})$.

We have found our group homomorphism. Indeed we can define a inverse map from $H \! \to \!M_4$,
\begin{equation}
\rho^{-1}(\bullet)
=
\frac{1}{2}\operatorname{Tr}\!\left[\bullet\,\bar{\sigma}^{\mu}\right] \, .
\end{equation}
Indeed,
\begin{equation}
\rho^{-1}(p^{\nu}\sigma_{\nu}) = \frac{1}{2}\operatorname{Tr}\!\left[p^{\nu}\sigma_{\nu}\bar{\sigma}^{\mu}\right]
=
\frac{1}{2}\operatorname{Tr}\!\left[\sigma^{\nu}\bar{\sigma}^{\mu}\right]p_{\nu}
=
\eta^{\mu\nu}p_{\nu}=p^{\mu}.
\end{equation}

Assembling everything together, we then get a map from Minkowski momentum space into itself via the following chain:
\begin{equation}
M_4
\xrightarrow{\ \rho\ }
H
\xrightarrow{\ \Gamma\ }
H
\xrightarrow{\ \rho^{-1}\ }
M_4,
\end{equation}
namely
\begin{equation}
p^{\nu}
\xrightarrow{\ \rho\ }
p^{\nu}\sigma_{\nu}
\xrightarrow{\ \Gamma\ }
\tau p^{\nu}\sigma_{\nu} \tau^{\dagger}
\xrightarrow{\ \rho^{-1}\ }
\frac{1}{2}\operatorname{Tr}\!\left[
\tau p^{\nu}\sigma_{\nu} \tau^{\dagger}\bar{\sigma}^{\mu}
\right]
=p'^{\mu}.
\end{equation}

This is nothing but a Lorentz transformation obtained by the \(SL(2,\mathbb{C})\) transformation \(\Gamma\), namely
\begin{equation}
\Lambda^{\mu}{}_{\nu}(\Gamma)
=\frac{1}{2}
\operatorname{Tr}\!\left[
\bar{\sigma}^{\mu}\tau\sigma_{\nu}\tau^{\dagger}
\right].
\end{equation}

It is now a trivial exercise to prove the homomorphism
\begin{equation}
SL(2,\mathbb{C}) \longrightarrow SO^{+}(1,3).
\end{equation}

Notice that the relation
\begin{equation}
\Lambda^{\mu}{}_{\nu}(\Gamma)
=
\frac{1}{2}
\operatorname{Tr}\!\left[
\bar{\sigma}^{\mu}\tau\sigma_{\nu}\tau^{\dagger}
\right]
\end{equation}
can in principle be inverted, in the sense that for a given \(\Lambda\) one can find
a corresponding \(\tau\in SL(2,\mathbb{C})\). However, the relation is not an
isomorphism, since it is double-valued,
\begin{equation}
\Lambda(\Gamma)=\Lambda(-\Gamma).
\end{equation}
This proves that 
\begin{equation}
  SO^+(1,3)  \cong SL(2, \mathbb{C})/\mathbb{Z}_2 \, ,
\end{equation}
and, by knowing that $SL(2, \mathbb{C})$ is topologically equivalent to $\mathbb{R}^3 \otimes S_3$, with $S_3$ the three-sphere, the Lorentz group is actually topologically equivalent to $\mathbb{R}^3 \otimes S_3/\mathbb{Z}_2$. So the non-trivial and compact topology of $SO^+(1,3)$ is the same as that of $SO(3)$. In conclusion, since the helicity is the quantum number associated to the rotations along the beam axis, \refeq{eq:Cap2_helicity}, which form a subgroup inside the rotation group whose double cover is \(SU(2)\), the allowed eigenvalues of \(h\) are the \(SU(2)\) weights, therefore
\begin{equation}
h \in \frac{1}{2}\mathbb{Z} \, .
\end{equation}

\begin{figure}[H]
\centering
\begin{tikzpicture}[scale=1.05, >=Latex]
  \coordinate (P) at (1.55,0.72);
  \coordinate (M) at (-1.55,-0.72);

  \shade[ball color=blue!8, opacity=0.55] (0,0) circle (2.15);
  \draw[thick] (0,0) circle (2.15);
  \draw[ gray!70] (-2.15,0) arc[start angle=180,end angle=360,x radius=2.15,y radius=0.45];
  \draw[dashed, gray!70] (2.15,0) arc[start angle=0,end angle=180,x radius=2.15,y radius=0.45];

  \draw[very thick, red!75!black,
    postaction={decorate},
    decoration={markings, mark=at position 0.55 with {\arrow{Latex}}}]
    (P) to[out=150,in=25] (-0.65,1.55) to[out=205,in=60] (M);

  \draw[dashed, thick, blue!70!black] (P) -- (M);
  \fill[blue!75!black] (P) circle (2.3pt);
  \fill[blue!75!black] (M) circle (2.3pt);

  \node[above right] at (P) {$p$};
  \node[below left] at (M) {$-p$};
  \node[align=center, font=\small] at (0,-2.85)
    {$S_3/\mathbb{Z}_2:\ p\sim -p$};
\end{tikzpicture}
\caption{Schematic representation of \(S_3/\mathbb{Z}_2\): the antipodal points \(p\) and \(-p\) are identified, therefore a continuous path which connects them on \(S_3\) is a closed path on this quotient space.}
\label{fig:Cap2:S3_Z2_closed_path}
\end{figure}

\section{Spinor-helicity formalism}\label{Cap2:sec:spinorhelicity}
In this section we introduce the spinor-helicity formalism, following mainly Ref.~\citeall{Badger_2024, elvang2014scatteringamplitudes,Henn:2014yza,Ochirov_2018}. This formalism provides an efficient description of massless on-shell particles interacting in four dimensions and is particularly useful when dealing with scattering amplitudes in gauge theory and gravity. The basic idea is to exploit the factorization of massless momenta into two dimensional
spinors, thereby replacing Lorentz vectors with spinorial variables.

This has two main advantages. First of all, it makes the little group properties of external massless states manifest, so that the helicities are directly encoded in the spinor variables. Second, it leads to very compact expressions for tree-level amplitudes, especially in Yang-Mills theory and gravity, where the
standard Feynman-diagram expansion doesn't make manifest most of the times the final on-shell simple result. Indeed, the traditional workflow in particle physics is to square the amplitude and sum over the polarisations of the external states when we compute the cross sections. The bottleneck in this procedure happens when we square the amplitude, as if we are left with $n$ Feynman diagrams, $n^2$ terms will appear in the computation of the cross section, therefore the calculation quickly becomes intractable as the number of external particles grows. However, if we fix the external polarisations the amplitude is just a complex number and different helicity configurations don't interfere, so the cross section would be simply the sum of $n$ squared contributions at fixed helicities.

Moreover, for massless fermions the helicity basis is particularly natural, because helicity and chirality coincide and chirality is conserved along massless fermion lines in gauge interactions. This reduces the number of non vanishing helicity configurations that have to be computed. A similar simplification occurs for external gluons and gravitons, because many helicity amplitudes vanish, while the non vanishing ones often take compact forms when expressed in spinor-helicity variables, like the maximally helicity violating (MHV) ones.

\subsection{Massless Weyl spinors}
We have seen that the proper-orthochronous Lorentz group $SO^+(1,3)$ is isomorphic to $SL(2,\mathbb{C})/\mathbb{Z}_2$. From the point of view of the associated Lie algebra, this means that we can work with the Lie algebra of its simply connected universal cover, namely $SL(2,\mathbb{C})$. Its Lie algebra is the complexification of the Lie algebra of $SU(2)$,
\begin{equation}
    \mathfrak{sl}(2,\mathbb{C}) \cong \mathfrak{su}(2)_{\mathbb{C}} \, ,
\end{equation}
consistent with $SL(2,\mathbb{C})$ being parametrized by three complex parameters or six real ones.

Equivalently, after complexification, the Lorentz algebra decomposes into two commuting copies of $\mathfrak{su}(2)$,
\begin{equation}
    \mathfrak{sl}(2,\mathbb{C})
    \cong
    \mathfrak{su}(2)_L \oplus \mathfrak{su}(2)_R \, .
\end{equation}

Therefore finite dimensional Lorentz representations are labelled with two spins, $(j_L,j_R)$, corresponding to the two chiral sectors. In particular, the two fundamental spinorial representations are
\begin{equation}
    \left(\frac{1}{2},0\right)
    \qquad \text{and} \qquad
    \left(0,\frac{1}{2}\right) \, .
\end{equation}
These are the left-handed (negative chirality) and right-handed (positive chirality) Weyl spinor representations, respectively. In Lorentzian signature they are complex conjugates of each other,
\begin{equation}
    \left(\frac{1}{2},0\right)^*
    =
    \left(0,\frac{1}{2}\right) \, .
\end{equation}
Using the chiral representation of the Dirac matrices,
\begin{equation}
    \gamma^\mu =
    \begin{pmatrix}
        0 & \sigma^\mu \\
        \bar{\sigma}^\mu & 0
    \end{pmatrix},
    \qquad
    \gamma^5 =
    \begin{pmatrix}
        -\mathds{1} & 0 \\
        0 & \mathds{1}
    \end{pmatrix},
\end{equation}
a four-component Dirac spinor can be decomposed into a couple of two Weyl spinors,
\begin{equation}
    u(p)=
    \begin{pmatrix}
        \xi_-(p) \\
        \xi_+(p)
    \end{pmatrix} \, .
\end{equation}
The subscripts refer to the chirality of the corresponding Weyl spinors. Indeed, defining the chiral projectors
\begin{equation}
    P_\pm = \frac{\mathds{1}\pm \gamma^5}{2} \, ,
\end{equation}
one finds
\begin{equation}
    u_+(p)=P_+u(p)=
    \begin{pmatrix}
        0 \\
        \xi_+(p)
    \end{pmatrix},
    \qquad
    u_-(p)=P_-u(p)=
    \begin{pmatrix}
        \xi_-(p) \\
        0
    \end{pmatrix} \, .
\end{equation}
In the massless case the Dirac equation becomes
\begin{equation}
    p_\mu\gamma^\mu u(p)=0 \, ,
\end{equation}
and, using the chiral form of the gamma matrices, it splits into two independent Weyl equations,
\begin{equation}
    p_\mu\sigma^\mu\xi_+(p)=0 \, ,
    \qquad
    p_\mu\bar{\sigma}^\mu\xi_-(p)=0 \, .
\end{equation}
This shows explicitly that the two chiral Lorentz representations, \(\xi_+(p)\) and \(\xi_-(p)\), transform in the two inequivalent Weyl spinor representations of the Lorentz group.

In light-cone coordinates, in $D=4$, we can rearrange a generic four momentum as
\begin{equation}
    \label{eq:Cap2:momenta_conventions}
    p^\pm=p^0\pm p^3 \, ,
    \qquad
    p_\perp=p^1+i p^2 \, ,
    \qquad
    p_\perp^*=p^1-i p^2 \, ,
\end{equation}
so that the two matrices entering the Weyl equations can be written as
\begin{equation}
    p\cdot\sigma=
    \begin{pmatrix}
        p^- & -p_\perp^* \\
        -p_\perp & p^+
    \end{pmatrix},
    \qquad
    p\cdot\bar{\sigma}=
    \begin{pmatrix}
        p^+ & p_\perp^* \\
        p_\perp & p^-
    \end{pmatrix} \, .
\end{equation}
Moreover, since $p^2=det( p\cdot\sigma)=det( p\cdot\bar{\sigma})$, in the massless case both matrices have vanishing determinant and rank one. Solving the Weyl equations, up to a phase which depends on our convention, one obtains for \(p^+\neq 0\)
\begin{equation}\label{eq:Cap2:pplus_non0_spinors}
    \xi_+(p)=\frac{1}{\sqrt{p^+}}
    \begin{pmatrix}
        p^+ \\
        p_\perp
    \end{pmatrix},
    \qquad
    \xi_-(p)=\frac{1}{\sqrt{p^+}}
    \begin{pmatrix}
        -p_\perp^* \\
        p^+
    \end{pmatrix} \, ,
\end{equation}
while for $p^+=0$ we get
\begin{equation}\label{eq:Cap2:pplus_0_spinors}
    \xi_+(p)=
    \begin{pmatrix}
        0\\
        \sqrt{p^-}
    \end{pmatrix},
    \qquad
    \xi_-(p)=
    \begin{pmatrix}
        -\sqrt{p^-}\\
        0
    \end{pmatrix} \, .
\end{equation}

The spinors are normalized as
\begin{equation}
    \xi_+^\dagger(p)\xi_+(p)
    =
    \xi_-^\dagger(p)\xi_-(p)
    =2E \, ,
\end{equation}
and, according to our phase convention they are related by charge conjugation,
\begin{equation}
    \xi_+(p)=i\sigma^2\xi_-^*(p) \, ,
    \qquad
    \xi_-^*(p)=-i\sigma^2\xi_+(p) \, .
\end{equation}
Notice that for massless (positive-energy) particles, helicity and chirality coincide.

The next step is to reinterpret these Weyl spinors as spinor-helicity variables. Several choices are found in the literature, so we are going to introduce the angle brackets and square brackets that,
according to Ref.~[\cite{elvang2014scatteringamplitudes}], are \textit{``the core of what we call the spinor-helicity formalism''}, in our convention.

We choose
\begin{equation}
    \ra{p} \equiv u_+(p)=  \begin{pmatrix}
        0 \\
        \xi_+(p)
    \end{pmatrix}, \qquad
     |p] \equiv u_{-}(p)=  \begin{pmatrix}
        \xi_-(p) \\
        0
    \end{pmatrix},
\end{equation} 
and 
\begin{equation}
    \la{p} \equiv \bar{u}_{-}(p)
    =
    \begin{pmatrix}
        0 & \xi_-^{\dagger}(p)
    \end{pmatrix},
    \qquad
    [p| \equiv \bar{u}_{+}(p)
    =
    \begin{pmatrix}
        \xi_+^{\dagger}(p) & 0
    \end{pmatrix} \, ,
\end{equation}
for fermions with positive energy. This implies that we can write the spinor products in a more compact way, i.e.
\begin{align}
    \lar{pq}
    &\equiv
    \langle p | q\rangle
    =
    \bar{u}_{-}(p)u_{+}(q)
    =
    \xi_-^{\dagger}(p)\xi_+(q) \, , \\[1mm]
    [p q]
    &\equiv
    [p|q]
    =
    \bar{u}_{+}(p)u_{-}(q)
    =
    \xi_+^{\dagger}(p)\xi_-(q) \, .
\end{align}
Using the explicit expressions in \refeq{eq:Cap2:pplus_non0_spinors}, for $p^+$ and $q^+ \neq 0$ one obtains
\begin{equation}
    \lar{pq}
    =
    -p_\perp\sqrt{\frac{q^+}{p^+}}
    +q_\perp\sqrt{\frac{p^+}{q^+}}
    =
    \frac{p^+q_\perp-q^+p_\perp}{\sqrt{p^+q^+}} \, ,
\end{equation}
and
\begin{equation}
    [p q]
    =
    p_\perp^*\sqrt{\frac{q^+}{p^+}}
    -q_\perp^*\sqrt{\frac{p^+}{q^+}}
    =
    \frac{p_\perp^*q^+-q_\perp^*p^+}{\sqrt{p^+q^+}} \, .
\end{equation}
The antisymmetry of the spinor products follows immediately,
\begin{equation}
    \lar{p q}=-\lar{q p} \, ,
    \qquad
    [p q]=-[q p] \, ,
    \qquad
    \lar{p p}=[p p]=0 \, .
\end{equation}
The two products are related by complex conjugation,
\begin{equation}
    \lar{p q}^{*}=\text{sign}(p^0q^0)[q p] \, .
\end{equation}
Finally, multiplying the two spinor products gives the Lorentz invariant scalar product of the two massless momenta,
\begin{equation}
    \lar{p q}[q p]=2p\cdot q \, .
\end{equation}

Notice that in the amplitude language we assume all the momenta to be outgoing, so we must perform an analytic continuation for the ingoing spinors by defining
\begin{equation}
|-p\rangle = i |p\rangle \, ,
    \qquad
    |-p] = i |p] \, ,
    \qquad
    \langle -p| = i \langle p| \, ,
    \qquad
    [-p| = i [p| \, .
\end{equation} 
Finally, this projection on the chirality basis implies that, a generic massless four momentum can be expanded as
\begin{equation}
\slashed{p}=\ra{p}[p| + |p]\la{p} \, .
\end{equation}

Another common notation, useful to make the Lorentz structure explicit, is the dotted-undotted spinor notation. In this language the two chiral spinor representations are distinguished by
\begin{equation}
    |p\rangle \longleftrightarrow \lambda_{\alpha}(p) \, ,
    \qquad
    |p] \longleftrightarrow \widetilde{\lambda}_{\dot\alpha}(p) \, ,
\end{equation}
while the corresponding dual spinors are obtained by raising indices with the invariant antisymmetric tensors of \(SL(2,\mathbb{C})\),
\begin{equation}
    \langle p| \longleftrightarrow \lambda^{\alpha}(p) \, ,
    \qquad
    [p| \longleftrightarrow \widetilde{\lambda}^{\dot\alpha}(p) \, .
\end{equation}
Our conventions for raising and lowering indices are
\begin{equation}
    \lambda^{\alpha}=\epsilon^{\alpha\beta}\lambda_{\beta} \, ,
    \qquad
    \lambda_{\alpha}=\epsilon_{\alpha\beta}\lambda^{\beta} \, ,
    \qquad
    \widetilde{\lambda}^{\dot\alpha}
    =\epsilon^{\dot\alpha\dot\beta}\widetilde{\lambda}_{\dot\beta} \, ,
    \qquad
    \widetilde{\lambda}_{\dot\alpha}
    =\epsilon_{\dot\alpha\dot\beta}\widetilde{\lambda}^{\dot\beta} \, .
\end{equation}
With this notation, a massless momentum is represented by the rank-one bi-spinor
\begin{equation}\label{eq:Cap2:massless_momentum_bispinor}
    p_{\alpha\dot\alpha}
    =p_{\mu}(\sigma^{\mu})_{\alpha\dot\alpha}
    =\lambda_{\alpha}(p)\widetilde{\lambda}_{\dot\alpha}(p) \, .
\end{equation}
This equation is just the dotted-undotted version of the factorisation of a null momentum into two spinors. Indeed,
\begin{equation}
    p^2=\det\left(p_{\alpha\dot\alpha}\right)=0 \, ,
\end{equation}
so \(p_{\alpha\dot\alpha}\) has rank one and can be written as an outer product of a left-handed and a right-handed spinor, as said previously.

More details on this formalism, such as the Gordon identity, the Fierz rearrangement and the Schouten identity, are reported in Ref.~[\cite{Dixon_1}],  while an interesting extension of this formalism to the massive case  can be found in Ref.~[\cite{arkanihamed2021scatteringamplitudesmassesspins}].

\subsection{Polarisation vectors}

We now extend the spinor-helicity description to massless particles of spin one. This is the case of photons and gluons, whose physical one-particle states are labelled by their momentum and helicity \(h=\pm1\). In the usual covariant description, an external vector particle is associated with a polarisation vector \(\varepsilon^\mu(p)\), satisfying the transversality condition
\begin{equation}
    p_\mu \varepsilon^\mu(p)=0 \, .
\end{equation}
However, this description contains a gauge redundancy, since the shift
\begin{equation}\label{eq:Cap2:GAUGETr}
    \varepsilon^\mu(p)\to \varepsilon^\mu(p)+\alpha p^\mu
\end{equation}
does not affect the gauge-invariant amplitudes. As we will se, spinor-helicity formalism makes this structure immediate. In order to write explicit polarisation vectors, we introduce an arbitrary null reference momentum \(q^\mu\), with
\begin{equation}
    q^2=0 \, ,
    \qquad
    p\cdot q\neq0 \, .
\end{equation}
Different choices of \(q\) just correspond to different gauge choices for the same external state.

In terms of spinor-helicity variables, the two polarisation vectors can be written as
\begin{equation}\label{eq:Cap2:polarisation_vectors_spinor_helicity}
    \varepsilon_-^\mu(p,q)
    =
    \frac{\langle p|\gamma^\mu|q]}{\sqrt{2}\,[p q]} \, ,
    \qquad
    \varepsilon_+^\mu(p,q)
    =-
    \frac{[ p|\gamma^\mu|q\rangle}{\sqrt{2}\,\langle  p q\rangle} \, ,
    \qquad
    \left(\varepsilon^{\pm}_\mu(p,q)\right)^* = \varepsilon^{\mp}_\mu(p,q) \, .
\end{equation}

Equivalently, using the dotted-undotted notation introduced above, a polarisation vector can be packaged into a bi-spinor as
\begin{equation}
    \varepsilon^{\pm}_{\alpha\dot\alpha}(p,q)
    =
    \varepsilon^{\pm}_{\mu}(p,q)
    (\sigma^\mu)_{\alpha\dot\alpha} \, .
\end{equation}
Then the two helicity states in \refeq{eq:Cap2:polarisation_vectors_spinor_helicity} become
\begin{equation}\label{eq:Cap2:polarisation_vectors_dotted_undotted}
    \varepsilon^-_{\alpha\dot\alpha}(p,q)
    =
    \sqrt{2}\,
    \frac{\lambda_{\alpha}(p)\widetilde{\lambda}_{\dot\alpha}(q)}{[pq]} \, ,
    \qquad
    \varepsilon^+_{\alpha\dot\alpha}(p,q)
    =
    -\sqrt{2}\,
    \frac{\lambda_{\alpha}(q)\widetilde{\lambda}_{\dot\alpha}(p)}{\langle pq\rangle} \, .
\end{equation}
In this form the Lorentz vector index has been indeed replaced by one undotted and one dotted spinor index. The spinors associated with the reference momentum \(q\) encode the gauge choice. Indeed changing \(q\) shifts \(\varepsilon^{\pm}_{\alpha\dot\alpha}\) by a term proportional to \(p_{\alpha\dot\alpha}\), which at the end amounts only in a gauge transformation, as in \refeq{eq:Cap2:GAUGETr}.

The reference momentum must be chosen so that the denominators do not vanish. Using the massless Dirac equations,
\begin{equation}
    \slashed{p}|p\rangle=0 \, ,
    \qquad
    \slashed{p}|p]=0 \, ,
\end{equation}
one immediately finds the transversality conditions
\begin{equation}
    p_\mu\varepsilon_\pm^\mu(p,q)=0 \, ,
    \qquad
    q_\mu\varepsilon_\pm^\mu(p,q)=0 \, .
\end{equation}
The first condition is the physical transversality with respect to the external momentum, while the second one reflects the gauge choice fixed by the reference momentum \(q\).

The polarisation vectors in \refeq{eq:Cap2:polarisation_vectors_spinor_helicity} obey the normalization identities
\begin{equation}
    \varepsilon_\pm(p_i,q)\cdot \varepsilon_\pm(p_j,q)=0 \, ,
    \qquad
    \varepsilon_+(p,q)\cdot \varepsilon_-(p,q)=-1 \, ,
\end{equation}
where the last sign follows from the mostly-minus metric convention. Furthermore, as said previously, changing the reference momentum corresponds to a gauge transformation. Indeed, for another null reference momentum \(q'\), one has
\begin{equation}
    \varepsilon_\pm^\mu(p,q')
    =
    \varepsilon_\pm^\mu(p,q)
    +
    \alpha_\pm(q,q')p^\mu \, ,
\end{equation}
for some coefficient \(\alpha_\pm(q,q')\). Therefore the dependence on \(q\) must drop out on any gauge invariant amplitude.

The little-group behaviour is also manifest. Under the scaling
\begin{equation}
    |p\rangle \to z|p\rangle \, ,
    \qquad
    |p] \to z^{-1}|p] \, ,
\end{equation}
for $z \in U(1)$, the polarisation vectors transform as
\begin{equation}
    \varepsilon_+^\mu(p,q)\to z^{-2}\varepsilon_+^\mu(p,q) \, ,
    \qquad
    \varepsilon_-^\mu(p,q)\to z^{2}\varepsilon_-^\mu(p,q) \, .
\end{equation}
This is precisely the expected little-group scaling for external massless particles of helicity \(h=\pm1\).

Since in the following we will also deal with gravitational amplitudes, it is useful to introduce the corresponding spin-two polarisation tensors. For a graviton of helicity \(h=\pm2\), they are given by the symmetric products
\begin{equation}\label{eq:Cap2:graviton_polarisation_tensor}
    \varepsilon_\pm^{\mu\nu}(p,q)
    =
    \varepsilon_\pm^\mu(p,q)\varepsilon_\pm^\nu(p,q) \, .
\end{equation}
Thus, graviton polarisation tensors carry twice the little-group weight of vector polarisation states, as expected for helicity \(\pm2\).

Combining this result with the little-group scaling of fermionic external states, we conclude that an \(n\)-point amplitude with external helicities \(h_i\) must transform as

\begin{equation}\label{eq:Cap2:amplitude_little_group_scaling}
    A_n\left(1^{h_1},\ldots,n^{h_n}\right)
    \xrightarrow{ \text{L.G.}} 
    \prod_{i=1}^{n} z_i^{-2h_i}\,
    A_n\left(1^{h_1},\ldots,n^{h_n}\right) \, .
\end{equation}

Before closing this section, let us stress one final point. Although the polarisation vectors in \refeq{eq:Cap2:polarisation_vectors_spinor_helicity} depend explicitly on the arbitrary reference momentum \(q\), physical amplitudes cannot depend on this choice, as said above. This independence is a consequence of gauge invariance. At the level of an amplitude with one external gauge boson, gauge invariance is encoded in the Ward identity
\begin{equation}\label{eq:Cap2:ward_identity_spin1}
    A_\mu(p,\ldots)\,p^\mu=0 \, .
\end{equation}

This provides a useful consistency check for spinor-helicity computations.

The analogous statement holds for external gravitons. In that case the amplitude carries two Lorentz indices and diffeomorphism invariance implies the gravitational Ward identity
\begin{equation}\label{eq:Cap2:ward_identity_spin2}
    A_{\mu\nu}(p,\ldots)\,p^\nu=0 \, .
\end{equation}
As a result, the replacement
\begin{equation}
    \varepsilon^{\mu\nu}(p,q)
    \to
    \varepsilon^{\mu\nu}(p,q)
    +p^\mu\xi^\nu+p^\nu\xi^\mu
\end{equation}
does not change the physical graviton amplitude. Thus, for both gauge theory and gravity, the spinor-helicity formalism separates the physical helicity information, encoded in little-group weights, from the unphysical gauge choice, encoded in the reference momentum.

\chapter{Tree-level  scattering amplitudes}\label{Cap3:Tree_level_scattering_amplitudes}
In this chapter we will provide a generic introduction to the modern view of  tree-level scattering amplitudes in both gauge and gravity theories.

We can now use the spinor-helicity formalism, introduced in Sec.~\ref{Cap2:sec:spinorhelicity}, in order to write gluon and graviton amplitudes at fixed external helicities in a more compact way. The main examples are 
the maximally helicity violating (MHV) amplitudes, whose helicity configuration, for a $n$-point amplitude, involves two negative helicity gluons or gravitons 
and $n-2$ positive helicity ones. We begin with gluon amplitudes in Yang--Mills theory. In particular, we present the
Parke--Taylor formula for colour-ordered tree-level MHV amplitudes. We then discuss
the corresponding gravitational amplitudes, whose compact form can be written through
the Hodges' formula.

We will then introduce colour-kinematics (CK) duality, which provides a consistent way to relate gauge-theory amplitudes to gravitational amplitudes through the double-copy construction.

After that, we will present the on-shell Britto--Cachazo--Feng--Witten (BCFW) recursion relations. These methods allow one to construct higher-point tree amplitudes recursively using lower-point on-shell building blocks.

Finally, we will discuss the universal behaviour of scattering amplitudes in soft and collinear limits, which is crucial if we are interested to study the infrared structure of gauge and gravity theories.

\section{Gluon and graviton MHV amplitudes}
\label{sec:gluon_graviton_mhv_amplitudes}

In this section I will introduce the compact expressions for the gluon and graviton MHV amplitudes. We start with the gluon case, following mainly Ref.~\citeall{Dixon_1,PT_formula}, and then we discuss the gravitational case using the Hodges' formula of Ref.~\citeall{hodges2012simpleformulagravitationalmhv}. 

\subsection{Parke--Taylor formula}
\label{subsec:parke_taylor_formula}
First of all we clarify the notion of colour-ordering in QCD amplitudes. At tree level, a generic \(n\)-gluon amplitude can be expanded using the Trace-based colour decomposition as
\begin{equation}\label{eq:Cap3:colour_decomposition_tree_gluon}
    \mathcal{A}_n^{\mathrm{tree}}(\{p_i,h_i,a_i\})
    =
    g^{n-2}
    \sum_{\sigma\in \Delta_{n-1}}
    \mathrm{Tr}\!\left(
    T^{a_{\sigma(1)}}\cdots T^{a_{\sigma(n)}}
    \right)
    A_n^{\mathrm{tree}}\left(
    \sigma(1^{h_1}),\ldots,\sigma(n^{h_n})
    \right) \, .
\end{equation}
The objects \(A_n^{\mathrm{tree}}\) are the colour-ordered partial amplitudes. They contain the kinematic and helicity dependence, while the explicit colour dependence is encoded in the trace. Each partial amplitude is associated with a fixed cyclic ordering of the external gluons, indeed in \refeq{eq:Cap3:colour_decomposition_tree_gluon} the sum runs over the $(n-1)!$ non cyclic permutations of the gluon labels. Let's also stress that the trace-based is particularly useful 
when dealing with gluon amplitudes, but of course it's not the only possible colour decomposition.

Before writing the first non-vanishing formula, it is useful to remark that some helicity configurations vanish identically. In particular, at tree level one has
\begin{equation}\label{eq:Cap3:all_plus_single_minus_vanish}
    A_n^{\mathrm{tree}}(1^\pm,2^+,\ldots,n^+)=0 \, .
\end{equation}
These identities show that the first non vanishing gluon amplitudes are not the all plus or single minus ones, but rather the amplitudes with two negative helicity gluons. These are the MHV ones.

For a colour-ordered \(n\)-gluon amplitude with only two negative helicity gluons located in positions \(i\) and \(j\), the Parke--Taylor formula reads
\begin{equation}\label{eq:Cap3:parke_taylor_formula}
    A_n^{\mathrm{tree}}
    \left(1^+,\ldots,i^-,\ldots,j^-,\ldots,n^+\right)
    =
    i\,\frac{\langle i j\rangle^4}
    {\langle 1 2\rangle\langle 2 3\rangle\cdots\langle n 1\rangle} \, .
\end{equation}
This result expresses in a compact way a \(n\)-point tree amplitude as a rational function of spinor products only, even if the number of Feynman diagrams contributing to the same process grows very rapidly with \(n\).

We can make some observations about \refeq{eq:Cap3:parke_taylor_formula}. First, the denominator reflects the cyclic ordering of the colour-ordered amplitude: only adjacent spinor products appear. Second, the formula has the correct little-group scaling. Indeed a positive helicity state \(k\neq i,j\) appears in two denominators and therefore carries weight \(z_k^{-2}\). A negative helicity leg, say \(i\), receives a factor \(z_i^4\) from the numerator and a factor \(z_i^{-2}\) from the denominator, giving a net weight \(z_i^2\), as required for helicity \(-1\).

Finally, the simple structure of the Parke--Taylor formula is also consistent with its factorisation properties. A multi-particle pole would correspond to factorizing the scattering process into two subprocesses,
each with at least four gluons,
\begin{equation}\label{eq:Cap3:multiparticle_factorisation_channel}
    A_n^{\mathrm{tree}}(\ldots)
    \xrightarrow{P^2\to0}
    A_{n-k+1}^{\mathrm{tree}}(\ldots,P^{\lambda_P},\ldots)
    \frac{i}{P^2}
    A_{k+1}^{\mathrm{tree}}(\ldots,(-P)^{-\lambda_P},\ldots) \, ,
    \qquad
    3\leq k\leq n-3 \, .
\end{equation}
In the MHV case, the two negative-helicity gluons are among the external legs appearing in the two subamplitudes on the right-hand side of \refeq{eq:Cap3:multiparticle_factorisation_channel}. The intermediate gluon carries one additional negative helicity on one side of the factorisation channel. Therefore we have only three negative helicity gluons to be distributed between the two lower point tree amplitudes. However, since all plus and single minus tree amplitudes vanish, each non vanishing gluon tree amplitude requires at least two negative helicities. The two subamplitudes would therefore require at least four negative helicity gluons in total. Hence we have found that the MHV amplitudes have no multi-particle factorisation poles. So, their remaining singularities are those ones in the soft and collinear regions, consistently with the adjacent bracket structure of \refeq{eq:Cap3:parke_taylor_formula}.

\subsection{Hodges formula}
\label{subsec:Hodges_formula}

We now turn to the gravitational analogue of the Parke--Taylor formula. While Yang--Mills partial amplitudes are colour-ordered and therefore  cyclically invariant, graviton amplitudes do not carry colour degrees of freedom and must be fully symmetric over $n$ gravitons. For this reason gravitational MHV amplitudes cannot have the same cyclic denominators as in \refeq{eq:Cap3:parke_taylor_formula}. 

For a $n$-graviton MHV amplitude with  gravitons with negative helicities in positions \(i\) and \(j\), one can pull out the helicity dependence as
\begin{equation}\label{eq:Cap3:graviton_mhv_hodges_factorisation}
    \mathcal{M}_n^{\mathrm{MHV}}
    \left(1^+,\ldots,i^-,\ldots,j^-,\ldots,n^+\right)
    =
    \lar{ij}^8\,
    \widetilde{\mathcal{M}}_n(1,2,\ldots,n) \, .
\end{equation}
If we compare this expression with the Yang--Mills case, the power \(\langle i j\rangle^8\) reflects the fact that a graviton has twice the helicity of a gluon. Indeed, the negative helicity legs must carry little group weight \(z_i^4\) and \(z_j^4\), while positive helicity gravitons must carry weight \(z_k^{-4}\).

As precisely shown in Ref.~\citeall{hodges2012simpleformulagravitationalmhv}, the helicity independent amplitude can be written in terms of a \(n\times n\) symmetric matrix \(\Phi\). Its off diagonal entries are

\begin{equation}\label{eq:Cap3:hodges_matrix_off_diagonal}
    \Phi^i{}_j
    =
    \frac{[ij]}{\lar{ij}} \, ,
    \qquad
    i\neq j \, .
\end{equation}
The diagonal entries are defined using two arbitrary reference spinors \(x\) and \(y\),
\begin{equation}\label{eq:Cap3:hodges_matrix_diagonal}
    \Phi^i{}_i
    =
    -\sum_{k\neq i}
    \frac{[ik]\lar{kx}\lar{ky}}
    {\lar{ik}\lar{ix}\lar{iy}} \, .
\end{equation}
The dependence on $x$ and $y$ is only auxiliary and drops out on the final amplitude.

In order to define the reduced determinant, we delete three rows \(i,j,k\) and three columns \(r,s,t\) from \(\Phi\). We denote the corresponding minor with
\begin{equation}
    |\Phi|^{ijk}_{rst} \, .
\end{equation}
Then the Hodges formula for the reduced MHV graviton amplitude reads
\begin{equation}\label{eq:Cap3:hodges_formula}
    \widetilde{\mathcal{M}}_n(1,2,\ldots,n)
    =
    (-1)^{n+1}
    \mathrm{sign}(ijk)\,\mathrm{sign}(rst)\,
    c_{ijk}\,c^{rst}\,
    |\Phi|^{ijk}_{rst} \, ,
\end{equation}
where $\text{sign}(ijk) \equiv \text{sign}(\sigma(i,j,k,1,2,\ldots,\slashed{i}, \slashed{j}, \slashed{k}, \ldots, n))$ is the signature of the permutation which
moves $i,j,k$ up front in the sequence, and 
\begin{equation}
c_{ijk} = \frac1{\langle ij \rangle \langle jk \rangle \langle ki \rangle } \, .
\end{equation}
Putting everything together, the full MHV graviton amplitude can therefore be written as
\begin{equation}\label{eq:Cap3:full_hodges_formula}
    \mathcal{M}_n^{\mathrm{MHV}}
    \left(1^+,\ldots,i^-,\ldots,j^-,\ldots,n^+\right)
    =
    \lar{ij}^8
    (-1)^{n+1}
    \mathrm{sign}(abc)\,\mathrm{sign}(rst)\,
    c_{abc}\,c^{rst}\,
    |\Phi|^{abc}_{rst} \, ,
\end{equation}
where $i$ and $j$ denote the two negative helicity gravitons, while $(a,b,c) $ and $ (r,s,t) $ are the indices of deleted rows and columns. So, we have derived the gravitational counterpart of Parke--Taylor. It gives all the tree-level MHV graviton amplitudes in a compact form using the spinor-helicity language, but replaces the cyclic structure of colour-ordered Yang--Mills amplitudes with a minor determinant that makes the bosonic permutation symmetry of $n$ gravitons manifest.

In the case of a four point graviton amplitude, at tree level, the MHV amplitude, according to \refeq{eq:Cap3:full_hodges_formula}, can be obtained for instance by choosing the minor 
\begin{equation}
 |\Phi|^{234}_{234}=\Phi^1{}_1=-\frac{[12]\lar{23}\lar{34}}{\lar{12}\lar{13}\lar{14}} \, ,
\end{equation}
leading to
\begin{equation}\label{eq:Cap3:four_graviton_mhv_amplitude}
    \mathcal{M}_4
    \left(1_h^-,2_h^-,3_h^+,4_h^+\right)
    =
    \kappa^2\,
    \frac{\lar{12}^{7}[12]}
    {\lar{13}\lar{14}\lar{23}\lar{24}\lar{34}^{2}} \, .
\end{equation}
All other helicity configurations of type \((-,-,+,+)\) are obtained by permuting the external labels, while the all-plus and single-minus configurations vanish at tree level.

For five gravitons instead, one convenient form is obtained by choosing the minor $|\Phi|^{123}_{345}$, reaching
\begin{equation}\label{eq:Cap3:five_graviton_mhv_amplitude}
    \mathcal{M}_5
    \left(1_h^-,2_h^-,3_h^+,4_h^+,5_h^+\right)
    =
    \kappa^3\,
    \lar{12}^{8}
    \frac{
    [15][24]\lar{14}\lar{25}
    -
    [14][25]\lar{24}\lar{15}
    }
    {
    \lar{12}\lar{23}\lar{31}\lar{34}\lar{45}\lar{53}
    \lar{14}\lar{25}\lar{24}\lar{15}
    } \, .
\end{equation}

We will encounter these expressions again later in this work.

\section{Colour-kinematics duality}\label{Cap3:sec:CKduality}

In this section we will discuss the relation between gravitational and Yang--Mills amplitudes. We will see that we can write gravity amplitudes as a double copy, or a squared, of Yang-Mills amplitudes, captured by the abstract expression: \textit{``gravity = (gauge theory)$^2$''}. We will follow mainly Ref.~\citeall{bern2019dualitycolorkinematicsapplications, Bern:2010ue, Elvang_Huang_2015}.

The starting point is that we can rearrange a $n$-point Yang--Mills tree amplitude in terms of diagrams with cubic vertices only. In this way the tree amplitude is written as a sum over all distinct trivalent diagrams,
\begin{equation}\label{eq:Cap3:BCJrepr}
    A_n^{tree}=\sum_{i \in trivalent}\frac{c_in_i}{\prod_{\alpha_i}p_{\alpha_i}^2} \, ,
\end{equation}
where the numerator factorizes into the colour part $c_i$ and a pure kinematic part $n_i$, while the denominator is given by the product of all propagators of a given diagram. As an example, the four-point gluon amplitude can be expanded into the three cubic channels as
\begin{equation}\label{eq:Cap3:four_gluon_cubic_decomposition}
\begin{aligned}
    \mathcal{A}_4^{\mathrm{tree}}(1,2,3,4)
    &=
    \vcenter{\hbox{%
    \begin{tikzpicture}[scale=0.58, baseline=-0.5ex]
        \coordinate (L) at (-0.65,0);
        \coordinate (R) at (0.65,0);
        \draw (L) -- (R);
        \draw (L) -- (-1.55,0.75) node[left] {$1$};
        \draw (L) -- (-1.55,-0.75) node[left] {$2$};
        \draw (R) -- (1.55,0.75) node[right] {$3$};
        \draw (R) -- (1.55,-0.75) node[right] {$4$};
    \end{tikzpicture}}}
    +
    \vcenter{\hbox{%
    \begin{tikzpicture}[scale=0.58, baseline=-0.5ex]
        \coordinate (T) at (0,0.65);
        \coordinate (B) at (0,-0.65);
        \draw (T) -- (B);
        \draw (T) -- (-0.95,1.35) node[left] {$1$};
        \draw (T) -- (0.95,1.35) node[right] {$3$};
        \draw (B) -- (-0.95,-1.35) node[left] {$2$};
        \draw (B) -- (0.95,-1.35) node[right] {$4$};
    \end{tikzpicture}}}
    +
    \vcenter{\hbox{%
    \begin{tikzpicture}[scale=0.58, baseline=-0.5ex]
        \coordinate (L) at (-0.55,0);
        \coordinate (R) at (0.55,0);
        \draw (L) -- (R);
        \draw (L) -- (1.45,0.95) node[right] {$3$};
        \draw (L) -- (-1.45,-0.95) node[left] {$2$};
        \draw (R) -- (-1.45,0.95) node[left] {$1$};
        \draw (R) -- (1.45,-0.95) node[right] {$4$};
    \end{tikzpicture}}}
    \\
    &=
    \frac{c_s n_s}{s}
    +
    \frac{c_t n_t}{t}
    +
    \frac{c_u n_u}{u} \, ,
\end{aligned}
\end{equation}
where the quadri-linear vertex has been absorbed into the cubic $s$, $t$ and $u$ channels. Note that, since we are expanding the perturbative series in terms of cubic Feynman diagrams only, the Jacobi identity must hold, so $c_s + c_t +c_u=0$, which can be generalized to any set of three trivalent diagrams
\begin{equation}\label{eq:Cap3:Jacobi_generalised}
    c_i + c_j +c_k = 0 \, .
\end{equation}
This implies that we can shift the kinematic factors $n_i$ without changing the total amplitude in several ways. For instance, one can apply a gauge transformation to the polarisation vectors, which of course does not change the total amplitude but changes the $n_i$ factors. A more non trivial transformation is obtained by the condition~(\ref{eq:Cap3:Jacobi_generalised}), which directly tells that, if we take the following changes,
\begin{equation}\label{eq:Cap3:BCJ_transf}
    n_i \to n_i +s_i\Delta \, , \qquad   n_j \to n_j +s_j\Delta \, , \qquad   n_k \to n_k +s_k\Delta \, , 
\end{equation}
the total amplitude stays invariant since the net deformation is proportional to $c_i + c_j +c_k$. Because of the similarity, \refeq{eq:Cap3:BCJ_transf} is often called \textbf{generalized gauge transformation}.

The idea of Bern, Carrasco and Johansson (BCJ) was that we can arrange the kinematic factors $n_i$ such that they should fulfill the same properties as the corresponding color factors $c_i$. More precisely, this \textbf{colour-kinematics duality} states that that scattering amplitudes of Yang–Mills theory, and its supersymmetric extensions, can be given in a representation where the numerators $n_i$ have the same algebraic properties as the corresponding color factors $c_i$. Using the representation~(\ref{eq:Cap3:BCJrepr}), the BCJ proposal is that one can always find a basis such that the following parallel relations hold, 
\begin{equation}\label{eq:Cap3:BCJ_duality}
    \begin{aligned}
        c_i = - c_j  \quad &\Longleftrightarrow \quad n_i = -n_j \, , \\
        c_i + c_j +c_k = 0 \quad &\Longleftrightarrow \quad n_i + n_j + n_k = 0 \, .
    \end{aligned}
\end{equation}

A remarkable consequence of this condition is that, if we replace the colour strcuture for our non abelian gauge theory with the gravitational charge, i.e. the 
the kinematic factors, we obtain directly the tree level gravitational amplitude,
\begin{equation}\label{eq:Cap3_BCJ_DC_relation}
    M_n^{tree}=\sum_{i \in cubic}\frac{n_i^2}{\prod_{\alpha_i}p_{\alpha_i}^2} \, .
\end{equation}
This relation is called \textbf{BCJ double-copy relation}, and this formula matches precisely all the poles that should appear in a gravitational amplitude.

Moreover, it turns out that we can generalize this expression. We can take indeed two distinct Yang--Mills numerators $n_i$ and $\widetilde{n_i}$, such that only one set of numerators, say $n_i$, satisfies the colour-kinematics duality~(\ref{eq:Cap3:BCJ_duality}), while the other copy $\widetilde{n_i}$ can be an arbitrary 
representation of the Yang--Mills amplitude, such that
\begin{equation}\label{eq:Cap3_BCJ_DC_relation_2}
    M_n^{tree}=\sum_{i \in cubic}\frac{n_i\widetilde{n_i}}{\prod_{\alpha_i}p_{\alpha_i}^2} \, .
\end{equation}
Indeed since $n_i$ and $\widetilde{n_i}$ are both valid representations of the same Yang--Mills amplitude, it follows that, if we define
\begin{equation}
\Delta_i = n_i - \widetilde{n_i} \, ,
\end{equation} 
we have, by gauge invariance,
\begin{equation}
\sum_{i \in cubic}\frac{c_i\Delta_i}{\prod_{\alpha_i}p_{\alpha_i}^2}  = 0 \, .
\end{equation}
Now, since the $n_i$'s satisfy the same algebraic properties as the $c_i$'s, we conclude that 
\begin{equation}
\sum_{i \in cubic}\frac{n_i\Delta_i}{\prod_{\alpha_i}p_{\alpha_i}^2}  = 0 \, ,
\end{equation}
which proves the corresponding generalized gauge invariance condition in the gravity case.

Now, it is useful to compare this result with another gauge-gravity relation, namely the Kawai--Lewellen--Tye (KLT) relations. These relations were originally derived in string theory, where closed string amplitudes can be written as sums of products of open string partial amplitudes. 

In the field-theory limit, open string amplitudes with external massless spin one states reduce to colour-ordered Yang--Mills amplitudes, while closed string amplitudes with external massless spin two states reduce to gravity amplitudes, and therefore KLT relations just express tree-level gravity amplitudes directly in terms of products of colour-ordered gauge-theory amplitudes. For instance, at four and five points one finds that
\begin{equation}\label{eq:Cap3:KLT_examples}
\begin{aligned}
    \mathcal{M}_4^{\mathrm{tree}}(1,2,3,4)
    &=
    -s_{12}\,
    A_4^{\mathrm{tree}}(1,2,3,4)\,
    A_4^{\mathrm{tree}}(1,2,4,3) \, ,
    \\
    \mathcal{M}_5^{\mathrm{tree}}(1,2,3,4,5)
    &=
    s_{23}s_{45}\,
    A_5^{\mathrm{tree}}(1,2,3,4,5)\,
    A_5^{\mathrm{tree}}(1,3,2,5,4)
    +(3\leftrightarrow4) \, .
\end{aligned}
\end{equation}

The BCJ double-copy representation can therefore be seen as another version of the gauge-gravity relation. However, unlike KLT which is written directly in terms of gauge-invariant colour-ordered amplitudes, BCJ works before that the sum over cubic diagrams has been performed. The individual numerators are not physical observables by themselves, but this is precisely what allows us to adjust their algebraic properties in order to mimic the Jacobi identities. Once such a representation is found, the passage from Yang--Mills theory to gravity is implemented by the replacement
\begin{equation}
    c_i \longrightarrow \widetilde n_i \, .
\end{equation}

A natural question is whether BCJ duality satisfying numerators can be obtained directly from a Lagrangian. In four dimensions this can be done for MHV amplitudes, while for other helicity configurations the construction becomes more difficult. In general, the ordinary Yang--Mills Feynman rules do not produce automatically numerators satisfying the Jacobi identity, even after distributing quadri-linear vertices among cubic diagrams. One possible way to make the duality manifest is to deform the Yang--Mills action by adding non local terms which vanish by the Jacobi identity,
\begin{equation}
    \mathcal{L}_{\mathrm{YM}}
    \longrightarrow
    \mathcal{L}_{\mathrm{YM}}
    +\mathcal{L}'_5+\mathcal{L}'_6+\cdots \, .
\end{equation}
Since the extra terms are proportional to algebraic identities, they do not change the theory; rather, they reorganise the Feynman rules so that the colour-kinematics structure becomes manifest.

This is the main advantage of the BCJ formulation over KLT for our purposes. KLT relations introduce the gauge-gravity duality at the level of tree amplitudes, while BCJ exposes it at the level of cubic diagrams, making the construction less manifestly gauge invariant term by term, but much closer to the diagrammatic structure of the perturbation theory. In particular, the same replacement of colour factors by kinematic numerators can be applied to loop integrands, where KLT relations do not directly generalise.

\section{BCFW recursion relations}
\label{sec:Cap3:BCFW}

In this section we introduce the Britto--Cachazo--Feng--Witten (BCFW) recursion relations, shown in Ref.~\citeall{BCFW_Tree_Level_Recursion}. The main idea is to consider a tree-level amplitude $M_n(p_1, \ldots, p_n)$ from its analytic structure  as a function of its complex momenta $p_1,\ldots,p_n$. The momenta are complexified by introducing a shift of the momenta, which preserves on-shellness and momentum
conservation, and which is linear in a complex variable $z$. In this way the amplitude $M_n(p_1, \ldots, p_n)$ becomes an analytic function of $z \in \mathbb{C}$. 

Let us shift a momentum $p_j\m$ by an arbitrary vector $q\m$,
\begin{equation}\label{eq:Cap3:BCFW_momentum_shift}
    \widehat p_j(z)=p_j+zq \, .
\end{equation}
In order to keep the momentum conservation, we must shift also another momentum $p_i\m$, with $i\neq j$,
\begin{equation}\label{eq:Cap3:BCFW_momentum_shift_2}
    \widehat p_i(z)=p_i-zq \, .
\end{equation}
Moreover, in order to preserve also the on-shell conditions for all values of \(z\), the shift vector must satisfy
\begin{equation}\label{eq:Cap3:BCFW_shift_conditions}
    q^2=0 \, ,
    \qquad
    q\cdot p_i=0 \, ,
    \qquad
    q\cdot p_j=0 \, .
\end{equation}
For massless particles these conditions can be implemented directly in spinor-helicity variables, obtaining
\begin{equation}\label{eq:Cap3:BCFW_spinor_shift}
    |\widehat j\rangle=|j\rangle+z|i\rangle \, ,
    \qquad
    |\widehat i]=|i]-z|j] \, ,
    \qquad
    |\widehat j]=|j] \, ,
    \qquad
    |\widehat i\rangle=|i\rangle \, .
\end{equation}
Equivalently, in two-component notation,
\begin{equation}\label{eq:Cap3:BCFW_momentum_shift_spinor_form}
    \widehat p_{i,\alpha\dot\alpha}(z)
    =p_{i,\alpha\dot\alpha}-z\,|i\rangle_{\alpha}[j|_{\dot\alpha} \, ,
    \qquad
    \widehat p_{j,\alpha\dot\alpha}(z)
    =p_{j,\alpha\dot\alpha}+z\,|i\rangle_{\alpha}[j|_{\dot\alpha} \, .
\end{equation}
In this convention the shift vector is
\begin{equation}
    q_{\alpha\dot\alpha}=q^\mu(\sigma_\mu)_{\alpha\dot\alpha}
    =|i\rangle[j| \, .
\end{equation}
This generalizes to the four vector in the full Dirac components,
\begin{equation}
    \slashed q
    =
    |i\rangle[j|+|j]\langle i| \, .
\end{equation}
The deformed amplitude \(M_n(z)\) has then become a meromorphic function of \(z\). 

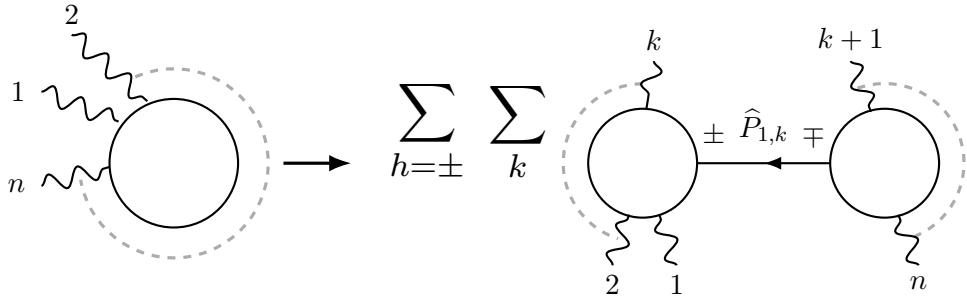
\begin{figure}[H]
\centering
\begin{tikzpicture}[scale=1.0, >=Latex]
    \coordinate (C) at (-4.2,0);
    \draw[thick] (C) circle (0.85);
    \draw[dashed, gray!65, very thick]
        (-4.71,1.14) arc[start angle=114,end angle=-170,radius=1.25];

    \draw[decorate, decoration={snake, amplitude=1.2mm, segment length=4mm}, thick]
        (-5.95,0.95) -- (-4.93,0.55) node[pos=0,left=2pt] {$1$};
    \draw[decorate, decoration={snake, amplitude=1.2mm, segment length=4mm}, thick]
        (-5.55,1.70) -- (-4.55,0.82) node[pos=0,above] {$2$};
    \draw[decorate, decoration={snake, amplitude=1.2mm, segment length=4mm}, thick]
        (-5.95,-0.30) -- (-5.05,-0.05) node[pos=0,left=2pt] {$n$};

    \draw[->, very thick] (-2.75,0) -- (-1.85,0);
    \node at (-0.85,0.42)
        {\scalebox{1.55}{$\displaystyle\sum_{h=\pm}$}};
    \node at (0.35,0.42)
        {\scalebox{1.55}{$\displaystyle\sum_{k}$}};

    \coordinate (L) at (2.0,0);
    \coordinate (R) at (5.2,0);
    \draw[thick] (L) circle (0.72);
    \draw[dashed, gray!65, very thick]
        (2.00,1.05) arc[start angle=90,end angle=252,radius=1.05];
    \draw[thick] (R) circle (0.72);
    \draw[dashed, gray!65, very thick]
        (4.84,0.99) arc[start angle=110,end angle=-70,radius=1.05];

    \draw[
        thick,
        postaction={decorate},
        decoration={markings,mark=at position 0.5 with {\arrow{>}}}
    ] (4.48,0) -- (2.72,0);
    \node[above] at (2.94,0.10) {$\pm$};
    \node[above] at (3.60,0.10) {$\widehat P_{1,k}$};
    \node[above] at (4.26,0.10) {$\mp$};

    \draw[decorate, decoration={snake, amplitude=1.2mm, segment length=4mm}, thick]
        (2.15,1.35) -- (2.05,0.72) node[pos=0,above] {$k$};
    \draw[decorate, decoration={snake, amplitude=1.2mm, segment length=4mm}, thick]
        (1.60,-1.35) -- (1.82,-0.68) node[pos=0,below] {$2$};
    \draw[decorate, decoration={snake, amplitude=1.2mm, segment length=4mm}, thick]
        (2.45,-1.35) -- (2.22,-0.68) node[pos=0,below] {$1$};

    \draw[decorate, decoration={snake, amplitude=1.2mm, segment length=4mm}, thick]
        (4.75,1.35) -- (5.02,0.68) node[pos=0,above] {$k+1$};
    \draw[decorate, decoration={snake, amplitude=1.2mm, segment length=4mm}, thick]
        (5.65,-1.35) -- (5.38,-0.68) node[pos=0,below] {$n$};
\end{tikzpicture}
\caption{Diagrammatic form of the BCFW recursion relation. The original \(n\)-point amplitude is reconstructed as a sum over factorisation channels and over the helicity \(h=\pm\) of the intermediate on-shell state.}
\label{fig:Cap3:BCFW_factorisation}
\end{figure}

We said that unitarity implies that the poles of amplitudes correspond
to the exchange of on-shell intermediate states. So the singularities of amplitudes stem from the
vanishing propagators in the Feynman diagrams. At tree level, this means that the amplitude, or
Feynman diagram, factorises into two smaller amplitudes connected by a vanishing propagator. Its poles occur when an internal momentum depending on \(z\) goes on shell. If the momentum flowing through a factorisation channel is
\begin{equation}\label{eq:Cap3:BCFW_internal_momentum}
    \widehat P_{1,k}(z)
    =p_1+\cdots+\widehat p_j(z)+\cdots+p_k \, ,
\end{equation}
then the pole is determined by
\begin{equation}
    \widehat P_{1,k}^2(z_k)=0 \, .
\end{equation}
Using \refeq{eq:Cap3:BCFW_momentum_shift} and \refeq{eq:Cap3:BCFW_momentum_shift_2}, we find
\begin{equation}\label{eq:Cap3:BCFW_pole_location}
    z_k=-\frac{P_{1,k}^2}{2P_{1,k}\cdot q} \, , \qquad z_k \neq 0,
\end{equation}
provided that the two shifted legs lie on opposite sides of the channel. If they lie on the same side, \(\widehat P_{1,k}\) is independent of \(z\) and no BCFW pole is generated in that channel. 
The factorisation at a BCFW pole has the diagrammatic form depicted in figure \ref{fig:Cap3:BCFW_factorisation}.
Notice that the poles $z_k$ cannot be located 
in the origin, which is the physical region in which the original amplitude were defined regularly.

Specifically, near \(z=z_k\) figure~\ref{fig:Cap3:BCFW_factorisation} shows that the amplitude behaves as
\begin{equation}\label{eq:Cap3:BCFW_factorisation_near_pole}
    iM_n(z)
    \xrightarrow{z\sim z_k}
    \sum_{h}
    iM_L\left(\ldots,\widehat p_j(z_k),-\widehat P^{-h}_{1,k}(z_k)\right)
    \frac{i}{\widehat P_{1,k}^2(z)}
    iM_R\left(\widehat P^{h}_{1,k}(z_k),\widehat p_i(z_k),\ldots\right) \, .
\end{equation}
The sum runs over the possible helicities of the intermediate on-shell state.

The recursion relation follows from Cauchy's theorem applied to \(M_n(z)/z\). Since the single pole at the origin gives the physical amplitude, one has
\begin{equation}\label{eq:Cap3:Cauchy_BCFW}
    \frac{1}{2\pi i}\oint_{\mathcal{C}_0} \frac{dz}{z}\,M_n(z)=M_n(0) \, 
\end{equation}
for any positive oriented path $\mathcal{C}_0$ close enough to the origin, which represents the physical region.

If we apply the global residue theorem, we find that
\begin{equation}\label{eq:Cap3:Cauchy_BCFW_2}
    \frac{1}{2\pi i}\oint_{\mathcal{C}} \frac{dz}{z}\,M_n(z)=0 \, 
\end{equation}
for any large contour $\mathcal{C}$ which surrounds all the BCFW poles $z_k$, provided that the amplitude vanishes at infinity, \(M_n(z)\to0\) for \(z\to\infty\), which is the 
case under consideration in many of the physical problems. Therefore we have
\begin{equation}\label{eq:Cap3:BCFW_residue_relation}
    M_n(0)
    =
    -\sum_k \underset{z=z_k}{\mathrm{Res}}\,\frac{M_n(z)}{z} \, .
\end{equation}
Using the factorisation formula into lower point amplitudes in \refeq{eq:Cap3:BCFW_factorisation_near_pole}, we can write the residue contribution explicitly as
\begin{equation}\label{eq:Cap3:BCFW_residue_factorisation}
\begin{aligned}
    iM_n(0)
    &=-\sum_{k=2}^{n-2}
    \underset{z=z_k}{\mathrm{Res}}\,
    \frac{iM_n(z)}{z}
    =-\sum_{h=\pm}
    \sum_{k=2}^{n-2}
    \underset{z=z_k}{\mathrm{Res}}\,
    \Biggl[
    iM_{k+1}\left(1,2,\ldots,\widehat j,\ldots,-\widehat P_{1,k}^{-h}\right)
    \frac{i}{z\,\widehat P_{1,k}^2(z)}
    \\
    &\qquad\times
    iM_{n-k+1}\left(\widehat P_{1,k}^{h},k+1,\ldots,\widehat i,\ldots,n\right)
    \Biggr] \, .
\end{aligned}
\end{equation}
Then, using the relation between \(z-z_k\) and \(\widehat P_{1,k}^2(z)\), the residue of the deformed propagator leads to the physical propagator \(1/P_{1,k}^2\). We finally obtain the BCFW recursion relation,
\begin{equation}\label{eq:Cap3:BCFW_recursion_general}
\begin{aligned}
    iM_n(p_1,\ldots,p_n)
    &=
    \sum_{h=\pm}
    \sum_{k=2}^{n-2}
    iM_{k+1}\left(1,2,\ldots,\widehat j,\ldots,-\widehat P_{1,k}^{-h}\right)
    \frac{i}{P_{1,k}^2} 
    \\
    &\times
    iM_{n-k+1}\left(\widehat P_{1,k}^{h},k+1,\ldots,\widehat i,\ldots,n\right) \, ,
\end{aligned}
\end{equation}
where all the hatted quantities in \refeq{eq:Cap3:BCFW_recursion_general} are evaluated at the corresponding pole \(z=z_k\). The sum over \(k\) runs over all allowed factorisation channels, which are basically given by all the partitions of the external legs for which the two shifted particles lie on opposite sides and both subamplitudes contain at least three external legs, otherwise at least one of the two subamplitudes identically vanish.

This completes the derivation of the BCFW recursion relations, up to one important assumption: the contribution from the contour at infinity must vanish. No theorem guarantees that for an arbitrary theory or for an arbitrary choice of shifted legs. Rather, it is a statement about the large \(z\) behaviour of the deformed amplitude \(M_n(z)\). When \(M_n(z)\to0\) as \(z\to\infty\), no boundary term is present and the physical amplitude is completely determined by the residues at its finite factorisation poles. In many physically relevant problems in Yang--Mills theory and in gravity there exist choices of shifts for which the required large \(z\) behaviour is good enough.

The practical importance of \refeq{eq:Cap3:BCFW_recursion_general} is that it is constructive. The amplitudes appearing on the right hand side have lower multiplicity than the original one, and therefore the recursion can be iterated. Starting from the three-point amplitudes, which are fixed by little group scaling, one can build higher point tree amplitudes from on-shell ``gluing'', making the BCFW recursion relations an alternative procedure to the standard perturbative diagrammatic expansion.

\section{Soft and collinear limits}

In this section we will study the factorisation properties of the amplitude in the case in which the emitted gluon or graviton is collinear or soft with respect another graviton or gauge boson or a parent massive scalar or quark. In the case of soft emissions, the historical work is presented in Ref.~\citeall{Weinberg_Thm}, while a more general result, using the BCFW recursion relations, has been shown in Ref.~\citeall{cachazo2014evidencenewsoftgraviton}.

\subsection{Soft theorems}
\label{subsec:soft_theorems}

We are ready to study the universal properties of amplitudes when one of the external massless particles becomes soft. We will mainly follow Ref.~\citeall{Bern_2014_2,Bern_2014,cachazo2014evidencenewsoftgraviton,Casali_2014}. The leading soft behaviour is dictated by the Weinberg's theorem, which basicaly states that when the momentum of one external gauge boson or graviton is taken to zero, the amplitude factorises into a universal soft factor multiplying the amplitude with the soft particle removed. In gauge theory and gravity, however, at tree level this universality extends beyond the leading pole. Gauge theory has a universal first subleading soft factor, while gravity has universal first and second subleading soft factors, as we will see soon.

Let us start with the gluonic case. We consider a colour-ordered tree amplitude in which a positive helicity gluon \(s\) is inserted between the hard legs \(a\) and \(b\),
\begin{equation}
    A_{n+1}^{\mathrm{tree}}(1,\ldots,a,s^+,b,\ldots,n) \, .
\end{equation}
Because the amplitude is colour ordered, the soft singularity is related only to the two adjacent hard legs \(a\) and \(b\). In principle one could take the usual soft limit by rescaling both spinors of the soft leg,
\begin{equation}
    \lambda_s\longrightarrow \sqrt{\delta}\,\lambda_s \, ,
    \qquad
    \widetilde\lambda_s\longrightarrow \sqrt{\delta}\,\widetilde\lambda_s \, ,
    \qquad
     p_s^{\alpha\dot\alpha}
    =\lambda_s^\alpha\widetilde\lambda_s^{\dot\alpha}
    \longrightarrow
    \delta p_s^{\alpha\dot\alpha}\, .
\end{equation}
However, for a gluon with positive helicity this limit is equivalent, up to a little group transformation, to a holomorphic soft limit. Indeed, using the L.G. scaling of an amplitude with a positive helicity gluon, one has
\begin{equation}\label{eq:Cap3:usual_vs_holomorphic_soft_gluon}
    A_{n+1}\!
    \left(\{\sqrt{\delta}\,\lambda_s,\sqrt{\delta}\,\widetilde\lambda_s,+1\},\ldots\right)
    =
    \delta\,
    A_{n+1}\!
    \left(\{\delta\lambda_s,\widetilde\lambda_s,+1\},\ldots\right) \, .
\end{equation}
Therefore it is convenient to present the subleading behaviour in terms of a holomorphic scaling of
the spinors,
\begin{equation}\label{eq:Cap3:holomorphic_soft_limit_gluon}
    \lambda_s \longrightarrow \delta\lambda_s \, ,
    \qquad
    \widetilde\lambda_s \longrightarrow \widetilde\lambda_s \, ,
    \qquad
    p_s^{\alpha\dot\alpha}
    \longrightarrow
    \delta p_s^{\alpha\dot\alpha} \, ,
\end{equation}
with \(\delta\to0\). This prescription is particularly convenient in spinor-helicity variables, because the subleading terms appear as definite powers of \(\delta\). The tree-level soft theorem for gluons reads
\begin{equation}\label{eq:Cap3:soft_gluon_theorem_spinor_helicity}
    \begin{aligned}
    A_{n+1}^{\mathrm{tree}}&(1,\ldots,a,s^+,b,\ldots,n)\\
    &\xrightarrow{\delta\to0}
    \left(
    \frac{1}{\delta^2}S_{\mathrm{YM}}^{(0)}(a,s^+,b)
    +
    \frac{1}{\delta}S_{\mathrm{YM}}^{(1)}(a,s^+,b)
    \right)
     A_n^{\mathrm{tree}}(1,\ldots,a,b,\ldots,n)
    +\mathcal{O}(\delta^0) \, .
    \end{aligned}
\end{equation}
The leading soft operator is the familiar \textbf{eikonal factor} which, written in  a spinor-helicity language, and suppressing all the gauge couplings, can be expanded as\footnote{In the following we will use the positive helicity form, since the negative helicity case follows directly by parity, according to the conventions in Appendix~\ref{app:parity_charge}.}
\begin{equation}\label{eq:Cap3:leading_soft_gluon_factor}
    S_{\mathrm{YM}}^{(0)}(a,s^+,b)
    =
    \frac{\langle ab\rangle}{\langle as\rangle\langle sb\rangle} \, .
\end{equation}
This is the spinorial version of the universal emission of a soft gluon from the two adjacent hard legs. It is independent of the helicities and particle species of the hard legs, so it's universal. The spin independence
arises from the classical nature of the soft gluon. Indeed, since its momentum vanishes, its wavelength is very
long, and it cannot resolve the details of the scattering process which has emitted it.

The first subleading correction\footnote{Strictly speaking, this is valid at tree-level, while at loop level, soft-collinear and non-factorising contributions
modify this tree-level relation, see Ref.~\citeall{Bern_2014,bonocore2015factorizationapproachnexttoleadingpowerthreshold,DELDUCA1990369}.} is a differential operator acting on the hard amplitude,
    \begin{equation}\label{eq:Cap3:subleading_soft_gluon_factor}
        S_{\mathrm{YM}}^{(1)}(a,s^+,b)
        =
        \frac{1}{\langle as\rangle}
        \widetilde\lambda_s^{\dot\alpha}
        \frac{\partial}{\partial\widetilde\lambda_a^{\dot\alpha}}
        +
        \frac{1}{\langle sb\rangle}
        \widetilde\lambda_s^{\dot\alpha}
        \frac{\partial}{\partial\widetilde\lambda_b^{\dot\alpha}} \, .
    \end{equation}

This operator is the spinor-helicity representation of the angular momentum operator. More explicitly, in spinor variables we can decompose the Lorentz generators as 
\begin{equation}\label{eq:Cap3:spinor_lorentz_generators}
    J_a^{\alpha\beta}
    =
    \lambda_a^{\alpha}\frac{\partial}{\partial\lambda_{a\beta}}
    +
    \lambda_a^{\beta}\frac{\partial}{\partial\lambda_{a\alpha}} \, ,
    \qquad
    \widetilde J_a^{\dot\alpha\dot\beta}
    =
    \widetilde\lambda_a^{\dot\alpha}
    \frac{\partial}{\partial\widetilde\lambda_{a\dot\beta}}
    +
    \widetilde\lambda_a^{\dot\beta}
    \frac{\partial}{\partial\widetilde\lambda_{a\dot\alpha}} \, .
\end{equation}
The subleading soft factor can be written in a  covariant form in terms of these generators as
\begin{equation}\label{eq:Cap3:subleading_soft_gluon_covariant}
    S_{\mathrm{YM}}^{(1)}(a,s^+,b)
    =
    \frac{k_{s,\mu}\varepsilon_{s,\nu}J_{a}^{\mu\nu}}
    {\sqrt{2}\,k_a\cdot k_s}
    -
    \frac{k_{s,\mu}\varepsilon_{s,\nu}J_{b}^{\mu\nu}}
    {\sqrt{2}\,k_b\cdot k_s} \, ,
\end{equation}

thus the differential operator in \refeq{eq:Cap3:subleading_soft_gluon_factor} is the explicit action of the angular momentum generator on the adjacent hard legs, written after choosing the holomorphic soft limit. Let us briefly explain how this form is obtained from BCFW recursion, following Ref.~\citeall{Casali_2014, Luo_2015}. Consider, for simplicity, the colour ordering
\begin{equation}
    A_{n+1}^{\mathrm{tree}}(s^+,1,\ldots,n) \, ,
\end{equation}
so that the soft leg \(s\) is adjacent to the hard legs \(1\) and \(n\). We perform the BCFW shift
\begin{equation}\label{eq:Cap3:Casali_BCFW_shift_soft_gluon}
    \lambda_s(z)=\lambda_s+z\lambda_n \, ,
    \qquad
    \widetilde\lambda_n(z)=\widetilde\lambda_n-z\widetilde\lambda_s \, .
\end{equation}
Now, after taking the holomorphic soft limit \(\lambda_s\to\delta\lambda_s\), the only singular BCFW channel is that one in which the soft gluon combines with the adjacent leg \(1\) into a three-point amplitude. Solving the on-shell condition for this pole gives a lower point amplitude in which the hard spinors change as
\begin{equation}\label{eq:Cap3:Casali_shifted_hard_amplitude}
\begin{aligned}
    \widetilde\lambda_1
    &\longrightarrow
    \widetilde\lambda_1
    +\delta\,\frac{\langle ns\rangle}{\langle n1\rangle}
    \widetilde\lambda_s \, ,
    \\
    \widetilde\lambda_n
    &\longrightarrow
    \widetilde\lambda_n
    +\delta\,\frac{\langle s1\rangle}{\langle n1\rangle}
    \widetilde\lambda_s \, .
\end{aligned}
\end{equation}
Therefore the singular part of the amplitude takes the form
\begin{equation}\label{eq:Cap3:Casali_soft_BCFW_singular_term}
\begin{aligned}
    A_{n+1}^{\mathrm{tree}}(s^+,1,\ldots,n)
    &\xrightarrow{\delta\to0}
    \frac{1}{\delta^2}
    \frac{\langle n1\rangle}{\langle ns\rangle\langle s1\rangle}
    \\
    &\quad\times
    A_n^{\mathrm{tree}}\left(
    \left\{\lambda_1,
    \widetilde\lambda_1+\delta\frac{\langle ns\rangle}{\langle n1\rangle}\widetilde\lambda_s\right\},
    \ldots,
    \left\{\lambda_n,
    \widetilde\lambda_n+\delta\frac{\langle s1\rangle}{\langle n1\rangle}\widetilde\lambda_s\right\}
    \right)
    +\mathcal{O}(\delta^0) \, .
\end{aligned}
\end{equation}
Expanding this lower point amplitude at first order in \(\delta\), one obtains
\begin{equation}\label{eq:Cap3:Casali_soft_expansion_hard_amplitude}
\begin{aligned}
    A_n^{\mathrm{tree}}\Big|_{\mathrm{shifted}}
    &
    =
    \left(
    1
    +\delta\frac{\langle ns\rangle}{\langle n1\rangle}
    \widetilde\lambda_s^{\dot\alpha}
    \frac{\partial}{\partial\widetilde\lambda_1^{\dot\alpha}}
    +\delta\frac{\langle s1\rangle}{\langle n1\rangle}
    \widetilde\lambda_s^{\dot\alpha}
    \frac{\partial}{\partial\widetilde\lambda_n^{\dot\alpha}}
    \right)
    A_n^{\mathrm{tree}}
    +\mathcal{O}(\delta^2) \, .
\end{aligned}
\end{equation}
Multiplying by the leading prefactor in \refeq{eq:Cap3:Casali_soft_BCFW_singular_term} gives
\begin{equation}
    S_{\mathrm{YM}}^{(0)}(n,s^+,1)
    =
    \frac{\langle n1\rangle}{\langle ns\rangle\langle s1\rangle} \, ,
    \qquad
    S_{\mathrm{YM}}^{(1)}(n,s^+,1)
    =
    \frac{1}{\langle s1\rangle}
    \widetilde\lambda_s^{\dot\alpha}
    \frac{\partial}{\partial\widetilde\lambda_1^{\dot\alpha}}
    +
    \frac{1}{\langle ns\rangle}
    \widetilde\lambda_s^{\dot\alpha}
    \frac{\partial}{\partial\widetilde\lambda_n^{\dot\alpha}} \, .
\end{equation}
Replacing \((n,1)\) by the two adjacent legs \((a,b)\) gives the general colour-ordered expression in \refeq{eq:Cap3:subleading_soft_gluon_factor}. The remaining BCFW channels are finite as \(\delta\to0\), and therefore they can only contribute starting at \(\mathcal{O}(\delta^0)\). This is why Yang--Mills theory has no universal sub-subleading divergent soft operators, while we will see that term in gravity.

We now turn to the gravitational case. Let us consider an \((n+1)\)-point graviton amplitude in which the soft leg is a positive helicity graviton \(s^+\). Also in this case we adopt the holomorphic soft limit
\begin{equation}\label{eq:Cap3:holomorphic_soft_limit_graviton}
    \lambda_s \longrightarrow \delta\lambda_s \, ,
    \qquad
    \widetilde\lambda_s \longrightarrow \widetilde\lambda_s \, ,
    \qquad
    k_s^{\alpha\dot\alpha}
    \longrightarrow
    \delta k_s^{\alpha\dot\alpha} \, .
\end{equation}
For a positive helicity graviton this is again related to the usual soft scaling using the L.G. transformations. Since a graviton has helicity \(+2\), the relation contains one extra power of \(\delta\) with respect to the gluon case,
\begin{equation}\label{eq:Cap3:usual_vs_holomorphic_soft_graviton}
    M_{n+1}\!
    \left(\{\sqrt{\delta}\,\lambda_s,\sqrt{\delta}\,\widetilde\lambda_s,+2\},\ldots\right)
    =
    \delta^2\,
    M_{n+1}\!
    \left(\{\delta\lambda_s,\widetilde\lambda_s,+2\},\ldots\right) \, .
\end{equation}
We will see that the holomorphic expansion leads to the gravitational soft theorem,
\begin{equation}\label{eq:Cap3:soft_graviton_theorem_spinor_helicity}
    M_{n+1}^{\mathrm{tree}}(1,\ldots,n,s^+)
    \xrightarrow{\delta\to0}
    \left(
    \frac{1}{\delta^3}S_{\mathrm{G}}^{(0)}
    +
    \frac{1}{\delta^2}S_{\mathrm{G}}^{(1)}
    +
    \frac{1}{\delta}S_{\mathrm{G}}^{(2)}
    \right)
    M_n^{\mathrm{tree}}(1,\ldots,n)
    +\mathcal{O}(\delta^0) \, .
\end{equation}

The leading soft factor is Weinberg's soft graviton factor, introduced in Ref.~\citeall{Weinberg_Thm}, which, suppressing the gravitational coupling, takes the form
\begin{equation}\label{eq:Cap3:leading_soft_graviton_factor_2}
    S_{\mathrm{G}}^{(0)}=\sum_{a=1}^{n}\frac{\varepsilon_{\mu\nu}k_a^{\mu}k_a^{\nu}}{k_s\cdot k_a} \, ,
\end{equation}
which is universal and independent of the helicities of the hard particles.

Introducing two arbitrary reference spinors \(x\) and \(y\) for the positive helicity graviton polarisation tensor, one can expand the leading soft factor as
\begin{equation}\label{eq:Cap3:leading_soft_graviton_factor}
    S_{\mathrm{G}}^{(0)}
    =
    -\sum_{a=1}^{n}
    \frac{[sa]}{\langle sa\rangle}
    \frac{\langle xa\rangle\langle ya\rangle}
    {\langle xs\rangle\langle ys\rangle} \, ,
\end{equation}
which we saw already in the Hodges' formula, \refeq{eq:Cap3:hodges_matrix_diagonal}. Remark that the dependence on the reference spinors \(x\) and \(y\) is a gauge choice and drops out of the physical amplitude by momentum conservation. Notice also that, differently from the gluonic soft factor, the sum runs over all hard legs, because there is no colour ordering in gravity.

The first subleading soft graviton operator is 
\begin{equation}\label{eq:Cap3:subleading_soft_graviton_factor}
    S_{\mathrm{G}}^{(1)}
    =
    -\frac{1}{2}
    \sum_{a=1}^{n}
    \frac{[sa]}{\langle sa\rangle}
    \left(
    \frac{\langle xa\rangle}{\langle xs\rangle}
    +
    \frac{\langle ya\rangle}{\langle ys\rangle}
    \right)
    \widetilde\lambda_s^{\dot\alpha}
    \frac{\partial}{\partial\widetilde\lambda_a^{\dot\alpha}} \, .
\end{equation}
This is the gravitational analogue of the subleading Yang--Mills operator. In covariant notation it is controlled by the total angular momentum of the hard particles,
\begin{equation}\label{eq:Cap3:subleading_soft_graviton_covariant}
    S_{\mathrm{G}}^{(1)}
    =
    \sum_{a=1}^{n}
    \frac{\varepsilon_{\mu\nu}k_a^{\mu}k_{s\rho}J_a^{\rho\nu}}
    {k_s\cdot k_a} \, ,
\end{equation}
where \(\varepsilon_{\mu\nu}\) is the polarisation tensor of the soft graviton and \(J_a^{\rho\nu}\) is the total angular momentum operator acting on the leg \(a\). Gauge invariance of this term follows from angular momentum conservation, in analogy with the way that Weinberg's leading factor is gauge invariant by using momentum conservation.

Gravity however has, at tree level, one further universal term with respect to the Yang--Mills case. In spinor-helicity variables the second subleading soft operator takes the form
\begin{equation}\label{eq:Cap3:subsubleading_soft_graviton_factor}
    S_{\mathrm{G}}^{(2)}
    =
    \frac{1}{2}
    \sum_{a=1}^{n}
    \frac{[sa]}{\langle sa\rangle}
    \widetilde\lambda_s^{\dot\alpha}
    \widetilde\lambda_s^{\dot\beta}
    \frac{\partial^2}
    {\partial\widetilde\lambda_a^{\dot\alpha}\partial\widetilde\lambda_a^{\dot\beta}} \, ,
\end{equation}
while using Lorentz indices it takes the form
\begin{equation}\label{eq:Cap3:subsubleading_soft_graviton_covariant}
    S_{\mathrm{G}}^{(2)}
    =
    \frac{1}{2}
    \sum_{a=1}^{n}
    \frac{\varepsilon_{\mu\nu}
    \left(k_{s\rho}J_a^{\rho\mu}\right)
    \left(k_{s\sigma}J_a^{\sigma\nu}\right)}
    {k_s\cdot k_a} \, .
\end{equation}
This term has no direct analogue in the divergent part of the Yang--Mills soft expansion. Notice also that, 
while $S_{\mathrm{G}}^{(0)}$ implies the universality of the gravitational coupling, $S_{\mathrm{G}}^{(1)}$ and $S_{\mathrm{G}}^{(2)}$ don't imply new physical relations because the 
little group properties are trivially satisfied.

\begin{figure}[ht]
\centering
\begin{tikzpicture}[scale=0.82, >=Latex]
    \draw[thick] (-5.20,0) circle (0.85);
    \node at (-5.20,0) {$M_{n+1}$};

    \draw[thick] (-5.80,0.58) -- (-6.52,1.02) node[left] {$1$};
    \draw[thick] (-6.04,-0.15) -- (-6.80,-0.28) node[left] {$2$};
    \draw[thick] (-4.65,-0.65) -- (-3.95,-1.10) node[right] {$n$};
    \draw[dashed, gray!65, very thick]
        (-6.45,-0.22) arc[start angle=190,end angle=310,radius=1.28];

    \draw[
        thick,
        decorate,
        decoration={snake, amplitude=1.0mm, segment length=3.0mm}
    ] (-5.20,1.65) -- (-5.20,0.85);
    \node[above] at (-5.20,1.65) {$s^+$};

    \draw[->, very thick] (-3.55,0) -- (-2.05,0);
    \node[above] at (-2.80,0.08) {$k_s\to0$};

    \node at (-1.35,0.35) {\scalebox{1.45}{$\displaystyle\sum_{a=1}^{n}$}};

    \draw[thick] (3.00,0) circle (0.95);
    \node at (3.00,0) {$M_n$};

    \draw[thick] (3.00,0.95) -- (3.00,1.55) node[above] {$1$};
    \draw[thick] (3.70,0.64) -- (4.32,1.08) node[right] {$2$};
    \draw[thick] (3.65,-0.74) -- (4.25,-1.22) node[right] {$n$};
    \draw[dashed, gray!65, very thick]
        (4.10,0.72) arc[start angle=32,end angle=-42,radius=1.26];

    \coordinate (V) at (0.50,0.45);
    \fill (V) circle (1.8pt);
    \draw[thick] (-0.35,1.05) -- (V);
    \node[above] at (-0.03,0.93) {$a$};
    \draw[thick] (V) -- (2.05,0.22);
    \draw[
        thick,
        decorate,
        decoration={snake, amplitude=1.0mm, segment length=3.0mm}
    ] (0.50,1.48) -- (V);
    \node[above] at (0.50,1.48) {$s^+$};
    \node[right] at (0.0,-0.15) {$S_{\mathrm G}^{(0,1,2)}$};

    \node at (4.85,0) {$+$};
    \node[right] at (5.18,0) {$\mathcal{O}(\delta^0)$};

    \node at (0.95,-2.35)
        {$\displaystyle
        S_{\mathrm G}^{(0,1,2)}(s,a)
        =
        \frac{S_{\mathrm G}^{(0)}(s,a)}{\delta^3}
        +
        \frac{S_{\mathrm G}^{(1)}(s,a)}{\delta^2}
        +
        \frac{S_{\mathrm G}^{(2)}(s,a)}{\delta}$};
\end{tikzpicture}
\caption{Diagrammatic form of the soft graviton factorisation. In the holomorphic soft limit the \((n+1)\)-point amplitude factorises into universal soft operators acting on the lower point hard amplitude, summed over all hard legs \(a\).}
\label{fig:Cap3:Weinberg_soft_factorisation}
\end{figure}

The proof is very similar to the previous case, however now there are important differences. First of all the amplitude is not colour-ordered, so we have to consider the contributions originating from all the intermediate hard legs. Moreover, as we said in Sec.~\ref{Cap3:sec:CKduality}, \textit{``gravity = (gauge theory)$^2$''}, so we have to account that also the non vanishing three-point amplitudes will be the ``double copy'' of the Yang--Mills ones.
Also in this case the only singular BCFW channel is the one in which the soft graviton combines with the hard leg $a$ into a three-point amplitude\footnote{In the following derivation we consider, just for simplicity, only graviton amplitudes.}, namely
\begin{equation}\label{eq:Cap3:CS_first_kind_terms}
    \sum_{a=1}^{n}\sum_{h}
    M_L\bigl(s(z^*),a,\widehat P_{s,a}^{h}\bigr)
    \frac{1}{(k_s+k_a)^2}
    M_R\bigl(-\widehat P_{s,a}^{-h},\ldots\bigr) \, .
\end{equation}
The other BCFW channels are regular in the soft limit and therefore cannot contribute to the divergent soft operators. The pole condition gives
\begin{equation}\label{eq:Cap3:CS_pole_data}
    z^*=-\frac{\langle as\rangle}{\langle an\rangle} \, ,
    \qquad
    \lambda_I=\lambda_a \, ,
    \qquad
    \widetilde\lambda_I
    =
    \widetilde\lambda_a
    +\frac{\langle ns\rangle}{\langle na\rangle}\widetilde\lambda_s \, .
\end{equation}
The three-point amplitude vanishes for \(h=h_a\), while the non vanishing contribution is given by \(h=-h_a\). Using the gravitational three-point amplitude, or equivalently the double copy of the Yang--Mills three-point amplitude, one obtains, up to the overall phases fixed by our spinor conventions,
\begin{equation}\label{eq:Cap3:CS_three_point_gravity_soft}
    M_L\bigl(s(z^*)^{+2},a,\widehat P_{s,a}^{h}\bigr)
    =
   - \frac{[sa]^2\langle na\rangle^2}
    {\langle ns\rangle^2} \, ,
\end{equation}
while the propagator can be expanded as
\begin{equation}
    \frac{1}{(k_s+k_a)^2}
    =
    \frac{1}{\langle sa\rangle[as]} \, .
\end{equation}
This gives the singular part of the amplitude, leading to 
\begin{equation}\label{eq:Cap3:CS_gravity_singular_BCFW_contribution}
\begin{aligned}
    M_{n+1}^{\mathrm{tree}}&(1,\ldots,n,s^+)
    \xrightarrow{\delta\to0}
    -\frac{1}{\delta^3}\sum_{a=1}^{n}
    \frac{[sa]}{\langle sa\rangle}
    \frac{\langle na\rangle^2}{\langle ns\rangle^2}
    \\
    &\quad\times
    M_n^{\mathrm{tree}}\Bigg(
    \ldots,
    \left\{\lambda_a,
    \widetilde\lambda_a
    +\delta\frac{\langle ns\rangle}{\langle na\rangle}\widetilde\lambda_s
    \right\},
    \ldots,
    \left\{\lambda_n,
    \widetilde\lambda_n
    +\delta\frac{\langle sa\rangle}{\langle an\rangle}\widetilde\lambda_s
    \right\}
    \Bigg)
    +\mathcal{O}(\delta^0) \, .
\end{aligned}
\end{equation}
Choosing the reference spinors in \refeq{eq:Cap3:leading_soft_graviton_factor} as \(x=y=n\), the prefactor in \refeq{eq:Cap3:CS_gravity_singular_BCFW_contribution} is exactly the contribution of the hard leg \(a\) to \(S_{\mathrm G}^{(0)}\).

The important point is that the internal state in the right subamplitude has become the hard particle \(a\), but with a shifted anti-holomorphic spinor. At the same time, momentum conservation of the lower point amplitude is guaranteed by another shift of \(\widetilde\lambda_n\). Indeed the BCFW residue is expressed as an \(n\)-point physical amplitude evaluated on shifted kinematics. Finally, expanding the lower point amplitude, we find
\begin{equation}\label{eq:Cap3:CS_gravity_Taylor_expansion}
\begin{aligned}
    M_n^{\mathrm{tree}}\Big|_{\mathrm{shifted}}
    &=
    \Bigg[
    1
    +\delta\left(
    \frac{\langle ns\rangle}{\langle na\rangle}
    \widetilde\lambda_s^{\dot\alpha}
    \frac{\partial}{\partial\widetilde\lambda_a^{\dot\alpha}}
    +
    \frac{\langle sa\rangle}{\langle an\rangle}
    \widetilde\lambda_s^{\dot\alpha}
    \frac{\partial}{\partial\widetilde\lambda_n^{\dot\alpha}}
    \right)
    \\
    &\quad+
    \frac{\delta^2}{2}
    \left(
    \frac{\langle ns\rangle}{\langle na\rangle}
    \widetilde\lambda_s^{\dot\alpha}
    \frac{\partial}{\partial\widetilde\lambda_a^{\dot\alpha}}
    +
    \frac{\langle sa\rangle}{\langle an\rangle}
    \widetilde\lambda_s^{\dot\alpha}
    \frac{\partial}{\partial\widetilde\lambda_n^{\dot\alpha}}
    \right)^2
    +\mathcal{O}(\delta^3)
    \Bigg]
    M_n^{\mathrm{tree}} \, .
\end{aligned}
\end{equation}
Since the gravitational prefactor in \refeq{eq:Cap3:CS_gravity_singular_BCFW_contribution} scales as \(\delta^{-3}\), the first three terms of this Taylor expansion are divergent, indeed they give  the leading, subleading and sub-subleading soft operators. Using the identities satisfied by stripped (off of the momentum conserving locus) amplitudes, reported in Ref.~\citeall{cachazo2014evidencenewsoftgraviton}, together with momentum and angular momentum conservation, the result can be rewritten in the symmetric form of \refeq{eq:Cap3:subleading_soft_graviton_factor} and \refeq{eq:Cap3:subsubleading_soft_graviton_factor}, in which the sum runs over all hard legs.

This also clarifies the difference from Yang--Mills theory. In the gluonic case the corresponding soft prefactor scales as \(\delta^{-2}\), so only the zeroth and first Taylor coefficients are divergent and universal. In gravity the three-point amplitude is effectively the square of the Yang--Mills one, giving a stronger \(\delta^{-3}\) singularity. Hence the quadratic Taylor coefficient remains divergent and produces the universal second subleading soft graviton operator.

\subsection{Collinear limit and splitting functions}
\label{Cap3:subsec:collinear_limit}

Next, we consider the limit in which two partons become collinear. The historical work on collinear factorisation in perturbative QCD is due to Altarelli and Parisi, Ref.~\citeall{ALTARELLI1977298}. For a modern approach in spinor-helicity variables we will follow mainly Ref.~\citeall{Bern:1994zx,Del_Duca_2020,Dixon_1,Henn:2014yza}.

In order to parametrise the approach to the collinear region, it is useful to introduce a light-cone, or Sudakov, decomposition for a massless momentum,
\begin{equation}\label{eq:Cap3:Sudakov}
    p_i^\mu
    =
    z_i\widetilde p^{\,\mu}
    +k_{i\perp}^\mu
    -
    \frac{k_{i\perp}^2}{2z_i\widetilde p\cdot n}\,n^\mu \, .
\end{equation}
Here \(\widetilde p^\mu\) and \(n^\mu\) are two light-like reference momenta, \(\widetilde p^2=n^2=0\), and $\widetilde{p}^{\, \mu}$ is aligned with the beam axis, while \(k_{i\perp}^\mu\) is transverse to both of them,
\begin{equation}
    k_{i\perp}\cdot \widetilde p=0 \, ,
    \qquad
    k_{i\perp}\cdot n=0 \, .
\end{equation}
The last term in \refeq{eq:Cap3:Sudakov} is fixed by the on-shell condition \(p_i^2=0\). In this parametrisation the collinear limit corresponds to sending the relative transverse momentum between two collinear particles to zero.

If we focus on two adjacent massless partons, labelled by \(a\) and \(b\), whose momenta become parallel, we can write in the strict collinear limit
\begin{equation}\label{eq:Cap3:collinear_momenta_partons}
    p_a \longrightarrow z P \, ,
    \qquad
    p_b \longrightarrow (1-z)P \, ,
    \qquad
    P^2=0 \, ,
\end{equation}
with \(0<z<1\). In spinor-helicity variables this implies, at leading order,
\begin{equation}\label{eq:Cap3:collinear_spinors_partons}
    |a\rangle \longrightarrow \sqrt z\,|P\rangle \, ,
    \qquad
    |b\rangle \longrightarrow \sqrt{1-z}\,|P\rangle \, ,
    \qquad
    |a] \longrightarrow \sqrt z\,|P] \, ,
    \qquad
    |b] \longrightarrow \sqrt{1-z}\,|P] \, .
\end{equation}
Therefore \(\langle ab\rangle\to0\) and \([ab]\to0\). More precisely, the invariant mass of the collinear pair is
\begin{equation}\label{eq:Cap3:collinear_invariant_mass}
    s_{ab}=(p_a+p_b)^2=\langle ab\rangle[ba]\longrightarrow0 \, .
\end{equation}
This is the origin of the collinear singularities of gauge-theory amplitudes.

At tree level, the singular behaviour is universal. For a colour-ordered amplitude, only adjacent partons in the colour decomposition can produce a collinear pole, as we can see by \refeq{eq:Cap3:parke_taylor_formula}. One has
\begin{equation}\label{eq:Cap3:partonic_collinear_factorisation}
    A_{n+1}^{\mathrm{tree}}(\ldots,a^{h_a},b^{h_b},\ldots)
    \xrightarrow{a\parallel b}
    \sum_{h=\pm}
    A_{n}^{\mathrm{tree}}(\ldots,P^h,\ldots)
    \mathrm{Split}^{P\to ab}_{-h}
    \left(z,a^{h_a},b^{h_b}\right) \, .
\end{equation}

\begin{figure}[ht]
    \centering
    \includegraphics[width=1.00\textwidth]{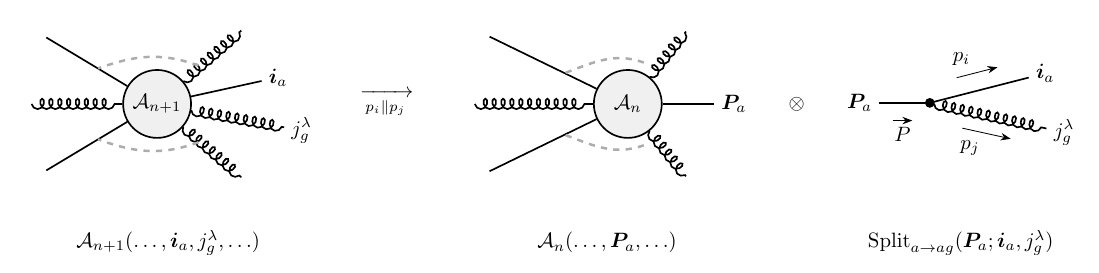}
    \caption{Diagrammatic representation of the universal collinear factorisation of a colour-ordered QCD amplitude into a lower point hard amplitude and a splitting amplitude.}
    \label{fig:Cap3:parton_gluon_splitting_factorisation}
\end{figure}

A schematic representation of this collinear factorisation is depicted in fig.~\ref{fig:Cap3:parton_gluon_splitting_factorisation}.

The label \(-h\) on the splitting amplitude follows the outgoing convention. Indeed the lower point amplitude contains a parent parton of helicity \(h\), while the three-point splitting amplitude contains the corresponding outgoing momentum \(-P\). The important point is that the splitting functions are universal. They depend only on the species and helicities of the two collinear partons and on the momentum fraction \(z\), but not on the remaining hard process.

Equivalently, the splitting amplitudes can be seen as the singular part of the three-point amplitude attached to the nearly on-shell propagator,
\begin{equation}\label{eq:Cap3:splitting_from_three_point}
    \mathrm{Split}^{P\to ab}_{-h}
    \left(z,a^{h_a},b^{h_b}\right)
    \sim
    A_3^{\mathrm{tree}}\left(a^{h_a},b^{h_b},(-P)^{-h}\right)
    \frac{i}{s_{ab}} \, .
\end{equation}
This formula explains why the splitting functions are singular when \(s_{ab}\to0\), and why their explicit spinor form is fixed by the three-point amplitudes.

Let's list now the gluon splitting amplitudes. Up to the overall gauge coupling, and choosing the holomorphic collinear branch, the independent non vanishing ones are given by
\begin{equation}\label{eq:Cap3:gluon_splitting_functions}
\begin{aligned}
    \mathrm{Split}^{g\to gg}_{-}(z,a^+,b^+)
    &=
    \frac{1}{\sqrt{z(1-z)}\,\langle ab\rangle} \, ,
    \\
    \mathrm{Split}^{g\to gg}_{+}(z,a^-,b^+)
    &=
    \frac{z^2}{\sqrt{z(1-z)}\,\langle ab\rangle} \, ,
    \\
    \mathrm{Split}^{g\to gg}_{+}(z,a^+,b^-)
    &=
    \frac{(1-z)^2}{\sqrt{z(1-z)}\,\langle ab\rangle} \, .
\end{aligned}
\end{equation}
Moreover,
\begin{equation}\label{eq:Cap3:vanishing_gluon_split}
    \mathrm{Split}^{g\to gg}_{+}(z,a^+,b^+)=0 \, ,
\end{equation}
which can be understood by a little group, or helicity, selection rule. 
The remaining helicity configurations are obtained by parity, namely by exchanging angle and square brackets and flipping all helicities, as explained in Appendix~\ref{app:parity_charge}. These expressions show explicitly that gluon amplitudes have a collinear singularity since the splitting functions contain a pole in \(\langle ab\rangle\).

The same factorisation applies when one or both of the collinear partons are quarks. In this case the possible lower point amplitudes are obtained by merging the two collinear particles according to the allowed QCD splittings,
\begin{equation}
    g\to gg \, ,
    \qquad
    g\to q\bar q \, ,
    \qquad
    q\to qg \, ,
    \qquad
    \bar q\to g\bar q \, .
\end{equation}
For example, for a quark and a gluon becoming collinear one finds\footnote{An exhaustive list of QCD splitting amplitudes is presented in Ref.~\citeall{Del_Duca_2000}.}
\begin{equation}\label{eq:Cap3:qg_splitting_functions}
\begin{aligned}
    \mathrm{Split}^{q\to qg}_{-}(z,q^+,g^+)
    &=
    \frac{1}{\sqrt{1-z}\,\langle qg\rangle} \, ,
    &
    \mathrm{Split}^{q\to qg}_{-}(z,q^+,g^-)
    &=
    -\frac{z}{\sqrt{1-z}\,[qg]} \, ,
    \\
    \mathrm{Split}^{\bar q\to g\bar q}_{-}(z,g^+,\bar q^{\, +})
    &=
    \frac{1}{\sqrt z\,\langle g\bar q\rangle} \, ,
    &
    \mathrm{Split}^{\bar q\to g\bar q}_{-}(z,g^-,\bar q^{\, +})
    &=
    -\frac{1-z}{\sqrt z\,[g\bar q]} \, .
\end{aligned}
\end{equation}
Notice that, unlike the soft factor, the splitting amplitude does depend on the particle species and on the helicities of the collinear partons. This is not in contradiction with universality since universality means that, once the collinear channel is fixed, the same splitting amplitude multiplies any hard process.

Squaring the factorisation formula and summing over the helicities of the collinear partons and of the parent parton gives the ordinary Altarelli--Parisi splitting kernels. For instance, from the two quark-gluon splitting amplitudes in \refeq{eq:Cap3:qg_splitting_functions}, one reconstructs the characteristic dependence for the unpolarised $P_{qq}(z)$ splitting
function,
\begin{equation}\label{eq:Cap3:Pqq_from_splitting_amplitudes}
    P_{qq}(z)
    \propto
    \left(\frac{1}{\sqrt{1-z}}\right)^2
    +
    \left(\frac{z}{\sqrt{1-z}}\right)^2
    =
    \frac{1+z^2}{1-z} \, ,
\end{equation}
up to the colour factor, the standard plus-prescription, and the virtual contribution proportional to \(\delta(1-z)\).

In the following we will first use these universal massless splitting amplitudes as a starter pack. In the next chapters we will study the collinear factorisation properties relevant for the gravitational amplitudes and for scalar--graviton emissions, where we will adapt the same logic to the corresponding kinematics, including the massive scalar case.

\chapter{Regge limit and BFKL equation}
\label{Cap4:Regge_limit_and_BFKL}

In this chapter we study the Regge, or high-energy, limit of scattering amplitudes in gauge theory and gravity. We begin with QCD, where amplitudes at fixed momentum transfer and large energy of the centre of mass factorise into universal building blocks in the multi-Regge kinematics (MRK). These are the impact factors (I.F.), which we denote by \(C\), and the central emission vertices (CEV), denoted by \(V\), which describe real emissions along the ladder. This factorisation leads to the Balitsky--Fadin--Kuraev--Lipatov (BFKL) equation, which is nothing but an evolution equation in rapidity which resums the logarithms generated by large rapidity intervals.

After introducing the QCD construction, we will discuss the extension to its double copy framework, the gravitational amplitudes. In this case the absence of colour and the double copy structure lead to some important differences, however the factorisation into impact factors and central emission vertices still holds in the multi-Regge kinematics regime. We will also emphasise the classical limit, which is the regime relevant for extracting long distance observables from scattering amplitudes, particularly useful for gravitational waves' physics.

\section{High-energy factorisation in QCD}
\label{sec:Cap4:high_energy_factorisation_QCD}

In this chapter we will introduce the Regge, or high-energy, limit in the QCD case. We will follow mainly the work presented in Ref.~\citeall{delduca1995introductionperturbativeqcdpomeron,delduca1996nexttoleadingcorrectionsbfklequation,Falcioni_2022,Ioffe:2010zz}, while the historical works related to high-energy factorisation in particle physics are introduced in Ref~\citeall{Kuraev:1976ge,Lipatov:1976zz}. 

For a generic partonic scattering process this limit is defined by
\begin{equation}\label{eq:Cap4:Regge_limit_definition}
    s \gg -t \, ,
    \qquad
    t \; \text{fixed} \, ,
\end{equation}
where \(\sqrt s\) is the energy of the partons in the centre of mass' frame and \(\sqrt{|t|}\) is the momentum transfer. In this regime perturbation theory cannot be organised by just power series of the coupling constant, because every additional emission, as we will see, can generate a large logarithm of the energy,
\begin{equation}
    L \equiv \log\left(\frac{s}{-t}\right) \, .
\end{equation}
The goal of the BFKL approach is precisely to resum the towers of high-energy logarithms which appear in this limit. Before deriving the BFKL equation, however, it is useful to understand the underlying tree-level factorisation structure.

Let's consider an elastic partonic scattering process,
\begin{equation}\label{eq:Cap4:two_to_two_partonic_scattering}
    i(p_1)+j(p_2)\longrightarrow i(p_3)+j(p_4) \, ,
\end{equation}
where \(i\) and \(j\) may be quarks or gluons. At leading power in the Regge limit, the dominant contribution is given by the exchange of a gluon in the crossed channel. In this limit the scattering is helicity conserving along each external line and the tree-level amplitude takes, after having suppressed the gauge coupling, the schematic form
\begin{equation}\label{eq:Cap4:tree_regge_two_to_two_factorisation}
    \mathcal A_{ij\to ij}^{\mathrm{tree}}
    \simeq
    2s\,
    C_i(p_1,p_3)
    \frac{1}{t}
    C_j(p_2,p_4) \, .
\end{equation}
The \textbf{impact factors} $C_i$ and \(C_j\) describe in an effective way the coupling of the external hard partons to the exchanged \(t\)-channel gluon, and contain the dependence on the particle species and colour representation of the external particles. 

The central propagator carries the momentum transfer, so in the physical momentum space we have
\begin{equation}
    q=p_1-p_3=p_4-p_2 \, ,
    \qquad
    q^2=t\simeq -|q_\perp|^2 \, .
\end{equation}
Thus, already at four points, the high-energy amplitude separates into two external impact factors connected by a transverse, or \textbf{Glauber}, exchange.

The terminology \emph{Glauber} refers to the momentum scaling of the exchanged particle. In the centre of mass frame of the two energetic incoming partons, the momentum exchanged in the \(t\)-channel has small light cone components compared with the large external energies, but has a finite transverse component,
\begin{equation}\label{eq:Cap4:Glauber_scaling}
    q^\mu \sim (q^+,q^-,q_\perp)
    \sim
    \left(\frac{|q_\perp|^2}{\sqrt{s}},\frac{|q_\perp|^2}{\sqrt{s}},q_\perp\right) \, .
\end{equation}
Consequently \(q^2\simeq -|q_\perp|^2\). This is the mode which mediates the transverse momentum exchange between particles separated by a large rapidity interval, and this still holds also for higher point amplitudes. In the following we will work in light cone coordinates, assuming that
the beam axis is aligned with the z-axis,
\begin{equation}
    p^{\pm}=p^0 \pm p^z \, ,
\end{equation}
while for the transverse part a useful convention is obtained by rotating the system such that, as in \refeq{eq:Cap2:momenta_conventions},
\begin{equation}
    p\p = p_x +ip_y\, ,  \qquad p^*\p = p_x -ip_y \, .
\end{equation}

The ladder structure of high-energy amplitudes can be understood as a sequence of such Glauber exchanges, separated by real emissions with a strong rapidity ordering.

The relevant multi-particle kinematics is the \textbf{multi-Regge kinematics}, or MRK. For the production of \(n\) additional gluons in a partonic scattering process,
\begin{equation}
    p_A+p_B\longrightarrow p_0+p_1+\cdots+p_n+p_{n+1} \, ,
\end{equation}
where MRK is defined by that strong ordering in rapidity together with comparable transverse momenta,
\begin{equation}\label{eq:Cap4:MRK_definition}
    y_0 \gg y_1 \gg \cdots \gg y_n \gg y_{n+1} \, ,
    \qquad
    |p_{0\perp}|\simeq |p_{1\perp}|\simeq \cdots \simeq |p_{n+1,\perp}| \, .
\end{equation}
Equivalently, we can say that the neighbouring invariant masses are large because the rapidity separations are large, while the transverse scales remain fixed. It is convenient to introduce the momenta exchanged in the \(t\)-channel, again in the physical momentum space,
\begin{equation}\label{eq:Cap4:t_channel_momenta_MRK}
    q_1=p_A-p_0 \, ,
    \qquad
    q_{r+1}=q_r-p_r \, ,
    \qquad
    t_r=q_r^2\simeq -|q_{r\perp}|^2 \, .
\end{equation}
In this notation the emitted gluon \(p_r\) is produced between two adjacent exchanged momenta, \(q_r\) and \(q_{r+1}\).

\begin{figure}[t]
    \centering
    \subfigure[Multigluon amplitude in MRK.\label{fig:Cap4:MRK_ladder}]{
        \includegraphics[width=0.36\textwidth]{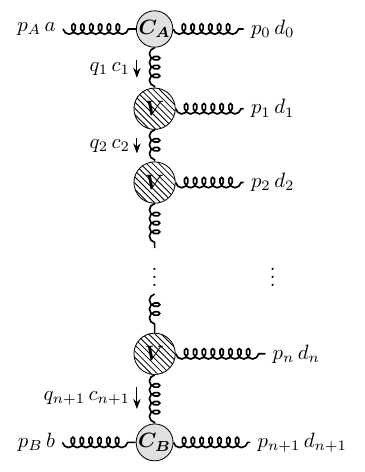}
    }
    \hfill
    \subfigure[Partonic interaction with gluon emission in MRK.\label{fig:Cap4:MRK_ladder_alternative}]{
        \includegraphics[width=0.36\textwidth]{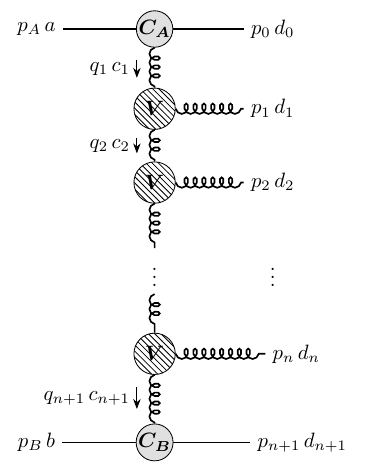}
    }
    \caption{Schematic representations of the Fadin--Kuraev--Lipatov ladder in multi-Regge kinematics. The impact factors are denoted by \(C_A\) and \(C_B\), while the central emission vertices are denoted by \(V\). Each parton or gluon is labelled by its momentum and its colour structure.}
    \label{fig:Cap4:MRK_ladders}
\end{figure}

In this limit the tree-level amplitude in MRK factorises into universal building blocks. Schematically, the ladder factorises in the the Fadin--Kuraev--Lipatov form,
\begin{equation}\label{eq:Cap4:FKL_tree_factorisation_schematic}
\begin{aligned}
    \mathcal A_{2\to n+2}^{\mathrm{tree}}
    &\simeq
    2s\,
    C_A(p_A,p_0)
    \frac{1}{t_1}
    \prod_{r=1}^{n}
    \left[
    V(q_r,p_r,q_{r+1})
    \frac{1}{t_{r+1}}
    \right]
    C_B(p_B,p_{n+1}) \, .
\end{aligned}
\end{equation}
Diagrammatic representations of this effective factorisation are shown in fig.~\ref{fig:Cap4:MRK_ladders}. In this convention the impact factors are denoted by \(C_A\) and \(C_B\), while the Lipatov currents, or \textbf{central emission vertices}, are denoted by \(V\), which describe the effective emission of an on-shell gluon from two off-shell \(t\)-channel Glauber gluons.

Unlike the impact factors, which are essentially phase factors and depend on the nature of the external particles, the Lipatov current is universal, indeed it is the same for any hard process in which the same high-energy ladder is embedded.

In a covariant form the Lipatov current can be written as
\begin{equation}\label{eq:Cap4:Lipatov_current_covariant}
\begin{aligned}
    V^\mu(q_r,q_{r+1})
    &={}
    (q_r+q_{r+1})_\perp^\mu
    +p_A^\mu
    \left(
    \frac{q_r^2}{p_r\cdot p_A}
    +
    \frac{p_r\cdot p_B}{p_A\cdot p_B}
    \right)
    -p_B^\mu
    \left(
    \frac{q_{r+1}^2}{p_r\cdot p_B}
    +
    \frac{p_r\cdot p_A}{p_A\cdot p_B}
    \right) \, ,
\end{aligned}
\end{equation}
up to conventions for the direction and for the signs of the exchanged momenta. This current, being related to a on-shell emitted gluon, must satisfy a gauge invariance,
\begin{equation}
    p_{r\mu}V^\mu(q_r,q_{r+1})=0 \, ,
    \qquad
    p_r=q_r-q_{r+1} \, .
\end{equation}
After contraction with a physical polarisation vector, the current gives the helicity dependent central emission vertex,
\begin{equation}\label{eq:Cap4:CEV_definition}
    V^{h}(q_r,p_r,q_{r+1})
    =
    V^\mu(q_r,q_{r+1})\,\varepsilon^h_\mu(p_r) \, ,
\end{equation}
For a gluon of positive helicity, in our complex transverse momentum convention, the CEV can be written as
\begin{equation}\label{eq:Cap4:gluon_CEV_positive_helicity}
    \vcenter{\hbox{\includegraphics[width=0.30\textwidth]{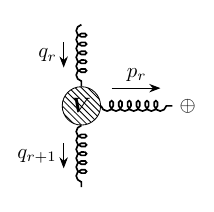}}}
    =
    V(q_r,p_r^{+},q_{r+1})
    =
    \frac{q_{r\perp}^{*}\,q_{r+1\perp}}{p_r} \, , 
\end{equation}
while the negative helicity CEV is obtained by parity, namely
\begin{equation}\label{eq:Cap4:gluon_CEV_parity}
    \left[V(q_r,p_r^{h},q_{r+1})\right]^{*}
    =
    V(q_r,p_r^{-h},q_{r+1}) \, .
\end{equation}
So, in the helicity formalism, this object takes a particularly compact form in terms of the complex transverse momenta, and this is one of the reasons why in the high-energy limit is convenient to adopt the spinor-helicity formalism described in the previous chapters.

The effective factorisation formula in \refeq{eq:Cap4:FKL_tree_factorisation_schematic} has a simple physical interpretation. The impact factors \(C_A\) and \(C_B\) create the two outgoing partons and couple them to the exchanged Glauber gluons. The Lipatov currents \(V\) insert real gluons along the ladder. Remark also that the propagators between neighbouring emissions are dominated by Glauber momenta, so they effectively depend only on transverse scales. The large energy dependence is instead carried by the rapidity intervals between successive emissions.

At loop level, virtual corrections dress the exchanged gluon and lead to \textbf{gluon Reggeisation}. This means that the elementary \(t\)-channel gluon exchange is promoted to the exchange of a Reggeised gluon, or Reggeon. This implies that the spin one gluon pole is replaced by a Regge pole whose position in the complex angular momentum plane is shifted by radiative corrections. The overall effect is that the tree-level amplitude for the process in \refeq{eq:Cap4:two_to_two_partonic_scattering} is multiplied by an energy dependent factor,
\begin{equation}\label{eq:Cap4:LL_gluon_reggeisation}
    \mathcal A_{ij\to ij}^{\mathrm{LL}}
    =
    \left(\frac{s}{-t}\right)^{C_{\mathrm A}\alpha_g(t,\mu^2)}
    \mathcal A_{ij\to ij}^{\mathrm{tree}} \, ,
\end{equation}
where \(C_{\mathrm A}\) is the quadratic Casimir in the adjoint representation, while \(\alpha_g(t,\mu^2)\) is the gluon Regge trajectory.

For an \(n\)-point amplitude in MRK, the same procedure can be applied on each \(t\)-channel exchange by dressing each gluon propagator with the corresponding Regge trajectory. Schematically, introducing a Regge transverse scale $\tau$, the Reggeised ladder takes the form
\begin{equation}\label{eq:Cap4:LL_reggeised_FKL_ladder}
\begin{aligned}
    &C_A\,
    \frac{1}{t_1}
    \left(\frac{s_1}{\tau}\right)^{C_{\mathrm A}\alpha_g(t_1)}
    V_1\,
    \frac{1}{t_2}
    \left(\frac{s_2}{\tau}\right)^{C_{\mathrm A}\alpha_g(t_2)}
    V_2
    \cdots
    V_n\,
    \frac{1}{t_{n+1}}
    \left(\frac{s_{n+1}}{\tau}\right)^{C_{\mathrm A}\alpha_g(t_{n+1})}
    C_B \, ,
\end{aligned}
\end{equation}

and hence the large logarithms generated by virtual corrections exponentiate into powers of \(s/(-t)\). 
In this sense the Regge trajectory plays the role of an anomalous dimension for the rapidity evolution of a single Reggeised gluon propagating in the \(t\)-channel. The exchange of one Reggeised gluon corresponds to an isolated Regge pole in the complex (global) angular momentum plane and gives the Reggeised form of the amplitude. However, if we are interested to go beyond the leading logarithm, the amplitude could receive contributions also from the exchange of two or more Reggeised gluons. These contributions cannot be absorbed into a single Regge trajectory and are associated with Regge branch cuts in the complex angular momentum plane rather than single poles\footnote{A more exhaustive description on the structure of the QCD amplitudes beyond the leading log is described in Ref~\citeall{Falcioni_2022}.}.

The MRK configuration is the starting point for the leading logarithmic BFKL equation. However, if we want to go beyond leading logarithmic accuracy, we have also to consider corrections to that strict multi-Regge kinematics. The real next-to-leading corrections arise when two emitted partons are not strongly ordered in rapidity.
The amplitude still factorises, but one of the elementary MRK building blocks is replaced by higher point effective vertices.
There are two qualitatively different possibilities. If the two partons close in rapidity are produced near one end of the ladder, one speaks of a forward-rapidity region. These corrections modify the impact factors,
\begin{equation}\label{eq:Cap4:NMRK_forward_replacement}
    C_A
    \quad\longrightarrow\quad
    C_A^{(2)} \, .
\end{equation}
An inclusive list, up to conventions,
of leading and next-to-leading impact factors for both gluons and quarks is presented in Ref.~\citeall{Del_Duca_2000}.

If instead the two gluons close in rapidity are produced in the middle of the ladder, one speaks of a central-rapidity region. These corrections modify the central emission vertices and enter the real NLL kernel of the BFKL equation, as we will see.

We will refer to these configurations as \textbf{next-to-multi-Regge kinematics}, or NMRK. In NMRK the strong ordering \refeq{eq:Cap4:MRK_definition} is relaxed for one pair of final-state particles. For instance, one may have
\begin{equation}\label{eq:Cap4:NMRK_definition}
    y_0 \gg \cdots \gg y_{r-1}
    \gg
    y_r \simeq y_{r+1}
    \gg
    y_{r+2}\gg\cdots\gg y_{n+1} \, ,
\end{equation}
while all transverse momenta remain of the same order. In the central region the replacement affects the central emission vertex,
\begin{equation}\label{eq:Cap4:NMRK_central_replacement}
    V(q_r,p_r,q_{r+1})
    \quad\longrightarrow\quad
    V(q_r,p_r,p_{r+1},q_{r+2}) \, .
\end{equation}
Up to conventions, and following Ref.~\citeall{Byrne_2022}, the two independent \textbf{two-gluon central emission vertices} can be written as
\begin{equation}\label{eq:Cap4:two_gluon_CEV_pp}
\begin{aligned}
    \vcenter{\hbox{\includegraphics[width=0.22\textwidth]{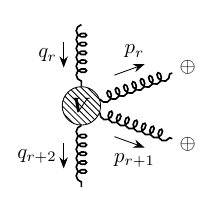}}}
    &={}
    V(q_r,p_r^{+},p_{r+1}^{+},q_{r+2})
    =
    \frac{q_{r\perp}^{*}\,q_{r+2\perp}}{p_{r\perp}}
    \sqrt{\frac{z_r}{z_{r+1}}}
    \frac{1}{\langle r\, (r+1)\rangle} \, ,
\end{aligned}
\end{equation}
\begin{equation}\label{eq:Cap4:two_gluon_CEV_pm}
\begin{aligned}
    &\vcenter{\hbox{\includegraphics[width=0.22\textwidth]{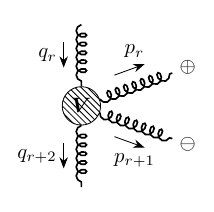}}}
    =
    V(q_r,p_r^{+},p_{r+1}^{-},q_{r+2})
    \\
    &\qquad =
    -\frac{p_{r\perp}^{*}}{p_{r\perp}}
    \Bigg\{
    -\frac{1}{s_{r,r+1}}
    \left[
    \frac{y_{r+1} p_{r+1\perp}|q_{r\perp}|^2}{p_{r+1\perp}^{*}}
    +
    \frac{z_r p_{r\perp}|q_{r+2\perp}|^2}{p_{r\perp}^{*}}
    +
    \frac{s_{r-2,r-1,r}\, p_{r\perp}p_{r+1\perp}}{p_r^-p_{r+1}^+}
    \right]
    \\
    &\qquad\quad
    +
    \frac{(q_{r+2\perp}+p_{r+1\perp})^2}{s_{r-2,r-1,r}}
    -
    \frac{q_{r+2\perp}+p_{r+1\perp}}{s_{r,r+1}}
    \left[
    \frac{p_{r\perp}}{y_r}
    -
    \frac{p_{r+1\perp}}{z_{r+1}}
    \right]
    \Bigg\} \, ,
\end{aligned}
\end{equation}
where we have used the longitudinal momentum fractions
\begin{equation}\label{eq:Cap4:NMRK_momentum_fractions}
    z_i=
    \frac{p_i^+}{p_r^+ + p_{r+1}^+}
    \, ,
    \qquad
    y_i=
    \frac{p_i^-}{p_r^- + p_{r+1}^-}
    \, ,
    \qquad
    i=r,r+1 \, ,
\end{equation}
and \(s_{ijk}\) denotes the multiparticle Mandelstam invariant, which in our all outgoing momentum convention is defined as\footnote{A schematic description of these kinematic configurations is given in the Appendix~\ref{app:kinematic_regions}.}
\begin{equation}
    s_{ijk}=(p_i+p_j+p_k)^2 \, .
\end{equation}

Note that both the impact factors and the central emission vertices (\refeq{eq:Cap4:two_gluon_CEV_pp} and \refeq{eq:Cap4:two_gluon_CEV_pm}) transform under parity into their complex conjugates,
\begin{equation}
    \left[V(q_r,p_r^{h_r},p_{r+1}^{h_{r+1}},q_{r+2})\right]^*=    V(q_r,p_r^{-h_r},p_{r+1}^{-h_{r+1}},q_{r+2}) \, .
\end{equation}
As a consistency check, the ordinary MRK factorisation is recovered when the two centrally emitted gluons are again strongly ordered in rapidity. To see this, let us introduce the ratio
\begin{equation}
    X_r = \frac{p_r^+}{p_{r+1}^+} \, .
\end{equation}
In the central NMRK region this quantity is kept finite, while the MRK limit corresponding to the ordering \(y_r\gg y_{r+1}\) is obtained by taking \(X_r\to\infty\). In this limit the two-gluon central emission vertex factorises into two ordinary Lipatov vertices connected by an intermediate \(t\)-channel propagator,
\begin{equation}
    V(q_r,p_r^{+},p_{r+1}^{+},q_{r+2})
    \xrightarrow{X_r\to\infty}
    V(q_r,p_r^{+},q_{r+1})
    \frac{1}{t_{r+1}}
    V(q_{r+1},p_{r+1}^{+},q_{r+2}) \, .
\end{equation}
This reproduces the Fadin--Kuraev--Lipatov factorisation in \refeq{eq:Cap4:FKL_tree_factorisation_schematic}. Diagrammatically, this corresponds exactly to the limit
\begin{equation}
    \vcenter{\hbox{\includegraphics[width=0.24\textwidth]{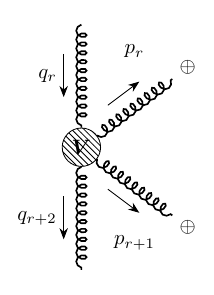}}}
    \xrightarrow{X_r\to\infty}
    \vcenter{\hbox{\includegraphics[width=0.24\textwidth]{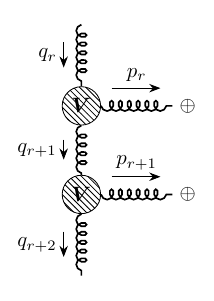}}} \, .
\end{equation}

A further consistency check is provided by the soft limit of either gluon in the tree-level two-gluon CEV. As discussed in Sec~\ref{subsec:soft_theorems}, a colour-ordered tree amplitude factorises universally when a positive helicity gluon of momentum \(p_s\) becomes soft,
\begin{equation}
    \mathcal A_{n+1}^{\mathrm{tree}}(\ldots,a,s^{+},b,\ldots)
    \xrightarrow{p_s\to 0}
    \frac{\langle a b\rangle}{\langle a s\rangle\langle s b\rangle}
    \mathcal A_n^{\mathrm{tree}}(\ldots,a,b,\ldots) \, .
\end{equation}
Accordingly, when \(p_r\to0\), the two-gluon CEV factorises into the NMRK limit of the same universal eikonal factor and an ordinary one-gluon CEV,
\begin{equation}\label{eq:Cap4:two_gluon_CEV_soft_limit}
    V(q_r,p_r^{h_r},p_{r+1}^{h_{r+1}},q_{r+2})
    \xrightarrow{p_r\to0}
    \operatorname{Soft}^{(0)}_{\mathrm A}(p_r^{h_r},p_{r+1})
    V(q_r,p_{r+1}^{h_{r+1}},q_{r+2}) \, .
\end{equation}
For a positive helicity soft gluon, we can expand the eikonal factor~(\ref{eq:Cap3:leading_soft_gluon_factor}) as
\begin{equation}
    \operatorname{Soft}^{(0)}_{\mathrm A}(p_r^{+},p_{r+1})
    =
    \frac{p_{r+1\perp}}{p_{r\perp}}
    \sqrt{\frac{p_r^+}{p_{r+1}^+}}
    \frac{1}{\langle r\,(r+1)\rangle} \, ,
\end{equation}
while the negative helicity factor follows immediately by complex conjugation. This provides direct agreement with the leading universal soft theorem derived in the previous chapter. Notice, however, that this check concerns only the leading soft singularity. The subleading Yang--Mills soft contribution is generated by an angular-momentum differential operator acting on the complete lower-point amplitude and therefore does not, in general, reduce to a multiplicative factor acting solely on the one-gluon central emission vertex.

Now, having established the factorised structure of high-energy amplitudes in QCD, we now turn to its resummation. In the next section we will show how the iteration of real emissions and virtual Regge-trajectory corrections leads to the BFKL equation, whose kernel governs the rapidity evolution of the ladder.

\section{The BFKL equation}
\label{sec:Cap4:BFKL_kernel}

The goal of the BFKL equation is to resum, to all orders in perturbation theory, the leading radiative corrections going as
\begin{equation}
    \alpha_s^n
    \log^n\!\left(\frac{s}{-t}\right) \, ,
\end{equation}
which become of order one in the Regge limit even if \(\alpha_s\ll1\). The BFKL equation is therefore an evolution equation in the rapidity interval
\begin{equation}\label{eq:Cap4:BFKL_rapidity_interval}
    Y = \log\!\left(\frac{s}{s_0}\right)\simeq L \equiv
    \log\!\left(\frac{s}{-t}\right) \, ,
\end{equation}
where \(s_0\) is another Regge factorisation scale of the order of the transverse scales. Its kernel describes how the state exchanged in the \(t\)-channel changes when the rapidity interval changes. In the following we will mainly follow Ref.~\citeall{Ioffe:2010zz,Kuraev:1976ge,Lipatov:1976zz, raj2025qcdgravitydoublecopyregge}.

\begin{figure}[t]
    \centering
    \includegraphics[width=0.78\textwidth]{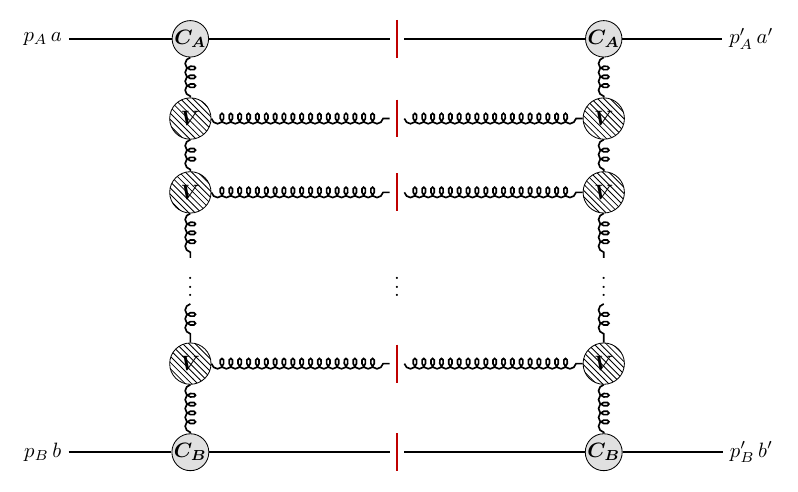}
    \caption{Schematic \(s\)-channel unitarity cut of two FKL amplitudes in MRK. The impact factors \(C_A\) and \(C_B\) couple the external particles to the ladder, while each pair of central emission vertices \(V\), one on either side of the cut, produces one real emission in the BFKL kernel. The Reggeised propagators between successive emissions contain the virtual corrections generated by the gluon Regge trajectory.}
    \label{fig:Cap4:BFKL_MRK_kernel}
\end{figure}

The equation is obtained most directly from unitarity in the \(s\)-channel. For the elastic process \(A+B\to A+B\), the discontinuity of the amplitude is given schematically by
\begin{equation}\label{eq:Cap4:s_channel_unitarity_BFKL}
    -i\,\operatorname{Disc}_s\mathcal A_{AB\to AB}
    =
    2\,\operatorname{Im}_s\mathcal A_{AB\to AB}
    =
    \sum_X
    \int d\Phi_X\,
    \mathcal A_{AB\to X}
    \mathcal A_{AB\to X}^{*} \, ,
\end{equation}
where the sum extends over all physical on-shell intermediate states and \(d\Phi_X\) denotes their Lorentz invariant phase space. At LL accuracy the relevant intermediate states are gluons produced in MRK. The amplitudes on the two sides of the cut can therefore be replaced by the FKL factorisation formula \refeq{eq:Cap4:FKL_tree_factorisation_schematic}.
The resulting cut is represented schematically in fig.~\ref{fig:Cap4:BFKL_MRK_kernel}.

The left-hand side of the cut is an amplitude for the production of an arbitrary number of gluons in MRK, while the right-hand side is its complex conjugate. The cut identifies the intermediate particles and requires a sum over their colours and helicities. Consequently, every emitted gluon is associated with a pair
\begin{equation}
    V\,V^{*}
\end{equation}
of one-gluon CEVs, which defines the part related to the real emissions in the BFKL approach. 

The first few perturbative orders illustrate the construction. With no central emissions, fig.~\ref{fig:Cap4:BFKL_MRK_kernel} reduces to the two-particle cut of the one-loop four-point amplitude. It describes the lowest order exchange of two \(t\)-channel gluons. After a projection on a colour-singlet state, this double Reggeon state has vacuum quantum numbers and it is often called the perturbative BFKL Pomeron\footnote{More details on the perturbative QCD Pomeron are described in Ref.~\citeall{delduca1995introductionperturbativeqcdpomeron}}. With one central emission, the same diagram becomes the three-particle cut of the two-loop four-point amplitude,
\begin{equation}\label{eq:Cap4:two_loop_three_particle_cut}
    \left.
    \operatorname{Disc}_s\mathcal A_{4}^{(2)}
    \right|_{3\text{-particle cut}}
    =
    i\sum_{\mathrm{states}}
    \int d\Phi_3\,
    \mathcal A_{2\to3}^{\mathrm{tree}}
    \left(\mathcal A_{2\to3}^{\mathrm{tree}}\right)^{*} \, .
\end{equation}
Each of the two five point amplitudes contains one CEV, and their product gives the first unitarity correction to the double Reggeon exchange. More generally, \(n\) central emissions correspond to an \((n+2)\)-particle cut of the \((n+1)\)-loop four-point amplitude.

Now we can also see where the logarithmic enhancement arises from. Looking at the phase space of those cut amplitudes, the MRK ordering reduces the longitudinal integrations to ordered rapidity integrals,
\begin{equation}\label{eq:Cap4:ordered_rapidity_integrals}
    \int_0^Ydy_1
    \int_0^{y_1}dy_2\cdots
    \int_0^{y_{n-1}}dy_n
    =
    \frac{Y^n}{n!}\simeq\frac{\log^n\!\left(\frac{s}{-t}\right)}{n!} \, ,
\end{equation}
and since each additional pair of CEVs provides one power of \(\alpha_s\), a ladder with \(n\) emitted gluons produces a contribution proportional to \((\alpha_sY)^n\). The BFKL equation resums the sum of all these ladders. Crossed ladders and configurations without a strict rapidity ordering do not generate the maximal power of \(Y\) and are therefore subleading at LL accuracy.

In fig.~\ref{fig:Cap4:BFKL_MRK_kernel} there are two impact factors \(C_A\) and \(C_B\) for each amplitude separately. After the two sides of the cut are combined, they give the cut impact factors
\begin{equation}\label{eq:Cap4:cut_impact_factors}
    \Phi_A
    \sim
    \sum_{\mathrm{colours,\,helicities}}C_A C_A^{*} \, ,
    \qquad
    \Phi_B
    \sim
    \sum_{\mathrm{colours,\,helicities}}C_B C_B^{*} \, .
\end{equation}
The quantities \(\Phi_A\) and \(\Phi_B\) contain all the dependence on the external particles, while the evolution inside the ladder is process independent.

We now describe the kernel more explicitly. Let the two Reggeised gluons carry transverse momenta \(q_{1\perp}\) and \(q_{2\perp}\), with
\begin{equation}
    q_{1\perp}+q_{2\perp}=q_\perp \, ,
\end{equation}
where \(q_\perp\) is the total momentum transfer. After one real emission their momenta become \(q'_{1\perp}\) and \(q'_{2\perp}\), and the emitted gluon carries orthogonal momentum
\begin{equation}\label{eq:Cap4:BFKL_emitted_momentum}
    p_\perp
    =q_{1\perp}-q'_{1\perp}
    =q'_{2\perp}-q_{2\perp} \, .
\end{equation}
Up to normalisation and choice of conventions, the real part of the kernel is obtained by sewing the two CEVs across the cut,
\begin{equation}\label{eq:Cap4:real_BFKL_kernel_from_CEV}
    \mathcal K_{\mathrm r}
    (q_{1\perp},q'_{1\perp};q_\perp)
    \propto{}
    \sum_{h,c}
    V^{h}(q_{1\perp},p_\perp,q'_{1\perp})
    \times
    \left[
    V^{h}(q_{2\perp},-p_\perp,q'_{2\perp})
    \right]^{*} \, .
\end{equation}
Thus the one-gluon CEV introduced in the previous section is precisely the object that, after \textit{``squaring''} it, builds the real BFKL kernel.

The kernel contains in principle also virtual contributions that we have introduced in the previous section,
\begin{equation}
    \omega(t)=C_{\mathrm A}\alpha_g(t,\mu^2) \, ,
\end{equation}
where the Regge trajectory dresses each of the two exchanged gluons between two consecutive emissions. Since these corrections do not change the transverse momenta of the double Reggeon state, the virtual kernel is diagonal in the transverse momentum space,
and, following Ref.~\citeall{Ioffe:2010zz}, we could derive that
\begin{equation}\label{eq:Cap4:virtual_BFKL_kernel}
    \mathcal K_{\mathrm v}
    (q_{1\perp},q'_{1\perp};q_\perp)
    ={}
    q_{1\perp}^{2}q_{2\perp}^{2}
    \delta^{(D-2)}(q_{1\perp}-q'_{1\perp})
    \times
    \left[
    \omega(-q_{1\perp}^{2})
    +
    \omega(-q_{2\perp}^{2})
    \right] \, .
\end{equation}
The factors \(q_{1\perp}^{2}q_{2\perp}^{2}\) compensate the two Reggeised propagators included in the transverse momentum measure of the BFKL equation, because, since the virtual contribution does not carry any additional real emission, the two propagators connecting the external legs in the ladder must appear only once.
The full leading order kernel is therefore
\begin{equation}\label{eq:Cap4:BFKL_kernel_real_virtual_decomposition}
    \mathcal K
    =
    \mathcal K_{\mathrm v}
    +
    \mathcal K_{\mathrm r} \, .
\end{equation}
Separately, both parts of the kernel are infrared divergent, because \(\mathcal K_{\mathrm r}\) is singular when the emitted gluon becomes soft, while \(\mathcal K_{\mathrm v}\) contains the corresponding divergence related to the Regge trajectories. At the end, in the cross section, the soft singularity due to the real emission cancels with the corresponding virtual divergence.

Let's moreover introduce
\begin{equation}
    G(Y;q_{1\perp},q'_{1\perp};q_\perp) \, ,
\end{equation}
which denotes the Green function describing the propagation of the two Reggeons with a rapidity interval \(Y\). Increasing \(Y\) by an infinitesimal amount either produces a virtual trajectory correction, without any real emission, or inserts one further real gluon through the product of CEVs in \refeq{eq:Cap4:real_BFKL_kernel_from_CEV}. This gives the BFKL equation,
which in the language of Green functions is described by the following convolution,
\begin{equation}\label{eq:Cap4:BFKL_evolution_equation}
    \frac{\partial}{\partial Y}
    G(Y;q_{1\perp},q'_{1\perp};q_\perp)
    ={}
    \int
    \frac{d^{D-2}k_\perp}
    {k_\perp^2(q_\perp-k_\perp)^2}
    \,\mathcal K(q_{1\perp},k_\perp;q_\perp)
    G(Y;k_\perp,q'_{1\perp};q_\perp) \, .
\end{equation}
Its initial condition describes the free propagation across a vanishing rapidity interval,
\begin{equation}\label{eq:Cap4:BFKL_initial_condition}
    G(0;q_{1\perp},q'_{1\perp};q_\perp)
    ={}
    q_{1\perp}^{2}(q_\perp-q_{1\perp})^{2}
    \times
    \delta^{(D-2)}(q_{1\perp}-q'_{1\perp}) \, .
\end{equation}
In operator notation, the equation takes the compact form
\begin{equation}\label{eq:Cap4:BFKL_operator_equation}
    \frac{d}{dY}\widehat G(Y)
    =
    \widehat{\mathcal K}\,\widehat G(Y) \, ,
    \qquad
    \widehat G(0)=\mathds{1} \, ,
\end{equation}
and hence
\begin{equation}
    \widehat G(Y)=e^{Y\widehat{\mathcal K}} \, .
\end{equation}
Equivalently, this differential equation states that increasing the rapidity interval by \(dY\) adds one elementary BFKL step, either a virtual correction or one real emission. The integral over \(k_\perp\) sums over the possible intermediate transverse momenta of the two Reggeons.
Expanding this exponential reproduces diagram by diagram the sequence of cut ladders in fig.~\ref{fig:Cap4:BFKL_MRK_kernel}.

Finally, the discontinuity of the elastic amplitude is obtained by convoluting this universal Green function with the cut impact factors,
\begin{equation}\label{eq:Cap4:BFKL_impact_factor_convolution}
\begin{aligned}
    \operatorname{Disc}_s\mathcal A_{AB}(s,t)
    \sim{}&
    is
    \int
    \frac{d^{D-2}q_{1\perp}}
    {q_{1\perp}^{2}(q_\perp-q_{1\perp})^{2}}
    \frac{d^{D-2}q'_{1\perp}}
    {q_{1\perp}^{\prime 2}(q_\perp-q'_{1\perp})^{2}}
    \\
    &\times
    \Phi_A(q_{1\perp},q_\perp)
    G(Y;q_{1\perp},q'_{1\perp};q_\perp)
    \Phi_B(q'_{1\perp},q_\perp) \, .
\end{aligned}
\end{equation}

At NLL the same diagrammatic construction is still valid, but of course the elementary building blocks must be evaluated at higher orders. For instance, the virtual kernel requires the two-loop gluon Regge trajectory. The one-particle real kernel instead could contain one-loop one-gluon CEVs. In addition, four-particle cuts of the elastic amplitude contain tree-level \(2\to4\) amplitudes in NMRK. If the two particles close in rapidity are emitted in the central-rapidity region, their squared two-particle CEV in \refeq{eq:Cap4:two_gluon_CEV_pp} and \refeq{eq:Cap4:two_gluon_CEV_pm} contributes to the real NLL kernel. If instead the two particles are produced in the forward-rapidity region, the corresponding correction affects the cut impact factors \(\Phi_A\) or \(\Phi_B\).

\section{Gravitational BFKL equation}
\label{sec:Cap4:Gravitational_BFKL_equation}

We now turn to the gravitational counterpart of the high-energy factorisation introduced in Sec.~\ref{sec:Cap4:high_energy_factorisation_QCD}. The construction is analogous to the QCD one where now, at fixed momentum transfer and large energy in the centre of mass'  frame, amplitudes are dominated by graviton exchange in the \(t\)-channel and factorise into impact factors, Glauber graviton propagators and central emission vertices regarding the real emissions along the ladder. However, the absence of colour and the spin two nature of the exchanged particle and the dimensionful gravitational coupling lead to important differences that we will discuss. In the following we will mainly follow Ref.~\citeall{barcaro2025centralemissionvertexgravitons,raj2025qcdgravitydoublecopyregge,rothstein2026systematiclagrangianformulationquantum,Rothstein_2016}.

At four points, a generic gravity amplitude in the Regge limit takes the form
\begin{equation}\label{eq:Cap4:gravity_four_point_Regge_factorisation}
    \mathcal M(1,2,3,4)
    \simeq
    -\left[\kappa C(2,3)\right]
    \frac{s^2}{t}
    \left[\kappa C(1,4)\right] \, .
\end{equation}
The additional power of \(s\), compared with the gauge-theory result in \refeq{eq:Cap4:tree_regge_two_to_two_factorisation}, reflects the spin two of the exchanged graviton. For graviton MHV amplitudes, this formula can be proven by using \refeq{eq:Cap3:four_graviton_mhv_amplitude}. For scalars and gravitons we can find that the impact factors are
\begin{equation}\label{eq:Cap4:gravity_impact_factors}
    C(i_\phi,j_\phi)=1 \, ,
    \qquad
    C(2_h^{-},3_h^{+})=1 \, ,
    \qquad
    C(1_h^{-},4_h^{+})
    =
    \left(\frac{p_{4\perp}^{*}}{p_{4\perp}}\right)^2 \, .
\end{equation}
As for the QCD case, at tree level these impact factors are pure phases and also their parity conjugates are obtained by complex conjugation.
The graviton impact factors in \refeq{eq:Cap4:gravity_impact_factors} are the double copy of the corresponding gluon impact factors, indeed the Yang--Mills phase is squared\footnote{We can expand the same impact factor in the Yang--Mills theory as \(\left(\frac{p_{4\perp}^{*}}{p_{4\perp}}\right)\), following the conventions in Ref.~\citeall{Del_Duca_2000}.}, while the colour factor has been removed.
Moreover, helicity is conserved along the direction of the beam, which means that
\begin{equation}
    C(i_h^{\nu},j_h^{\nu'})
    \propto
    \delta_{\nu,-\nu'} \, .
\end{equation}
Helicity-flip impact factors are power suppressed in the Regge limit by \(t/s\). 

\begin{figure}[t]
    \centering
    \subfigure[
\label{fig:five_point_MRK_graviton}]{
        \includegraphics[width=0.36\textwidth]{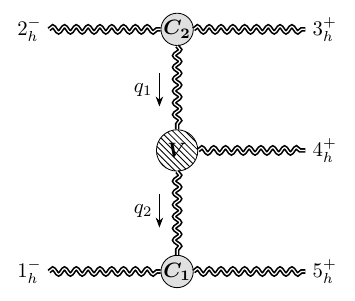}
    }
    \hfill
    \subfigure[]{
        \includegraphics[width=0.36\textwidth]{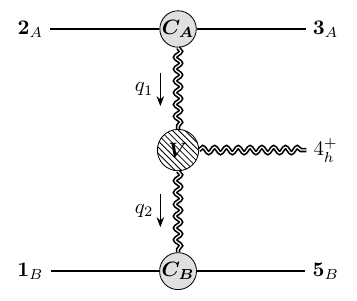}
    }
    \caption{Regge factorisation of five point amplitudes in MRK.}
    \label{fig:Cap4:MRK_5_point_diagrams}
\end{figure}

The same factorisation extends to amplitudes with one graviton emitted in the central region coupled to two external gravitons or massive scalars. In MRK, a five-point amplitude takes the form
\begin{equation}\label{eq:Cap4:gravity_five_point_MRK_factorisation}
\begin{aligned}
    \mathcal M(1,2,3,p^{h},5)
    \simeq{}&
    -s^2
    \left[\kappa C_A\right]
    \frac{1}{t_1}
    \left[\kappa V(q_1,p^{h},q_{2})\right]
    \frac{1}{t_{2}}
    \left[\kappa C_B\right] \, .
\end{aligned}
\end{equation}
A schematic representation is shown in fig.~\ref{fig:Cap4:MRK_5_point_diagrams}.

For a positive-helicity graviton, the one-graviton CEV could be derived by \refeq{eq:Cap3:five_graviton_mhv_amplitude}, reaching
\begin{equation}\label{eq:Cap4:one_graviton_CEV_positive_helicity}
    \vcenter{\hbox{\includegraphics[width=0.25\textwidth]{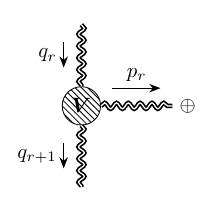}}}
    =
    V(q_r,p_r^{+},q_{r+1})
    =
    \frac{q_{r\perp}^{*}q_{r+1\perp}}{p_{r\perp}^{2}}
    \left(
    q_{r\perp}^{*}q_{r+1\perp}
    -
    q_{r\perp}q_{r+1\perp}^{*}
    \right) \, .
\end{equation}
The negative-helicity vertex follows as usual by parity.
The double-copy structure is manifest by comparing \refeq{eq:Cap4:one_graviton_CEV_positive_helicity} with the gluon CEV in \refeq{eq:Cap4:gluon_CEV_positive_helicity}. The first factor is the gauge-theory Lipatov vertex, while the second factor supplies its gravitational kinematic copy, namely
\begin{equation}\label{eq:Cap4:one_graviton_CEV_kinematic_copy}
    \mathcal F_{ h}(q_r,p_r^{+},q_{r+1})
    =
    \frac{
    q_{r\perp}^{*}q_{r+1\perp}
    -q_{r\perp}q_{r+1\perp}^{*}}
    {p_{r\perp}} \, .
\end{equation}
We can appreciate that the gravitational vertex is not just the naive square of the gluonic expression because, in a covariant description, it is the difference between the strict double copy of the QCD Lipatov current and a bremsstrahlung subtraction term. The latter is  
required by Steinmann relations, to avoid overlapping singularities of Mandelstam invariants, see Ref.~\citeall{Caron_Huot_2016,Lipatov:1982er}. As for QCD, in the soft limit of the emitted graviton, this vertex reduces to the Regge limit of Weinberg's soft graviton current in \refeq{eq:Cap3:leading_soft_graviton_factor}.
 We can indeed write the Weinberg theorem in covariant form as in \refeq{eq:Cap3:leading_soft_graviton_factor_2},
\begin{equation}\label{eq:Cap4:Weinberg_soft_current_covariant}
\begin{aligned}
    \mathcal M_{n+1}
    (k_1,\ldots,k_n,p_r^{h_r})
    \xrightarrow{p_r\to0}{}&
    \varepsilon_{\mu\nu}^{h_r}(p_r)
    S_{\mathrm{G}}^{(0)\, \mu\nu}(p_r)\,
    \mathcal M_n(k_1,\ldots,k_n),
    \\
    S_{\mathrm{G}}^{(0)\, \mu\nu}(p_r)
    ={}&
    \sum_{a=1}^{n}
    \frac{k_a^\mu k_a^\nu}{k_a\cdot p_r} \, .
\end{aligned}
\end{equation}
 Since
\(\varepsilon_{\mu\nu}^{h}=\varepsilon_\mu^{h}\varepsilon_\nu^{h}\),
the helicity projection of the current is the sum of the eikonal factors
\((k_a\cdot\varepsilon^{h})^2/(k_a\cdot p_r)\).

We may now specialise the hard momenta in
\refeq{eq:Cap4:Weinberg_soft_current_covariant} to Regge kinematics and
retain the leading power in \(|t|/s\). The same projection can be obtained
directly by matching the soft limit of the five-point Regge factorisation
onto the four-point one. Since
\(q_{r+1}=q_r-p_r\to q_r\) and \(t_{r+1}\to t_r\), one finds
\begin{equation}\label{eq:Cap4:graviton_CEV_Weinberg_matching}
\begin{aligned}
    \frac{1}{t_r}
    V&(q_r,p_r^+,q_{r+1})
    \frac{1}{t_{r+1}}
    \xrightarrow{p_r\to0}{}
    \operatorname{Soft}_{\mathrm G,\mathrm{Regge}}^{(0)}
    (p_r^+;q_r)\,
    \frac{1}{t_r},
    \\
    \operatorname{Soft}&_{\mathrm G,\mathrm{Regge}}^{(0)}
    (p_r^+;q_r)
    ={}
    \left.
    \varepsilon_{\mu\nu}^{+}(p_r)
    S_{\mathrm{G}}^{(0)\, \mu\nu}(p_r)
    \right|_{\mathrm{Regge}}
    \\
    ={}&
    \lim_{p_r\to0}
    \frac{V(q_r,p_r^+,q_r-p_r)}{t_r}
    ={}
    \frac{
    q_{r\perp}^{*}p_{r\perp}
    -q_{r\perp}p_{r\perp}^{*}}
    {p_{r\perp}^{2}} \, .
\end{aligned}
\end{equation}
Indeed, \(t_r\simeq-|q_{r\perp}|^2\), while the two factors entering the
CEV behave as
\(q_{r\perp}^*q_{r+1\perp}\to|q_{r\perp}|^2\) and
\(q_{r\perp}^*q_{r+1\perp}
-q_{r\perp}q_{r+1\perp}^*
\to q_{r\perp}p_{r\perp}^*
-q_{r\perp}^*p_{r\perp}\).
The contributions of each incoming--outgoing hard pair cancel at zero
momentum transfer, while the first surviving term is linear in \(q_r\) and gives
\refeq{eq:Cap4:graviton_CEV_Weinberg_matching}.

The general tree-level gravity amplitude in MRK therefore has the same iterative structure as its QCD counterpart,
\begin{equation}\label{eq:Cap4:gravity_FKL_tree_factorisation}
\begin{aligned}
    \mathcal M_{2\to n+2}^{\mathrm{tree}}
    \simeq{}&
    -s^2
    \left[\kappa C_A\right]
    \frac{1}{t_1}
    \prod_{r=1}^{n}
    \left[
    \kappa V(q_r,p_r^{h_r},q_{r+1})
    \frac{1}{t_{r+1}}
    \right]
    \left[\kappa C_B\right] \, ,
\end{aligned}
\end{equation}
where the exchanged momenta have the usual Glauber scaling and are slightly off shell, while the gravitons emitted by the CEVs are on shell and strongly ordered in rapidity. 

Moreover, as in gauge theory, the first real corrections beyond strict MRK arise in NMRK, when two neighbouring gravitons are emitted with comparable rapidities, as in \refeq{eq:Cap4:NMRK_definition}.

In the central-rapidity region, the two consecutive one-graviton CEVs are replaced by a two-graviton CEV,
\begin{equation}\label{eq:Cap4:gravity_NMRK_central_replacement}
    V(q_r,p_r,q_{r+1})
    \frac{1}{t_{r+1}}
    V(q_{r+1},p_{r+1},q_{r+2})
    \quad\longrightarrow\quad
    V(q_r,p_r,p_{r+1},q_{r+2}) \, ,
\end{equation}
and, accordingly, the relevant six-point amplitude factorises as
\begin{equation}\label{eq:Cap4:gravity_six_point_NMRK_factorisation}
\begin{aligned}
    \mathcal M_{2\to4}^{\mathrm{tree}}
    \simeq{}&
    -s^2
    \left[\kappa C_A\right]
    \frac{1}{t_r}
    \left[
    \kappa^2
    V(q_r,p_r^{h_r},p_{r+1}^{h_{r+1}},q_{r+2})
    \right]
    \frac{1}{t_{r+2}}
    \left[\kappa C_B\right] \, .
\end{aligned}
\end{equation}
There are two independent helicity configurations. The configuration with same helicities can be extracted from a six graviton MHV amplitude. Introducing again 
the parameter $X_r$ which encodes the rapidity separation between two neighbouring gravitons,
\begin{equation}\label{eq:Cap4:X_r}
    X_r=\frac{p_r^{+}}{p_{r+1}^{+}} \, ,
\end{equation}
we can expand
\begin{equation}\label{eq:Cap4:two_graviton_CEV_pp}
\begin{aligned}
    \vcenter{\hbox{\includegraphics[width=0.20\textwidth]{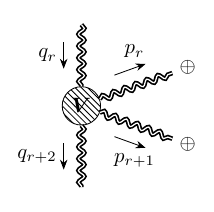}}}
    &={}
    V(q_r,p_r^{+},p_{r+1}^{+},q_{r+2})
    \\
    &={}
    \sqrt{X_r}
    \frac{-p_{r+1\perp}}{\langle r\,(r+1)\rangle}
    \frac{q_{r\perp}^{*}q_{r+2\perp}}
    {p_{r\perp}^{2}p_{r+1\perp}^{2}}
    \\
    &\quad\times
    \left(
    q_{r\perp}^{*}p_{r\perp}
    -q_{r\perp}p_{r\perp}^{*}
    \right)
    \left(
    q_{r+2\perp}^{*}p_{r+1\perp}
    -q_{r+2\perp}p_{r+1\perp}^{*}
    \right)
    \\
    &\quad+
    (r\leftrightarrow r+1) \, .
\end{aligned}
\end{equation}
In order to make manifest the double-copy structure of this result, let
\(q_{r+1}=q_r-p_r\), so that
\(p_r=q_r-q_{r+1}\) and
\(p_{r+1}=q_{r+1}-q_{r+2}\).
Denoting the two-gluon CEV and the two-graviton CEV by
\(V_{g^{+}g^{+}}\) and \(V_{h^{+}h^{+}}\), respectively,
\refeq{eq:Cap4:two_graviton_CEV_pp} takes the form
\begin{equation}\label{eq:Cap4:two_graviton_CEV_double_copy}
\begin{aligned}
    V_{h^{+}h^{+}}(q_r,p_r^{+},p_{r+1}^{+},q_{r+2})
    ={}&
    V_{g^{+}g^{+}}(q_r,p_r^{+},p_{r+1}^{+},q_{r+2})
    \mathcal F_h(q_r,p_r^{+},q_{r+1})
    \mathcal F_h(q_{r+1},p_{r+1}^{+},q_{r+2})
    \\
    &{}+
    \left(
    \begin{array}{c}
        q_{r+1}\rightarrow q_r-q_{r+1}+q_{r+2}
        \\
        X_r\rightarrow 1/X_r
    \end{array}
    \right) \, .
\end{aligned}
\end{equation}
The presence of two factors \(\mathcal F_h\) reflects the emission of
two gravitons. Indeed, each emitted graviton provides the same
kinematic factor that appears in the double-copy representation of
the one-graviton CEV. Therefore, for a fixed ordering, the factor
\(\mathcal F_h(q_r,p_r^{+},q_{r+1})\) in \refeq{eq:Cap4:one_graviton_CEV_kinematic_copy} is associated with the emission
of the first graviton, while
\(\mathcal F_h(q_{r+1},p_{r+1}^{+},q_{r+2})\) is associated with the
emission of the second one. The additional term in
\refeq{eq:Cap4:two_graviton_CEV_double_copy} exchanges the two emitted
gravitons, ensuring that the complete vertex is symmetric under their
interchange, as required by Bose statistics, while the same term is not present in QCD because of the colour ordering.

By contrast, the opposite helicity two-graviton CEV follows from the
NMRK limit of a six-graviton NMHV amplitude. Its rational
structure is considerably more involved. We report its expression below just for completeness.
\begin{equation}\label{eq:Cap4:two_graviton_CEV_pm}
\begin{aligned}
    \vcenter{\hbox{
    \includegraphics[width=0.20\textwidth]
    {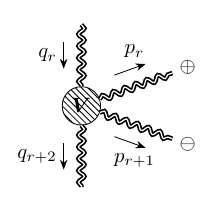}}}
    &\;=\;
    V_{h^{+}h^{-}}(q_r,q_{r+1},q_{r+2};X_r)
    =
    V_{(\oplus\ominus,1)}
    +V_{(\oplus\ominus,2)}
    +V_{(\oplus\ominus,3)} \, .
\end{aligned}
\end{equation}
The three contributions are given explicitly below,
\begin{equation}\label{eq:Cap4:two_graviton_CEV_pm_1}
\begin{aligned}
V_{(\oplus\ominus,1)}
={}&
-\frac{
q_{r\perp}^{*}p_{r\perp}q_{r+2\perp}q_{r+2\perp}^{*}}
{(p_{r\perp}+p_{r+1\perp})X_r(1+X_r)
(p_{r\perp}^{*}-p_{r+1\perp}^{*}X_r)}
\\
&-
\frac{
q_{r\perp}^{*}p_{r\perp}^{5}p_{r\perp}^{*}q_{r+2\perp}^{*}}
{q_{r+1\perp}^{2}X_r
(p_{r\perp}p_{r\perp}^{*}
+q_{r+1\perp}q_{r+1\perp}^{*}X_r)
(p_{r\perp}p_{r\perp}^{*}
+q_{r+1\perp}p_{r+1\perp}^{*}X_r)}
\\
&+
\frac{
q_{r\perp}q_{r\perp}^{*}p_{r\perp}^{6}q_{r+2\perp}^{*}
(p_{r\perp}^{*}-p_{r+1\perp}^{*}X_r)}
{(p_{r\perp}+p_{r+1\perp})p_{r+1\perp}^{2}X_r
(p_{r\perp}p_{r\perp}^{*}
+p_{r+1\perp}p_{r+1\perp}^{*}X_r)^{2}
(p_{r\perp}-p_{r+1\perp}X_r)}
\\
&\quad +
\frac{
q_{r\perp}q_{r\perp}^{*}p_{r\perp}q_{r+2\perp}^{2}
(p_{r+1\perp}^{*})^{6}X_r^{4}}
{(p_{r\perp}^{*})^{2}
(p_{r\perp}p_{r\perp}^{*}
+q_{r+1\perp}p_{r+1\perp}^{*}X_r)}
\\[-1pt]
&\qquad {}\times
\frac{1}{
(p_{r\perp}p_{r\perp}^{*}
+p_{r+1\perp}p_{r+1\perp}^{*}X_r)^{2}
(p_{r\perp}-p_{r+1\perp}^{*}X_r)}
\\
&+
\frac{
q_{r\perp}q_{r\perp}^{*}p_{r\perp}
q_{r+2\perp}q_{r+2\perp}^{*}
(q_{r+1\perp}^{*}p_{r+1\perp}
-q_{r+2\perp}p_{r+1\perp}^{*})}
{q_{r+1\perp}(p_{r\perp}^{*})^{2}p_{r+1\perp}^{2}}
\\
&-
\frac{
q_{r\perp}q_{r\perp}^{*}p_{r\perp}
q_{r+2\perp}q_{r+2\perp}^{*}
\bigl[q_{r+1\perp}(p_{r\perp}-p_{r+1\perp})
+p_{r\perp}p_{r+1\perp}\bigr]}
{q_{r+1\perp}^{2}p_{r\perp}^{*}p_{r+1\perp}^{2}X_r}
\, ,
\end{aligned}
\end{equation}
\begingroup
\setlength{\jot}{1pt}
\allowdisplaybreaks[4]
\begin{align}
V_{(\oplus\ominus,2)}
&={}
\frac{
X_r^{3}q_{r\perp}q_{r+2\perp}q_{r+2\perp}^{*}p_{r\perp}}
{(1+X_r)(p_{r\perp}-X_rq_{r+1\perp})
(p_{r\perp}-X_rp_{r+1\perp})}
\notag\\
&+
\frac{
p_{r\perp}q_{r\perp}q_{r\perp}^{*}
q_{r+2\perp}q_{r+2\perp}^{*}
(q_{r+1\perp}p_{r+1\perp}^{*}
-q_{r\perp}^{*}p_{r+1\perp})}
{q_{r+1\perp}p_{r\perp}^{*}p_{r+1\perp}^{2}
(q_{r\perp}^{*}-q_{r+1\perp}^{*}+q_{r+2\perp}^{*})}
 -
\frac{
p_{r+1\perp}^{2}q_{r\perp}^{2}q_{r+2\perp}^{*}}
{q_{r+1\perp}(p_{r\perp}-X_rq_{r+1\perp})}
\notag\\[-1pt]
&\times
\frac{1}{
\bigl[p_{r\perp}q_{r\perp}^{*}
-q_{r+1\perp}^{*}(p_{r\perp}-X_rq_{r+1\perp})\bigr]\bigl[p_{r\perp}q_{r\perp}^{*}
-p_{r+1\perp}^{*}(p_{r\perp}-X_rq_{r+1\perp})\bigr]}
\notag\\
& +
\frac{
X_r^{3}q_{r\perp}q_{r+2\perp}^{2}p_{r\perp}
(p_{r+1\perp}^{*})^{5}}
{q_{r\perp}^{*}-q_{r+1\perp}^{*}+q_{r+2\perp}^{*}}
\notag\\[-1pt]
&\times
\frac{1}{
\bigl[
(q_{r\perp}-q_{r+1\perp}+q_{r+2\perp})
(q_{r\perp}^{*}-q_{r+1\perp}^{*}+q_{r+2\perp}^{*})
+X_rp_{r+1\perp}p_{r+1\perp}^{*}
\bigr]}
\notag\displaybreak[4]\\[-1pt]
&\times
\frac{1}{
\bigl[
p_{r\perp}q_{r\perp}^{*}
-p_{r+1\perp}^{*}(p_{r\perp}-X_rq_{r+1\perp})
\bigr]
\bigl[
p_{r\perp}q_{r\perp}^{*}
-p_{r+1\perp}^{*}(p_{r\perp}-X_rp_{r+1\perp})
\bigr]}
\notag\\
& +
\frac{
q_{r\perp}q_{r\perp}^{*}q_{r+2\perp}^{*}
p_{r\perp}^{5}p_{r+1\perp}^{*}}
{p_{r+1\perp}^{2}(p_{r\perp}-X_rp_{r+1\perp})}
\notag\\[-1pt]
&\times
\frac{1}{
(p_{r\perp}p_{r\perp}^{*}
+X_rp_{r+1\perp}p_{r+1\perp}^{*})
\bigl[
p_{r\perp}
(q_{r\perp}^{*}-q_{r+1\perp}^{*}+q_{r+2\perp}^{*})
+X_rp_{r+1\perp}p_{r+1\perp}^{*}
\bigr]}
\, ,
\label{eq:Cap4:two_graviton_CEV_pm_2}
\end{align}
\nopagebreak[4]
\begin{equation}\label{eq:Cap4:two_graviton_CEV_pm_3}
\begin{aligned}
V_{(\oplus\ominus,3)}
={}&
-\frac{
X_r^{3}q_{r\perp}q_{r+2\perp}q_{r+2\perp}^{*}
(p_{r\perp}^{*}-X_rp_{r+1\perp}^{*})}
{(1+X_r)^{2}(p_{r\perp}^{*}+p_{r+1\perp}^{*})
(p_{r\perp}-X_rp_{r+1\perp})}
\\
&+
\frac{
(q_{r\perp}^{*})^{2}q_{r+2\perp}q_{r+2\perp}^{*}}
{(1+X_r)^{2}(p_{r\perp}^{*}-X_rp_{r+1\perp}^{*})
(q_{r\perp}^{*}-q_{r+1\perp}^{*}
+q_{r+2\perp}^{*}-X_rp_{r+1\perp}^{*})}
\\
&+
\frac{
q_{r\perp}q_{r\perp}^{*}q_{r+2\perp}q_{r+2\perp}^{*}
p_{r+1\perp}^{*}
\bigl[
(q_{r\perp}^{*}-q_{r+1\perp}^{*}+q_{r+2\perp}^{*})
p_{r\perp}^{*}
+2X_rp_{r+1\perp}p_{r+1\perp}^{*}
\bigr]}
{p_{r+1\perp}^{2}(p_{r\perp}^{*})^{2}
(q_{r\perp}^{*}-q_{r+1\perp}^{*}+q_{r+2\perp}^{*})}
\\
&-
\frac{
q_{r\perp}q_{r\perp}^{*}
q_{r+2\perp}q_{r+2\perp}^{*}}
{(p_{r\perp}^{*})^{2}p_{r+1\perp}
(q_{r\perp}^{*}-q_{r+1\perp}^{*}+q_{r+2\perp}^{*})^{2}}
\\
&\times\Big[q_{r\perp}^{*}p_{r\perp}^{*}
(q_{r\perp}^{*}-q_{r+1\perp}^{*}+q_{r+2\perp}^{*})
-X_r\bigl(q_{r\perp}^{*}q_{r+1\perp}^{*}
-4(p_{r+1\perp}^{*})^{2}\bigr)
(q_{r\perp}^{*}-q_{r+1\perp}^{*}+q_{r+2\perp}^{*})
\\[-1pt]
&{}+X_r(p_{r+1\perp}^{*})^{2}
(p_{r\perp}^{*}+p_{r+1\perp}^{*})
\Big] +
\frac{
X_r^{3}q_{r\perp}q_{r+2\perp}(p_{r+1\perp}^{*})^{5}
}
{(q_{r\perp}^{*}-q_{r+1\perp}^{*}+q_{r+2\perp}^{*})^{2}}
\\[-1pt]
&\times
\frac{1}{
\bigl[
(q_{r\perp}-q_{r+1\perp}+q_{r+2\perp})
(q_{r\perp}^{*}-q_{r+1\perp}^{*}+q_{r+2\perp}^{*})
+X_rp_{r+1\perp}p_{r+1\perp}^{*}
\bigr]}
\\[-1pt]
&\times
\frac{1}{
(p_{r\perp}^{*}-X_rp_{r+1\perp}^{*})
(q_{r\perp}^{*}-q_{r+1\perp}^{*}
+q_{r+2\perp}^{*}-X_rp_{r+1\perp}^{*})}
\\
&-
\frac{
X_r^{3}q_{r\perp}q_{r\perp}^{*}q_{r+2\perp}
(p_{r+1\perp}^{*})^{5}}
{(p_{r\perp}^{*})^{2}
(p_{r\perp}^{*}+p_{r+1\perp}^{*})}
\times
\frac{1}{
(p_{r\perp}^{*}-X_rp_{r+1\perp}^{*})
(p_{r\perp}p_{r\perp}^{*}
+X_rp_{r+1\perp}p_{r+1\perp}^{*})}
\, .
\end{aligned}
\end{equation}
\endgroup
Unlike the MHV case, the expression above does not exhibit a simple double-copy structure. The reason is that in the six-graviton NMHV amplitude, several factorisation
channels and gauge-theory orderings contribute at the same leading power,
and their NMRK limit cannot be reorganised as a single two-gluon CEV
multiplied by two \(\mathcal F_h\) factors. However in the MRK
limit, the vertex instead factorises into two one-graviton CEVs, making
the double-copy structure manifest again.

Indeed, as in QCD, if we sistematically send $X_r \to \infty$ in \refeq{eq:Cap4:two_graviton_CEV_pm}, we find the exact factorisation into two one-graviton CEV, 
\begin{equation}\label{eq:Cap4:two_graviton_CEV_MRK_limit}
    V(q_r,p_r^{h_r},p_{r+1}^{h_{r+1}},q_{r+2})
    \xrightarrow{X_r\to\infty}
    V(q_r,p_r^{h_r},q_{r+1})
    \frac{1}{t_{r+1}}
    V(q_{r+1},p_{r+1}^{h_{r+1}},q_{r+2}) \, .
\end{equation}
Diagrammatically,
\begin{equation}
    \vcenter{\hbox{\includegraphics[width=0.23\textwidth]{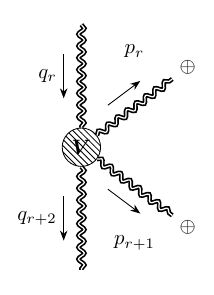}}}
    \xrightarrow{X_r\to\infty}
    \vcenter{\hbox{\includegraphics[width=0.25\textwidth]{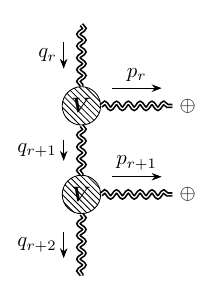}}} \, .
\end{equation}
This reproduces the gravitational FKL factorisation in \refeq{eq:Cap4:gravity_FKL_tree_factorisation}.

The tree-level one-graviton CEV provides the building
block of the real contribution at leading order of the  gravitational BFKL kernel, whose construction is analogous to the QCD case\footnote{See Ref.~\citeall{Lipatov:1982er,LIPATOV1982411,raj2025qcdgravitydoublecopyregge,rothstein2026systematiclagrangianformulationquantum}.}. As in QCD, this
contribution follows from \(s\)-channel unitarity by sewing two
tree-level CEVs and summing over the helicity of the intermediate cut graviton,
\begin{equation}\label{eq:Cap4:gravity_BFKL_real_kernel}
    \mathcal K_h^{(0),\mathrm{real}}(q_1,q_2;q)
    =
    \sum_{h=\pm2}
    V(q_1,p^{h},q_2)
    \left[
    V(q-q_1,-p^{h},q-q_2)
    \right]^{*} \, .
\end{equation}
Together with the virtual contribution generated by the one-loop
graviton Regge trajectory, this defines the leading logarithmic
gravitational BFKL kernel and reproduces the 
convolutive structure of the BFKL equation and, even in this case, if we are interested to go beyond leading logarithmic
accuracy, one must also include loop corrections to the one-graviton
CEV and the Regge trajectory, as well as the tree-level two-graviton CEV
in NMRK. 

Its physical interpretation is nevertheless different from the QCD
one. To sum up, in QCD, at fixed Reggeon number, the emission of an on-shell
\(s\)-channel gluon and the virtual correction encoded in the gluon
Regge trajectory contribute at the same logarithmic accuracy. Each
insertion carries a factor of \(\alpha_s\), while the corresponding
ordered rapidity integral supplies a factor
\(L=\log(s/|t|)\), therefore real and virtual contributions must be 
 combined, and their iteration generates the dominant series
\((\alpha_s L)^n\).

In gravity we have the same organisation for the real and virtual emissions, however with a
high-energy power counting. To see this, we recall that in the eikonal
description of small scattering angles, with \(s\gg |t|\), the 
momentum transfer \(q_\perp\) is the Fourier conjugate to the impact
parameter \(b\). We define the impact parameter amplitude by
\begin{equation}\label{eq:Cap4:eikonal_impact_parameter_definition}
\begin{aligned}
    \widetilde{\mathcal M}(s,b)
    &\equiv
    \frac{1}{2s}
    \int\frac{d^2q_\perp}{(2\pi)^2}\,
    e^{i b\cdot q_\perp}
    \mathcal M(s,-q_\perp^2),
    \\
    \widetilde S(s,b)
    &=1+i\widetilde{\mathcal M}(s,b)
    \simeq e^{2i\delta(s,b)} \, ,
\end{aligned}
\end{equation}
where $\widetilde S(s,b)$ is the classical $S$-matrix in the impact parameter space. Here we have introduced the eikonal phase \(2\delta(s,b)\),
which, in the classical limit, plays the role of the scattering action, indeed its derivative with
respect to \(b\) determines the transverse impulse and therefore the
deflection angle, through the relation
\(Q_\perp^\mu=2\hbar\,\partial\delta/\partial b_\mu\). Following the
historical work by Amati, Ciafaloni and Veneziano, in Ref.~\citeall{Amati:1987uf,Amati:1990xe}, the leading phase
\(\delta_0\) is obtained from the tree-level exchange of one Glauber graviton and, in $D=4$ dimensions, using \(\kappa^2/(8\pi)=G\), it can be written as
\begin{equation}\label{eq:Cap4:leading_eikonal_phase}
    2\delta_0(s,b)\equiv\chi(b,s)
    =
    \frac{\kappa^2s}{2\hbar}
    \int\frac{d^2q}{(2\pi)^2}\,
    \frac{e^{i b\cdot q}}{q^2}
   =-\frac{2Gs}{\hbar}
    \log\!\left(\frac{\lvert b\rvert}{L_{\mathrm{IR}}}\right) \, ,
\end{equation}
where \(L_{\mathrm{IR}}\) is an infrared cutoff, distinct from the
Regge logarithm \(L\)\footnote{For a modern description of the eikonal phase look at Ref.~\citeall{raj2025qcdgravitydoublecopyregge,divecchia2024gravitationaleikonalparticlestring}.}.

The appearance of \(Gs\) reflects the energy growth of the effective
gravitational charge. By contrast, the correction associated with the
graviton Regge trajectory is proportional to \(GtL\), so that, schematically,
\begin{equation}\label{eq:Cap4:eikonal_regge_ratio}
    \frac{\text{Reggeisation}}{\text{eikonal}}
    \sim \frac{|t|}{s}\,L \, ,
\end{equation}
which shows that, in the strict Regge limit, parallel eikonal exchanges dominate over
vertical iterations that dress a single graviton into a Reggeised
one\footnote{A more elegant description using the Wilson lines' approach is reported in Ref.~\citeall{melville2013wilsonlinegravity}.}.

In momentum
space, the exchange of \(n\) independent Glauber gravitons gives an
\(n\)-fold transverse convolution of the one-graviton amplitude. The
Fourier transform in
\refeq{eq:Cap4:eikonal_impact_parameter_definition} turns this
convolution into an ordinary product, while the permutations of the
identical exchanges supply the factor \(1/n!\), consequently,
\begin{equation}\label{eq:Cap4:leading_eikonal_exponentiation}
\begin{aligned}
    i\widetilde{\mathcal M}_{\mathrm{parallel}}^{(n-1)}(s,b)
    &=\frac{\bigl(2i\delta_0(s,b)\bigr)^n}{n!},
    \\
    \widetilde S_{\mathrm{eik}}(s,b)
    &=1+i\sum_{n=1}^{\infty}
    \widetilde{\mathcal M}_{\mathrm{parallel}}^{(n-1)}(s,b)
    =e^{2i\delta_0(s,b)} \, ,
\end{aligned}
\end{equation}
which can be represented diagrammatically as
\begin{equation}\label{eq:Cap4:ACV_eikonal_resummation_diagram}
    \vcenter{\hbox{\includegraphics[width=0.62\textwidth]
    {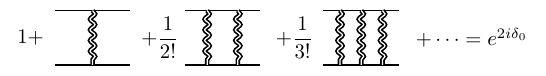}}} \, .
\end{equation}
In particular, the logarithm of \(b\) contained
in \(\delta_0\) is a transverse infrared logarithm and does not yield a new rapidity logarithm.

\begin{figure}[htbp]
    \centering
    \includegraphics[width=0.82\textwidth]
    {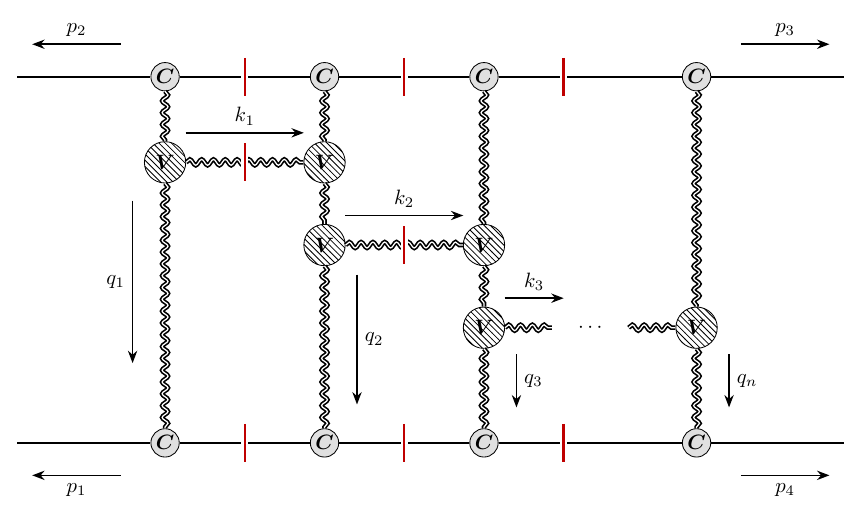}
    \caption{Classical multi-\(H\) sequence participating to the gravitational kernel obtained by three-particle \(s\)-channel cuts.}
    \label{fig:Cap4:multi_H_gravity}
\end{figure}

To generate a classical correction with an explicit dependence on
the rapidity, distinct Glauber exchanges must instead be connected by a soft
graviton. In the \(s\)-channel discontinuity this graviton is on shell,
and its phase-space integral provides \(\int dy\simeq L\). The first such
connected configuration is the \(H\) diagram, and if we repeat the construction
we find the multi-\(H\) sequence, depicted in fig.~\ref{fig:Cap4:multi_H_gravity}. These terms are subleading in the PM
expansion relative to the leading eikonal, but they provide the leading
classical Regge logarithms at their respective PM orders.

This distinction becomes bigger when one restricts the attention to
classical observables. Restoring Planck's constant according to the
standard classical scaling\footnote{A more detailed description of the classical limit and KMOC formalism for computing classical G.R. observables is presented in Appendix~\ref{app:Classical_limit}.},
\begin{equation}\label{eq:Cap4:classical_hbar_scaling}
    q^\mu=\hbar\,\bar q^\mu,
    \qquad
    p^\mu=\hbar\,\bar p^\mu,
    \qquad
    \kappa\longrightarrow\frac{\kappa}{\sqrt{\hbar}} \, ,
\end{equation}
the hard external momenta and the impact parameter are kept fixed, while
the transferred and radiated momenta are soft. Classical terms are those ones which
 survive in the \(\hbar\to0\) limit, after the
superclassical eikonal iterations have been combined or subtracted. Moreover, following Ref.~\citeall{Holstein_2004}, the loop
order alone is therefore not a criterion because infrared graviton exchange
can make a loop contribution classical, whereas a closed loop made only
of soft gravitons is quantum. This is precisely why the two-loop \(H\)
diagram may contribute to a classical observable.

Finally, to summarize, there are two conceptually different ways in which the gravitational
kernel enters the rapidity evolution. The first is the ordinary BFKL
iteration at fixed number of Reggeons. In this case the longitudinal phase space leads therefore to the same ordered rapidity
integral encountered in QCD and yields \(L^n/n!\), as shown in
\refeq{eq:Cap4:ordered_rapidity_integrals}. This generates the maximal
power of \(L\) at every order and hence the usual leading logarithmic
BFKL tower, as in fig.~\ref{fig:Cap4:BFKL_MRK_kernel}.

This iteration leaves unchanged the number of independent
Glauber exchanges between the hard particles and, so, provides
additional rapidity logarithms, but no further eikonal enhancement by
\(Gs\). Up to transverse momentum factors, the resulting terms scale as
\begin{equation}\label{eq:Cap4:gravity_BFKL_scaling}
    \mathcal M^{(n)}_{\mathrm{BFKL}}
    \sim G^2s^3\bigl(GL\bigr)^n \, .
\end{equation}
For \(n=1\), the three-particle cut reconstructs the \(H\) diagram and
gives the classical 3PM term \(G^3s^3L\). For \(n>1\), further
iterations between the same Reggeons introduce additional soft graviton
loops without the extra power of \(s\) required by the classical
counting. Moreover, these terms, as well as insertions of the graviton Regge
trajectory, are quantum under the scaling in
\refeq{eq:Cap4:classical_hbar_scaling}, therefore they cannot give a classical contribution.

The second iteration increases instead the number of Reggeons, so
the number of parallel eikonal graviton exchanges in the \(t\) channel.
An additional disconnected Reggeon would only reproduce the expansion of
the eikonal phase and hence a superclassical, two-particle reducible,
iteration. The connected classical contribution is obtained by joining
neighbouring Reggeons through the real kernel and selecting the maximal
sequence of three-particle \(s\)-channel cuts. In this way every new
\(H\) unit contributes not only a rapidity logarithm, but also the
additional high-energy factor \(Gs\) that was absent from the ordinary
BFKL iteration. The resulting topology is the multi-\(H\) sequence shown
in Fig.~\ref{fig:Cap4:multi_H_gravity}.

With \(N+1\) Reggeons connected by \(N\) kernels, the classical leading
logarithm appears at \((2N+1)\)PM and obeys
\begin{equation}\label{eq:Cap4:multi_H_classical_scaling}
    \mathcal M_{H^N}\sim
    Gs^2\bigl(G^2sL\bigr)^N\sim
    G^{2N+1}s^{N+2}L^N \, .
\end{equation}
The first member, \(N=1\), is the \(H\) diagram at 3PM; \(N=2\) is the
double-\(H\) diagram at 5PM and scales as \(G^5s^4L^2\). In contrast to
the terms with several emissions in \refeq{eq:Cap4:gravity_BFKL_scaling},
every extra Reggeon in \refeq{eq:Cap4:multi_H_classical_scaling} brings
the eikonal enhancement required to preserve the classical \(\hbar\)
counting. The multi-\(H\) sequence therefore gives the highest classical
power \(L^N\) at each odd PM order\footnote{A new interesting systematic approach to classical gravity using SCET is provided in Ref.~\citeall{rothstein2026systematiclagrangianformulationquantum}.}.

Hence the gravitational kernel has two distinct roles. At fixed number of Reggeons, its rapidity iteration yields the BFKL tower, which is quantum
beyond the first \(H\) term. Between states with progressively more
Reggeons, it instead constructs the classical multi-\(H\) series. At
next-to-leading logarithmic accuracy, the tree-level two-graviton CEVs in
\refeq{eq:Cap4:two_graviton_CEV_pp} and
\refeq{eq:Cap4:two_graviton_CEV_pm} provide the corresponding
two-emission blocks, provided that the emitted gravitons are not closed
into another loop. Moreover, when dealing with corrections to the leading logarithm, we must also account for contributions originating form collinear regions, that we will discuss in the next chapter. Gravity thus retains the BFKL kernel and its
double-copy origin, but changes the QCD hierarchy, because the Reggeised ladder
is dominant in QCD, whereas classical high-energy gravity is organised by
eikonal exchange and the multi-\(H\) expansion.

\chapter{Gravitational amplitudes in the collinear limit}
\label{cap:Gravitational_amp_in_the_coll_limit}

In this chapter we will describe the interplay between high-energy and
collinear factorisation in gravity. We begin with the all-graviton MHV
amplitudes introduced in Sec.~\ref{subsec:Hodges_formula}. Their NMRK limit
will allow us to extract the forward correction to the gravitational
impact factor defined in Sec.~\ref{sec:Cap4:Gravitational_BFKL_equation}.
We  will then discuss the collinear regime and derive the corresponding
graviton splitting amplitudes, making explicit the relation with the
general factorisation formula discussed in
Sec.~\ref{Cap3:subsec:collinear_limit}.

In the second part of the chapter, the BCFW recursion relations reviewed
in Sec.~\ref{sec:Cap3:BCFW} will be used to construct the tree-level
five-point amplitude for two massive scalars scattering with the emission
of one graviton. Its high-energy and collinear limits will provide the
massive-scalar analogue of the impact-factor correction. Finally, the
formalism of Appendix~\ref{app:Classical_limit} will be used to extract the
tree-level waveform in the high-energy collinear regime.

\section{Graviton amplitudes in the collinear region}
\label{sec:Cap5:graviton_amplitudes_collinear_region}

We first consider the five-graviton MHV amplitude in a forward NMRK
configuration. In this region, the two final-state gravitons \(3\) and
\(4\) have comparable rapidities and are produced close to the direction
of the incoming graviton \(2\). The same pair will subsequently be taken
to the collinear region, providing a direct connection between the
NMRK impact factor and the universal splitting amplitude.

\subsection{The all-graviton impact factor in NMRK}
\label{subsec:Cap5:all_graviton_NMRK_impact_factor}

Our starting point is the five-graviton MHV amplitude derived in
\refeq{eq:Cap3:five_graviton_mhv_amplitude},
\begin{equation}\label{eq:Cap5:five_graviton_MHV_starting_point}
    \mathcal M_5
    \left(1_h^-,2_h^-,3_h^+,4_h^+,5_h^+\right)
    \quad=
    \kappa^3\langle12\rangle^8
    \frac{
    [15][24]\langle14\rangle\langle25\rangle
    -[14][25]\langle24\rangle\langle15\rangle}
    {\langle12\rangle\langle23\rangle\langle31\rangle
    \langle34\rangle\langle45\rangle\langle53\rangle
    \langle14\rangle\langle25\rangle
    \langle24\rangle\langle15\rangle} \, .
\end{equation}
It is useful to denote the two terms in the numerator by
\(\mathcal M_5^{(I)}\) and \(\mathcal M_5^{(II)}\), respectively, so that
\begin{equation}\label{eq:Cap5:M5_two_contributions}
    \mathcal M_5=\mathcal M_5^{(I)}-\mathcal M_5^{(II)} \, .
\end{equation}

We choose the incoming momenta along the two light-cone directions and
take \(p_3^+\sim p_4^+\), while all transverse momenta remain of the same
order. The finite and positive longitudinal ratio
\begin{equation}\label{eq:Cap5:NMRK_X_definition}
    X=\frac{p_3^+}{p_4^+}
\end{equation}
therefore, as said in the previous chapter, parametrises this forward NMRK region. Using the
spinor-helicity conventions of Chap.~\ref{Cap2:scatteringamp},
the two contributions in \refeq{eq:Cap5:M5_two_contributions} reduce at
leading power to
\begin{equation}\label{eq:Cap5:M5_NMRK_two_contributions}
\begin{aligned}
    \mathcal M_5^{(I)}
    ={}&
    \kappa^3s^2
    \frac{
    p_{4\perp}^{*}|p_{5\perp}|^2(1+X)}
    {p_{3\perp}(p_{3\perp}-Xp_{4\perp})
    (-p_{5\perp}^{2})p_{4\perp}} \, ,
    \\
    \mathcal M_5^{(II)}
    ={}&
    \kappa^3s^2
    \frac{
    p_{4\perp}(p_{5\perp}^{*})^2(1+X)}
    {p_{3\perp}(p_{3\perp}-Xp_{4\perp})
    (-p_{5\perp}^{2})p_{4\perp}} \, .
\end{aligned}
\end{equation}
Combining them gives
\begin{equation}\label{eq:Cap5:five_graviton_NMRK_amplitude}
    \mathcal M_5^{\mathrm{NMRK}}
    ={}
    \frac{\kappa^3s^2}{t}
    \left(\frac{p_{5\perp}^{*}}{p_{5\perp}}\right)^2
    \frac{
    p_{5\perp}
    (p_{4\perp}^{*}p_{5\perp}
    -p_{4\perp}p_{5\perp}^{*})(1+X)}
    {p_{3\perp}(p_{3\perp}-Xp_{4\perp})p_{4\perp}} \, ,
\end{equation}
where
\begin{equation}
    t=-|p_{5\perp}|^2 \, .
\end{equation}
The extra power of \(s\) compared with the gauge-theory case is the same spin two
enhancement encountered in the four-point Regge factorisation in
\refeq{eq:Cap4:gravity_four_point_Regge_factorisation}.

\begin{figure}[H]
    \centering
    \includegraphics[width=0.34\textwidth]
    {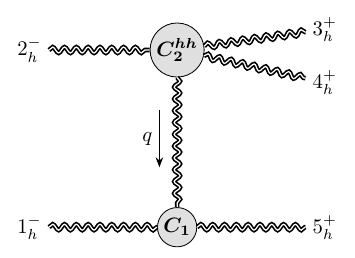}
    \caption{Forward NMRK factorisation of the five-graviton amplitude.
    The upper building block \(C_2^{hh}\) describes the transition
    \(2_h^-\to3_h^++4_h^+\), while \(C_1\) is the ordinary lower impact
    factor for \(1_h^-\to5_h^+\).}
    \label{fig:Cap5:five_point_graviton_NMRK}
\end{figure}

In the forward NMRK region, the two nearby gravitons are absorbed into a
higher point impact factor. We adopt the notation displayed in
fig.~\ref{fig:Cap5:five_point_graviton_NMRK} and write
\begin{equation}\label{eq:Cap5:NMRK_impact_factor_factorisation}
\begin{aligned}
    \mathcal M_5^{\mathrm{NMRK}}
    ={}&
    -\frac{\kappa^3s^2}{t}
    C_2^{hh}(2_h^-;3_h^+,4_h^+)
    C_1(1_h^-,5_h^+) \, .
\end{aligned}
\end{equation}

The ordinary lower impact factor is the phase already found in
\refeq{eq:Cap4:gravity_impact_factors},
\begin{equation}\label{eq:Cap5:lower_graviton_impact_factor}
    C_1(1_h^-,5_h^+)
    =
    \left(\frac{p_{5\perp}^{*}}{p_{5\perp}}\right)^2 \, ,
\end{equation}
whereas comparison with
\refeq{eq:Cap5:five_graviton_NMRK_amplitude} determines the corrected
upper impact factor,
\begin{equation}\label{eq:Cap5:two_graviton_NMRK_impact_factor}
\begin{aligned}
    C_2^{hh}(2_h^-;3_h^+,4_h^+)
    ={}&-
    \frac{
    p_{5\perp}
    (p_{4\perp}^{*}p_{5\perp}
    -p_{4\perp}p_{5\perp}^{*})(1+X)}
    {p_{3\perp}(p_{3\perp}-Xp_{4\perp})p_{4\perp}} \, .
\end{aligned}
\end{equation}
Although three transverse momenta appear in this expression, only two
are independent because the transverse momentum conservation gives
\(p_{3\perp}+p_{4\perp}+p_{5\perp}=0\), because of our choice of incoming momenta.

A useful consistency check is obtained by restoring a strong rapidity
ordering between gravitons \(3\) and \(4\). Remark that the
MRK limit corresponds to send \(X\to\infty\). The corrected impact factor must
then resolve into an ordinary impact factor and a one-graviton CEV joined
by an additional Glauber propagator,
\begin{equation}\label{eq:Cap5:NMRK_impact_factor_MRK_limit}
    C_2^{hh}(2_h^-;3_h^+,4_h^+)
    \xrightarrow{X\to\infty}
    C_2(2_h^-,3_h^+)
    \frac{1}{t_1}
    V(q_1,p_4^+,q_2) \, ,
\end{equation}
where \(C_2(2_h^-,3_h^+)=1\) and
\(t_1=-|p_{3\perp}|^2\). Thus
\refeq{eq:Cap5:NMRK_impact_factor_factorisation} continuously reduces to
the MRK factorisation of
\refeq{eq:Cap4:gravity_five_point_MRK_factorisation}, which diagrammatically is depicted in fig.~\ref{fig:five_point_MRK_graviton}.

\subsection{Collinear limit of the five-graviton amplitude}
\label{subsec:Cap5:five_graviton_collinear_limit}

We now let the two nearby  gravitons become collinear. In the
strict collinear limit,
\begin{equation}\label{eq:Cap5:strict_collinear_momenta}
    p_3=zP \, ,
    \qquad
    p_4=(1-z)P \, ,
    \qquad
    P^2=0 \, ,
\end{equation}
where \(0<z<1\). In terms of the NMRK variable introduced above,
\begin{equation}\label{eq:Cap5:z_X_relation}
    z=\frac{p_3^+}{p_3^++p_4^+}
    =\frac{X}{1+X} \, .
\end{equation}
The universal tree-level factorisation reviewed in
Sec.~\ref{Cap3:subsec:collinear_limit} becomes
\begin{equation}\label{eq:Cap5:five_graviton_collinear_factorisation}
    \mathcal M_5
    (1_h^{\nu_1},2_h^{\nu_2},3_h^{\nu_3},
    4_h^{\nu_4},5_h^{\nu_5})
   \xrightarrow{\;3\parallel4 \;}
    \sum_{\nu=\pm2}
    \mathcal M_4
    (1_h^{\nu_1},2_h^{\nu_2},P_h^{\nu},5_h^{\nu_5})
    \mathrm{Split}_{-\nu}
    (z,3_h^{\nu_3},4_h^{\nu_4}) \, .
\end{equation}
For the MHV configuration in
\refeq{eq:Cap5:five_graviton_MHV_starting_point}, only the positive helicity
parent contributes, because the alternative reduced amplitude contains just one 
positive helicity gravitons and vanishes at tree level.

The strict collinear spinors would be
\begin{equation}\label{eq:Cap5:strict_collinear_spinors}
    |3\rangle=\sqrt z\,|P\rangle \, ,
    \qquad
    |4\rangle=\sqrt{1-z}\,|P\rangle \, .
\end{equation}
However we will see now why another parametrisation is required. Substituting
\refeq{eq:Cap5:strict_collinear_spinors} into the two pieces of the
five point amplitude and extracting
\(\mathcal M_4(1_h^-,2_h^-,P_h^+,5_h^+)\) gives
\begin{equation}\label{eq:Cap5:naive_collinear_M5I}
\begin{aligned}
    \mathcal M_5^{(I)}
    \xrightarrow{3\parallel4}{}&
    \mathcal M_4(1_h^-,2_h^-,P_h^+,5_h^+)
    \frac{[15]\langle25\rangle[2P]}
    {\langle2P\rangle\langle34\rangle[12]}
    \frac{1}{z\sqrt z\sqrt{1-z}}
    \\
    ={}&
    \mathcal M_4(1_h^-,2_h^-,P_h^+,5_h^+)
    \frac{[1P][P2]}{\langle34\rangle[12]}
    \frac{1}{z\sqrt z\sqrt{1-z}} \, .
\end{aligned}
\end{equation}
Here we used the outgoing momentum conservation,
\(p_1+p_2+P+p_5=0\), which implies
\(-[15]\langle52\rangle=[1P]\langle P2\rangle\).
The second contribution similarly gives
\begin{equation}\label{eq:Cap5:naive_collinear_M5II}
\begin{aligned}
    \mathcal M_5^{(II)}
    \xrightarrow{3\parallel4}{}&
    \mathcal M_4(1_h^-,2_h^-,P_h^+,5_h^+)
    \frac{[1P][25]\langle15\rangle}
    {\langle1P\rangle\langle34\rangle[12]}
    \frac{1}{z\sqrt z\sqrt{1-z}}
    \\
    ={}&
    \mathcal M_4(1_h^-,2_h^-,P_h^+,5_h^+)
    \frac{[1P][P2]}{\langle34\rangle[12]}
    \frac{1}{z\sqrt z\sqrt{1-z}} \, ,
\end{aligned}
\end{equation}
after having used the same momentum conservation that, in this case, implied \(-[25]\langle51\rangle=[2P]\langle P1\rangle\).          
The two expressions are identical and would appear to cancel in
\(\mathcal M_5^{(I)}-\mathcal M_5^{(II)}\). There is however an elephant in the room,
because the same strict substitution also sets
\(\langle34\rangle\propto\langle PP\rangle=0\), so the substitution in \refeq{eq:Cap5:strict_collinear_spinors} is actually ill defined. The cancellation is
therefore of the indeterminate form \(0/0\), and the first displacement
away from exact collinearity must be retained.

\begin{figure}[H]
    \centering
    \includegraphics[width=0.48\textwidth]{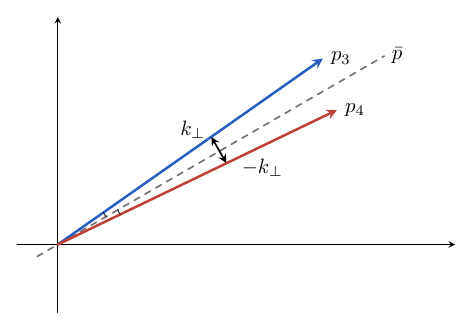}
    \caption{Sudakov parametrisation of the nearly collinear momenta
    \(p_3\) and \(p_4\). Their common null direction is \(\bar p\), while
    the relative transverse momenta are \(k_\perp\) and \(-k_\perp\).}
    \label{fig:Cap5:Sudakov_collinear_momenta}
\end{figure}

We regulate the limit through a two-step Sudakov decomposition. Let
\(\widetilde p^\mu\) and \(n^\mu\) be null reference vectors along the plus
and minus directions respectively, and let \(v_\perp^\mu\) be transverse to both. A
generic on-shell momentum along the common collinear direction is
\begin{equation}\label{eq:Cap5:Sudakov_pbar}
    \bar p^\mu
    =
    y\widetilde p^{\,\mu}+v_\perp^\mu
    -\frac{v_\perp^2}{2y\widetilde p\cdot n}\,n^\mu \, ,
    \qquad
    \bar p^2=0 \, .
\end{equation}
The direction is generic with respect to the beam, so
\(y\), \(v_\perp\), \(\widetilde p\) and \(n\) are all kept fixed in the
collinear limit. The small variable is instead the relative transverse
momentum, \(k_\perp\), shown in
fig.~\ref{fig:Cap5:Sudakov_collinear_momenta}.

The two on-shell momenta are parametrised as
\begin{equation}\label{eq:Cap5:Sudakov_p3_p4}
\begin{aligned}
    p_3^\mu
    ={}&
    zy\widetilde p^{\,\mu}+zv_\perp^\mu+k_\perp^\mu
    -\frac{(zv_\perp+k_\perp)^2}
    {2zy\widetilde p\cdot n}\,n^\mu \, ,
    \\
    p_4^\mu
    ={}&
    (1-z)y\widetilde p^{\,\mu}+(1-z)v_\perp^\mu-k_\perp^\mu
   -
    \frac{((1-z)v_\perp-k_\perp)^2}
    {2(1-z)y\widetilde p\cdot n}\,n^\mu \, .
\end{aligned}
\end{equation}
The components proportional to \(n^\mu\) ensure the on-shell conditions
\(p_3^2=p_4^2=0\). Using \refeq{eq:Cap5:Sudakov_pbar}, the same relations
can be reorganised as
\begin{equation}\label{eq:Cap5:Sudakov_p3_p4_pbar}
\begin{aligned}
    p_3^\mu
    ={}&z\bar p^\mu+k_\perp^\mu
    -\frac{v_\perp\cdot k_\perp}{y\widetilde p\cdot n}\,n^\mu
    -\frac{k_\perp^2}{2zy\widetilde p\cdot n}\,n^\mu \, ,
    \\
    p_4^\mu
    ={}&(1-z)\bar p^\mu-k_\perp^\mu
    +\frac{v_\perp\cdot k_\perp}{y\widetilde p\cdot n}\,n^\mu
    -
    \frac{k_\perp^2}{2(1-z)y\widetilde p\cdot n}\,n^\mu \, .
\end{aligned}
\end{equation}
Consequently, the momentum of the pair is off shell only at quadratic
order in the relative transverse momentum,
\begin{equation}\label{eq:Cap5:collinear_parent_off_shellness}
\begin{aligned}
    P^\mu=p_3^\mu+p_4^\mu
    ={}&
    \bar p^\mu
    -\frac{k_\perp^2}
    {2z(1-z)y\widetilde p\cdot n}\,n^\mu \, ,
    \\
    P^2={}&-\frac{k_\perp^2}{z(1-z)}
    =\mathcal O(k_\perp^2) \, .
\end{aligned}
\end{equation}
Thus \(\bar p\), rather than \(P\), is the on-shell null momentum that enters the
reduced four-point amplitude at the order required below.

The spinor expansion can be obtained directly from our 
convention. For \(p_3^+=zy\widetilde p^+\) and
\(p_{3\perp}=zv_\perp+k_\perp\), one has
\begin{equation}\label{eq:Cap5:p3_spinor_Sudakov_decomposition}
\begin{aligned}
    \xi_+(p_3)
    ={}&
    \frac{1}{\sqrt{p_3^+}}
    \binom{p_3^+}{p_{3\perp}}
    =
    \frac{1}{\sqrt{zy\widetilde p^+}}
    \binom{zy\widetilde p^+}{zv_\perp+k_\perp}
    \\
    ={}&
    \frac{1}{\sqrt{zy\widetilde p^+}}
    \binom{zy\widetilde p^+}{zv_\perp}
    +
    \frac{1}{\sqrt{zy\widetilde p^+}}
    \binom{0}{k_\perp}
    \\
    ={}&
    \frac{\sqrt z}{\sqrt{y\widetilde p^+}}
    \binom{y\widetilde p^+}{v_\perp}
    +
    \frac{k_\perp}
    {\sqrt z\sqrt{y\widetilde p^+n^-}}
    \binom{0}{\sqrt{n^-}}
    \\
    ={}&
    \sqrt z\,\xi_+(\bar p)
    +
    \frac{k_\perp}
    {\sqrt z\sqrt{y\widetilde p^+n^-}}\,
    \xi_+(n) \, .
\end{aligned}
\end{equation}
This shows explicitly that \(\xi_+(p_3)\) is decomposed on the two massless
spinors associated with \(\bar p\) and \(n\). Since \(n^\mu\) has only a
non vanishing minus light-cone component,
\(\sqrt{y\widetilde p^+n^-}
=\sqrt{\bar p^+n^-}=\langle\bar p n\rangle\), after fixing the phase of
the reference spinors so that this bracket is real. Applying the same
decomposition to \(p_4\), and then taking the conjugate spinors, gives
\begin{equation}\label{eq:Cap5:Sudakov_angle_spinors}
\begin{aligned}
    |3\rangle
    ={}&
    \sqrt z\,|\bar p\rangle
    +\frac{k_\perp}{\sqrt z\langle\bar p n\rangle}|n\rangle \, ,
    \\
    |4\rangle
    ={}&
    \sqrt{1-z}\,|\bar p\rangle
    -\frac{k_\perp}{\sqrt{1-z}\langle\bar p n\rangle}|n\rangle \, ,
\end{aligned}
\end{equation}
and, consequently,
\begin{equation}\label{eq:Cap5:Sudakov_square_spinors}
\begin{aligned}
    |3]
    ={}&
    \sqrt z\,|\bar p]
    +\frac{k_\perp^{*}}{\sqrt z\langle\bar p n\rangle}|n] \, ,
    \\
    |4]
    ={}&
    \sqrt{1-z}\,|\bar p]
    -\frac{k_\perp^{*}}{\sqrt{1-z}\langle\bar p n\rangle}|n] \, .
\end{aligned}
\end{equation}
In particular,
\begin{equation}\label{eq:Cap5:angle_34_collinear_expansion}
    \langle34\rangle
    =-\frac{k_\perp}{\sqrt{z(1-z)}}
    +\mathcal O(k_\perp^2) \, ,
\end{equation}
so the small bracket that was ill defined in the strict limit is now
under perturbative control.

We next insert \refeq{eq:Cap5:Sudakov_angle_spinors} and
\refeq{eq:Cap5:Sudakov_square_spinors} into the two terms of the MHV
amplitude. At first order in \(k_\perp\), corrections from the common
denominators are identical in
\(\mathcal M_5^{(I)}\) and \(\mathcal M_5^{(II)}\) and therefore cancel in
their difference. The non trivial contribution is supplied by the
numerators. Factoring out the reduced four-point amplitude gives
\begin{equation}\label{eq:Cap5:M5I_Sudakov_expansion}
\begin{aligned}
    \mathcal M_5^{(I)}
    ={}&
    \mathcal M_4(1_h^-,2_h^-,\bar p_h^+,5_h^+)
    \frac{[1\bar p][\bar p2]}
    {\langle34\rangle[12]}
    \frac{1}{z\sqrt z\sqrt{1-z}}
    \\
    &\times\left[
    1-
    \frac{k_\perp\langle1n\rangle}
    {(1-z)\langle\bar p n\rangle\langle1\bar p\rangle}
    -
    \frac{k_\perp^*[2n]}
    {(1-z)\langle\bar p n\rangle[2\bar p]}
    +\mathcal O(k_\perp^2)
    \right] \, ,
\end{aligned}
\end{equation}
while
\begin{equation}\label{eq:Cap5:M5II_Sudakov_expansion}
\begin{aligned}
    \mathcal M_5^{(II)}
    ={}&
    \mathcal M_4(1_h^-,2_h^-,\bar p_h^+,5_h^+)
    \frac{[1\bar p][\bar p2]}
    {\langle34\rangle[12]}
    \frac{1}{z\sqrt z\sqrt{1-z}}
    \\
    &\times\left[
    1-
    \frac{k_\perp\langle2n\rangle}
    {(1-z)\langle\bar p n\rangle\langle2\bar p\rangle}
    -
    \frac{k_\perp^*[1n]}
    {(1-z)\langle\bar p n\rangle[1\bar p]}
    +\mathcal O(k_\perp^2)
    \right] \, .
\end{aligned}
\end{equation}
The leading term cancels between the two contributions, while the terms
linear in \(k_\perp\) combine with the pole in
\refeq{eq:Cap5:angle_34_collinear_expansion}. The finite result is therefore
\begin{equation}\label{eq:Cap5:M5_collinear_Sudakov_result}
\begin{aligned}
    \mathcal M_5
    \xrightarrow{3\parallel4}{}&
    \mathcal M_4(1_h^-,2_h^-,\bar p_h^+,5_h^+)
    \frac{[1\bar p][2\bar p]}{[12]}
    \frac{1}{z(1-z)k_\perp}
    \\
    &\times\Bigg[
    \frac{k_\perp}{\langle\bar p n\rangle}
    \left(
    \frac{\langle2n\rangle}{\langle2\bar p\rangle}
    -\frac{\langle1n\rangle}{\langle1\bar p\rangle}
    \right)
    -
    \frac{k_\perp^*}{\langle\bar p n\rangle}
    \left(
    \frac{[2n]}{[2\bar p]}
    -\frac{[1n]}{[1\bar p]}
    \right)
    \Bigg] \, .
\end{aligned}
\end{equation}
Comparison with
\refeq{eq:Cap5:five_graviton_collinear_factorisation} identifies
\begin{equation}\label{eq:Cap5:graviton_split_redundant_form}
\begin{aligned}
    \mathrm{Split}_{-}(z,3_h^+,4_h^+)
    ={}&
    \frac{[1\bar p][2\bar p]}{[12]}
    \frac{1}{z(1-z)k_\perp}
    \\
    &\times\Bigg[
    \frac{k_\perp}{\langle\bar p n\rangle}
    \left(
    \frac{\langle2n\rangle}{\langle2\bar p\rangle}
    -\frac{\langle1n\rangle}{\langle1\bar p\rangle}
    \right)
   -
    \frac{k_\perp^*}{\langle\bar p n\rangle}
    \left(
    \frac{[2n]}{[2\bar p]}
    -\frac{[1n]}{[1\bar p]}
    \right)
    \Bigg] \, .
\end{aligned}
\end{equation}
This intermediate expression seems to depend on the hard momenta \(p_1\)
and \(p_2\), as well as on the auxiliary vector \(n\). However universality
requires all such dependence to cancel, and indeed if we expand the spinor products in
light-cone components we reach the compact result
\begin{equation}\label{eq:Cap5:graviton_split_pp_compact}
\begin{aligned}
    \mathrm{Split}_{-}(z,3_h^+,4_h^+)
    ={}&-\frac{1}{z(1-z)}
    \frac{\bar p_\perp^*k_\perp-k_\perp^*\bar p_\perp}
    {\bar p_\perp k_\perp}
    \\
    ={}&-\frac{1}{z(1-z)}
    \left(
    \frac{\bar p_\perp^*}{\bar p_\perp}
    -\frac{k_\perp^*}{k_\perp}
    \right) \, ,
\end{aligned}
\end{equation}
which is of course symmetric under the swap $z \iff 1-z$ because the two emitted gravitons have the same helicities and our choice of  $z$ rather than $1-z$ in \refeq{eq:Cap5:Sudakov_p3_p4} was arbitrary.

The remaining independent helicity configurations follow without
repeating the Sudakov expansion. The Hodges representation implies that (\refeq{eq:Cap3:graviton_mhv_hodges_factorisation})
an MHV amplitude can be written as
\begin{equation}\label{eq:Cap5:Hodges_helicity_factor}
    \mathcal M_n(1,\ldots,n)
    =\langle ij\rangle^8
    \widetilde{\mathcal M}_n(1,\ldots,n) \, ,
\end{equation}
where \(i\) and \(j\) label the negative-helicity gravitons and
\(\widetilde{\mathcal M}_n\) is independent of their choice. Changing the
helicity assignment from
\((1^-,2^-,3^+,4^+,5^+)\) to
\((1^-,2^+,3^-,4^+,5^+)\) therefore gives the chain
\begin{equation}\label{eq:Cap5:Hodges_collinear_helicity_chain}
\begin{aligned}
    &\mathcal M_5(1_h^-,2_h^+,3_h^-,4_h^+,5_h^+)
    \\
    &\quad=
    \frac{\langle13\rangle^8}{\langle12\rangle^8}
    \mathcal M_5(1_h^-,2_h^-,3_h^+,4_h^+,5_h^+)
    \\
    &\quad\xrightarrow{3\parallel4}
    \frac{\langle13\rangle^8}{\langle12\rangle^8}
    \mathcal M_4(1_h^-,2_h^-,\bar p_h^+,5_h^+)
    \times
    \mathrm{Split}_{-}(z,3_h^+,4_h^+)
    \\
    &\quad=
    \frac{\langle13\rangle^8}{\langle12\rangle^8}
    \frac{\langle12\rangle^8}{\langle1\bar p\rangle^8}
    \mathcal M_4(1_h^-,2_h^+,\bar p_h^-,5_h^+)
    \times
    \mathrm{Split}_{-}(z,3_h^+,4_h^+) \, .
\end{aligned}
\end{equation}
The last equality follows by applying the same Hodges relation to the
four-point amplitude. On the other hand, direct collinear factorisation
of the helicity configuration on the first line gives
\begin{equation}\label{eq:Cap5:direct_mp_collinear_factorisation}
    \mathcal M_5(1_h^-,2_h^+,3_h^-,4_h^+,5_h^+)
    \xrightarrow{3\parallel4}
    \mathcal M_4(1_h^-,2_h^+,\bar p_h^-,5_h^+)
    \mathrm{Split}_{+}(z,3_h^-,4_h^+) \, .
\end{equation}
Comparison of the two factorised expressions therefore yields
\begin{equation}\label{eq:Cap5:graviton_split_mp_from_pp}
\begin{aligned}
    \mathrm{Split}_{+}(z,3_h^-,4_h^+)
    ={}&
    \frac{\langle13\rangle^8}
    {\langle1\bar p\rangle^8}
    \mathrm{Split}_{-}(z,3_h^+,4_h^+)
    \\
    ={}&z^4\,
    \mathrm{Split}_{-}(z,3_h^+,4_h^+)
    \\
    ={}&-\frac{z^3}{1-z}
    \left(
    \frac{\bar p_\perp^*}{\bar p_\perp}
    -\frac{k_\perp^*}{k_\perp}
    \right) \, .
\end{aligned}
\end{equation}
Here we used that 
\(p_3^+=z\bar p^+\). The
 configuration with all positive-helicity gravitons and a negative-helicity parent instead
vanishes,
\begin{equation}\label{eq:Cap5:vanishing_graviton_split}
    \mathrm{Split}_{+}(z,3_h^+,4_h^+)=0 \, ,
\end{equation}
as a consequence of the vanishing of all plus and single-minus graviton
tree amplitudes, which is just a direct consequence of the angular momentum conservation in gravity. The parity conjugate splitting amplitudes are obtained
by complex conjugation.

Finally, the transverse components of
\refeq{eq:Cap5:Sudakov_p3_p4_pbar} can be inverted to express the result
directly in terms of the daughter momenta,
\begin{equation}\label{eq:Cap5:k_pbar_from_p3_p4}
    k_\perp=(1-z)p_{3\perp}-zp_{4\perp} \, ,
    \qquad
    \bar p_\perp=p_{3\perp}+p_{4\perp} \, .
\end{equation}
The independent splitting amplitudes therefore take the form
\begin{equation}\label{eq:Cap5:final_graviton_splitting_amplitudes}
\begin{aligned}
    \mathrm{Split}_{-}(z,3_h^+,4_h^+)
    ={}&-\frac{1}{z(1-z)}
    \Bigg[
    \frac{p_{3\perp}^*+p_{4\perp}^*}
    {p_{3\perp}+p_{4\perp}}
    -
    \frac{(1-z)p_{3\perp}^*-zp_{4\perp}^*}
    {(1-z)p_{3\perp}-zp_{4\perp}}
    \Bigg] \, ,
    \\
    \mathrm{Split}_{+}(z,3_h^-,4_h^+)
    ={}&-\frac{z^3}{1-z}
    \Bigg[
    \frac{p_{3\perp}^*+p_{4\perp}^*}
    {p_{3\perp}+p_{4\perp}}
   -
    \frac{(1-z)p_{3\perp}^*-zp_{4\perp}^*}
    {(1-z)p_{3\perp}-zp_{4\perp}}
    \Bigg] \, ,
    \\
    \mathrm{Split}_{+}(z,3_h^+,4_h^+)
    ={}&0 \, .
\end{aligned}
\end{equation}
Unlike the gauge-theory splitting amplitudes in
\refeq{eq:Cap3:gluon_splitting_functions}, these expressions contain no
pole in the magnitude of the relative transverse momentum. The ratios
in brackets are pure phases, so the direction of the collinear approach
is retained but the splitting amplitudes remain finite as
\(|k_\perp|\to0\). Gravity therefore has no collinear divergence at tree
level. This gives the gravitational collinear limit a slightly different
interpretation from its gauge-theory counterpart since the universal
splitting term is finite, so different terms can survive when we send \(s_{34}\to0\). Nevertheless,
the universal contribution can still be isolated, indeed under a rotation of
the two daughter momenta around their sum \(P\), the splitting amplitude
acquires different phases, according to the expressions displayed above. Thus the directional dependence is not a breakdown of
universality, but rather the property that distinguishes the distinct universal
collinear contributions in the finite terms, as pointed out in Ref.~\citeall{Bern_1999}. 

\section{The five-point massive-scalar amplitude from on-shell recursion}
\label{sec:Cap5:five_point_massive_scalar_BCFW}

\begin{figure}[H]
    \centering
    \includegraphics[width=0.44\textwidth]
    {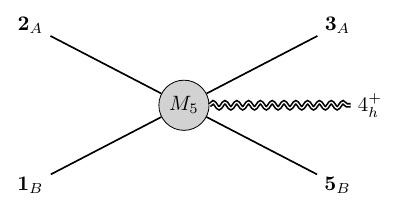}
    \caption{Generic five-point tree amplitude involving two massive
    scalars of flavour \(A\), two massive scalars of flavour \(B\), and
    one positive helicity graviton.}
    \label{fig:Cap5:generic_massive_five_point_amplitude}
\end{figure}

We now turn to the tree-level process involving two distinguishable massive
scalars and the emission of one graviton,
\begin{equation}\label{eq:Cap5:massive_five_point_process}
    \mathcal M_5^{(0)}
    \left(
    \mathbf 1_B,\mathbf 2_A,\mathbf 3_A,4_h^+,\mathbf 5_B
    \right) \, .
\end{equation}
The corresponding on-shell diagram is shown in
fig.~\ref{fig:Cap5:generic_massive_five_point_amplitude}.
Using the outgoing convention throughout, we have
\begin{equation}\label{eq:Cap5:massive_five_point_kinematics}
\begin{gathered}
    p_1+p_2+p_3+p_4+p_5=0 \, ,
    \qquad
    p_1^2=p_5^2=m_B^2 \, ,
    \\
    p_2^2=p_3^2=m_A^2 \, ,
    \qquad
    p_4^2=0 \, .
\end{gathered}
\end{equation}
Bold labels will be used to distinguish the massive scalar legs from the
massless graviton. The amplitude was constructed through on-shell recursion in
Ref.~\citeall{Britto_2022}, which we will follow. 

\subsection{Equal-mass shift and factorisation channels}
\label{subsec:Cap5:equal_mass_shift}

The standard BCFW construction reviewed in Sec.~\ref{sec:Cap3:BCFW}
preserves the on-shell condition of each massless shifted state, while here there is only one
massless leg, with two massive scalars \(A\) and $B$. A useful deformation is therefore the three-line equal-mass
shift
\begin{equation}\label{eq:Cap5:equal_mass_shift}
\begin{aligned}
    &\mathcal M_5^{(0)}
    (\mathbf 1_B,\mathbf 2_A,\mathbf 3_A,4_h^+,\mathbf 5_B)
    \longrightarrow
    \mathcal M_5^{(0)}
    (\mathbf 1_B,\widehat{\mathbf 2}_A,
    \widehat{\mathbf 3}_A,\widehat 4_h^+,\mathbf 5_B) \, ,
    \\
    &\widehat p_4
    =
    |\widehat 4\rangle[4|
    =
    \bigl(|4\rangle+z(p_2-p_3)|4]\bigr)[4| \, ,
    \\
    &\widehat p_2
    =
    p_2+z\,p_3|4][4| \, ,
    \qquad
    \widehat p_3
    =
    p_3-z\,p_2|4][4| \, .
\end{aligned}
\end{equation}
The deformation preserves total momentum. It does not keep the two
scalar masses separately fixed at their physical value for \(z\neq0\);
instead, it preserves their equality,
\begin{equation}\label{eq:Cap5:equal_shifted_masses}
    \widehat p_2^{\,2}
    =
    p_2^2+z[4|32|4]
    =
    m_A^2-z[4|23|4]
    =
    \widehat p_3^{\,2}
    \equiv
    \widehat m_A^{\,2} \, .
\end{equation}
Consequently, every lower-point object entering the recursion remains an
equal-mass on-shell amplitude, with the common mass regarded as a
shifted kinematic variable. The holomorphic deformation of
\(|4\rangle\) in \refeq{eq:Cap5:equal_mass_shift} has no boundary
contribution for this five-point amplitude. However the corresponding
anti-holomorphic deformation would have instead generated a boundary term, because, in that case, the polarisation tensor would not be power suppressed with $z$.

\begin{figure}[H]
    \centering
    \includegraphics[width=\textwidth]
    {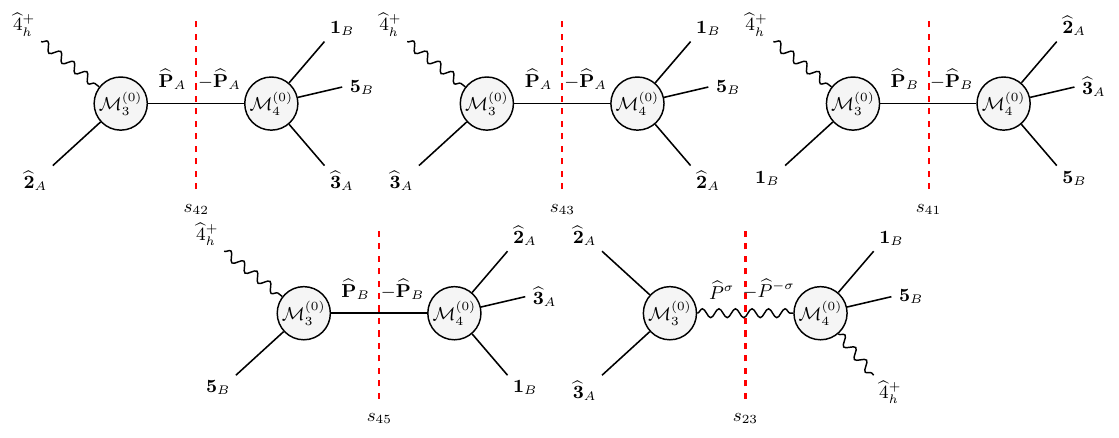}
    \caption{The five factorisation channels generated by the equal-mass
    shift.
    The first four channels exchange a massive scalar of flavour \(A\)
    or \(B\); the last channel exchanges a graviton and therefore
    includes a sum over \(\sigma=\pm2\).}
    \label{fig:Cap5:massive_five_point_BCFW_channels}
\end{figure}

The shifted amplitude has five factorisation channels. They are respectively \(s_{42}\), \(s_{43}\), \(s_{41}\),
\(s_{45}\), and \(s_{23}\), as displayed in
fig.~\ref{fig:Cap5:massive_five_point_BCFW_channels}.

Using the recursion relation in
\refeq{eq:Cap3:BCFW_recursion_general}, the two channels with an
intermediate scalar of flavour \(A\) are
\begin{equation}\label{eq:Cap5:BCFW_residues_A}
\begin{aligned}
    \left.i\mathcal M_5^{(0)}\right|_{42}
    ={}&
    i\mathcal M_3^{(0)}
    (\widehat{\mathbf 2}_A,\widehat{\mathbf P}_A,
    \widehat 4_h^+)
    \frac{i}{s_{42}-m_A^2}
    i\mathcal M_4^{(0)}
    (-\widehat{\mathbf P}_A,\mathbf 1_B,
    \widehat{\mathbf 3}_A,\mathbf 5_B) \, ,
    \\
    \left.i\mathcal M_5^{(0)}\right|_{43}
    ={}&
    i\mathcal M_3^{(0)}
    (\widehat{\mathbf 3}_A,\widehat{\mathbf P}_A,
    \widehat 4_h^+)
    \frac{i}{s_{43}-m_A^2}
    i\mathcal M_4^{(0)}
    (-\widehat{\mathbf P}_A,\mathbf 1_B,
    \widehat{\mathbf 2}_A,\mathbf 5_B) \, .
\end{aligned}
\end{equation}
The two analogous channels with an intermediate scalar of flavour \(B\)
are
\begin{equation}\label{eq:Cap5:BCFW_residues_B}
\begin{aligned}
    \left.i\mathcal M_5^{(0)}\right|_{41}
    ={}&
    i\mathcal M_3^{(0)}
    (\mathbf 1_B,\widehat{\mathbf P}_B,\widehat 4_h^+)
    \frac{i}{s_{41}-m_B^2}
    i\mathcal M_4^{(0)}
    (-\widehat{\mathbf P}_B,\widehat{\mathbf 3}_A,
    \mathbf 5_B,\widehat{\mathbf 2}_A) \, ,
    \\
    \left.i\mathcal M_5^{(0)}\right|_{45}
    ={}&
    i\mathcal M_3^{(0)}
    (\mathbf 5_B,\widehat{\mathbf P}_B,\widehat 4_h^+)
    \frac{i}{s_{45}-m_B^2}
    i\mathcal M_4^{(0)}
    (-\widehat{\mathbf P}_B,\widehat{\mathbf 3}_A,
    \mathbf 1_B,\widehat{\mathbf 2}_A) \, .
\end{aligned}
\end{equation}
Finally, the \(s_{23}\) channel contains an intermediate graviton,
\begin{equation}\label{eq:Cap5:BCFW_residue_graviton}
\begin{aligned}
    \left.i\mathcal M_5^{(0)}\right|_{23}
    ={}&
    \sum_{\sigma=\pm2}
    i\mathcal M_3^{(0)}
    (\widehat{\mathbf 2}_A,\widehat{\mathbf 3}_A,
    \widehat P^\sigma)
    \frac{i}{s_{23}}
    i\mathcal M_4^{(0)}
    (\mathbf 1_B,\mathbf 5_B,\widehat 4_h^+,
    -\widehat P^{-\sigma}) \, .
\end{aligned}
\end{equation}
All hatted quantities in these expressions are evaluated at the pole of
the corresponding channel and their sum reconstructs the physical
amplitude at \(z=0\). Although the individual residues contain spurious
poles, these cancel after the five terms are combined.

\subsection{The reconstructed five-point amplitude}
\label{subsec:Cap5:reconstructed_massive_five_point_amplitude}

It is convenient to introduce
\begin{equation}\label{eq:Cap5:massive_five_point_shorthand}
\begin{gathered}
    K_A=p_2-p_3 \, ,
    \qquad
    K_B=p_1-p_5 \, ,
    \qquad
    s_{ij}=(p_i+p_j)^2 \, ,
    \\
    D_A=(s_{42}-m_A^2)(s_{43}-m_A^2) \, ,
    \qquad
    D_B=(s_{41}-m_B^2)(s_{45}-m_B^2) \, .
\end{gathered}
\end{equation}
Because all momenta are outgoing, \(K_A\) and \(K_B\) are not the
physical \(t\)-channel momentum transfers. We also use the compact
spinor strings
\begin{equation}\label{eq:Cap5:massive_five_point_spinor_strings}
\begin{gathered}
    [4|23|4]
    \equiv
    [4|\slashed p_2\slashed p_3|4] \, ,
    \qquad
    [4|51|4]
    \equiv
    [4|\slashed p_5\slashed p_1|4] \, ,
    \\
    [4|K_AK_B|4]
    \equiv
    [4|\slashed K_A\slashed K_B|4] \, .
\end{gathered}
\end{equation}

To present the result in a compact way, it's useful to organise its
rational structure into six terms. The amplitude, obtained just by BCFW recursion relations, is
\begin{equation}\label{eq:Cap5:massive_five_point_amplitude_compact}
\begin{aligned}
    &\mathcal M_5^{(0)}
    (\mathbf 1_B,\mathbf 2_A,\mathbf 3_A,4_h^+,\mathbf 5_B)
    \\
    &\quad=
    -\frac{\kappa^3}{2}
    \left\{
    \left[
    \sum_{a=1}^{6}\mathcal R_a
    \right]
    +
    \bigl(
    2,3,K_A,m_A
    \leftrightarrow
    1,5,K_B,m_B
    \bigr)
    \right\} \, .
\end{aligned}
\end{equation}
The first two contributions are
\begin{equation}\label{eq:Cap5:massive_five_point_R1_R2}
\begin{aligned}
    \mathcal R_1
    ={}&
    \frac{
    p_5\cdot p_1\,[4|23|4]^2}
    {s_{15}D_A} \, ,
    \\
    \mathcal R_2
    ={}&
    \frac{
    [4|K_AK_B|4]^2-8[4|23|4]^2}
    {16s_{23}s_{15}} \, ,
\end{aligned}
\end{equation}
while the term proportional to the sum of the masses is
\begin{equation}\label{eq:Cap5:massive_five_point_R3}
    \mathcal R_3
    ={}
    \frac{
    (m_A^2+m_B^2)[4|23|4]}
    {8D_AD_B}
    \left[
    2(s_{23}-s_{15})[4|23|4]
    +
    [4|K_B|4\rangle[4|K_AK_B|4]
    \right] \, .
\end{equation}
The next two terms are
\begin{equation}\label{eq:Cap5:massive_five_point_R4}
    \mathcal R_4
    ={}
    -\frac{
    K_A\cdot K_B\,
    (s_{15}-s_{23})^2[4|23|4]}
    {32s_{23}s_{15}D_AD_B}
    \left[
    4s_{15}[4|51|4]
    -
    [4|K_A|4\rangle[4|K_AK_B|4]
    \right] \, ,
\end{equation}
and
\begin{equation}\label{eq:Cap5:massive_five_point_R5}
\begin{aligned}
    \mathcal R_5
    &={}
    -\frac{
    K_A\cdot K_B\,[4|51|4]}
    {8s_{23}s_{15}D_B}
    \left[
    [4|K_B|4\rangle[4|K_AK_B|4]
    -
    4(s_{23}+s_{15})[4|23|4]
    \right]
    \\
    &-
    \frac{
    K_A\cdot K_B\,
    [4|K_A|4\rangle[4|K_B|4\rangle}
    {64s_{23}D_AD_B}
    \left(
    [4|K_AK_B|4]^2
    -
    8[4|51|4]^2
    \right) \, .
\end{aligned}
\end{equation}
For the last contribution, let's define
\begin{equation}\label{eq:Cap5:massive_five_point_trace_factor}
    \mathcal T
    ={}
    \operatorname{tr}_{2}
    (\slashed K_A\slashed K_B\slashed K_A\slashed K_B)
    +
    2[4|K_A|4\rangle^2
    +
    2[4|K_B|4\rangle^2
    -
    2s_{23}^2
    -
    2s_{15}^2 \, ,
\end{equation}
where \(\operatorname{tr}_{2}\) denotes the trace over the chiral
Weyl spinor indices, rather than the full Dirac trace. Then
\begin{equation}\label{eq:Cap5:massive_five_point_R6}
\begin{aligned}
    \mathcal R_6
    ={}&
    \mathcal T
    \Bigg[
    \frac{
    [4|K_A|4\rangle[4|K_B|4\rangle
    [4|23|4][4|51|4]}
    {64s_{23}s_{15}D_AD_B}
    \\
    &+
    \frac{
    [4|51|4]
    \left[
    2(s_{23}-s_{15})[4|51|4]
    -
    [4|K_A|4\rangle[4|K_AK_B|4]
    \right]}
    {64s_{23}D_AD_B}
    \Bigg] \, .
\end{aligned}
\end{equation}
The expression collected in
\refeq{eq:Cap5:massive_five_point_amplitude_compact}, with the six
contributions specified in
\refeq{eq:Cap5:massive_five_point_R1_R2},
\refeq{eq:Cap5:massive_five_point_R3},
\refeq{eq:Cap5:massive_five_point_R4},
\refeq{eq:Cap5:massive_five_point_R5}, and
\refeq{eq:Cap5:massive_five_point_R6}, reproduces the result of
Ref.~\citeall{Britto_2022}. The exchange operation in
\refeq{eq:Cap5:massive_five_point_amplitude_compact} interchanges the
two scalar lines, including their masses and the associated denominator
factors. This five-point expression will
be the starting point for the high-energy and collinear limits discussed
in the following sections.

\section{The massive-scalar impact factor and its collinear limit}
\label{sec:Cap5:massive_scalar_impact_factor_collinear_limit}

We now expand the five-point amplitude derived above in the forward
high-energy region. We take the outgoing scalar \(\mathbf 3_A\) and the
emitted graviton \(4_h^+\) to have comparable rapidities, while the
\(B\)-scalar remains in the opposite rapidity region. As in the
pure graviton case, their longitudinal separation is parametrised by
\begin{equation}\label{eq:Cap5:massive_NMRK_X_definition}
    X=\frac{p_3^+}{p_4^+} \, ,
    \qquad 0<X<\infty \, .
\end{equation}
The Regge expansion is performed at fixed \(X\), transverse momenta and
masses, with \(s\gg |t|,m_A^2,m_B^2\). Keeping the leading term, of order
\(s^2\), gives the following factorisation,
\begin{equation}\label{eq:Cap5:massive_five_point_NMRK_factorisation}
\begin{aligned}
    \mathcal M_5^{(0),\mathrm{NMRK}}
    ={}&
    -\frac{\kappa^3s^2}{t}\,
    C_1^{\phi}(\mathbf 1_B,\mathbf 5_B)\,
    C_2^{\phi h}(\mathbf 2_A;\mathbf 3_A,4_h^+)
    \\
    ={}&
    -\frac{\kappa^3s^2}{t}\,
    C^{\phi\to\phi h}
    (p_{3\perp},p_{4\perp},m_A^2,X) \, ,
\end{aligned}
\end{equation}
where the ordinary scalar impact factor is
\(C_1^{\phi}=1\), whereas the upper building block is
\(C_2^{\phi h}\equiv C^{\phi\to\phi h}\).

\begin{figure}[H]
    \centering
    \includegraphics[width=0.28\textwidth]
    {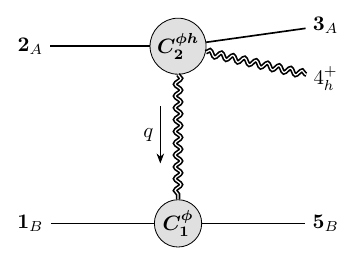}
    \caption{Forward NMRK factorisation of the five-point amplitude with
    massive scalars and one emitted graviton.}
    \label{fig:Cap5:massive_five_point_NMRK}
\end{figure}

To display the result compactly, we introduce
\begin{equation}\label{eq:Cap5:massive_impact_factor_denominators}
\begin{gathered}
    A_X=4X(X+2)+5 \, ,
    \\
    \mathcal D_3=m_A^2+(1+X)^2|p_{3\perp}|^2 \, ,
    \qquad
    \mathcal D_4=m_A^2+(1+X)^2|p_{4\perp}|^2\, ,
    \\
    \mathcal D_{34}
    =m_A^2+
    |(p_{3\perp}-Xp_{4\perp})|^2\, .
\end{gathered}
\end{equation}
The corrected NMRK impact factor then reads
\begin{equation}\label{eq:Cap5:massive_scalar_NMRK_impact_factor}
    C^{\phi\to\phi h}
    (p_{3\perp},p_{4\perp},m_A^2,X)
    =
    \frac{\mathcal N^{\phi\to\phi h}}
    {4(1+X)^2\mathcal D_3\mathcal D_4\mathcal D_{34}} \, ,
\end{equation}
where it is useful to separate its numerator according to the powers of
the conjugate transverse momentum $p_{3\perp}^*$,
\begin{equation}\label{eq:Cap5:massive_impact_factor_numerator_split}
    \mathcal N^{\phi\to\phi h}
    =\mathcal N_3+\mathcal N_2+\mathcal N_1+\mathcal N_0 \, .
\end{equation}
The four contributions are
\begin{equation}\label{eq:Cap5:massive_impact_factor_N3}
\begin{aligned}
    \mathcal N_3
    ={}&
    X^2(1+X)A_X(p_{3\perp}+p_{4\perp})
    (p_{3\perp}^*)^3\mathcal D_4 \, ,
\end{aligned}
\end{equation}
\begin{equation}\label{eq:Cap5:massive_impact_factor_N2}
\begin{aligned}
    \mathcal N_2
    ={}&-(p_{3\perp}^*)^2
    \Bigg\{
    (1+X)p_{4\perp}^*
    \Bigg[
    m_A^2X
    \Bigg(
    \bigl[X(2X+1)A_X+2\bigr]p_{3\perp}
    \\
    &
    +X\bigl\{2X[2X(2X+5)+9]+3\bigr\}p_{4\perp}
    \Bigg)
    \\
    &
    +(1+X)^2p_{4\perp}^*
    \Bigg(
    (1+X)^2(4X^2+1)p_{3\perp}^2
    +X^2(XA_X-2)p_{4\perp}^2
    \\
    &
    +X\bigl\{X[X(4X(X+3)+13)+3]+1\bigr\}
    p_{3\perp}p_{4\perp}
    \Bigg)
    \Bigg]
    +m_A^4X^2A_X
    \Bigg\} \, .
\end{aligned}
\end{equation}
The term proportional to \(p_{3\perp}^*p_{4\perp}^*\) is
\begin{equation}\label{eq:Cap5:massive_impact_factor_N1}
\begin{aligned}
    \mathcal N_1
    ={}&-p_{3\perp}^*p_{4\perp}^*
    \Bigg\{
    (1+X)p_{4\perp}^*
    \Bigg[
    m_A^2
    \Bigg(
    \bigl\{X[X[X(4X(X+4)+23)+13]+7]+2\bigr\}p_{3\perp}
    \\
    &
    +X(2X+1)\bigl\{X[X(2X+7)+9]+1\bigr\}p_{4\perp}
    \Bigg)
    \\
    &
    -X(1+X)^2p_{4\perp}^*
    \Bigg(
    (1+X)^2(4X^2+1)p_{3\perp}^2
    \\
    &
    +\bigl\{[4X(X+2)+3]X^2+X-1\bigr\}
    p_{3\perp}p_{4\perp}
    +X(1-2X)p_{4\perp}^2
    \Bigg)
    \Bigg]
    \\
    &\hspace{1.0cm}
    +2m_A^4X\bigl\{2X[2X(X+2)+3]+1\bigr\}
    \Bigg\} \, .
\end{aligned}
\end{equation}
Finally,
\begin{equation}\label{eq:Cap5:massive_impact_factor_N0}
\begin{aligned}
    \mathcal N_0
    ={}&(p_{4\perp}^*)^2
    \Bigg\{
    Xp_{4\perp}^*
    \Bigg[
    m_A^2(2X+1)(1+X)
    \Bigg(
    (2X+1)(X^2+X-1)p_{4\perp}
    \\
    &
    +\bigl[X(1+X)(2X+1)+1\bigr]p_{3\perp}
    \Bigg)
    +X(1+X)^3p_{4\perp}p_{4\perp}^*
    (p_{3\perp}-Xp_{4\perp})
    \Bigg]
    \\
    &\hspace{1.0cm}
    -m_A^4\bigl[2X(1+X)+1\bigr]^2
    \Bigg\} \, .
\end{aligned}
\end{equation}

The strong rapidity ordering limit provides a useful check on this expression.
Expanding the corrected result for \(X\to\infty\) gives
\begin{equation}\label{eq:Cap5:massive_impact_factor_large_X_expansion}
\begin{aligned}
    C^{\phi\to\phi h}
    ={}&\mathcal C^{(0)}
    +\frac{\mathcal C^{(1)}}{X}
    +\frac{\mathcal C^{(2)}}{X^2}
    +\mathcal O\!\left(\frac{1}{X^3}\right) \, ,
\end{aligned}
\end{equation}
with
\begin{equation}\label{eq:Cap5:massive_impact_factor_large_X_C0_C1}
\begin{aligned}
    \mathcal C^{(0)}
    ={}&
    \frac{(p_{3\perp}+p_{4\perp})
    (p_{3\perp}p_{4\perp}^*
    -p_{4\perp}p_{3\perp}^*)}
    {p_{3\perp}p_{4\perp}^2} \, ,
    \\
    \mathcal C^{(1)}
    ={}&
    \frac{(p_{3\perp}+p_{4\perp})
    (p_{3\perp}-p_{4\perp})
    (p_{3\perp}p_{4\perp}^*
    -p_{4\perp}p_{3\perp}^*)}
    {p_{3\perp}p_{4\perp}^3} \, .
\end{aligned}
\end{equation}
The first mass dependent correction occurs at the next order,
\begin{equation}\label{eq:Cap5:massive_impact_factor_large_X_C2}
\begin{aligned}
    \mathcal C^{(2)}
    ={}&
    \frac{p_{3\perp}^3p_{4\perp}^*}{p_{4\perp}^4}
    -\frac{p_{3\perp}^2p_{3\perp}^*}{p_{4\perp}^3}
    +\frac{p_{3\perp}p_{4\perp}^*}{4p_{4\perp}^2}
    -\frac{2m_A^2p_{3\perp}}{p_{4\perp}^3}
    -\frac{p_{3\perp}^*}{4p_{4\perp}}
    +\frac{3p_{4\perp}^*}{4p_{4\perp}}
    -\frac{3m_A^2}{p_{4\perp}^2}
    \\
    &
    -\frac{5p_{3\perp}^*}{4p_{3\perp}}
    -\frac{p_{4\perp}^*}{2p_{3\perp}}
    -\frac{(p_{4\perp}^*)^2}
    {4p_{3\perp}p_{3\perp}^*}
    +\frac{m_A^2}{p_{3\perp}^2} \, .
\end{aligned}
\end{equation}
Now, using
\begin{equation}\label{eq:Cap5:massive_NMRK_to_MRK_transverse_momenta}
    p_{3\perp}=-q_{1\perp} \, ,
    \qquad
    p_{4\perp}=q_{1\perp}-q_{2\perp} \, ,
    \qquad
    p_{5\perp}=q_{2\perp} \, ,
\end{equation}
the leading term reproduces an ordinary scalar impact factor (which is trivial) followed by a
one-graviton CEV,
\begin{equation}\label{eq:Cap5:massive_impact_factor_MRK_limit}
    \lim_{X\to\infty}C^{\phi\to\phi h}
    =\frac{V(q_1,p_4^+,q_2)}{t_1} \, ,
    \qquad
    t_1=-|q_{1\perp}|^2 \, .
\end{equation}
Thus the corrected NMRK building block has the required MRK limit,
while the inverse powers of \(X\) retain the information lost under a strict MRK ordering. Notice also that $C^{\phi\to\phi h}$ depends just on the mass of the upper 
massive scalar, close in rapidity with the emitted graviton.

For later use in the computation of classical observables, it is also
useful to take the classical limit of the impact factor itself. Following
the scaling introduced in Appendix~\ref{app:Classical_limit}, we keep the
mass and the rapidities fixed and write the transverse momenta transferred
to the massive particles and carried by the emitted graviton as
\(p_{i\perp}^{\mathrm{phys}}=\hbar p_{i\perp}\). In this limit the NMRK
variable is not held fixed, but scales according to
\begin{equation}\label{eq:Cap5:massive_impact_factor_classical_X_scaling}
    X
    =\frac{m_A}{\hbar|p_{4\perp}|}
    e^{y_2-y_4}-1
    =\frac{m_A}{\hbar|p_{4\perp}|}\,\Sigma-1 \, ,
    \qquad
    \Sigma\equiv e^{y_2-y_4} \, ,
\end{equation}
therefore in this classical soft expansion \(X\sim\hbar^{-1}\), while the
transverse momenta in the full impact factor are simultaneously soft.
If we expand \refeq{eq:Cap5:massive_scalar_NMRK_impact_factor} in $\hbar$ the leading order, which is the finite classical result, is
\begin{align}
    C_{\mathrm{cl}}^{\phi\to\phi h}
    ={}&
    \lim_{\hbar\to0}
    C^{\phi\to\phi h}
    \left(
    \hbar p_{3\perp},\hbar p_{4\perp},m_A^2,
    \frac{m_A\Sigma}{\hbar|p_{4\perp}|}-1
    \right)\label{eq:Cap5:massive_impact_factor_classical_limit}
    \\
    ={}&\Sigma^2
    \Bigg[
    \frac{\Sigma^2p_{3\perp}p_{4\perp}^*}
    {(\Sigma^2+1)^2p_{4\perp}^2}
    +\frac{(\Sigma^2-1)p_{4\perp}^*
    -(\Sigma^2+2)p_{3\perp}^*}
    {(\Sigma^2+1)^2p_{4\perp}}
    -\frac{(p_{3\perp}^*)^2}
    {\Sigma^2p_{3\perp}p_{3\perp}^*
    +p_{4\perp}p_{4\perp}^*}
    \Bigg] \, ,\notag
\end{align}

and, also in this case, we recover the MRK limit by sending $\Sigma \to \infty$, finding out again the usual Lipatov 
current in \refeq{eq:Cap5:massive_impact_factor_MRK_limit}.

\subsection{Massive scalar--graviton collinear limit}
\label{subsec:Cap5:massive_scalar_graviton_collinear_limit}

\begin{figure}[H]
    \centering
    \includegraphics[width=\textwidth]
    {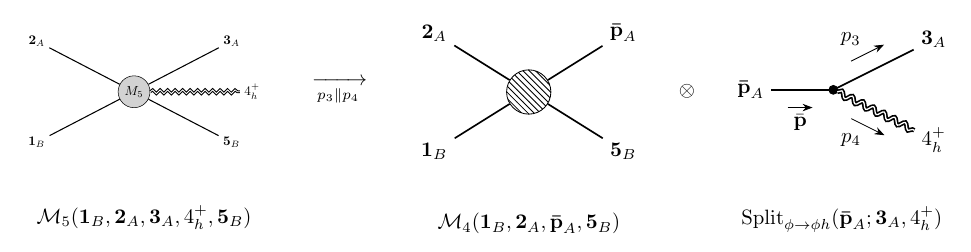}
    \caption{Collinear factorisation of the five-point amplitude when
    the outgoing scalar \(\mathbf 3_A\) and graviton \(4_h^+\) approach
    the common massive direction \({\bf{\bar {\mathbf p}}}_A\).}
    \label{fig:Cap5:massive_scalar_splitting_factorisation}
\end{figure}

We next take the outgoing scalar \(\mathbf 3_A\) and graviton \(4_h^+\)
to the collinear region. The factorisation relevant to the leading
high-energy term is represented in
fig.~\ref{fig:Cap5:massive_scalar_splitting_factorisation} and reads
\begin{equation}\label{eq:Cap5:massive_scalar_collinear_factorisation}
\begin{aligned}
    &\mathcal M_5^{(0)}
    (\mathbf 1_B,\mathbf 2_A,\mathbf 3_A,4_h^+,\mathbf 5_B)
    \\
    &\quad\xrightarrow{\,\mathbf 3_A\parallel4_h^+\,}
    \mathcal M_4^{(0)}
    (\mathbf 1_B,\mathbf 2_A,\bar{\mathbf p}_A,\mathbf 5_B)
    \,\mathrm{Split}_{\phi\to\phi h}
    (z,\mathbf 3_A,4_h^+) \, .
\end{aligned}
\end{equation}

At leading power in the Regge expansion, the reduced four-point
amplitude is
\begin{equation}\label{eq:Cap5:massive_four_point_high_energy_limit}
    \left.\mathcal M_4^{(0)}
    (\mathbf 1_B,\mathbf 2_A,\bar{\mathbf p}_A,\mathbf 5_B)
    \right|_{\mathcal O(s^2)}
    =-\kappa^2\frac{s^2}{s_{15}} \, .
\end{equation}
Consequently, after stripping the additional common power of \(\kappa\)
as in our convention used previously and using the results in Appendix~\ref{app:five_point_MRK_NMRK}, the high-energy
splitting function is obtained directly from
\(C^{\phi\to\phi h}\).

The massless Sudakov parametrisation in
\refeq{eq:Cap5:Sudakov_pbar}--(\ref{eq:Cap5:Sudakov_p3_p4_pbar})
must now be adapted because the scalar and its parent have mass \(m_A\).
We define the on-shell parent direction by
\begin{equation}\label{eq:Cap5:massive_Sudakov_parent}
    \bar p^\mu
    =y\widetilde p^{\,\mu}+\bar p_{\perp}^{\, \mu}
    +\frac{m_A^2-\bar p_{\perp}^{\, 2}}
    {2y\widetilde p\cdot n}\,n^\mu \, ,
    \qquad
    \bar p^2=m_A^2 \, ,
\end{equation}
and write the two daughter momenta as
\begin{equation}\label{eq:Cap5:massive_scalar_graviton_Sudakov}
\begin{aligned}
    p_3^\mu
    ={}&z\bar p^\mu+k_\perp^\mu
    +\frac{(1-z^2)m_A^2
    -2z\bar p_{\perp}\cdot k_\perp-k_\perp^2}
    {2zy\widetilde p\cdot n}\,n^\mu \, ,
    \\
    p_4^\mu
    ={}&(1-z)\bar p^\mu-k_\perp^\mu
    +\frac{-(1-z)^2m_A^2
    +2(1-z)\bar p_{\perp}\cdot k_\perp-k_\perp^2}
    {2(1-z)y\widetilde p\cdot n}\,n^\mu \, .
\end{aligned}
\end{equation}
These relations enforce \(p_3^2=m_A^2\) and \(p_4^2=0\). Their plus
and transverse components imply
\begin{equation}\label{eq:Cap5:massive_collinear_X_and_transverse_relations}
\begin{gathered}
    X=\frac{p_3^+}{p_4^+}=\frac{z}{1-z} \, ,
    \\
    p_{3\perp}=z\bar p_\perp+k_\perp \, ,
    \qquad
    p_{4\perp}=(1-z)\bar p_\perp-k_\perp \, .
\end{gathered}
\end{equation}
Substituting these relations into
\refeq{eq:Cap5:massive_scalar_NMRK_impact_factor} and expanding about
\(k_\perp=0\) gives
\begin{equation}\label{eq:Cap5:massive_scalar_graviton_splitting_function}
    \mathrm{Split}_{\phi\to\phi h}
    (z,\mathbf 3_A,4_h^+)
    ={}
    \frac{[z(5z-2)+1](\bar p_\perp^*)^2
    \bigl[z\bar p_\perp\bar p_\perp^*
    -m_A^2(z-1)\bigr]}
    {4(z-1)
    (m_A^2+\bar p_\perp\bar p_\perp^*)
    \bigl[m_A^2(z-1)^2
    +z^2\bar p_\perp\bar p_\perp^*\bigr]}
    +\mathcal O(k_\perp) \, .
\end{equation}
The leading term is therefore of order \(k_\perp^0\): no inverse power
of the relative transverse momentum is generated. As in the pure graviton
case, the gravitational splitting function remains finite as
\(|k_\perp|\to0\) for generic \(0<z<1\), while the scalar mass modifies
its finite coefficient. The pole at
the endpoint \(z=1\), where the graviton becomes soft, is instead a soft
singularity and should not be confused with a collinear divergence.

\section{Tree-level waveform in the high-energy collinear limit}
\label{sec:Cap5:tree_waveform_high_energy_collinear}

We finally use the classical massive impact factor derived above to obtain the
tree-level waveform in the high-energy collinear region. The general waveform map has been reviewed in
Appendix~\ref{app:Classical_limit}. In particular, after the two
mass-shell constraints have been used to perform the longitudinal
integrations, the radiation kernel reduces to the transverse transform
in \refeq{eq:app:tree_radiation_kernel_reduced}. This is the same step
that produces the Bessel functions in the waveform calculation of
Sec.~8.3.1 of
Ref.~\citeall{divecchia2024gravitationaleikonalparticlestring}.

We denote the momentum of the emitted on-shell graviton by
\(k\equiv p_4\), in order to face the conventions of
Appendix~\ref{sec:app:waveform_tree_amplitude}, while \(q_{1\perp}\) denotes the transverse integration
variable. In the upper NMRK cluster,
\begin{equation}\label{eq:Cap5:waveform_NMRK_momentum_assignment}
    p_{3\perp}=-q_{1\perp},
    \qquad
    p_{4\perp}=k_\perp,
    \qquad
    t=-|k_\perp-q_{1\perp}|^2 \, .
\end{equation}
Combining \refeq{eq:Cap5:massive_five_point_NMRK_factorisation} with
\refeq{eq:app:spectral_metric_waveform}, the positive-helicity spectral
waveform takes the form
\begin{equation}\label{eq:Cap5:collinear_spectral_waveform}
    \widehat h_+^{(0)}(k;b)
    =
    \frac{\kappa^4s^2}
    {16\pi r\sqrt{\Delta_{12}}}\,
    \mathcal W_+(k_\perp,b;\Sigma) \, ,
\end{equation}
where the minus sign in the amplitude cancels the one in \(t\), and
\begin{equation}\label{eq:Cap5:collinear_waveform_transverse_transform}
    \mathcal W_+(k_\perp,b;\Sigma)
    =
    \int\frac{\mathrm d^2q_{1\perp}}{(2\pi)^2}\,
    e^{i b\cdot q_{1\perp}}
    \frac{
    C_{\mathrm{cl}}^{\phi\to\phi h}
    (-q_{1\perp},k_\perp;\Sigma)}
    {|k_\perp-q_{1\perp}|^2} \, .
\end{equation}
As we said in the previous sections, all the dependence on the
high-energy collinear correction is contained in
\(C_{\mathrm{cl}}^{\phi\to\phi h}\).

We now evaluate the transverse transform. A rotation of the transverse
plane allows us to choose the following parametrisation, up to 
conventios associated with the chosen
orientation of the impact parameter,
\begin{equation}\label{eq:Cap5:waveform_polar_coordinates}
\begin{gathered}
    k_\perp=k_\perp^*=P>0,
    \qquad
    q_{1\perp}=Qe^{i\theta},
    \qquad
    q_{1\perp}^*=Qe^{-i\theta},
    \\
    B\equiv\sqrt{-b^2}>0,
    \qquad
    b\cdot q_{1\perp}
    =BQ\cos(\theta-\beta),
\end{gathered}
\end{equation}
where \(\beta\) is the angle between the transverse vectors \(b\) and
\(k_\perp\). Substitution of
\refeq{eq:Cap5:waveform_NMRK_momentum_assignment} into
\refeq{eq:Cap5:massive_impact_factor_classical_limit} gives
\begin{equation}\label{eq:Cap5:classical_impact_factor_harmonics}
    \frac{
    C_{\mathrm{cl}}^{\phi\to\phi h}(-q_{1\perp},k_\perp;\Sigma)}
    {|k_\perp-q_{1\perp}|^2}
    =
    \frac{\Sigma^2}
    {P^2+Q^2-2PQ\cos\theta}
    \sum_{n=-2}^{1}c_n(Q)e^{in\theta} \, ,
\end{equation}
with
\begin{equation}\label{eq:Cap5:classical_impact_factor_harmonic_coefficients}
\begin{gathered}
    c_{-2}(Q)
    =-\frac{Q^2}{P^2+\Sigma^2Q^2},
    \qquad
    c_{-1}(Q)
    =\frac{\Sigma^2+2}{(1+\Sigma^2)^2}\frac{Q}{P},
    \\
    c_0(Q)
    =\frac{\Sigma^2-1}{(1+\Sigma^2)^2},
    \qquad
    c_1(Q)
    =-\frac{\Sigma^2}{(1+\Sigma^2)^2}\frac{Q}{P} \, .
\end{gathered}
\end{equation}

The shifted massless propagator is expanded separately inside and
outside the circle \(Q=P\),
\begin{equation}\label{eq:Cap5:Poisson_kernel_expansion}
\begin{aligned}
    \frac{1}{P^2+Q^2-2PQ\cos\theta}
   =
    \begin{cases}
    \displaystyle
    \frac{1}{P^2-Q^2}
    \sum_{m\in\mathbb Z}
    \left(\dfrac{Q}{P}\right)^{|m|}e^{im\theta},
    &0<Q<P,
    \\[4mm]
    \displaystyle
    \frac{1}{Q^2-P^2}
    \sum_{m\in\mathbb Z}
    \left(\dfrac{P}{Q}\right)^{|m|}e^{im\theta},
    &Q>P.
    \end{cases}
\end{aligned}
\end{equation}
The angular integral is then elementary. The Jacobi--Anger expansion
decomposes a plane wave into angular harmonics according to
\begin{equation}\label{eq:Cap5:Jacobi_Anger_expansion}
    e^{iz\cos\varphi}
    =
    \sum_{j\in\mathbb Z}
    i^jJ_j(z)e^{ij\varphi} \, .,
\end{equation}
where $J_j(z)$ are the Bessel functions of the first kind.
In the present case, \(z=BQ\) and
\(\varphi=\theta-\beta\). Using
the orthogonality of the Fourier modes, and setting $\ell = m+n$, we find
\begin{equation}\label{eq:Cap5:Jacobi_Anger_angular_integral}
\begin{aligned}
    &\int_0^{2\pi}\mathrm d\theta\,
    e^{iBQ\cos(\theta-\beta)}e^{i\ell\theta}=
    e^{i\ell\beta}
    \sum_{j\in\mathbb Z}i^jJ_j(BQ)
    \int_0^{2\pi}\mathrm d\varphi\,
    e^{i(j+\ell)\varphi}
    \\
    &\quad=
    2\pi e^{i\ell\beta}i^{-\ell}J_{-\ell}(BQ)
    =2\pi i^\ell e^{i\ell\beta}J_\ell(BQ) \, .
\end{aligned}
\end{equation}
In the last equality we used
\(J_{-\ell}(z)=(-1)^\ell J_\ell(z)\) and
\((-1)^\ell i^{-\ell}=i^\ell\), valid for integer \(\ell\).
Consequently, the original two dimensional integral is reduced exactly
to a series of one dimensional radial Bessel transforms,
\begin{equation}\label{eq:Cap5:collinear_waveform_radial_series}
    \mathcal W_+
    ={}\frac{\Sigma^2}{2\pi}
    \sum_{\ell\in\mathbb Z}
    i^\ell e^{i\ell\beta}
    \Bigg[
    \int_0^P\mathrm dQ\,
    \frac{QJ_\ell(BQ)}{P^2-Q^2}
    \mathcal A_\ell\!\left(Q,\frac{Q}{P}\right)
    +\int_P^\infty\mathrm dQ\,
    \frac{QJ_\ell(BQ)}{Q^2-P^2}
    \mathcal A_\ell\!\left(Q,\frac{P}{Q}\right)
    \Bigg] \, .
\end{equation}
Introducing \(\rho\), the radial expansion parameter,
\begin{equation}\label{eq:Cap5:waveform_radial_expansion_parameter}
    \rho(Q)
    \equiv
    \begin{cases}
        Q/P, & 0<Q<P, \\[1mm]
        P/Q, & Q>P \, ,
    \end{cases}
\end{equation}
\(0<\rho<1\), as required for the convergence of the
Poisson expansion in \refeq{eq:Cap5:Poisson_kernel_expansion}, the
coefficient entering each radial harmonic is
\begin{equation}\label{eq:Cap5:collinear_waveform_radial_coefficient}
\begin{aligned}
    \mathcal A_\ell(Q,\rho)
    \equiv{}&
    \sum_{n=-2}^{1}c_n(Q)\rho^{|\ell-n|}
    \\
    ={}&
    -\frac{Q^2}{P^2+\Sigma^2Q^2}\rho^{|\ell+2|}
    +\frac{\Sigma^2+2}{(1+\Sigma^2)^2}
    \frac{Q}{P}\rho^{|\ell+1|}
+\frac{\Sigma^2-1}{(1+\Sigma^2)^2}\rho^{|\ell|}
    -\frac{\Sigma^2}{(1+\Sigma^2)^2}
    \frac{Q}{P}\rho^{|\ell-1|} \, .
\end{aligned}
\end{equation}
Equation~\eqref{eq:Cap5:collinear_waveform_radial_series} is the desired
analytic result for the transverse transform, in which each azimuthal harmonic
of the waveform has been expanded as a radial Hankel transform, and the factors
\(e^{i\ell\beta}\) retain its angular dependence.

The denominators in the two radial regions appear singular at \(Q=P\).
This is an artefact of the harmonic expansion. Indeed in the original
two dimensional integrand the propagator becomes singular only when
\(Q=P\) and \(\theta=0\), namely for \(q_{1\perp}=k_\perp\). At precisely
this point the complete impact factor has a second order zero and cancels
the propagator. In the radial representation the same cancellation is
encoded in \(\sum_{n=-2}^{1}c_n(P)=0\), and becomes manifest only after
the four contributions have been combined into \(\mathcal A_\ell\).
The two radial expressions in
\refeq{eq:Cap5:collinear_waveform_radial_series} therefore approach the
same finite limit as \(Q\to P\), whereas treating the four contributions
separately would introduce spurious poles that are absent from the
original waveform.

Finally,  the signal in 
time domain could be obtained by inserting
\refeq{eq:Cap5:collinear_waveform_radial_series} into
\refeq{eq:Cap5:collinear_spectral_waveform} and performing the frequency
transform in \refeq{eq:app:time_domain_metric_waveform}. The absence of inverse powers of the small relative transverse momentum in
\refeq{eq:Cap5:massive_scalar_graviton_splitting_function} implies that no
separate collinear divergence arises. The waveform remains finite in the
external collinear limit, while its soft endpoint is governed by the
Weinberg singularity.

\chapter{Gravity Kernels}
\label{cap:Gravity_kernels}

The high-energy building blocks derived in the previous chapters become
physical kernels once they are sewn across an \(s\)-channel unitarity cut.
In this chapter we first construct the real gravitational kernel associated
with the \(H\) diagram. We then extend the construction to a radiative
\(H\) diagram and to the NMRK region in which a massive scalar and an intermediate
on-shell graviton have comparable rapidities. Finally, we integrate the
elastic kernel and recover the classical logarithmic contribution of the
\(H\) diagram. We follow the conventions of Chap.~\ref{Cap4:Regge_limit_and_BFKL}.

\section{The gravitational \texorpdfstring{\(H\)}{H} kernel}
\label{sec:Cap6:gravitational_H_kernel}

The gravitational kernel is obtained by sewing two one-graviton CEVs
across an \(s\)-channel unitarity cut and summing over the helicity of
the intermediate on-shell graviton. Following the conventions of
Appendix~C of Ref.~\citeall{barcaro2025centralemissionvertexgravitons}, we denote its momentum
by
\begin{equation}\label{eq:Cap6:H_kernel_momentum_routing}
\begin{gathered}
    k=q_1-q_2=q_4-q_3 \, ,
    \\
    q_3=q-q_1 \, ,
    \qquad
    q_4=q-q_2=q_3+k \, .
\end{gathered}
\end{equation}
Here \(q\) is the total momentum transfer in the corresponding four
point amplitude. The momentum \(k\) is outgoing from the left CEV and
\(-k\) is outgoing from the right one. Since the kernel depends only on
transverse momenta, we will denote
\(\boldsymbol q_i^2=-|q_{i\perp}|^2\).

In the notation of the one-graviton CEV introduced
in \refeq{eq:Cap4:one_graviton_CEV_positive_helicity} and of its parity
conjugate, the gravity kernel is obtained by summing the two helicity configurations,
\begin{equation}\label{eq:Cap6:H_kernel_helicity_sum}
    \mathcal K^h(q_1,q_2;q)
    ={}
    \mathcal K^h_{(h^+,h^-)}(q_1,q_2;q)
    +
    \mathcal K^h_{(h^-,h^+)}(q_1,q_2;q) \, .
\end{equation}

The opposite helicities on the two sides arise because reversing the
momentum of the graviton crossing the cut is equivalent to a parity transformation, hence each CEV turns into its
complex conjugate. Equation~\ref{eq:Cap6:H_kernel_helicity_sum} is
therefore the gravitational analogue of the sum over physical polarisation
entering the real BFKL kernel. Parity exchanges $\mathcal K^h_{(h^+,h^-)}$ and $\mathcal K^h_{(h^-,h^+)}$,
so their sum is real even though each term is not.
Notice moreover that \(\mathcal K^h\) denotes the kernel obtained just after sewing
the two CEVs, so the four Reggeised graviton propagators are not included
and will be restored when defining the complete \(H\)-diagram
integrand.

Diagrammatically,
\begin{equation}\label{eq:Cap6:H_kernel_diagrammatic_sum}
    \vcenter{\hbox{\includegraphics[width=0.3\textwidth]
    {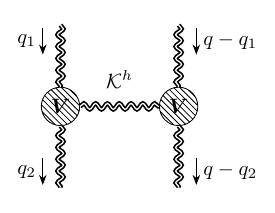}}}
    =
    \vcenter{\hbox{\includegraphics[width=0.3\textwidth]
    {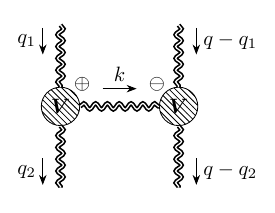}}}
    +
    \vcenter{\hbox{\includegraphics[width=0.3\textwidth]
    {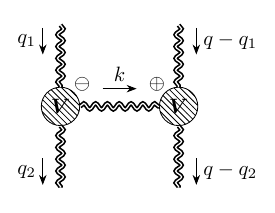}}} \, .
\end{equation}
The two terms in \refeq{eq:Cap6:H_kernel_helicity_sum} are just
\begin{subequations}\label{eq:Cap6:H_kernel_helicity_terms}
\begin{align}
    \mathcal K^h_{(h^+,h^-)}(q_1,q_2;q)
    ={}&
    V_{h^+}(q_1,q_2)\,
    V_{h^-}(q_3,q_4) \, ,
    \\
    \mathcal K^h_{(h^-,h^+)}(q_1,q_2;q)
    ={}&
    V_{h^-}(q_1,q_2)\,
    V_{h^+}(q_3,q_4) \, .
\end{align}
\end{subequations}

Combining the two contributions and using \(q_4=q_3+k\), and knowing that $k=q_1-q_2$, the kernel
takes the form
\begin{equation}\label{eq:Cap6:H_kernel_explicit}
\begin{aligned}
    \mathcal K^h(q_1,q_2;q)
    ={}&
    \frac{
    (q_{1\perp}^*q_{2\perp}-q_{1\perp}q_{2\perp}^*)
    (q_{3\perp}k_\perp^*-q_{3\perp}^*k_\perp)}
    {k_\perp^2(k_\perp^*)^2}
    \\
    &\times
    \Bigl[
    q_{1\perp}^*q_{2\perp}q_{3\perp}
    (k_\perp^*+q_{3\perp}^*)
    +q_{1\perp}q_{2\perp}^*q_{3\perp}^*
    (k_\perp+q_{3\perp})
    \Bigr] \, .
\end{aligned}
\end{equation}

For completeness, its relation with the Yang--Mills case can be made
explicit by expanding the corresponding Yang--Mills kernel using the one-gluon CEV in \refeq{eq:Cap4:CEV_definition},
\begin{equation}\label{eq:Cap6:H_YM_kernel}
    \mathcal K^g(q_1,q_2;q)
    =
    \boldsymbol q^2
    -
    \frac{
    \boldsymbol q_1^2\boldsymbol q_4^2
    +\boldsymbol q_2^2\boldsymbol q_3^2}
    {\boldsymbol k^2} \, .
\end{equation}
In this way, we can expand the gravity kernel via a double copy relation as
\begin{equation}\label{eq:Cap6:H_kernel_compact}
    \mathcal K^h(q_1,q_2;q)
    ={}
    \left[\mathcal K^g(q_1,q_2;q)\right]^2+
    \frac{4}{(\boldsymbol k^2)^2}
    \Bigl[
    \boldsymbol q_1^2\boldsymbol q_2^2
    \boldsymbol q_3^2\boldsymbol q_4^2
    -(\boldsymbol q_1\cdot\boldsymbol q_2)^2
    \boldsymbol q_3^2\boldsymbol q_4^2
    -(\boldsymbol q_3\cdot\boldsymbol q_4)^2
    \boldsymbol q_1^2\boldsymbol q_2^2
    \Bigr] \, ,
\end{equation}
The terms beyond \(\left[\mathcal K^g\right]^2\) originate from the bremsstrahlung subtraction in the gravitational Lipatov current discussed in Chap~\ref{Cap4:Regge_limit_and_BFKL}. They remove simultaneous singularities in overlapping channels and ensure the correct analytic structure of the gravitational kernel.

In order to connect with the transverse integrand inside the \(H\) diagram, we must include also the four Glauber propagators,
\begin{equation}\label{eq:Cap6:H_kernel_complete_integrand}
    \mathcal I_H(q_1,q_2;q)
    =
    \frac{\mathcal K^h(q_1,q_2;q)}
    {\boldsymbol q_1^2\boldsymbol q_2^2
     \boldsymbol q_3^2\boldsymbol q_4^2} \, .
\end{equation}

\section{The radiative \texorpdfstring{\(H\)}{H} kernel}
\label{sec:Cap6:radiative_H_kernel}

\begin{figure}[htbp]
    \centering
    \setlength{\abovecaptionskip}{5pt}
    \subfigure[
    Two-graviton CEV on the left.
    \label{fig:Cap6:H_emission_left}]{
        \includegraphics[width=0.29\textwidth]
        {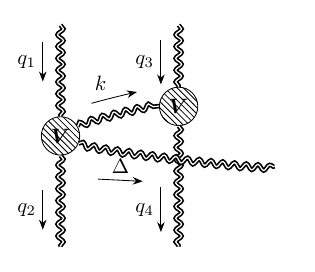}
    }
    \hfill
    \subfigure[
    Two-graviton CEV on the right.
    \label{fig:Cap6:H_emission_right}]{
        \includegraphics[width=0.29\textwidth]
        {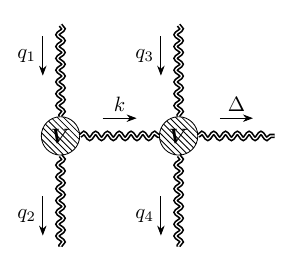}
    }
    \caption{Radiative \(H\) kernels. One graviton crosses the unitarity
    cut, while the second graviton, of helicity \(\nu\), remains in the
    final state.}
    \label{fig:Cap6:radiative_H_kernels}
\end{figure}

We next insert an additional on-shell graviton into the \(H\) diagram.
At tree level, this amounts to replacing either one of the one-graviton
CEVs entering \refeq{eq:Cap6:H_kernel_helicity_sum} by the two-graviton
CEV derived in \refeq{eq:Cap4:two_graviton_CEV_pp} and
\refeq{eq:Cap4:two_graviton_CEV_pm}. The two possible placements give
the left and right radiative kernels shown in
Fig.~\ref{fig:Cap6:radiative_H_kernels}.

We denote by \(k\) the graviton crossing the cut and by \(\Delta\) the
additional graviton in the final state. For the diagram with a double emission in the right side the
momentum conservation reads
\begin{equation}\label{eq:Cap6:H_radiative_right_routing}
    k=q_1-q_2 \, ,
    \qquad
    q_3=q-k \, ,
    \qquad
    q_4=q-\Delta \, ,
\end{equation}
thus the two-graviton CEV on the right is
\(V(q-k,q,q-\Delta;X)\). In agreement with the convention used in
Chap.~\ref{Cap4:Regge_limit_and_BFKL}, its single longitudinal variable is
\begin{equation}\label{eq:Cap6:H_radiative_X_definition}
    X
    \equiv
    \frac{k^+}{\Delta^+}
    =
    \frac{|k_\perp|}{|\Delta_\perp|}
    e^{y_k-y_\Delta} \, .
\end{equation}
The right kernel therefore is
\begin{equation}\label{eq:Cap6:H_radiative_right}
    \mathcal H_\nu^R(q_1,q_2;q,\Delta;X)
    =
    \sum_{h=\pm2}
    \left[V(q_1,k^h,q_2)\right]^*
    V(q-k,(-k)^h,\Delta^\nu,q-\Delta;X) \, .
\end{equation}
The left contribution is obtained by interchanging the two sides of the
cut, therefore its two-graviton CEV depends on the same ratio \(X\), and the full
radiative kernel is
\begin{equation}\label{eq:Cap6:H_radiative_full_kernel}
    \mathcal H_\nu^{\mathrm{rad}}
    =\mathcal H_\nu^L+\mathcal H_\nu^R \, .
\end{equation}

For a final graviton with positive helicity, the helicity sum in
\refeq{eq:Cap6:H_radiative_right} contains two configurations. We use
the first entry in each subscript for the one-graviton CEV and the last
two for the two-graviton CEV:
\begin{equation}\label{eq:Cap6:H_radiative_positive_helicity}
\begin{aligned}
    \mathcal H_+^R
    ={}&
    \mathcal H^R_{(-,++)}
    +\mathcal H^R_{(+,-+)} \, ,
    \\
    \mathcal H^R_{(-,++)}
    ={}&
    V_{h^-}(q_1,q_2)\,
    V_{h^+h^+}(q-k,q,q-\Delta;X) \, ,
    \\
    \mathcal H^R_{(+,-+)}
    ={}&
    V_{h^+}(q_1,q_2)\,
    V_{h^-h^+}(q-k,q,q-\Delta;X) \, .
\end{aligned}
\end{equation}
The term in the radiative kernel containing a two-graviton CEV with same helicities can be displayed in closed form as
\begingroup
\setlength{\jot}{1pt}
\begin{equation}\label{eq:Cap6:H_radiative_equal_helicity_explicit}
\begin{aligned}
\mathcal H^R_{(-,++)}
={}&
\frac{
q_{1\perp}(q_\perp-\Delta_\perp)
(q_{1\perp}^*-k_\perp^*)(k_\perp^*-q_\perp^*)
(q_{1\perp}k_\perp^*-q_{1\perp}^*k_\perp)
}
{\Delta_\perp^2 k_\perp^2(k_\perp^*)^2
(k_\perp+X\Delta_\perp)}
\\
&\times\Bigl[
\bigl((q_\perp^*)^2\Delta_\perp k_\perp
+q_\perp^2k_\perp^*\Delta_\perp^*\bigr)
(k_\perp+X\Delta_\perp)
\\
&\qquad
-q_\perp q_\perp^*
(\Delta_\perp k_\perp^*+\Delta_\perp^*k_\perp)
(k_\perp+X\Delta_\perp)
\\
&\qquad
+\bigl(q_\perp^*(\Delta_\perp-k_\perp)
-q_\perp(\Delta_\perp^*-k_\perp^*)\bigr)
(\Delta_\perp k_\perp^*-\Delta_\perp^*k_\perp)k_\perp
\\
&\qquad
-(\Delta_\perp k_\perp^*-\Delta_\perp^*k_\perp)^2k_\perp
\Bigr] \, .
\end{aligned}
\end{equation}
\endgroup

However the NMHV term is much longer because it involves the rational
structure of \(V_{h^-h^+}\). It is useful to organise it as
\begin{equation}\label{eq:Cap6:H_radiative_mixed_helicity_sum}
    \mathcal H^R_{(+,-+)}
    =
    \sum_{i=0}^{7}\sum_{j=0}^{i}a_{ij} \, ,
    \qquad
    a_{ij}=\frac{\operatorname{num}(a_{ij})}
    {\operatorname{den}(a_{ij})} \, ,
\end{equation}
where the indices make the powers of \(X\) transparent,
\begin{equation}\label{eq:Cap6:H_radiative_aij_scaling}
    \operatorname{den}(a_{ij})\sim X^i \, ,
    \qquad
    \operatorname{num}(a_{ij})\sim X^j \, .
\end{equation}
Only
\begin{equation}\label{eq:Cap6:H_radiative_aij_nonzero}
\begin{gathered}
    a_{00},\ a_{10},\ a_{11},\ a_{20},\ a_{30},\ a_{33},
    \\
    a_{40},\ a_{41},\ a_{44},\ a_{76},\ a_{77}
\end{gathered}
\end{equation}
are non vanishing. Two representative terms are
\begin{equation}\label{eq:Cap6:H_radiative_a00}
\begin{aligned}
a_{00}
={}&
\frac{
q_{1\perp}^*(q_{1\perp}-k_\perp)(q_\perp-k_\perp)
(q_\perp^*-k_\perp^*)(q_\perp-\Delta_\perp)
(q_\perp^*-\Delta_\perp^*)
}
{q_\perp\Delta_\perp^2 k_\perp k_\perp^*
(q_\perp^*-k_\perp^*-\Delta_\perp^*)}
\\
&\times
(q_{1\perp}k_\perp^*-q_{1\perp}^*k_\perp)
(q_\perp\Delta_\perp^*-q_\perp^*\Delta_\perp
+\Delta_\perp k_\perp^*) \, ,
\end{aligned}
\end{equation}
and
\begin{equation}\label{eq:Cap6:H_radiative_a41}
\begin{aligned}
a_{41}
={}&
\frac{
q_{1\perp}^*k_\perp^4(q_\perp-k_\perp)
(q_{1\perp}-k_\perp)(q_\perp^*-k_\perp^*)
(q_\perp^*-\Delta_\perp^*)
}
{X\Delta_\perp^2(\Delta_\perp-k_\perp)
(k_\perp+X\Delta_\perp)
(k_\perp k_\perp^*+X\Delta_\perp\Delta_\perp^*)^2}
\\
&\times
(q_{1\perp}k_\perp^*-q_{1\perp}^*k_\perp)
(k_\perp^*+X\Delta_\perp^*) \, .
\end{aligned}
\end{equation}
The remaining terms \(a_{ij}\) are obtained by evaluating the
NMHV CEV of \refeq{eq:Cap4:two_graviton_CEV_pm} for
\(q_r=q-k\), \(q_{r+1}=q\), \(q_{r+2}=q-\Delta\) and \(X_r=X\),
multiplying it by \(V_{h^+}(q_1,q_2)\), and collecting the resulting
terms according to their powers of \(X\). This organisation is
particularly convenient for extracting the two strongly ordered limits.

At the level of the kernel one finds that, when \(X\to\infty\), the cut graviton stands at a much larger rapidity than
the observed graviton, and consequently the two-graviton CEV factorises into two one-graviton CEVs,
\begin{equation}\label{eq:Cap6:H_radiative_X_infinity}
    \lim_{X\to\infty}\mathcal H_+^R
    =
    \mathcal K^h(q_1,q_2;q-k)\,
    \mathcal P_{\mathrm R}(q)\,
    V_{h^+}(q,q-\Delta) \, ,
\end{equation}
where
\begin{equation}\label{eq:Cap6:reggeized_propagator_right}
    \mathcal P_{\mathrm R}(p)
    =-\frac{1}{|p_\perp|^2} \, .
\end{equation}

Diagrammatically this factorisation corresponds to 
\begingroup
\setlength{\abovedisplayskip}{6pt}
\setlength{\belowdisplayskip}{6pt}
\begin{equation}\label{eq:Cap6:H_radiative_X_infinity_diagram}
    \lim_{X\to\infty}
    \vcenter{\hbox{\includegraphics[width=0.27\textwidth]
    {Diagrams_PDF/H_emission_right_nomassive.pdf}}}
    =
    \vcenter{\hbox{\includegraphics[width=0.33\textwidth]
    {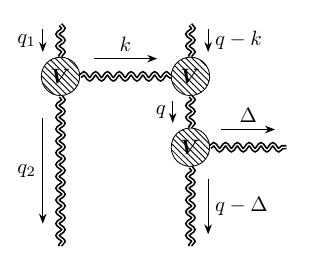}}} \, .
\end{equation}
\endgroup

Conversely, for \(X\to0\), the observed graviton is emitted first and
the intermediate Reggeised momentum is \(q-\Delta-k\):
\begin{equation}\label{eq:Cap6:H_radiative_X_zero}
    \lim_{X\to0}\mathcal H_+^R
    ={}
    \mathcal K^h(q_1,q_2;q-\Delta-k)\,
    \mathcal P_{\mathrm R}(q-\Delta-k)
    V_{h^+}(q-\Delta-k,q-\Delta) \, .
\end{equation}
The complementary factorisation is represented diagrammatically by
\begingroup
\setlength{\abovedisplayskip}{6pt}
\setlength{\belowdisplayskip}{6pt}
\begin{equation}\label{eq:Cap6:H_radiative_X_zero_diagram}
    \lim_{X\to0}
    \vcenter{\hbox{\includegraphics[width=0.27\textwidth]
    {Diagrams_PDF/H_emission_right_nomassive.pdf}}}
    =
    \vcenter{\hbox{\includegraphics[width=0.33\textwidth]
    {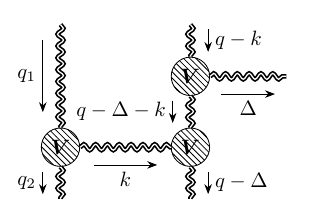}}} \, .
\end{equation}
\endgroup
The two limits are therefore the two MRK orderings of the same NMRK
kernel.

The departure from strict rapidity ordering is obtained by expanding
the same function at the two ends of the \(X\) interval,
\begin{subequations}\label{eq:Cap6:H_radiative_asymptotic_series}
\begin{align}
    \mathcal H_{\infty,+}^R
    ={}&
    \sum_{n=0}^{\infty}\frac{1}{X^n}
    f_\infty^{(n)}(\boldsymbol q_\perp) \, ,
    \\
    \mathcal H_{0,+}^R
    ={}&
    \sum_{n=0}^{\infty}X^n
    f_0^{(n)}(\boldsymbol q_\perp) \, ,
\end{align}
\end{subequations}
\begingroup
\widowpenalty=10000
where the leading orders are the MRK coefficients, reproducing
\refeq{eq:Cap6:H_radiative_X_infinity} and
\refeq{eq:Cap6:H_radiative_X_zero}, respectively.
Here \(\boldsymbol q_\perp\) collectively denotes all transverse
momenta. The first correction in the large \(X\) region is
\par
\endgroup
\begingroup
\setlength{\jot}{1pt}
\setlength{\abovedisplayskip}{6pt}
\setlength{\belowdisplayskip}{6pt}
\begin{align}
\mathcal H_{\infty,+}^{R(1)}
={}&
\frac{1}{X}
\frac{
q_{1\perp}^*(q_\perp-\Delta_\perp)(q_\perp-k_\perp)
(q_{1\perp}-k_\perp)
(k_\perp^*q_{1\perp}-k_\perp q_{1\perp}^*)
}
{\Delta_\perp^3 k_\perp^2 k_\perp^*q_\perp^2}
\notag\\
&\times\Bigl[
k_\perp q_\perp^*
\Bigl\{k_\perp(\Delta_\perp^2q_\perp^*
-\Delta_\perp^*q_\perp^2)
\notag\\
&\qquad\qquad
+q_\perp(\Delta_\perp+k_\perp)
(\Delta_\perp q_\perp^*-\Delta_\perp^*q_\perp)
\Bigr\}
\notag\displaybreak[2]\\
&\qquad
-k_\perp^*q_\perp^2
\Bigl\{\Delta_\perp^*(\Delta_\perp+k_\perp-q_\perp)^2
-\Delta_\perp^*(\Delta_\perp^2+k_\perp^2+q_\perp^2)
\notag\\
&\qquad\qquad
+(\Delta_\perp+k_\perp)
(\Delta_\perp q_\perp^*-\Delta_\perp^*q_\perp)
+\Delta_\perp^*q_\perp^2+\Delta_\perp^2q_\perp^*
\Bigr\}
\Bigr]
\notag\\
&+
\frac{1}{X}
\frac{
q_{1\perp}(q_\perp-\Delta_\perp)
(q_\perp^*-k_\perp^*)(q_{1\perp}^*-k_\perp^*)
}
{\Delta_\perp^3k_\perp(k_\perp^*)^2}
\notag\\
&\times
\Bigl\{\Delta_\perp^*k_\perp-\Delta_\perp k_\perp^*
+q_\perp(k_\perp^*-\Delta_\perp^*)
-q_\perp^*(k_\perp-\Delta_\perp)\Bigr\}
\notag\\
&\times
(k_\perp\Delta_\perp^*-\Delta_\perp k_\perp^*)
(k_\perp^*q_{1\perp}-k_\perp q_{1\perp}^*) \, .
\label{eq:Cap6:H_radiative_first_correction_infinity}
\end{align}
\endgroup
In the small \(X\) region the corresponding first correction is
\begingroup
\setlength{\jot}{1pt}
\setlength{\abovedisplayskip}{6pt}
\setlength{\belowdisplayskip}{6pt}
\begin{equation}\label{eq:Cap6:H_radiative_first_correction_zero}
\begin{aligned}
\mathcal H_{0,+}^{R(1)}
={}&
\frac{X}{\Delta_\perp k_\perp^3(k_\perp^*)^3}
\Biggl[
-k_\perp^*q_{1\perp}(q_\perp-\Delta_\perp)
(q_{1\perp}^*-k_\perp^*)(k_\perp^*-q_\perp^*)
\\
&\quad\times
(\Delta_\perp k_\perp^*-k_\perp\Delta_\perp^*)
(k_\perp^*q_{1\perp}-k_\perp q_{1\perp}^*)
\\
&\quad\times
\Bigl\{k_\perp\Delta_\perp^*-\Delta_\perp k_\perp^*
+q_\perp(k_\perp^*-\Delta_\perp^*)
-q_\perp^*(k_\perp-\Delta_\perp)\Bigr\}
\\
&\quad+
\frac{
k_\perp q_{1\perp}^*(q_{1\perp}-k_\perp)
(q_\perp^*-k_\perp^*)(q_\perp^*-\Delta_\perp^*)
(k_\perp q_{1\perp}^*-k_\perp^*q_{1\perp})
}
{(k_\perp^*+\Delta_\perp^*-q_\perp^*)^2}
\\
&\quad\times\Biggl(
\Delta_\perp(q_\perp^*-k_\perp^*)^2
\Bigl\{2(k_\perp^*)^2(\Delta_\perp-q_\perp)
+k_\perp(q_\perp^*)^2+k_\perp k_\perp^*q_\perp^*\Bigr\}
\\
&\qquad-
\Delta_\perp^*(q_\perp^*-k_\perp^*)
\Bigl\{-(\Delta_\perp^*)^2
\bigl[2\Delta_\perp(k_\perp-2\Delta_\perp)
+q_\perp^2+q_\perp(3\Delta_\perp-k_\perp)\bigr]
\\
&\qquad\qquad
+k_\perp^*q_\perp^*
\bigl[-\Delta_\perp^2+3\Delta_\perp k_\perp-k_\perp^2
+q_\perp(\Delta_\perp+k_\perp)\bigr]
+\Delta_\perp k_\perp(q_\perp^*)^2
\Bigr\}
\\
&\qquad+
(\Delta_\perp^*)^2
\Bigl\{(k_\perp^*)^2
(\Delta_\perp^2-6\Delta_\perp k_\perp+k_\perp^2
-3q_\perp^2+2q_\perp(\Delta_\perp+k_\perp))
\\
&\qquad\qquad
+k_\perp^*q_\perp^*
(-\Delta_\perp^2+5k_\perp(\Delta_\perp-k_\perp)
+q_\perp^2+4k_\perp q_\perp)
\\
&\qquad\qquad
-k_\perp(q_\perp^*)^2
(\Delta_\perp-2k_\perp+2q_\perp)
\Bigr\}
\\
&\qquad+
(\Delta_\perp^*)^3
\Bigl\{k_\perp q_\perp^*
(\Delta_\perp-4k_\perp+4q_\perp)
\\
&\qquad\qquad
-k_\perp^*
\bigl[2k_\perp(\Delta_\perp-2k_\perp)
+q_\perp^2-q_\perp(\Delta_\perp-3k_\perp)\bigr]
\Bigr\}
\\
&\qquad+
2k_\perp(\Delta_\perp^*)^4(k_\perp-q_\perp)
\Biggr)
\Biggr] \, .
\end{aligned}
\end{equation}
\endgroup
These terms retain the first finite informations which arise if we relax the strict MRK configuration. The same kernels in the negative helicity case follows by
complex conjugation, while the left contribution follows by reflecting
the cut diagram. Remark that, since \(\Delta\) remains an external on-shell graviton,
the radiative kernel is not closed into an additional graviton loop and
can contribute to a classical radiative observable.

\section{Classical NMRK kernel from the massive impact factor}
\label{sec:Cap6:classical_NMRK_kernel}

In this section we will apply the same unitarity construction of Sec~\ref{sec:Cap6:gravitational_H_kernel} to the massive impact
factor in the classical limit derived in
Sec.~\ref{sec:Cap5:massive_scalar_impact_factor_collinear_limit}.
We retain the momentum routing of
\refeq{eq:Cap6:H_kernel_momentum_routing},
\begin{equation}\label{eq:Cap6:NMRK_H_routing}
    k=q_1-q_2=q_4-q_3 \, ,
    \qquad
    q_3=q-q_1 \, ,
    \qquad
    q_4=q-q_2 \, .
\end{equation}
For the left impact factor, the transverse momenta in
\refeq{eq:Cap5:massive_NMRK_to_MRK_transverse_momenta} are therefore
\begin{equation}\label{eq:Cap6:NMRK_left_routing}
\begin{aligned}
    p_{3L}&=-q_1 \, ,
    &
    p_{4L}&=q_1-q_2=k \, ,
    &
    p_{5L}&=q_2 \, .
\end{aligned}
\end{equation}
The right impact factor is obtained by the replacements
\(q_1\to q_3\), \(q_2\to q_4\) and, since the momenta crossing the cut
are outgoing from the left amplitude and incoming into the right one, momentum conservation implies
\begin{equation}\label{eq:Cap6:NMRK_massive_routing}
\begin{aligned}
    p_{3R}&=-q_3=q_1-q \, ,
    &
    p_{4R}&=q_3-q_4=-k=q_2-q_1 \, ,
    &
    p_{5R}&=q_4=q-q_2 \, ,
\end{aligned}
\end{equation}
and notice that \(p_{3R}\neq -p_{3L}\), since each impact factor is defined in
a local frame where the incoming massive state has zero transverse momentum, therefore we denote the intermediate massive legs 
with $\ell$ and $\ell_{B}$ to avoid ambiguity.

\begin{figure}[htbp]
    \centering
    \includegraphics[width=0.66\textwidth]
    {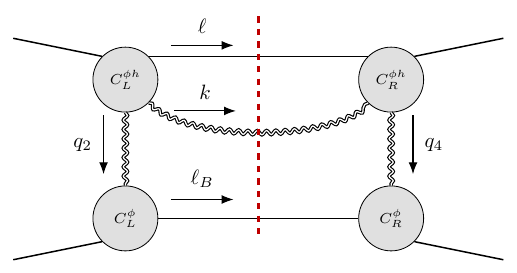}
    \caption{Three-particle cut defining the classical NMRK \(H\)
    kernel. The upper scalar and the on-shell graviton form an NMRK
    cluster, while \(q_2\) and \(q_4\) are the two Reggeised momenta.}
    \label{fig:Cap6:H_NMRK_massive_cut}
\end{figure}

The corresponding three-particle cut is represented in
fig.~\ref{fig:Cap6:H_NMRK_massive_cut}. The graviton \(k\) belongs to
the same rapidity building block as the upper massive scalar \(\ell\), indeed
it is emitted by \(C_L^{\phi h}\) and absorbed by
\(C_R^{\phi h}\). The lower scalar crosses the cut between the two
ordinary scalar impact factors.

As we know at this point, the negative helicity expression follows by parity,
\begin{equation}\label{eq:Cap6:NMRK_impact_factor_helicity}
\begin{aligned}
    C_{-h,a}^{\mathrm{cl}}
    ={}&
    \left[C_{h,a}^{\mathrm{cl}}\right]^* \, ,
    \qquad a=L,R \, ,
\end{aligned}
\end{equation}
and remark that the rapidity separation, as in Sec.~\ref{sec:Cap5:massive_scalar_impact_factor_collinear_limit},  is encoded in the variable 
\begin{equation}
\Sigma=e^{y_\ell-y_k}>0 \, .
\end{equation} 
Using the momentum routings in \refeq{eq:Cap6:NMRK_left_routing} and
\refeq{eq:Cap6:NMRK_massive_routing} in the two classical impact
factors gives the following explicit kernel in the more general NMRK
configuration,
\begingroup
\setlength{\jot}{1pt}
\begin{align}
\mathcal K_{\mathrm{NMRK}}^{\mathrm{cl}}
\equiv{}&
\sum_{\lambda=\pm2}
C_{\lambda,L}^{\mathrm{cl}}
\left[C_{\lambda,R}^{\mathrm{cl}}\right]^*
\notag\\
={}&
\frac{\Sigma^4}
{(\Sigma^2+1)^2
\bigl[
(q_{1\perp}-q_{2\perp})
(q_{1\perp}^*-q_{2\perp}^*)
+\Sigma^2q_{1\perp}q_{1\perp}^*
\bigr]}
\notag\\[-2pt]
\times&\Bigg\{\frac{1}{(q_{1\perp}^*-q_{2\perp}^*)^2}
\Biggl[
-\Sigma^4q_{1\perp}^2(q_{2\perp}^*)^2
\notag\\[-2pt]
+{}&
\Sigma^2q_{1\perp}q_{2\perp}q_{2\perp}^*
\bigl[
q_{1\perp}^*(\Sigma^2+3)-3q_{2\perp}^*
\bigr]
-{}
q_{2\perp}^2(q_{1\perp}^*-q_{2\perp}^*)
\bigl[
q_{1\perp}^*+q_{2\perp}^*(\Sigma^2-1)
\bigr]
\Biggr]
\notag\\[-2pt]
\times{}&
\Biggl[
-\frac{(q_\perp^*-q_{1\perp}^*)^2}
{\Sigma^2(q_\perp-q_{1\perp})
(q_\perp^*-q_{1\perp}^*)
+(q_{1\perp}-q_{2\perp})
(q_{1\perp}^*-q_{2\perp}^*)}
\notag\\[-2pt]
+{}&
\frac{
\Sigma^2(q_\perp-q_{1\perp})
(q_{1\perp}^*-q_{2\perp}^*)}
{(\Sigma^2+1)^2(q_{1\perp}-q_{2\perp})^2}
+\frac{
-\Sigma^2(q_\perp^*-2q_{1\perp}^*+q_{2\perp}^*)
-2q_\perp^*+q_{1\perp}^*+q_{2\perp}^*}
{(\Sigma^2+1)^2(q_{1\perp}-q_{2\perp})}
\Biggr]
\notag\displaybreak[2]\\[2pt]
&+\frac{1}{(q_{1\perp}-q_{2\perp})^2}
\Biggl[
\Sigma^4q_{2\perp}q_{1\perp}^*
\bigl(q_{1\perp}q_{2\perp}^*
-q_{2\perp}q_{1\perp}^*\bigr)
\notag\\[-2pt]
+{}&
\Sigma^2q_{2\perp}q_{2\perp}^*
(q_{2\perp}-q_{1\perp})
(q_{2\perp}^*-3q_{1\perp}^*)
-(q_{2\perp}^*)^2
(q_{1\perp}-q_{2\perp})^2
\Biggr]
\notag\\[-2pt]
\times{}&
\Biggl[
-\frac{(q_\perp-q_{1\perp})^2}
{\Sigma^2(q_\perp-q_{1\perp})
(q_\perp^*-q_{1\perp}^*)
+(q_{1\perp}-q_{2\perp})
(q_{1\perp}^*-q_{2\perp}^*)}
\notag\\[-2pt]
+{}&
\frac{
-\Sigma^2(q_\perp-2q_{1\perp}+q_{2\perp})
-2q_\perp+q_{1\perp}+q_{2\perp}}
{(\Sigma^2+1)^2
(q_{1\perp}^*-q_{2\perp}^*)}
+
\frac{
\Sigma^2(q_{1\perp}-q_{2\perp})
(q_\perp^*-q_{1\perp}^*)}
{(\Sigma^2+1)^2
(q_{1\perp}^*-q_{2\perp}^*)^2}
\Biggr]
\Bigg\} \, .
\label{eq:Cap6:classical_NMRK_kernel}
\end{align}
\endgroup
The corresponding
transverse integrand also contains the two Reggeised propagators that
remain outside the impact factors,
\begin{equation}\label{eq:Cap6:classical_NMRK_integrand}
\begin{aligned}
    \mathcal I_{\mathrm{NMRK}}^{\mathrm{cl}}
    (q_{1\perp},q_{2\perp};q_\perp)
    ={}&
    \frac{
    \mathcal K_{\mathrm{NMRK}}^{\mathrm{cl}}
    (q_{1\perp},q_{2\perp};q_\perp)}
    {\boldsymbol q_2^2\boldsymbol q_4^2} \, ,
\end{aligned}
\end{equation}
in which we have to keep in mind the momentum conservation in \refeq{eq:Cap6:NMRK_H_routing}.

If we include the reduced three-particle phase space, derived in
Appendix~\ref{app:NMRK_phase_space}, we must remind that \(X=m_A\Sigma/(\hbar|k_\perp|)-1\sim\hbar^{-1}\), so that, in the classical expansion, 
\refeq{eq:app:NMRK_longitudinal_phase_space_result} reduces to $\frac{1}{s}\left(1+\mathcal O(\hbar)\right)$. Therefore, the phase-space in \refeq{eq:app:five_point_NMRK_reduced_phase_space} reduces in the classical limit to
\begin{equation}\label{eq:Cap6:NMRK_phase_space_factor}
    \mathrm d\mathcal P_{3,\mathrm{cl}}^{\mathrm{NMRK}}
    ={}
    \frac{1}{4s}\,
    \widehat{\mathrm d}y_k\,
    \widehat{\mathrm d}^{\,d}\boldsymbol\ell\,
    \widehat{\mathrm d}^{\,d}\boldsymbol k\,
    \widehat{\mathrm d}^{\,d}\boldsymbol\ell_B
   \, \widehat{\delta}^{(d)}
    \left(
    \boldsymbol\ell+\boldsymbol k+\boldsymbol\ell_B
    \right)
    +\mathcal O(\hbar) \, .
\end{equation}
Multiplying the transverse integrand by this measure yields the
classical cut contribution,
\begin{equation}\label{eq:Cap6:classical_NMRK_scaling}
    \mathrm d\mathcal P_{3,\mathrm{cl}}^{\mathrm{NMRK}}\,
    \mathcal I_{\mathrm{NMRK}}^{\mathrm{cl}}
    ={}
    \frac{1}{4s}
    \frac{
    \mathcal K_{\mathrm{NMRK}}^{\mathrm{cl}}
    (q_{1\perp},q_{2\perp};q_\perp)}
    {\boldsymbol q_2^2\boldsymbol q_4^2}
    \widehat{\mathrm d}y_k \,
    \widehat{\mathrm d}^{\,d}\boldsymbol\ell\,
    \widehat{\mathrm d}^{\,d}\boldsymbol k\,
    \widehat{\mathrm d}^{\,d}\boldsymbol\ell_B\,
    \widehat{\delta}^{(d)}
    \left(
    \boldsymbol\ell+\boldsymbol k+\boldsymbol\ell_B
    \right)
    +\mathcal O(\hbar) \, .
\end{equation}
Consequently, up to the overall \(S\)-matrix normalisation,
\begin{equation}\label{eq:Cap6:NMRK_cut}
\begin{aligned}
    2\,\operatorname{Im}
    \mathcal M_{\mathrm{NMRK}}^{\mathrm{cl}}
    ={}&
    \frac{\kappa^6s^3}{4}
    \int
    \widehat{\mathrm d}y_k\,
    \widehat{\mathrm d}^{\,d}\boldsymbol\ell\,
    \widehat{\mathrm d}^{\,d}\boldsymbol k\,
    \widehat{\mathrm d}^{\,d}\boldsymbol\ell_B
    \\
    &\times
    \widehat{\delta}^{(d)}
    \left(
    \boldsymbol\ell+\boldsymbol k+\boldsymbol\ell_B
    \right)
    \frac{
    \mathcal K_{\mathrm{NMRK}}^{\mathrm{cl}}
    (q_{1\perp},q_{2\perp};q_\perp)}
    {\boldsymbol q_2^2\boldsymbol q_4^2}
    +\mathcal O(\hbar) \, .
\end{aligned}
\end{equation}
The phase-space factor \(1/(4s)\) converts the product of the two
\(s^2\) five-point amplitudes into the \(s^3\) behaviour characteristic
of the classical \(H\) sector, defined in \refeq{eq:Cap4:multi_H_classical_scaling}. The finite rapidity separation survives
through the dependence on \(\Sigma\), whereas the factor
\(1+1/X\) in the exact NMRK measure contributes only beyond the leading
classical order.

In the strict MRK limit, $\Sigma \to \infty$, \refeq{eq:Cap6:classical_NMRK_integrand} reproduces exactly the MRK \(H\) integrand in \refeq{eq:Cap6:H_kernel_complete_integrand}.
In the next section we will show how this MRK amplitude automatically generates the leading logarithmic accuracy.

\section{Integration of the \texorpdfstring{\(H\)}{H} diagram}
\label{sec:Cap6:H_diagram_integration}

We now integrate the elastic kernel of
\refeq{eq:Cap6:H_kernel_complete_integrand}, following the procedure in
Ref.~\citeall{alessio2026gravitationalamplitudesregge}. Restoring
\(\hbar\) and working in \(D=4-2\epsilon\), we have
\begin{equation}\label{eq:Cap6:H_diagram_unintegrated}
\begin{aligned}
    \operatorname{Im}\mathcal M_{\mathrm{MRK}}^{(2)}(s,q^2)
    \simeq{}&
    \frac{(8\pi G)^3s^3}{8\hbar^3}
    \mu^{4\epsilon}
    \int_{y_1}^{y_2}
    \widehat{\mathrm d}y
    \int_{\boldsymbol q_1,\boldsymbol q_2}
    \mathcal I_H(q_1,q_2;q) \, ,
    \\
    \int_{\boldsymbol\ell}
    \equiv{}&
    \int\frac{\mathrm d^{\,2-2\epsilon}\boldsymbol\ell}
    {(2\pi)^{2-2\epsilon}} \, .
\end{aligned}
\end{equation}
The interval \(y_1<y<y_2\) is the rapidity available to the cut
graviton. At leading power, its integration gives
\begin{equation}\label{eq:Cap6:H_rapidity_logarithm}
    \int_{y_1}^{y_2}\widehat{\mathrm d}y
    \simeq
    \frac{1}{2\pi}
    \log\left(\frac{s}{|t|}\right) \, ,
\end{equation}
thus the
longitudinal phase space produces the Regge logarithm. It is convenient
to denote the remaining transverse convolution by
\begin{equation}\label{eq:Cap6:H_transverse_convolution_definition}
    \mathcal J_H(q^2)
    =
    \mu^{4\epsilon}
    \int_{\boldsymbol q_1,\boldsymbol q_2}
    \mathcal I_H(q_1,q_2;q)
    =
    \mu^{4\epsilon}
    \int_{\boldsymbol q_1,\boldsymbol q_2}
    \frac{\mathcal K^h(q_1,q_2;q)}
    {\boldsymbol q_1^2\boldsymbol q_2^2
     \boldsymbol q_3^2\boldsymbol q_4^2} \, .
\end{equation}
We obtain
\begin{equation}\label{eq:Cap6:H_diagram_transverse_convolution}
    \operatorname{Im}\mathcal M_{\mathrm{MRK}}^{(2)}(s,q^2)
    \simeq
    \frac{(8\pi G)^3}{16\pi\hbar^3}
    s^3
    \log\left(\frac{s}{|t|}\right)
    \mathcal J_H(q^2) \, ,
\end{equation}
where the transverse integration has been performed in the \(d=D-2=2-2\epsilon\)
dimensional space orthogonal to the two light-cone directions. 

To reduce the integral, we first expand
\refeq{eq:Cap6:H_kernel_compact} and express every transverse scalar
product through inverse propagators. The resulting
two-loop integrals belong to the following family of denominators,
\begin{equation}\label{eq:Cap6:H_integral_denominator_family}
    \mathcal P_H=\Bigl\{
    (\boldsymbol q-\boldsymbol q_1)^2,
    \boldsymbol q_1^2,
    (\boldsymbol q-\boldsymbol q_2)^2,
    (\boldsymbol q_1-\boldsymbol q_2)^2,
    \boldsymbol q_2^2
    \Bigr\} \, .
\end{equation}
Factors generated in the numerator cancel some elements of
\(\mathcal P_H\). Integration by parts identities then reduce all the
remaining terms to two master topologies\footnote{More exhaustive details of the integration are reported in the Appendix B of Ref~\citeall{alessio2026gravitationalamplitudesregge}.},
\begin{subequations}\label{eq:Cap6:H_master_integrals}
\begin{align}
    \mathcal I_{01110}(q)
    \equiv{}&
    \int_{\boldsymbol q_1,\boldsymbol q_2}
    \frac{1}
    {\boldsymbol q_1^2
    (\boldsymbol q-\boldsymbol q_2)^2
    (\boldsymbol q_1-\boldsymbol q_2)^2} \, ,
    \\
    \mathcal I_{11101}(q)
    \equiv{}&
    \int_{\boldsymbol q_1,\boldsymbol q_2}
    \frac{1}
    {(\boldsymbol q-\boldsymbol q_1)^2
    \boldsymbol q_1^2
    (\boldsymbol q-\boldsymbol q_2)^2
    \boldsymbol q_2^2} \, .
\end{align}
\end{subequations}
The first integral is a two-loop chain and can be evaluated by applying
the one-loop bubble formula twice. The second factorises directly into
the product of two one-loop bubbles. More explicitly, the IBP reduction
of the transverse convolution reads
\begin{equation}\label{eq:Cap6:H_IBP_reduction}
\begin{aligned}
    \mathcal J_H(q^2)
    =\mu^{4\epsilon}\Bigg[
    -\frac{2}{3}(d+1)\boldsymbol q^2
    \mathcal I_{01110}(q)
    +(\boldsymbol q^2)^2
    \mathcal I_{11101}(q)
    \Bigg] \, .
\end{aligned}
\end{equation}

Both master integrals follow from the standard massless
\(d\)-dimensional convolution
\begin{equation}\label{eq:Cap6:massless_convolution_formula}
\begin{aligned}
    &\int
    \frac{\mathrm d^d\boldsymbol\ell}{(2\pi)^d}
    \frac{1}
    {(\boldsymbol\ell^2)^\alpha
    [(\boldsymbol q-\boldsymbol\ell)^2]^\beta}
    \\
    &\quad=
    \frac{1}{(4\pi)^{d/2}}
    (\boldsymbol q^2)^{d/2-\alpha-\beta}
    \frac{
    \Gamma(\alpha+\beta-d/2)
    \Gamma(d/2-\alpha)
    \Gamma(d/2-\beta)}
    {\Gamma(\alpha)\Gamma(\beta)
    \Gamma(d-\alpha-\beta)} \, .
\end{aligned}
\end{equation}
For
\(\mathcal I_{01110}\), the first integration changes the power of the
propagator appearing in the second bubble, and applying
\refeq{eq:Cap6:massless_convolution_formula} once more completes the
chain. For \(\mathcal I_{11101}\), the integrations over
\(\boldsymbol q_1\) and \(\boldsymbol q_2\) are independent and each
gives the case \(\alpha=\beta=1\). In the adopted
\(\overline{\mathrm{MS}}\) normalisation, the resulting Gamma functions
combine into
\begin{equation}\label{eq:Cap6:H_transverse_convolution_integrated}
    \mathcal J_H(q^2)
    ={}
    \frac{1}{(4\pi)^2}
    \left(
    \frac{4\pi\mu^2e^{\gamma_E}}
    {\boldsymbol q^2}
    \right)^{2\epsilon}
    \Bigg[
    \frac{
    \Gamma(-\epsilon)^4
    \Gamma(1+\epsilon)^2}
    {\Gamma(-2\epsilon)^2}
    +
    \frac{
    2(2\epsilon-3)
    \Gamma(-\epsilon)^3
    \Gamma(1+2\epsilon)}
    {3\Gamma(-3\epsilon)}
    \Bigg] \, .
\end{equation}
Expanding around \(\epsilon=0\), its imaginary part becomes
\begin{equation}\label{eq:Cap6:H_diagram_epsilon_expansion}
\begin{aligned}
    \operatorname{Im}\mathcal M_{\mathrm{MRK}}^{(2)}(s,q^2)
    \simeq{}&
    4\left(\frac{G}{\hbar}\right)^3
    s^3
    \log\left(\frac{s}{|t|}\right)
    \left(
    \frac{4\pi\mu^2}{\boldsymbol q^2}
    \right)^{2\epsilon}
    \left[
    -\frac{1}{\epsilon^2}
    +\frac{2}{\epsilon}
    +\zeta_2
    +\mathcal O(\epsilon)
    \right] \, .
\end{aligned}
\end{equation}
This reproduces the two-loop \(H\)-diagram result originally obtained
by Amati, Ciafaloni and Veneziano in Ref.~\citeall{Amati:1990xe}.

The leading \(1/\epsilon^2\) pole has a direct soft interpretation. In
the limit \(k^0=\omega\to0\), the Lipatov current reduces to the
high-energy expansion of the Weinberg soft current,
\begin{equation}\label{eq:Cap6:Lipatov_Weinberg_soft_limit}
    -\frac{\kappa^3s^2}{t_1t_2}
    J^{\mu\nu}(q_1,k,q_2)
    \xrightarrow{\omega\to0}
    \left.
    S_{\mathrm{G}}^{(0)\, \mu\nu}(k)
    \right|_{\mathrm{Regge}}\,
    \mathcal M_{2\to2}^{(0)}(s,q^2) \, ,
\end{equation}
\begin{samepage}
where \(\mathcal M_{2\to2}^{(0)}(s,q^2)\) is the tree-level elastic
amplitude which, after restoring \(\hbar\), in the Regge limit reads
\begin{equation}\label{eq:Cap6:tree_elastic_Regge_factorisation}
    \mathcal M_{2\to2}^{(0)}(s,q^2)
    \simeq
    -\frac{1}{\hbar}
    \left[\kappa C_i(p_1,p_4)\right]
    \frac{s^2}{t}
    \left[\kappa C_j(p_2,p_3)\right] \, ,
\end{equation}
\end{samepage}
in agreement with \refeq{eq:Cap4:gravity_four_point_Regge_factorisation}.
The high-energy Weinberg current appearing in
\refeq{eq:Cap6:Lipatov_Weinberg_soft_limit} is
\begin{equation}\label{eq:Cap6:Weinberg_current_Regge}
     \left.
    S_{\mathrm{G}}^{(0)\, \mu\nu}(k)
    \right|_{\mathrm{Regge}}\,
    =\kappa\Bigg[
    (q\cdot k)
    \left(
    \frac{p_1^\mu p_1^\nu}{(p_1\cdot k)^2}
    -
    \frac{p_2^\mu p_2^\nu}{(p_2\cdot k)^2}
    \right)
    -
    \frac{p_1^\mu q^\nu+p_1^\nu q^\mu}{p_1\cdot k}
    +
    \frac{p_2^\mu q^\nu+p_2^\nu q^\mu}{p_2\cdot k}
    \Bigg] \, ,
\end{equation}
where the momentum transfer is \(q=-p_1-p_4=p_2+p_3\).
It coincides with the leading expansion of
the Weinberg soft factor up to quadratic terms going as $\mathcal O(q^2)$. At the level of the three-particle unitarity
cut, both tree-level five-point amplitudes therefore factorise in the
soft region, therefore, after summing over the polarisation of the cut graviton,
their product becomes schematically
\begin{equation}\label{eq:Cap6:H_cut_Weinberg_factorisation}
\begin{aligned}
    \sum_{\lambda=\pm2}
    \mathcal M_{2\to3}^{(0),\lambda}
    \left[\mathcal M_{2\to3}^{(0),\lambda}\right]^*
    \xrightarrow{\omega\to0}{}&
    \left|\mathcal M_{2\to2}^{(0)}\right|^2
    S_{\mathrm{G}}^{(0)\, \mu\nu} \,
    \Pi_{\mu\nu,\rho\sigma}
    \left[
    S_{\mathrm{G}}^{(0)\, \mu\nu}
    \right]^* \, ,
\end{aligned}
\end{equation}
where \(\Pi_{\mu\nu,\rho\sigma}\) is the
projector over the graviton's physical polarisations. Integrating this soft result over the on-shell
 phase space reproduces exactly the coefficient of the leading
\(1/\epsilon^2\) pole in
\refeq{eq:Cap6:H_diagram_epsilon_expansion}. Hence the double pole is the
universal Weinberg contribution to the cut, whereas the subleading pole
and the finite terms probe the full transverse Lipatov kernel.

The scaling \(G^3s^3\log(s/|t|)\) identifies this result as the first
classical Regge logarithm of the multi-\(H\) sequence discussed in
Chap.~\ref{Cap4:Regge_limit_and_BFKL}. Since the elastic \(H\) diagram
corrects the four-point amplitude, its classical part contributes to the
impulse and hence to the scattering angle through the KMOC relation
reviewed in Appendix~\ref{app:Classical_limit}. If we add real radiative emissions to the 
\(H\) diagrams, as discussed in Sec~\ref{sec:Cap6:radiative_H_kernel}, the corresponding five-point kernel enters the expectation value of the metric field
as described in Sec.~\ref{sec:app:waveform_tree_amplitude}.
Going beyond the leading logarithmic accuracy requires also to look at the region in
which the cut graviton and a neighbouring massive particle have
comparable rapidities. Its classical contribution is described by the
NMRK kernel \(\mathcal K_{\mathrm{NMRK}}^{\mathrm{cl}}\) constructed in
Sec.~\ref{sec:Cap6:classical_NMRK_kernel}.

\chapter{Conclusions}
\label{cap:Conclusions}

In this thesis, we have investigated the high-energy behaviour of
gravitational scattering amplitudes and its relation to classical
observables. The main purpose was to extend the familiar factorisation of
amplitudes in multi-Regge kinematics to regions in which two particles have
comparable rapidities, and to determine which of the resulting building
blocks survive the classical limit. This required combining on-shell
methods, high-energy factorisation, unitarity and
maps from amplitudes to observables.

We first reviewed the spinor-helicity formalism, on shell recursion relations and the
soft and collinear limits of tree-level amplitudes. These tools were then
connected with the Regge description of gauge theories and gravity. In strict
MRK, gravitational amplitudes factorise into impact factors, Reggeised
propagators and one-graviton central emission vertices. The latter display a
double-copy structure inherited from a gauge theory and reduce to the
high-energy form of the Weinberg current in the soft limit. The extension to
two emitted gravitons showed that this structure remains manifest for equal
helicities, whereas the opposite-helicity result, extracted from an NMHV
amplitude, involves several rational contributions and does not admit the
same simple representation.

We next studied the first corrections to that strict rapidity ordering. Starting
from the five-graviton MHV amplitude, we extracted the all-graviton NMRK
impact factor and verified its reduction to the expected sequence of MRK
vertices. Its external collinear limit yielded the gravitational splitting
functions. Although these functions retain a non trivial complex phase,
they remain finite in the magnitude of the relative transverse momentum, in
agreement with the absence of tree-level collinear divergences in gravity.

The analysis was then extended to massive matter. Using BCFW recursion relations, we
reconstructed the tree-level five-point amplitude describing the scattering
of two massive scalars with the emission of one graviton. Its high-energy
limit provided the corrected scalar--graviton impact factor in NMRK. We
checked both its strict MRK limit and its correlated classical limit, in
which the transverse momenta scale with \(\hbar\) while the rapidity ratio
becomes large. The corresponding massive scalar--graviton splitting function
is again finite in the collinear limit, with the scalar mass modifying only
its finite coefficient.

The same five-point amplitude has then been used to construct the leading classical
waveform. After the longitudinal constraints in the radiation kernel have been
performed, the remaining transverse Fourier transform was decomposed into
azimuthal harmonics. The angular integration led to a series of radial Hankel
transforms, making explicit how the impact parameter and collinear
dependences are encoded in Bessel functions. 

Finally, we constructed the gravitational kernels obtained by sewing the
high-energy vertices across unitarity cuts. The one-graviton CEV produced the
elastic \(H\) kernel, at LL accuracy, while the two-graviton CEV generated its radiative
extension. Sewing two classical massive impact factors also yielded an NMRK
kernel that retains the finite rapidity separation between the cut graviton
and the neighbouring massive particle, leading to the NLL accuracy. Its strict MRK limit agrees with the
ordinary \(H\) kernel, while its soft limit is governed by the Weinberg
current. The transverse integration of the elastic kernel reproduced the
classical two-loop \(H\) diagram result of Amati, Ciafaloni and Veneziano. In
particular, the leading infrared double pole was identified with the
universal soft contribution, whereas the subleading terms probe the full
Lipatov kernel.

These results also clarify the physical role of a BFKL like evolution in
gravity. Unlike QCD, where a Reggeised ladder organises the dominant
high-energy logarithms, classical gravity is dominated by eikonal
exchanges. New connected classical corrections arise when distinct Glauber
exchanges are linked by on-shell gravitons, producing the \(H\) and
multi-\(H\) sequence. The kernels derived here are therefore the building blocks of selected logarithmic
corrections to classical scattering and radiation observables.

Several natural developments of this work remain open:
\begin{itemize}
    \item The tree-level waveform can be extended for loops. This
    requires the computation of the classical part of five-point amplitudes at loop level, providing the first 
    corrections to the high-energy waveform;

    \item The radiative \(H\) kernel constructed in this thesis can be
    integrated to obtain the corresponding NLL contribution. This is more
    involved than the elastic calculation because the two-graviton CEV has a
    lengthy rational structure. In particular, its opposite-helicity sector
    contains several factorisation channels which make the computation really harder;

    \item On the same line, one can construct and evaluate the next
    multi-\(H\) contribution, denoted by \(H_2\), by sewing amplitudes across
    two successive three-particle unitarity cuts. This would provide the
    first explicit test of the iteration of gravitational kernels beyond a
    single \(H\) diagram and would help determine how the classical
    multi-\(H\) sequence organises the higher-order Regge logarithms.
\end{itemize}

Together, these directions would extend the factorised description developed
here from leading tree-level radiation and the first classical \(H\) diagram
to a systematic treatment of loop waveforms and higher multi-Regge
corrections to classical gravity.

\chapter*{Acknowledgements}
\phantomsection
\addcontentsline{toc}{chapter}{Acknowledgements}
\markboth{Acknowledgements}{Acknowledgements}

I would like to thank Prof.~Paolo Pani for his support and for enabling
my participation in the 7th School of Analytic Computing in High-Energy
and Gravitational Theoretical Physics in Atrani through funding provided
by the Italian Ministry of University and Research (MUR) under the FIS2
Advanced Grant ET-NOW (CUP:~B53C25001080001).

I would also like to express my sincere gratitude to Dr.~Vittorio Del Duca for
his guidance, encouragement and continuous support throughout this work.

I am also grateful to Francesco Alessio, Emmet Byrne, Emanuele Rosi and
Victor Sanz Sanchis for the many valuable discussions and for their
collaboration during this project. Finally, I would like to thank the
theoretical physics group at Sapienza University of Rome and at the INFN
Laboratori Nazionali di Frascati for providing a stimulating and welcoming
scientific environment.

\appendix
\chapter{Conventions and useful formulae}
\label{app:conventions}

\section{Poincaré algebra}
\label{app:poincare_algebra}

We collect here the conventions for the generators of the Poincaré algebra.
The generators of spacetime translations are denoted by \(P_\mu\), while the
Lorentz generators are denoted by \(M_{\mu\nu}=-M_{\nu\mu}\). With the metric
\(\eta_{\mu\nu}=\mathrm{diag}(+,-,-,-)\), their commutation relations are
\begin{equation}
    [P_\mu,P_\nu]=0 \, ,
\end{equation}
\begin{equation}
    [M_{\mu\nu},P_\rho]
    =
    i\left(
    \eta_{\nu\rho}P_\mu
    -
    \eta_{\mu\rho}P_\nu
    \right) \, ,
\end{equation}
\begin{equation}
    [M_{\mu\nu},M_{\rho\sigma}]
    =
    i\left(
    \eta_{\nu\rho}M_{\mu\sigma}
    +
    \eta_{\mu\sigma}M_{\nu\rho}
    -
    \eta_{\mu\rho}M_{\nu\sigma}
    -
    \eta_{\nu\sigma}M_{\mu\rho}
    \right) \, .
\end{equation}

It is useful to introduce the rotation and boost generators as 
\begin{equation}
    J_i=\frac{1}{2}\epsilon_{ijk}M^{jk} \, ,
    \qquad
    K_i=M_{0i}=-M^{0i} \, .
\end{equation}

In terms of these generators, the Lorentz algebra becomes
\begin{equation}
    [J_i,J_j]=i\epsilon_{ijk}J_k \, ,
    \qquad
    [J_i,K_j]=i\epsilon_{ijk}K_k \, ,
    \qquad
    [K_i,K_j]=-i\epsilon_{ijk}J_k \, .
\end{equation}

Finally, the action of rotations and boosts on the momentum generators is
\begin{equation}
    [J_i,P_0]=0 \, ,
    \qquad
    [J_i,P_j]=i\epsilon_{ijk}P_k \, ,
\end{equation}
\begin{equation}
    [K_i,P_0]=-iP_i \, ,
    \qquad
    [K_i,P_j]=-i\delta_{ij}P_0 \, .
\end{equation}

\section{Discrete symmetries}
\label{app:parity_charge}

In this section we describe the action of the discrete symmetries \(\mathcal{C}\) and \(\mathcal{P}\) on spinors and on the full amplitudes, following mainly Ref.~[\cite{Dixon_1}].

Let us first consider parity.  Since the helicity is the projection of the angular momentum along the beam axis, parity flips the helicity of a massless particle,
\begin{equation}
    \mathcal{P}: \qquad h\longrightarrow -h \, .
\end{equation}
In spinor-helicity variables this corresponds to exchanging the two chiral spinor representations. With our conventions, this amounts to the replacement
\begin{equation}\label{eq:app:parity_spinor_brackets}
    \mathcal{P}: \qquad
    \langle i j\rangle \longleftrightarrow [j i] \, ,
    \qquad
    [i j] \longleftrightarrow \langle j i\rangle \, .
\end{equation}
Therefore, at the level of a colour-ordered helicity amplitude,
\begin{equation}\label{eq:app:parity_amplitude}
    A_n(1^{h_1},2^{h_2},\ldots,n^{h_n})
    \xrightarrow{\,\mathcal{P}\,}
    A_n(1^{-h_1},2^{-h_2},\ldots,n^{-h_n})
    =
    A_n(1^{h_1},2^{h_2},\ldots,n^{h_n})
    \bigg|_{\langle ij\rangle \leftrightarrow  [ji]} \, .
\end{equation}
This rule implies that, for instance, a MHV amplitude written only in terms of angle brackets, under parity is mapped into its anti-MHV amplitude, written only in terms of square brackets.

Charge conjugation exchanges particles and antiparticles. Along a fermion line, it can be implemented by interchanging the labels of the fermion and antifermion while keeping track of the ordering and of possible signs fixed by the colour-ordered Feynman rules. Schematically,
\begin{equation}\label{eq:app:charge_conjugation_fermion_line}
    \mathcal{C}: \qquad
    f_i^{h_i}\,\bar f_j^{h_j}
    \longleftrightarrow
    \bar f_i^{h_i}\, f_j^{h_j} \, .
\end{equation}
Since the  gauge interactions preserve chirality along massless fermion lines, the allowed helicity configurations are strongly constrained, indeed for a massless fermion-antifermion pair coupled to a vector current, the two helicities on the same fermion line must be opposite.

As a simple illustration, consider a four-fermion amplitude of the form
\begin{equation}
    A_4(1^+_{\bar e},2^-_{e},3^+_{q},4^-_{\bar q}) \, .
\end{equation}
Acting with charge conjugation on the electron line exchanges the labels \(1\leftrightarrow 2\), while acting on the quark line exchanges \(3\leftrightarrow 4\). Instead, parity flips all helicities and replaces angle brackets by square brackets according to \refeq{eq:app:parity_spinor_brackets}. Hence the four non-vanishing helicity configurations can be generated from one representative amplitude by combining \(\mathcal{P}\) and \(\mathcal{C}\) transformations.

In practice, the rules used throughout the thesis are the following:
\begin{itemize}
    \item \(\mathcal{P}\) flips every external helicity and exchanges angle and square spinor brackets as in \refeq{eq:app:parity_spinor_brackets};
    \item \(\mathcal{C}\) exchanges fermions with antifermions on the chosen fermion line, which in colour-ordered amplitudes is implemented by the corresponding relabelling of the external legs, possibly accompanied by the sign dictated by the ordering convention;
    \item since chirality is conserved on massless fermion lines in gauge interactions, many helicity configurations vanish identically, and the remaining ones are related by \(\mathcal{P}\) and \(\mathcal{C}\).
\end{itemize}

Ref.~[\cite{Dixon_1}] illustrates, as an example, the application of these discrete simmetries to the Drell-Yan process in order to clarify the picture.

\chapter{Classical limit and waveforms from amplitudes}
\label{app:Classical_limit}

Scattering amplitudes are quantum objects, but suitable expectation values
constructed from them admit a well defined classical limit. In this appendix we discuss the relation
between amplitudes and classical observables in
the framework developed by Kosower, Maybee and O'Connell (KMOC)
\citeall{Kosower:2018adc,kosower2022sagexreviewscatteringamplitudes}.
We summarize the restoration of \(\hbar\), the associated
classical power counting and the extraction of a gravitational waveform from
a tree-level five-point amplitude following mainly Ref.~\citeall{Cristofoli:2022waveforms,divecchia2024gravitationaleikonalparticlestring}.

\section{Restoring \texorpdfstring{\(\hbar\)}{hbar} and the classical scaling}
\label{sec:app:hbar_classical_scaling}

We work in relativistic units, \(c=1\), while keeping \(\hbar\) explicit.
The classical limit is taken as \(\hbar\to0\) at fixed masses, velocities,
impact parameter and Newton's constant. Restoring \(\hbar\) requires
distinguishing a physical momentum \(p^\mu\) from its associated wavevector,
\begin{equation}\label{eq:app:momentum_wavevector}
    \bar p^\mu\equiv\frac{p^\mu}{\hbar} \, .
\end{equation}
The two quantities have dimensions of mass and inverse length, respectively.

Factors of \(\hbar\) enter both through the couplings and through the
kinematics. In gauge theories and gravity the couplings appearing in an
amplitude must be replaced according to
\begin{equation}\label{eq:app:hbar_couplings}
    e\longrightarrow\frac{e}{\sqrt{\hbar}} \, ,
    \qquad
    \kappa\longrightarrow\frac{\kappa}{\sqrt{\hbar}}
    =
    \sqrt{\frac{8\pi G}{\hbar}} \, .
\end{equation}

For the scattering of massive bodies, the external momenta remain fixed
while momentum transfers and the momenta of emitted massless particles
scale as
\begin{equation}\label{eq:app:classical_momentum_scaling}
    p_i^\mu=m_i u_i^\mu=\mathcal O(\hbar^0) \, ,
    \qquad
    q_i^\mu=\hbar\bar q_i^\mu \, ,
    \qquad
    k^\mu=\hbar\bar k^\mu \, ,
\end{equation}
where $k\m$ is the momentum of the emitted  soft graviton.
This rescaling produces additional powers of \(\hbar\) from integration
measures and on-shell delta functions.
For example,
\begin{equation}\label{eq:app:hbar_measure_scaling}
    \widehat{\mathrm d}^{\,4}q
    =
    \hbar^4\widehat{\mathrm d}^{\,4}\bar q \, , \quad   
    \widehat{\delta}
    \bigl((p\pm q)^2-m^2\bigr)
    =
    \frac{1}{\hbar}
    \widehat{\delta}
    \bigl(\pm2p\cdot\bar q+\hbar\bar q^2\bigr) \, ,  \quad  
    e^{\pm i b\cdot q/\hbar}
    =
    e^{\pm i b\cdot\bar q} \, ,
\end{equation}
where
\begin{equation}
    \widehat{\mathrm d}^{\,4}q
    \equiv
    \frac{\mathrm d^4q}{(2\pi)^4} \, ,
    \qquad
    \widehat{\delta}^{(n)}(q)
    \equiv
    (2\pi)^n\delta^{(n)}(q) \, .
\end{equation}
The factors in \refeq{eq:app:hbar_measure_scaling} are only part of the
power counting because the overall scaling must also include the couplings,
numerators and propagators in each relevant momentum region.
For a soft momentum transfer \(q=\hbar\bar q\), for example, a massless
propagator and a massive propagator off-shell behave as
\begin{equation}\label{eq:app:hbar_propagator_scaling}
    \frac{1}{q^2}
    =
    \frac{1}{\hbar^2}
    \frac{1}{\bar q^2} \, ,
    \qquad
    \frac{1}{(p+q)^2-m^2}
    =
    \frac{1}{\hbar}
    \frac{1}{2p\cdot\bar q+\hbar\bar q^2} \, ,
\end{equation}
where \(p^2=m^2\). The two propagators therefore supply different powers
of \(\hbar\), and propagators belonging to other momentum regions can
scale differently again. Thus, loop order alone does not determine the
classical content of a contribution. All factors must be combined at the
level of the complete observable, whose small-\(\hbar\) expansion, as described in Ref~\citeall{Holstein_2004,Kosower:2018adc}, has the
schematic form 
\begin{equation}\label{eq:app:observable_hbar_expansion}
    \Delta O
    =
    \sum_{r<0}\hbar^r
    \Delta O_{\mathrm{supercl}}^{(r)}
    +
    \Delta O_{\mathrm{cl}}
    +
    \mathcal O(\hbar) \, .
\end{equation}
Terms with positive powers of \(\hbar\) vanish in the classical limit, while
the terms with negative powers are usually called \emph{superclassical}. In a properly defined observable they cancel between virtual
and unitarity-cut contributions, or reorganise into the exponentiation of
the leading interaction. The finite \(\hbar^0\) term is instead our
classical contribution.

\section{Classical observables in the KMOC formalism}
\label{sec:app:KMOC_observables}

The KMOC formalism, described in Ref.~\citeall{Kosower:2018adc}, does not identify an amplitude by itself with a classical
observable. Instead, it embeds amplitudes into expectation values evaluated
on states that describe the incoming classical configuration. Since plane
waves are completely delocalised, two normalised wave packets are used to
prepare massive particles with a prescribed impact parameter,
\begin{equation}\label{eq:app:KMOC_incoming_state}
    \lvert\psi\rangle_{\mathrm{in}}
    ={}
    \int
    \mathrm d\Phi_{m_1}(p_1)
    \mathrm d\Phi_{m_2}(p_2)
    \times
    \phi_1(p_1)\phi_2(p_2)
    e^{i b_1\cdot p_1/\hbar}
    e^{i b_2\cdot p_2/\hbar}
    \lvert p_1,p_2\rangle_{\mathrm{in}} \, ,
\end{equation}
where
\begin{equation}\label{eq:app:bmuperpendic}
\begin{aligned}
    \mathrm d\Phi_m(p)
    &\equiv
    \widehat{\mathrm d}^{\,4}p\,
    \widehat{\delta}^{(+)}(p^2-m^2) \, ,
    &
    b^\mu
    &\equiv b_1^\mu-b_2^\mu \, ,
    \\
    p_1\cdot b
    &=p_2\cdot b=0 \, .
\end{aligned}
\end{equation}
The wave packets must satisfy the ``Goldilocks'' hierarchy
\begin{equation}\label{eq:app:KMOC_goldilocks}
    \ell_{\mathrm c}^{(i)}
    =
    \frac{\hbar}{m_i}
    \ll
    \ell_{\mathrm w}^{(i)}
    \ll
    \sqrt{-b^2} \, .
\end{equation}
They are therefore much wider than the Compton wavelengths, suppressing
quantum localisation effects, but much narrower than the characteristic
scattering scale. The classical result is obtained by extracting the leading
term in this hierarchy.
The wave packet average then localises smooth functions of the massive
momenta around \(p_i^\mu=m_i u_i^\mu\), while the phase in
\refeq{eq:app:KMOC_incoming_state} retains the dependence on the impact
parameter.

Let \(O\) be an operator associated with a measurable quantity. Its
integrated change during the scattering is
\begin{equation}\label{eq:app:KMOC_observable}
    \Delta O
    =
    \langle\psi|
    S^\dagger O S-O
    |\psi\rangle \, .
\end{equation}
Writing \(S=1+iT\) and using unitarity gives the exact identity
\begin{equation}\label{eq:app:KMOC_virtual_cut}
    \Delta O
    =
    i\langle\psi|[O,T]|\psi\rangle
    +
    \langle\psi|T^\dagger[O,T]|\psi\rangle \, .
\end{equation}
The first term is linear in a scattering amplitude and is conventionally
called the virtual contribution. The second contains products of amplitudes
 and has the interpretation of a
unitarity cut. Beyond the leading order the two pieces are not separately
physical, because only their sum has a well defined classical limit and the
superclassical iterations cancel  between them.

Important examples are the impulse imparted to the first body and the
total radiated momentum,
\begin{equation}\label{eq:app:KMOC_impulse_radiation}
\begin{aligned}
    \Delta p_1^\mu
    &=
    \langle\psi|
    S^\dagger P_1^\mu S-P_1^\mu
    |\psi\rangle \, ,
    \\
    K_{\mathrm{rad}}^\mu
    &=
    \langle\psi|
    S^\dagger P_{\mathrm{rad}}^\mu S
    |\psi\rangle \, ,
\end{aligned}
\end{equation}
with
\begin{equation}
    P_{\mathrm{rad}}^\mu
    =
    \sum_{h=\pm2}
    \int\mathrm d\Phi_0(k)\,
    k^\mu a_h^\dagger(k)a_h(k) \, .
\end{equation}
These observables probe amplitudes in different ways. At leading order, the first
commutator term in \refeq{eq:app:KMOC_virtual_cut} gives the conservative
impulse, and hence the scattering angle, from the tree-level four-point
amplitude. Higher-loop four-point amplitudes can supply further classical
corrections whenever their complete \(\hbar\) scaling contains a finite
term. The radiated momentum behaves differently, because the incoming state
contains no gravitons and \(P_{\mathrm{rad}}^\mu\) is quadratic in creation
and annihilation operators, so its first non vanishing contribution is
quadratic in the tree-level five-point radiation kernel. A waveform instead
measures a local field operator, which is linear in creation and annihilation
operators, and is
therefore linear in the same radiation kernel at leading order. The elastic
four-point amplitude alone cannot produce a waveform, since it contains no
on-shell emitted graviton.

Classical radiation is described by a coherent state,
\begin{equation}\label{eq:app:radiation_coherent_state}
    \lvert\alpha\rangle
    =
    \mathcal N_\alpha
    \exp\left[
    \sum_{h=\pm2}
    \int\mathrm d\Phi_0(k)\,
    \alpha_h(k)a_h^\dagger(k)
    \right]\lvert0\rangle \, ,
\end{equation}
whose occupation number becomes large in the classical limit. The
appropriate scaling is made explicit by setting
\(k=\hbar\bar k\) and
\(\bar\alpha_h(\bar k)=\hbar^{3/2}\alpha_h(\hbar\bar k)\), so that the mean graviton number  is
\begin{equation}\label{eq:app:coherent_state_occupation}
    N_{\mathrm{gr}}
    =
    \sum_{h=\pm2}
    \int\mathrm d\Phi_0(k)\,
    \lvert\alpha_h(k)\rvert^2
    =
    \frac{1}{\hbar}
    \sum_{h=\pm2}
    \int\mathrm d\Phi_0(\bar k)\,
    \lvert\bar\alpha_h(\bar k)\rvert^2 \, .
\end{equation}
A classical gravitational wave is not described by a state with a fixed
number of gravitons, but by a coherent superposition of states with different
occupation numbers. Indeed, a coherent state satisfies
\(a_h(k)\lvert\alpha\rangle=\alpha_h(k)\lvert\alpha\rangle\), so that the
expectation value of the linearised gravitational field is non vanishing
and is determined directly by the profile \(\alpha_h(k)\). Moreover,
\(N_{\mathrm{gr}}\sim\hbar^{-1}\), and therefore the relative fluctuations
scale as
\(\Delta N_{\mathrm{gr}}/N_{\mathrm{gr}}\sim N_{\mathrm{gr}}^{-1/2}\to0\)
in the classical limit, which ensure the classical nature of the gravitational wave's field, so that correlation functions factorise into products
of classical fields. At leading order, the one-graviton emission amplitude,
through the radiation kernel, determines \(\alpha_h(k)\), because the latter gives
the spectral content of the waveform, whereas
\(\lvert\alpha_h(k)\rvert^2\) determines the radiated energy and momentum.

\section{The waveform from a tree-level amplitude}
\label{sec:app:waveform_tree_amplitude}

Following Sec.~8.3.1 of
Ref.~\citeall{divecchia2024gravitationaleikonalparticlestring}, we start
from the canonically normalised free graviton field,
\begin{equation}\label{eq:app:free_graviton_field}
    \widehat H_{\mu\nu}(x)
    ={}
    \sum_{\lambda=\pm}
    \int\mathrm d\Phi_0(k)
    \Bigl[
    \varepsilon_{\mu\nu}^{(2\lambda)}(k)
    a_{2\lambda}(k)e^{-ik\cdot x}+
    \varepsilon_{\mu\nu}^{(2\lambda)*}(k)
    a_{2\lambda}^{\dagger}(k)e^{ik\cdot x}
    \Bigr] \, ,
\end{equation}
where
\(\varepsilon_{\mu\nu}^{(2\lambda)}
=\varepsilon_\mu^{(\lambda)}\varepsilon_\nu^{(\lambda)}\).
With the normalisation \(\kappa=\sqrt{8\pi G}\), the physical metric
perturbation and the corresponding classical waveform are
\begin{equation}\label{eq:app:classical_metric_waveform}
\begin{aligned}
    h_{\mu\nu}^{\mathrm{cl}}(x)
    &\equiv
    \kappa\lim_{\hbar\to0}
    \langle\psi|
    S^\dagger\widehat H_{\mu\nu}(x)S
    |\psi\rangle \, ,
    \\
    g_{\mu\nu}(x)
    &=
    \eta_{\mu\nu}+2h_{\mu\nu}^{\mathrm{cl}}(x) \, .
\end{aligned}
\end{equation}
Thus, the waveform is obtained directly as a one-point function of the
metric field in the outgoing state.

Since the incoming state contains no gravitons, the expectation value of
an outgoing annihilation operator can be written as
\begin{equation}\label{eq:app:outgoing_annihilation_operator}
\begin{aligned}
    \langle a_{2\lambda}(k)\rangle_{\mathrm{out}}
    ={}&
    i\langle\psi|a_{2\lambda}(k)T|\psi\rangle
    +
    \langle\psi|T^\dagger a_{2\lambda}(k)T|\psi\rangle \, .
\end{aligned}
\end{equation}
At leading order the second term is of higher order in the gravitational
coupling, while the first one contains the tree-level five-point amplitude,
\begin{equation}
    \langle p_1',p_2',k^{2\lambda}|T|p_1,p_2\rangle
    ={}
    \mathcal M_5^{(0)}
    (p_1,p_2\to p_1',p_2',k^{2\lambda})\,
    \widehat{\delta}^{(4)}
    (p_1+p_2-p_1'-p_2'-k) \, ,
\end{equation}
therefore the leading waveform is completely determined by the classical
limit of \(\mathcal M_5^{(0)}\).

After performing the \(\hbar\) rescaling described in
Sec.~\ref{sec:app:hbar_classical_scaling}, we return to natural units and
define \(p_i'=p_i-q_i\), so that \(q_1+q_2=k\).
The 
radiation kernel in the impact parameter space is
\begin{equation}\label{eq:app:tree_radiation_kernel}
\begin{aligned}
    \mathcal R_{\lambda}^{(0)}(k;b_1,b_2)
    ={}&
    \int
    \widehat{\mathrm d}^{\,4}q_1
    \widehat{\mathrm d}^{\,4}q_2\,
    \widehat{\delta}(2p_1\cdot q_1)
    \widehat{\delta}(2p_2\cdot q_2)
    \times
    \widehat{\delta}^{(4)}(q_1+q_2-k)\,
    e^{i b_1\cdot q_1+i b_2\cdot q_2}
    \\
    &\times
    \mathcal M_5^{(0)}
    \bigl(
    p_1,p_2
    \to
    p_1-q_1,p_2-q_2,k^{2\lambda}
    \bigr) \, .
\end{aligned}
\end{equation}
The on-shell conditions in
\refeq{eq:app:tree_radiation_kernel} have been expanded to leading order
in \(\hbar\), and the wave packet has localised
\(p_i^\mu\) onto \(m_i u_i^\mu\). Their longitudinal integrations can be
performed explicitly following Ref.~\citeall{De_Angelis_2024}. Let's define
\begin{equation}\label{eq:app:waveform_longitudinal_decomposition}
\begin{gathered}
    \Delta_{12}
    \equiv
    (p_1\cdot p_2)^2-m_1^2m_2^2 = m_1^2m_2^2 \, \sinh^2(Y)\, ,
    \qquad
    q_1=q_{1\perp}+\alpha p_1+\beta p_2 \, ,
    \\
    p_1\cdot q_{1\perp}
    =p_2\cdot q_{1\perp}=0 \, .
\end{gathered}
\end{equation}
After using momentum conservation to set \(q_2=k-q_1\), the two
mass-shell constraints fix \(\alpha\) and \(\beta\), hence,
\begin{equation}\label{eq:app:waveform_on_shell_transfers}
\begin{aligned}
    q_1^{\star}(q_{1\perp})
    ={}&
    q_{1\perp}
    +\frac{p_2\cdot k}{\Delta_{12}}
    \left[
    (p_1\cdot p_2)p_1-m_1^2p_2
    \right] \, ,
    \\
    q_2^{\star}(q_{1\perp})
    ={}&k-q_1^{\star}(q_{1\perp}) \, .
\end{aligned}
\end{equation}
The corresponding Jacobian reduces the original eight-dimensional
measure to the two transverse components of \(q_1\),
\begin{equation}\label{eq:app:waveform_reduced_measure}
\begin{aligned}
    &\int
    \widehat{\mathrm d}^{\,4}q_1
    \widehat{\mathrm d}^{\,4}q_2\,
    \widehat{\delta}(2p_1\cdot q_1)
    \widehat{\delta}(2p_2\cdot q_2)
    \widehat{\delta}^{(4)}(q_1+q_2-k)
    \, f(q_1,q_2)
    \\
    &\hspace{2.0cm}=
    \frac{1}{4\sqrt{\Delta_{12}}}
    \int
    \widehat{\mathrm d}^{\,2}q_{1\perp}\,
    f(q_1^{\star},q_2^{\star}) \, .
\end{aligned}
\end{equation}
Using \(q_2^{\star}=k-q_1^{\star}\),
\(b=b_1-b_2\), and \refeq{eq:app:bmuperpendic}, the phase can be written as
\begin{equation}\label{eq:app:waveform_impact_parameter_phase}
    e^{i b_1\cdot q_1^{\star}
    +i b_2\cdot q_2^{\star}}
    ={}
    e^{i b_2\cdot k}
    e^{i b\cdot q_1^{\star}}    
    ={}
    e^{i b_2\cdot k}
    e^{i b\cdot q_{1\perp}} \, .
\end{equation}
The first factor only translates the origin of retarded time according to
\(u\rightarrow u-n\cdot b_2\), therefore we can choose
\(b_2=0\) and \(b_1=b\). With this choice, the radiation kernel takes the
reduced form
\begin{equation}\label{eq:app:tree_radiation_kernel_reduced}
    \mathcal R_{\lambda}^{(0)}(k;b)
    ={}
    \frac{1}{4\sqrt{\Delta_{12}}}
    \int
    \widehat{\mathrm d}^{\,2}q_{1\perp}\,
    e^{i b\cdot q_{1\perp}} 
    \times
    \mathcal M_5^{(0)}
    \bigl(
    p_1,p_2
    \to
    p_1-q_1^{\star},p_2-q_2^{\star},k^{2\lambda}
    \bigr) \, .
\end{equation}

Consider now an asymptotic observer at
\(x^\mu=(x^0,r\boldsymbol n)\), with
\(r\gg\sqrt{-b^2}\), and define
\(u=x^0-r\) and \(n^\mu=(1,\boldsymbol n)\). For a graviton of momentum
\(k^\mu=\omega n^\mu\), the  waveform in the frequency domain projected onto the
helicity \(2\lambda\) is
\begin{equation}\label{eq:app:spectral_metric_waveform}
    \widehat h_{\lambda}^{(0)}
    (k;b)
    =
    \frac{\kappa}{4\pi r}
    \mathcal R_{\lambda}^{(0)}(k;b) \, .
\end{equation}
Its form in the time domain is obtained just through a Fourier transform,
\begin{equation}\label{eq:app:time_domain_metric_waveform}
\begin{aligned}
    h_{\lambda}^{(0)}
    (u,r,\boldsymbol n)
    ={}&
    \int_{-\infty}^{+\infty}
    \widehat{\mathrm d}\omega\,
    e^{-i\omega u}
    \widehat h_{\lambda}^{(0)}
    \bigl(\omega n;b\bigr)
    +\mathcal O(r^{-2}) \, ,
\end{aligned}
\end{equation}
where
\(\widehat{\mathrm d}\omega\equiv\mathrm d\omega/(2\pi)\), and the
negative frequency  is fixed by the reality of
\(h_{\mu\nu}^{\mathrm{cl}}\).

The Lorentz indices do not appear in
\refeq{eq:app:spectral_metric_waveform} because
\(\mathcal M_5^{(0)}(\ldots,k^{2\lambda})\) is already contracted with the
polarisation tensor of the emitted graviton. The calculation therefore consists of taking
the classical limit of the helicity amplitude, performing the transform in
\refeq{eq:app:tree_radiation_kernel_reduced}, multiplying by the asymptotic factor
in \refeq{eq:app:spectral_metric_waveform}, and finally Fourier transforming
in \(\omega\).

In the high-energy regime studied in Chap.~\ref{Cap4:Regge_limit_and_BFKL},
the five-point amplitude entering \refeq{eq:app:tree_radiation_kernel_reduced}
factorises according to
\refeq{eq:Cap4:gravity_five_point_MRK_factorisation}. The helicity waveform
is consequently built from the two gravitational impact factors, the two
\(t\)-channel propagators and the one-graviton CEV. In this regime, the
Lipatov vertex derived in Chap.~\ref{Cap4:Regge_limit_and_BFKL} is precisely
the tree-level radiative building block required by the radiation kernel.
Loop five-point amplitudes and the second term in
\refeq{eq:app:outgoing_annihilation_operator} enter only when the waveform
is computed beyond leading order.

\begin{figure}[H]
    \centering
    \resizebox{0.80\textwidth}{!}{%
    \begin{tikzpicture}[
        >=Latex,
        massive/.style={
            draw=black,
            line width=0.65pt,
            -{Latex[length=2.2mm]}
        },
        graviton/.style={
            draw=black,
            line width=0.65pt,
            decorate,
            decoration={
                snake,
                amplitude=1.1pt,
                segment length=5.2pt
            },
            double
        },
        amplitude/.style={
            circle,
            draw=blue!70!black,
            fill=blue!58,
            minimum size=10mm,
            inner sep=0pt
        }
    ]

    \node[anchor=west] at (0,2.7) {
        \(\displaystyle
        \widehat h_{\lambda}^{(0)}(k;b)
        =
        \frac{\kappa}{16\pi r\sqrt{\Delta_{12}}}
        \int
        \widehat{\mathrm d}^{\,2}q_{1\perp}\,
        e^{i b\cdot q_{1\perp}}
        \)
    };

    \begin{scope}[xshift=-4.5cm]

    \node at (8.5,0) {\(\displaystyle\times\)};

    \node[amplitude] (M) at (11.3,0) {};
    \node at (11.3,0.85) {
        \(\mathcal M_5^{(0)}\)
    };

    \draw[massive]
        (9.25,1.25) -- (M.north west)
        node[pos=0.10,above left=-1pt] {\(p_1\)};

    \draw[massive]
        (9.25,-1.25) -- (M.south west)
        node[pos=0.10,below left=-1pt] {\(p_2\)};

    \draw[massive]
        (M.north east) -- (13.35,1.25)
        node[pos=0.92,above right=-1pt] {\(p_1-q_1^{\star}\)};

    \draw[massive]
        (M.south east) -- (13.35,-1.25)
        node[pos=0.90,below right=-1pt] {\(p_2-q_2^{\star}\)};

    \draw[graviton]
        (M.east) -- (15.05,0);

    \draw[-{Latex[length=2.2mm]},line width=0.55pt]
        (12.05,0.32) -- (14.45,0.32)
        node[midway,above=2pt] {\(k^{2\lambda}\)};

    \node at (15.05,0.68) {
        \(r\rightarrow\infty\)
    };

    \end{scope}

    \end{tikzpicture}%
    }
    \caption{
        Diagrammatic representation of the leading
        gravitational waveform in the frequency domain.  }
    \label{fig:app:tree_waveform_diagram}
\end{figure}
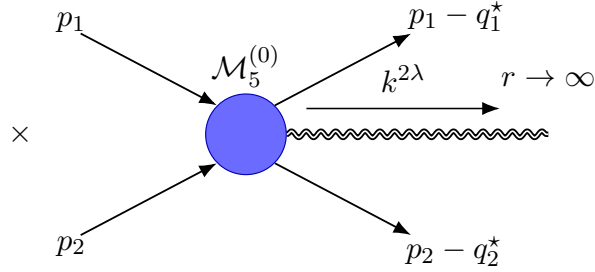

\begingroup
\let\clearpage\relax
\let\cleardoublepage\relax
\chapter{Kinematic regions}
\label{app:kinematic_regions}
\endgroup

This appendix collects the five-point kinematic formulae used in
Chap.~\ref{cap:Gravitational_amp_in_the_coll_limit}. The general
multi-particle construction is discussed in
Ref.~\citeall{barcaro2025centralemissionvertexgravitons}, while here the discussion is
specialised to the amplitude with four massive scalars and one graviton.
This makes explicit the high-energy expansion underlying the massive
impact factor that was given in
Sec.~\ref{sec:Cap5:massive_scalar_impact_factor_collinear_limit}.

\section{Five-point light-cone kinematics}
\label{app:five_point_light_cone_kinematics}

We use the light-cone conventions introduced in
\refeq{eq:Cap2:momenta_conventions},
\begin{equation}\label{eq:app:light_cone_mass_shell}
    p^\mu=(p^+,p^-;p_\perp,p_\perp^*) \, ,
    \qquad
    p^2=p^+p^- -p_\perp p_\perp^*=m^2 \, .
\end{equation}
Consequently,
\begin{equation}\label{eq:app:light_cone_scalar_product}
    2p_i\cdot p_j
    =p_i^+p_j^-+p_i^-p_j^+
    -p_{i\perp}p_{j\perp}^*
    -p_{i\perp}^*p_{j\perp} \, .
\end{equation}
For a positive-energy outgoing momentum it is useful to write
\begin{equation}\label{eq:app:massive_rapidity_parametrisation}
\begin{gathered}
    p_i^+=m_{i\perp}e^{y_i} \, ,
    \qquad
    p_i^-=m_{i\perp}e^{-y_i} \, ,
    \\
    m_{i\perp}^2=m_i^2+|p_{i\perp}|^2 \, .
\end{gathered}
\end{equation}
For a massless particle, \(m_{i\perp}=|p_{i\perp}|\). The incoming
particles have negative energy in the outgoing convention and their
light-cone components therefore carry the opposite sign.

Consider the process in \refeq{eq:Cap5:massive_five_point_process}. The
legs \(1\) and \(5\) have mass \(m_B\), the legs \(2\) and \(3\) have
mass \(m_A\), and \(p_4^2=0\). We introduce the positive longitudinal
scales
\begin{equation}\label{eq:app:incoming_longitudinal_scales}
    P_A^+\equiv-p_2^+>0 \, ,
    \qquad
    P_B^-\equiv-p_1^->0 \, .
\end{equation}
A convenient parametrisation of the five momenta is
\begin{equation}\label{eq:app:five_point_massive_momenta}
\begin{aligned}
    p_1^\mu
    &={}
    \left(-\frac{m_B^2}{P_B^-},-P_B^-;0,0\right) \, ,
    \\
    p_2^\mu
    &={}
    \left(-P_A^+,-\frac{m_A^2}{P_A^+};0,0\right) \, ,
    \\
    p_3^\mu
    &={}
    \left(
    p_3^+,\frac{m_A^2+|p_{3\perp}|^2}{p_3^+};
    p_{3\perp},p_{3\perp}^*
    \right) \, ,
    \\
    p_4^\mu
    &={}
    \left(
    p_4^+,\frac{|p_{4\perp}|^2}{p_4^+};
    p_{4\perp},p_{4\perp}^*
    \right) \, ,
    \\
    p_5^\mu
    &={}
    \left(
    \frac{m_B^2+|p_{5\perp}|^2}{p_5^-},p_5^-;
    p_{5\perp},p_{5\perp}^*
    \right) \, .
\end{aligned}
\end{equation}
The leading components of momentum conservation give
\begin{equation}\label{eq:app:five_point_leading_momentum_conservation}
\begin{gathered}
    P_A^+\simeq p_3^++p_4^+ \, ,
    \qquad
    P_B^-\simeq p_5^- \, ,
    \\
    p_{3\perp}+p_{4\perp}+p_{5\perp}=0 \, .
\end{gathered}
\end{equation}
The components omitted in the first line are fixed by exact momentum
conservation and affect only power suppressed terms in the Regge expansion.

\section{MRK and NMRK at five points}
\label{app:five_point_MRK_NMRK}

For the five-point process, multi-Regge kinematics corresponds to the
strong rapidity ordering
\begin{equation}\label{eq:app:five_point_MRK_ordering}
\begin{gathered}
    y_3\gg y_4\gg y_5 \, ,
    \\
    p_3^+\gg p_4^+\gg p_5^+ \, ,
    \qquad
    p_3^-\ll p_4^-\ll p_5^- \, ,
\end{gathered}
\end{equation}
while the transverse masses remain of the same order. Thus the graviton
is separated by a large rapidity interval from both scattered scalars.
In the NMRK region relevant to the upper massive impact factor, the
scalar \(\mathbf 3_A\) and the graviton instead form a rapidity cluster,
\begin{equation}\label{eq:app:five_point_NMRK_ordering}
\begin{gathered}
    y_3\simeq y_4\gg y_5 \, ,
    \\
    p_3^+\sim p_4^+\gg p_5^+ \, ,
    \qquad
    p_3^-\sim p_4^-\ll p_5^- \, .
\end{gathered}
\end{equation}
Here \(y_3-y_4\) is held finite while the total rapidity interval becomes
large. Accordingly, the invariant mass of the pair \((p_3,p_4)\) remains
finite and must not be expanded as a large invariant. The analogous
region with a cluster next to the lower scalar line follows by exchanging
the two ends of the ladder.

The finite separation inside the upper cluster is encoded by the
variable introduced in \refeq{eq:Cap5:massive_NMRK_X_definition},
\begin{equation}\label{eq:app:NMRK_X_rapidity}
    X=\frac{p_3^+}{p_4^+}
    =\frac{m_{3\perp}}{|p_{4\perp}|}
    e^{y_3-y_4} \, ,
    \qquad
    p_3^+=\frac{X}{1+X}P_A^+ \, ,
    \qquad
    p_4^+=\frac{1}{1+X}P_A^+ \, .
\end{equation}
NMRK is obtained by taking the high-energy limit at fixed
\(X=\mathcal O(1)\). The further limit \(X\to\infty\) separates legs
\(3\) and \(4\) and recovers the corresponding MRK ordering.

Defining the large expansion parameter by
\begin{equation}\label{eq:app:five_point_large_s}
    s=P_A^+P_B^-
    \simeq(p_3^++p_4^+)p_5^- \, ,
\end{equation}
the Mandelstam invariants \(s_{ij}=(p_i+p_j)^2\) become, at the
first non vanishing order in the Regge expansion,
\begin{equation}\label{eq:app:five_point_NMRK_invariants}
\begin{aligned}
    s_{12}&=s+m_A^2+m_B^2 \, ,
    &
    s_{13}&=-\frac{X}{1+X}s+m_A^2+m_B^2 \, ,
    \\
    s_{14}&=-\frac{s}{1+X}+m_B^2 \, ,
    &
    s_{15}&=-|p_{5\perp}|^2 \, ,
    \\
    s_{23}&=-\frac{m_A^2}{X(1+X)}
    -\frac{1+X}{X}|p_{3\perp}|^2 \, ,
    &
    s_{24}&=\frac{X}{1+X}m_A^2
    -(1+X)|p_{4\perp}|^2 \, ,
    \\
    s_{25}&=-s+m_A^2+m_B^2 \, ,
    &
    s_{35}&=\frac{X}{1+X}s+m_A^2+m_B^2 \, ,
    \\
    s_{45}&=\frac{s}{1+X}+m_B^2 \, ,
    &
    s_{34}&=\frac{1+X}{X}m_A^2
    +X|p_{4\perp}|^2
    +\frac{|p_{3\perp}|^2}{X}
    \\
    &&&
    -p_{3\perp}p_{4\perp}^*
    -p_{3\perp}^*p_{4\perp} \, .
\end{aligned}
\end{equation}
Corrections to these expressions are suppressed by inverse powers of
\(s\). At fixed \(X\), the invariants
\(s_{12},s_{13},s_{14},s_{25},s_{35}\), and \(s_{45}\) are of order
\(s\), whereas \(s_{15},s_{23},s_{24}\), and \(s_{34}\) remain of order
\(|t|\) or \(m_i^2\). 

\section{Massive momenta and the spinor expansion}
\label{app:massive_momenta_spinor_expansion}
\enlargethispage{3\baselineskip}

The compact amplitude in
\refeq{eq:Cap5:massive_five_point_amplitude_compact} contains spinor
strings built from massive momenta. To expand them with ordinary
massless spinors, every massive momentum \(p_i\), with
\(i\in\{1,2,3,5\}\), can be written as the sum of two null vectors,
\begin{equation}\label{eq:app:massive_momentum_null_decomposition}
\begin{aligned}
    p_i^\mu
    &={}
    \widetilde p_i^{\,\mu}+p_{i,m}^{\mu} \, ,
    \\
    \widetilde p_i^{\,\mu}
    &={}
    \left(
    p_i^+,\frac{|p_{i\perp}|^2}{p_i^+};
    p_{i\perp},p_{i\perp}^*
    \right),
    &
    p_{i,m}^{\mu}
    &={}
    \left(
    0,\frac{m_i^2}{p_i^+};0,0
    \right),
    && i\in\{2,3\},
    \\
    \widetilde p_i^{\,\mu}
    &={}
    \left(
    \frac{|p_{i\perp}|^2}{p_i^-},p_i^-;
    p_{i\perp},p_{i\perp}^*
    \right),
    &
    p_{i,m}^{\mu}
    &={}
    \left(
    \frac{m_i^2}{p_i^-},0;0,0
    \right),
    && i\in\{1,5\}.
\end{aligned}
\end{equation}
Indeed, \(\widetilde p_i^{\,2}=p_{i,m}^2=0\), while their sum obeys
\(p_i^2=m_i^2\). The split depends on the chosen light-cone direction,
but it is particularly convenient here because it separates the
massless kinematics from the term that restores the massive on-shell
condition.

Let
\(\lvert\widetilde i\rangle,\lvert\widetilde i]\) and
\(\lvert i_m\rangle,\lvert i_m]\) denote the massless spinors associated
with \(\widetilde p_i\) and \(p_{i,m}\), respectively. Linearity then gives
\begin{equation}\label{eq:app:single_massive_spinor_string}
    [4|p_i|4\rangle
    =
    [4\widetilde i]\langle\widetilde i4\rangle
    +[4i_m]\langle i_m4\rangle \, .
\end{equation}
For example, using \(K_B=p_1-p_5\) from
\refeq{eq:Cap5:massive_five_point_shorthand}, one obtains
\begin{equation}\label{eq:app:KB_massive_spinor_decomposition}
\begin{aligned}
    [4|K_B|4\rangle
    ={}&
    [4\widetilde 1]\langle\widetilde 1 4\rangle
    +[4 1_m]\langle1_m4\rangle
    \\
    &{}-
    [4\widetilde 5]\langle\widetilde 5 4\rangle
    -[4 5_m]\langle5_m4\rangle \, .
\end{aligned}
\end{equation}
No massive spinor-helicity variables are required in this step, all the
objects on the right-hand side are spinor products of null momenta.

The same decomposition applies to strings containing two momenta. In a
compact notation, with \(p_{i,0}\equiv\widetilde p_i\), it reads
\begin{equation}\label{eq:app:double_massive_spinor_string}
    [4|p_ip_j|4]
    =
    \sum_{\alpha,\beta\in\{0,m\}}
    [4i_\alpha]
    \langle i_\alpha j_\beta\rangle
    [j_\beta4] \, .
\end{equation}
This formula, together with linearity in \(K_A\) and \(K_B\), expands all
the structures in
\refeq{eq:Cap5:massive_five_point_spinor_strings}, including
\([4|23|4]\), \([4|51|4]\), and \([4|K_AK_B|4]\).

To extract the NMRK impact factor, one first substitutes
\refeq{eq:app:five_point_NMRK_invariants} and the decomposed spinor
strings into the full five-point amplitude of
Ref.~\citeall{Britto_2022}. The limit \(s\to\infty\) is then taken at
fixed \(X\), masses, and transverse momenta. Terms belonging to different
pieces of the amplitude must be combined before the expansion, since
non factorising leading powers and spurious poles cancel only in their
sum. The coefficient of the surviving \(s^2/t\) term defines
\begin{equation}\label{eq:app:massive_impact_factor_extraction}
\begin{aligned}
    C^{\phi\to\phi h}
    (p_{3\perp},p_{4\perp},m_A^2,X)
    ={}&
    -\lim_{s\to\infty}
    \frac{t}{\kappa^3s^2}
    \mathcal M_5^{(0)}
    (\mathbf 1_B,\mathbf 2_A,
    \mathbf 3_A,4_h^+,\mathbf 5_B)
    \\
    &\hspace{1.7cm}
    \text{at fixed }X,p_{i\perp},m_A,m_B \, .
\end{aligned}
\end{equation}
This is precisely the building block displayed in
\refeq{eq:Cap5:massive_five_point_NMRK_factorisation}. Keeping \(X\)
fixed during the expansion preserves the finite rapidity separation
inside the upper cluster; taking \(X\to\infty\) only afterwards gives
its MRK limit.

\chapter{The NMRK phase-space integral}
\label{app:NMRK_phase_space}

In this appendix  we derive the phase-space factor used in
\refeq{eq:Cap6:NMRK_phase_space_factor}. The correction with respect to
the strict MRK measure originates from keeping the finite longitudinal
momentum carried by the graviton in the upper NMRK cluster. We will follow Ref~\citeall{alessio2026gravitationalamplitudesregge}, 
extending the same procedure to the NMRK case.

\section{The multi-particle on-shell measure}
\label{app:general_on_shell_phase_space}

We use the hatted measure and delta function
\begin{equation}\label{eq:app:hatted_measure_definition}
    \widehat{\mathrm d}^{\,D}k
    \equiv
    \frac{\mathrm d^Dk}{(2\pi)^D} \, ,
    \qquad
    \widehat{\delta}^{(D)}(k)
    \equiv
    (2\pi)^D\delta^{(D)}(k) \, .
\end{equation}
For two massive particles and \(n\) gravitons, the on-shell phase space
can be written as
\begin{subequations}\label{eq:app:phase_space_2_plus_n}
\begin{align}
    \int\mathrm d\mathcal P_{2+n}
    ={}&
    \prod_{i=0}^{n+1}
    \int\mathrm d\Phi(k_i)\,
    \widehat{\delta}^{(D)}
    \left(
    p_1+p_2+\sum_{j=0}^{n+1}k_j
    \right) \, ,
    \\
    \mathrm d\Phi(k_0)
    ={}&
    \widehat{\mathrm d}^{\,D}k_0\,
    \widehat{\delta}^{+}(k_0^2-m_B^2) \, ,
    \\
    \mathrm d\Phi(k_{n+1})
    ={}&
    \widehat{\mathrm d}^{\,D}k_{n+1}\,
    \widehat{\delta}^{+}(k_{n+1}^2-m_A^2) \, ,
    \\
    \mathrm d\Phi(k_r)
    ={}&
    \widehat{\mathrm d}^{\,D}k_r\,
    \widehat{\delta}^{+}(k_r^2) \, ,
    \qquad r=1,\ldots,n \, .
\end{align}
\end{subequations}
The momenta are all outgoing in the amplitude convention. Let
\(d=D-2\) denote the dimension of the transverse space, solving each
positive energy mass-shell constraint gives
\begin{equation}\label{eq:app:rapidity_phase_space_measure}
    \mathrm d\Phi(k_i)
    =
    \frac{1}{2}\,
    \widehat{\mathrm d}y_i\,
    \widehat{\mathrm d}^{\,d}\boldsymbol k_i \, .
\end{equation}
Moreover, the light-cone decomposition of the momentum-conservation
delta function is
\begin{equation}\label{eq:app:light_cone_delta_decomposition}
    \widehat{\delta}^{(D)}(k)
    =
    2\,
    \widehat{\delta}^{(d)}(\boldsymbol k)\,
    \widehat{\delta}(k^+)\,
    \widehat{\delta}(k^-) \, .
\end{equation}
Combining these relations yields
\begin{equation}\label{eq:app:phase_space_rapidity_form}
    \int\mathrm d\mathcal P_{2+n}
    ={}
    \frac{1}{2^{n+1}}
    \left(
    \prod_{i=0}^{n+1}
    \int\widehat{\mathrm d}y_i
    \int
    \widehat{\mathrm d}^{\,d}\boldsymbol k_i
    \right)
    \widehat{\delta}^{(d)}
    \left(
    \sum_{j=0}^{n+1}\boldsymbol k_j
    \right)
    \times
    \prod_{\lambda=\pm}
    \widehat{\delta}
    \left(
    p_1^\lambda+p_2^\lambda
    +\sum_{j=0}^{n+1}k_j^\lambda
    \right) \, .
\end{equation}

\section{Reduction of the five-point NMRK phase space}
\label{app:five_point_NMRK_phase_space}

We now specialise to the final state
\(\mathbf 3_A+4_h+\mathbf 5_B\). It is convenient to parametrise the
incoming rapidities by
\begin{equation}\label{eq:app:incoming_rapidity_variables}
\begin{gathered}
    y_1=-\frac{\Delta y_{21}}{2}+\bar y_{21} \, ,
    \qquad
    y_2=\frac{\Delta y_{21}}{2}+\bar y_{21} \, ,
    \\
    x_1=e^{-\bar y_{21}}\sqrt{\frac{m_B}{m_A}} \, ,
    \qquad
    x_2=\frac{1}{x_1} \, .
\end{gathered}
\end{equation}
The positive longitudinal scales defined in
\refeq{eq:app:incoming_longitudinal_scales} are then
\begin{equation}\label{eq:app:incoming_scales_x1_x2}
\begin{gathered}
    \sqrt{s}
    =
    \sqrt{P_A^+P_B^-}
    =
    \sqrt{m_Am_B}\,
    e^{\Delta y_{21}/2} \, ,
    \\
    P_A^+=x_2\sqrt{s} \, ,
    \qquad
    P_B^-=x_1\sqrt{s} \, .
\end{gathered}
\end{equation}

In the NMRK region \(y_3\simeq y_4\gg y_5\), the two longitudinal delta
functions reduce at leading power to
\begin{subequations}\label{eq:app:NMRK_longitudinal_delta_functions}
\begin{align}
    \widehat{\delta}_+
    ={}&
    \widehat{\delta}
    \left(
    P_A^+-p_3^+-p_4^+
    \right) \, ,
    \\
    \widehat{\delta}_-
    ={}&
    \widehat{\delta}
    \left(
    P_B^--p_5^-
    \right) \, .
\end{align}
\end{subequations}
Using the rapidity parametrisation of
\refeq{eq:app:massive_rapidity_parametrisation}, the first one becomes
\begin{equation}\label{eq:app:NMRK_plus_delta}
\begin{aligned}
    \widehat{\delta}_+
    ={}&
    \widehat{\delta}
    \left(
    x_2\sqrt{s}
    -m_{3\perp}e^{y_3}
    -|p_{4\perp}|e^{y_4}
    \right)
    \\
    ={}&
    \frac{
    \widehat{\delta}
    \left[
    y_3-y_4
    -\log
    \left(
    \frac{
    x_2\sqrt{s}\,e^{-y_4}
    -|p_{4\perp}|}
    {m_{3\perp}}
    \right)
    \right]}
    {
    \left(
    x_2\sqrt{s}\,e^{-y_4}
    -|p_{4\perp}|
    \right)e^{y_4}} \, .
\end{aligned}
\end{equation}
Similarly,
\begin{equation}\label{eq:app:NMRK_minus_delta}
\begin{aligned}
    \widehat{\delta}_-
    ={}&
    \widehat{\delta}
    \left(
    x_1\sqrt{s}
    -m_{5\perp}e^{-y_5}
    \right)
    \\
    ={}&
    \frac{
    \widehat{\delta}
    \left[
    y_5-\log
    \left(
    \frac{m_{5\perp}}{x_1\sqrt{s}}
    \right)
    \right]}
    {x_1\sqrt{s}} \, .
\end{aligned}
\end{equation}
The integrations over the two massive-particle rapidities therefore
give
\begin{equation}\label{eq:app:NMRK_longitudinal_integral_intermediate}
\begin{aligned}
    &\int
    \widehat{\mathrm d}y_3\,
    \widehat{\mathrm d}y_5\,
    \widehat{\delta}_+\widehat{\delta}_-
    \\
    &\quad=
    \frac{1}{
    x_1\sqrt{s}
    \left(
    x_2\sqrt{s}\,e^{-y_4}
    -|p_{4\perp}|
    \right)e^{y_4}}
    \\
    &\quad=
    \frac{1}{s}
    \left[
    1-
    \frac{x_1}{\sqrt{s}}
    |p_{4\perp}|e^{y_4}
    \right]^{-1} \, .
\end{aligned}
\end{equation}
The remaining ratio is fixed by the longitudinal momentum fraction of
the graviton,
\begin{equation}\label{eq:app:NMRK_graviton_longitudinal_fraction}
    \frac{x_1}{\sqrt{s}}
    |p_{4\perp}|e^{y_4}
    =
    \frac{|p_{4\perp}|}{m_A}e^{-(y_2-y_4)}
    =
    \frac{p_4^+}{P_A^+}
    =
    \frac{1}{1+X} \, .
\end{equation}
Hence the required correction is
\begin{equation}\label{eq:app:NMRK_longitudinal_phase_space_result}
    \int
    \widehat{\mathrm d}y_3\,
    \widehat{\mathrm d}y_5\,
    \widehat{\delta}_+\widehat{\delta}_-
    =
    \frac{1}{s}
    \left(
    1+\frac{1}{X}
    \right) \, .
\end{equation}

For one emitted graviton, \(n=1\), the reduced phase space can finally
be written as
\begin{equation}\label{eq:app:five_point_NMRK_reduced_phase_space}
\begin{aligned}
    \mathrm d\mathcal P_3^{\mathrm{NMRK}}
    ={}&
    \frac{1}{4s}
    \left(
    1+\frac{1}{X}
    \right)
    \widehat{\mathrm d}y_4
    \prod_{i=3}^{5}
    \widehat{\mathrm d}^{\,d}\boldsymbol p_i\,
    \widehat{\delta}^{(d)}
    \left(
    \boldsymbol p_3+\boldsymbol p_4+\boldsymbol p_5
    \right) \, .
\end{aligned}
\end{equation}
In strict MRK, \(X\to\infty\), this reduces to the usual
\(1/(4s)\) measure in Ref~\citeall{alessio2026gravitationalamplitudesregge}.

\backmatter
\cleardoublepage


\newpage
\phantomsection
\addcontentsline{toc}{chapter}{\bibname}
\nocite{*}
\bibliography{bibiografia}

\chapter*{Ringraziamenti}
\phantomsection
\addcontentsline{toc}{chapter}{Ringraziamenti}
\markboth{Ringraziamenti}{Ringraziamenti}

Inizio ringraziando nuovamente il gruppo dei colleghi teorici a Frascati per tutti i fantastici momenti passati ai Labs in questo semestre. Mi mancheranno i pranzi alla mensa alle prese di qualche discussione illuminante ma allo stesso tempo stravagante usualmente intavolata da Federico. Ringrazio Emanuele per le varie discussioni in questi mesi ma soprattutto a lui va una forte stima per avermi sopportato per due settimane intere ad Atrani. Ringrazio in particolar modo il gruppo di Amplitudes per quella partita leggendaria di Dixit a Stoccolma.

Ringrazio (in ordine di nome) i miei compagni di corso: Angiolino Pio per la sua innata abilità nel saper inventare le misure mancanti di laboratorio, Damiano per avermi 
concesso ripetutamente un goccio della sua acqua ad ogni pranzo al parco di fronte la Sapienza, Eugenio per le nostre pacate e amichevoli discussioni e per le sue controverse opinioni politiche, Francesco 
per aver replicato in C\texttt{++} i codici del CERN, Simone per non avermi mai dato buca ad una manifestazione popolare. Mi mancheranno i pomeriggi interi passati in laboratorio a tentare di racimolare più misure possibili, confidando nella statistica poissoniana affinché alcune di queste si rivelassero davvero buone. Ammetto che senza di voi questo percorso sarebbe stato molto più ostico e tortuoso.

Ringrazio ChatGPT perché funziona (\verb|\cite{Bonvini}|). 

Nonostante ne pianga ancora la scomparsa, ci tengo anche a ringraziare il Maestro Francesco Guccini per avermi accompagnato in ogni singolo viaggio in macchina fatto con mio padre, e per avermi insegnato che a volte è meglio essere una pecora nera che unirmi al resto della schiera. ``Ma nella fantasia ho l'immagine sua, gli eroi son tutti giovani e belli''. 

Ringrazio il maestro Yoda per avermi insegnato il motto Jedi per il quale nella vita esiste solo fare o non fare, non c'è provare.

Ringrazio anche Goku per essere stato una mia totale fonte di ispirazione fin dall'infanzia e per avermi insegnato a non mollare mai.

Una dedica speciale va a mio padre, il quale mi ha sempre sostenuto e supportato verso il mio percorso, nonostante tutte le difficoltà che si sono presentate in questi anni. Un caloroso abbraccio è rivolto verso Camilla,
la quale mi ha sollevato anche nei momenti di più totale buio e che non ha mai smesso di credere in me, nemmeno per un momento, riportandomi sempre a far vedere quel barlume di speranza oltre l'oscurità.

Ringrazio tutte le mie zie e i miei zii, le mie cugine e i miei cugini, per il continuo aiuto nel momento della necessità e per i pranzi familiari ricchi di allegria e ``caciaronaggine''.
Un grande bacio è rivolto ai miei nonni, ai quali spero di aver regalato una gioia con questa laurea.

Un grazie di cuore spetta a Valeria per la sua costante presenza nella mia vita e a Claudia per le stupende serate passate al bar del Disagio. (Non) ringrazio invece Ginevra per quelle giornate dedicate a preparare il TOLC ($\heartsuit$). Ringrazio altresì Mirko per avermi fatto scoprire la passione per il tennis
e per le sue battute infantili capaci di risollevarmi l'animo anche nelle serate più mosce.

Un sentito abbraccio è rivolto ai miei amici di Oriolo (escluso Pasquali al quale sono rivolte solo malparole per battermi continuamente a tennis), giovani compagni di squadra e accaniti giocatori di calcetto, e ai miei amici di Vejano, abili nel rallegrarmi tramite qualche cena ``acchittata'' improvvisamente. Non so onestamente come avrei potuto passare i sabati sera senza di voi.

Infine un ultimo ma sincero ringraziamento è rivolto a Giulia e Isabella, alle quali vorrei ribadire il forte affetto che mi lega a loro, nonostante tutto.
Avevo davvero sperato in un epilogo diverso delle cose, ma non sempre si può essere totalmente padroni del destino. Vi voglio bene.

\end{document}